\documentclass[prd,aps,twocolumn,a4paper,showkeys,nofootinbib,showpacs]{revtex4-1}

\usepackage[colorlinks=true,linkcolor=blue,linktocpage=true,anchorcolor=black,citecolor=blue,filecolor=black,menucolor=black,urlcolor=blue]{hyperref}

\usepackage{graphicx,psfrag}
\usepackage{mathrsfs}
\usepackage{amsmath,amsfonts,amssymb}
\usepackage{comment}
\usepackage{ulem}
\usepackage[top=2cm,bottom=2cm,left=1.4cm,right=1.4cm]{geometry}

\usepackage[T1]{fontenc}

\newcommand{\be}{\begin{equation}}
\newcommand{\ee}{\end{equation}}
\newcommand{\bea}{\begin{eqnarray}}
\newcommand{\eea}{\end{eqnarray}}
\newcommand{\bel}{\begin{align}}
\newcommand{\eel}{\end{align}}

\def\i{{\rm i}}

\def\Msun{M_{\odot}}
\def\GMc2{G M_{\odot} c^{-2}}

\def\lm{{\ell m}}

\def\lm{{\ell m}}

\def\de{\partial}
\def\lm{{\ell m}}

\def\ii{{\rm i}}
\def\l{{\ell }}

\def\Msun{M_\odot}

\def\TEOB{\texttt{TEOBResumS}}
\def\TEOBResumS{\texttt{TEOBResumS}}

\def\TEOBResumSvfour{\texttt{TEOBResumSv4.1.4}}
\def\TEOBResumSvfourtwo{\texttt{TEOBResumSv4.2.0}}
\def\TEOBResumSvfourthree{\texttt{TEOBResumSv4.3.0}}
\def\TEOBResumSvfourthreeone{\texttt{TEOBResumSv4.3.1}}
\def\TEOBResumSvfourthreetwo{\texttt{TEOBResumSv4.3.2}}

\def\nrsurqeight{{\texttt{NRHybSur3dq8}}}
\def\nrsurqfifteen{{\texttt{NRHybSur2dq15}}}

\def\l{{\ell}}
\def\lm{{\ell m}}

\def\tA22{{t_{A_{22}}^{\rm peak}}}

\def\d2Amx{{\ddot{A}_{\rm peak}}}

\def\Amatch{{A_\tmatch}}
\def\dAmatch{{\dot{A}_\tmatch}}

\def\omgmatch{{\omega_{\tmatch}}}

\def\tmatch{{t_\lm^{\rm match}}}
\def\hbar{{\bar{h}}}

\DeclareSymbolFontAlphabet{\mathrsfs}{rsfs}
\DeclareMathAlphabet{\mathcal}{OMS}{cmsy}{m}{n}
\DeclareSymbolFontAlphabet{\mathrsfs}{rsfs}
\DeclareMathAlphabet\mathbfcal{OMS}{cmsy}{b}{n}

\usepackage{color}
\definecolor{cyan}{rgb}{0,0.9,0.9}
\definecolor{orange}{rgb}{0.9,0.5,0}
\definecolor{magenta}{rgb}{1,0,1}
\definecolor{purple}{rgb}{0.8,0.4,0.8}
\definecolor{gray}{rgb}{0.8242,0.8242,0.8242}
\definecolor{dodgerblue}{rgb}{0.12, 0.56, 1.0}

\begin{document}
        
\title{Analytic systematics in next-generation of effective-one-body gravitational waveform models for future observations}

\author{Alessandro \surname{Nagar}${}^{1,2}$}
\author{Piero \surname{Rettegno}${}^{1}$}
\author{Rossella \surname{Gamba}${}^{3}$}
\author{Simone \surname{Albanesi}${}^1$}
\author{Angelica \surname{Albertini}${}^{4,5}$}
\author{Sebastiano \surname{Bernuzzi}${}^{3}$}

\affiliation{${}^1$INFN Sezione di Torino, Via P. Giuria 1, 10125 Torino, Italy}
\affiliation{${}^2$ Institut des Hautes Etudes Scientifiques, 91440 Bures-sur-Yvette, France}
\affiliation{${}^3$Theoretisch-Physikalisches Institut, Friedrich-Schiller-Universit{\"a}t Jena, 07743, Jena, Germany}  
\affiliation{${}^4$Astronomical Institute of the Czech Academy of Sciences,
Bo\v{c}n\'{i} II 1401/1a, CZ-141 00 Prague, Czech Republic}
\affiliation{${}^4$Faculty of Mathematics and Physics, Charles University in Prague, 18000 Prague, Czech Republic}

\begin{abstract}
The success of analytic waveform modeling within the effective-one-body (EOB) approach 
relies on the precise understanding of the physical importance of each technical element included
in the model. The urgency of constructing progressively more sophisticated and complete waveform 
models (e.g. including spin precession and eccentricity) partly defocused the research from a careful
comprehension of each building block (e.g. Hamiltonian, radiation reaction,
ringdown attachment). Here we go back to the spirit of the first EOB works.  We focus first 
on nonspinning, quasi-circular, black hole binaries and analyze systematically the mutual synergy between 
numerical relativity (NR) informed functions and the high post-Newtonian corrections (up to 5PN) to 
the EOB potentials. Our main finding is that it is essential to correctly control the noncircular 
part of the dynamics during the late plunge up to merger. When this happens, either using 
NR-informed non-quasi-circular corrections to the waveform (and flux) or high-PN corrections 
in the radial EOB potentials $(D,Q)$, it is easy to obtain EOB/NR unfaithfulness $\sim 10^{-4}$ 
with the noise of either Advanced LIGO or 3G detectors.
We then improve the {\tt TEOBResumS-GIOTTO} waveform model (dubbed \TEOBResumSvfourthreetwo{})
for quasi-circular, spin-aligned black hole binaries. We obtain maximal 
EOB/NR unfaithfulness ${\bar{\cal F}}^{\rm max}_{\rm EOBNR}\sim 10^{-3}$ 
(with Advanced LIGO noise and in the total mass range $10-200M_\odot$) for the dominant 
$\ell=m=2$ mode all over the 534 spin-aligned configurations available through the Simulating 
eXtreme Spacetime catalog. The model performance, also including higher modes, is then explored 
using the NR surrogates \nrsurqeight{} and \nrsurqfifteen, to validate \TEOBResumSvfourthreetwo{} 
up to mass ratio $m_1/m_2=15$. We find that, over the set of configurations considered, more than $98\%$ 
of the total-mass-maximized unfaithfulness lie below the $3\%$ threshold when comparing to the surrogate models.

\end{abstract}

\maketitle

\section{Introduction}
With the ever-growing sensitivities of gravitational wave (GWs) observatories, the issue of 
systematic errors due to modeling is at the forefront of GW astronomy. This topic has 
received some attention in the recent literature~\cite{Littenberg:2012uj,Purrer:2019jcp,Gamba:2020wgg,Owen:2023mid,Read:2023hkv},
and such errors are typically quantified in terms of their impact on parameter estimation, and studied by comparing
the posterior distributions obtained with different models.
Unfortunately, so far only few studies have attempted to link the observed differences to the specifics of the models themselves,
in large part due to their extreme complexity, which stems from the need of describing a large class of systems and
physical effects.

In this respect, the effective one body (EOB) \cite{Buonanno:2000ef,Damour:2000we,Damour:2001tu,Damour:2015isa} approach to the general relativistic two body problem is currently
the only formalism sufficiently flexible and accurate to generate reliable waveforms for any kind 
of coalescing binaries, from quasi-circular and eccentric black holes~\cite{Akcay:2020qrj,Schmidt:2020yuu,Nagar:2020xsk,Ossokine:2020kjp,Riemenschneider:2021ppj,Gamba:2021ydi,Gamba:2020ljo,Gamba:2021gap,Ramos-Buades:2021adz,Bonino:2022hkj},  
to neutron stars~\cite{Lackey:2018zvw,Godzieba:2020bbz,Gamba:2020ljo,Gamba:2020wgg,Breschi:2022ens,Gamba:2022mgx} or mixed binaries~\cite{Matas:2020wab,Gonzalez:2022prs}. 
This framework is characterized by few, well defined, building blocks (i.e. Hamiltonian, radiation reaction, waveform, ringdown description) 
that can incorporate a variable amount of analytic information -- usually post-Newtonian results 
repackaged in some resummed fashion -- suitably augmented by Numerical Relativity (NR) information.

Within the EOB approach, the issue of understanding waveform systematics rephrases as 
the need of evaluating the impact that each (sub)building block of the model may have on 
waveform accuracy\footnote{This waveform accuracy is typically evaluated using phasing comparisons with
NR waveform data, considered as exact for any practical purpose.}.
This conceptual approach was followed in e.g. Ref.~\cite{Damour:2007yf, Damour:2007vq,Damour:2008te,Damour:2009kr}, 
and was essential to identify the best analytical waveform description through plunge up to merger.
Note that this effort was nontrivial, as the flexibility of the framework implies some degree of arbitrariness in e.g. resummation choices. 
We then define ``analytic'' waveform systematics all errors stemming from the choices in the analytic structure of a model.
In recent years, the rush of constructing physically complete EOB waveform models, to satisfy the needs of 
gravitational wave data analysis, partly defocused the current EOB research from the scope of minimizing analytic systematics, and deeply 
understanding the interplay of the various analytical elements with each other and with the procedure of {\it calibrating}
the analytical model. 
Here we make an effort to go back to the original EOB philosophy~\cite{Damour:2007yf,Damour:2007vq,Damour:2008te,Damour:2009kr}. 
We attempt to understand, one by one, the influence of various building elements of the Hamiltonian,
or other well-defined and established procedure, like the determination of next-to-quasi-circular (NQC)
to the waveform and radiation reaction. This knowledge allows then to better clarify which directions
to follow aiming at improved waveform accuracy for next generation detectors 
like Einstein Telescope~\cite{Hild:2008ng,Hild:2009ns, Hild:2010id} and Cosmic Explorer~\cite{Evans:2021gyd}.

The paper is organized as follows. In Sec.~\ref{new:a6c} we present an improved nonspinning sector
of \TEOBResumS{} via a more precise NR-informed effective-5PN function $a_6^c$.
Using this model as reference baseline, in Sec.~\ref{sec:systematics} we explore various
analytic systematics, like the effect of changing the ringdown matching point and the inclusion 
of high-order (noncircular) terms in the EOB potentials. This allows the construction of different
quasi-circular EOB waveform models with different analytic content and related levels of 
EOB/NR unfaithfulness. To enlarge our battery of EOB models, and related analytic 
systematics, in Sec.~\ref{sec:nc} we present an upgrade of the eccentric EOB model of
Ref.~\cite{Nagar:2021xnh}, that offers a better performance either for quasi-circular 
binaries or for scattering angle. In Sec.~\ref{teob:best} we complement the newly determined
$a_6^c(\nu)$ function of Sec.~\ref{new:a6c} with progressively different NR-informed spin-orbit
sectors. This eventually brings us to the construction of a new, spin-aligned, waveform model 
for quasi-circular binaries (including higher modes) that performs better than the state-of-the-art \TEOBResumS{}.
In particular, here we also incorporate within \TEOBResumS{} the description of the ringdown of
the $\ell=2$, $m=1$ mode that is part of the {\tt SEOBNRv5} model~\cite{Pompili:2023tna}. 
This allows us to eliminate long-standing issues due to the bad modelization of this mode
in the standard \TEOBResumS{} construction~\cite{Nagar:2020pcj}.
The notation and analytical information of this paper strongly builds upon 
Refs.~\cite{Nagar:2020pcj,Riemenschneider:2021ppj,Nagar:2021xnh},
and we assume the reader to be familiar with the content and notation of those papers.
We only recall a few notational elements: $(m_1,m_2)$ are the masses of the two black holes, 
with $q=m_1/m_2\geq 1$ the mass ratio, $M\equiv m_1+m_2$ the total mass and $\nu\equiv m_1 m_2/M^2$
the symmetric mass ratio and $X_i\equiv m_i/M$ with $i=1,2$. The dimensionless spin magnitudes are 
$\chi_i\equiv S_i/m_1^2$ with $i=1,2$, and we indicate with $\tilde{a}_0\equiv X_1\chi_1+X_2\chi_2$
the effective spin, usually indicated as $\chi_{\rm eff}$ in the literature. 
If not stated otherwise,  we use geometric units with $c=G=1$.
\begin{table*}[t]
 \caption{\label{tab:a6c}Nonspinning SXS simulations used in the first part of the paper to construct and check
 various the various EOB nonspinning models characterized in Table~\ref{tab:models}. They are selected as 
 nonspinning because the initial effective spin of the system, $\tilde{a}_0$ is smaller than $10^{-6}$. Only some datasets are 
 used to determine the first-guess values of $a_6^c$, then fitted with functional forms and coefficients also 
 listed in Table~\ref{tab:models}. All datasets are then used to validate the models, either with time-domain
 phasing comparisons or EOB/NR unfaithfulness calculations. The fifth column reports the nominal NR phasing
 uncertainty at merger $\delta\phi^{\rm NR}_{\rm mrg}$, obtained by taking the phase difference between 
 the highest and second highest resolutions available.}
   \begin{center}
     \begin{ruledtabular}
\begin{tabular}{c c c c c |c c c c c c }
        $\#$ & SXS & $q$ & $\nu$ & $\delta\phi^{\rm NR}_{\rm mrg}$[rad]  & \multicolumn{5}{c}{ $a_6^c$ first-guess values}\\ 
 & & & & & {\tt D3Q3\_NQC}  & {\tt D5Q5\_NQC} & {\tt D5Q5} &  {\tt D3Q3} & {\tt D5Q3}\\
  \hline
  \hline
  1 & SXS:BBH:0180   & 1      & 0.25  & $+0.42$ &$-36.0$ &$-98.0$ & $-5$ & 11& $-13$\\  
  2  &  SXS:BBH:0007 &  1.5 &  0.24           & $-0.0186$ &$-41.5$ &$-92.0$ & $-15$ &13 & $-13.5$\\
  3  &  SXS:BBH:0169  & 2 &  $0.\bar{2}$               &  $-0.0271$ &$-48.0$ & $-81.5$ & $-29$ & 15& $-15$\\
  4   & SXS:BBH:0259     & 2.5 & 0.2041                   & $+0.0080$ &$\dots$   &$\dots$ &$-31$ & $\dots$ & $\dots$ \\
  5  &  SXS:BBH:0168  & 3  &  0.1875             & $+0.1144$&$-51.0$  & $-68.0$ & $-30.2$ & 17 & $-19$ \\
  6    & SXS:BBH:0294  & 3.5 & 0.1728   & $+1.325$ &$\dots$ & $\dots$ & $-29.2$ &$\dots$ & $\dots$ \\
  7  &  SXS:BBH:0295  & 4.5 & 0.1488            & $-0.240$ &$-45.0$ & $-54.0$ &$-28.5$ &14.5 & $-17$ \\
  8    & SXS:BBH:0056          & 5   & 0.1389  & $-0.439$& $\dots$ & $\dots$ & $-27.7$ & $\dots$ & $\dots$ \\
  9    &  SXS:BBH:0296                            & 5.5 &0.1302 & $-0.443$ &$\dots$ & $\dots$ &$\dots$ &$\dots$ &$\dots$ \\
  10  & SXS:BBH:0166                             & 6 &0.1224 & $\dots$ & $\dots$& $\dots$& $\dots$& $\dots$& $\dots$\\
  11  &  SXS:BBH:0297  & 6.5     &  0.1156 & $-0.053$ &$-34.5$ & $-42.0$ & $-26$ &11 & $-12$ \\
  12  &  SXS:BBH:0298  & 7      &  0.1094 & $+0.078$ &$-31.5$ & $-39.5$ & $-25$ &10& $-11$\\ 
  13 &  SXS:BBH:0299  & 7.5 & 0.1038& $+0.050$ & $\dots$& $\dots$& $\dots$& $\dots$& $\dots$\\
   14 &  SXS:BBH:0063  & 8 & 0.0988 & $-1.009$ & $\dots$& $\dots$& $\dots$& $\dots$& $\dots$\\
15&     SXS:BBH:0301   & 9 & 0.090 & $+0.16$ & $\dots$& $\dots$& $\dots$& $\dots$& $\dots$\\
  16 &  SXS:BBH:0302  & 9.5   & 0.0826& $-0.020$ & $-19.5$ & $-25.9$ & $-14$ &8.5& $-8$
  \end{tabular}
 \end{ruledtabular}
 \end{center}
 \end{table*}
 
\begin{figure*}[t]
	\center
	\includegraphics[width=0.32\textwidth]{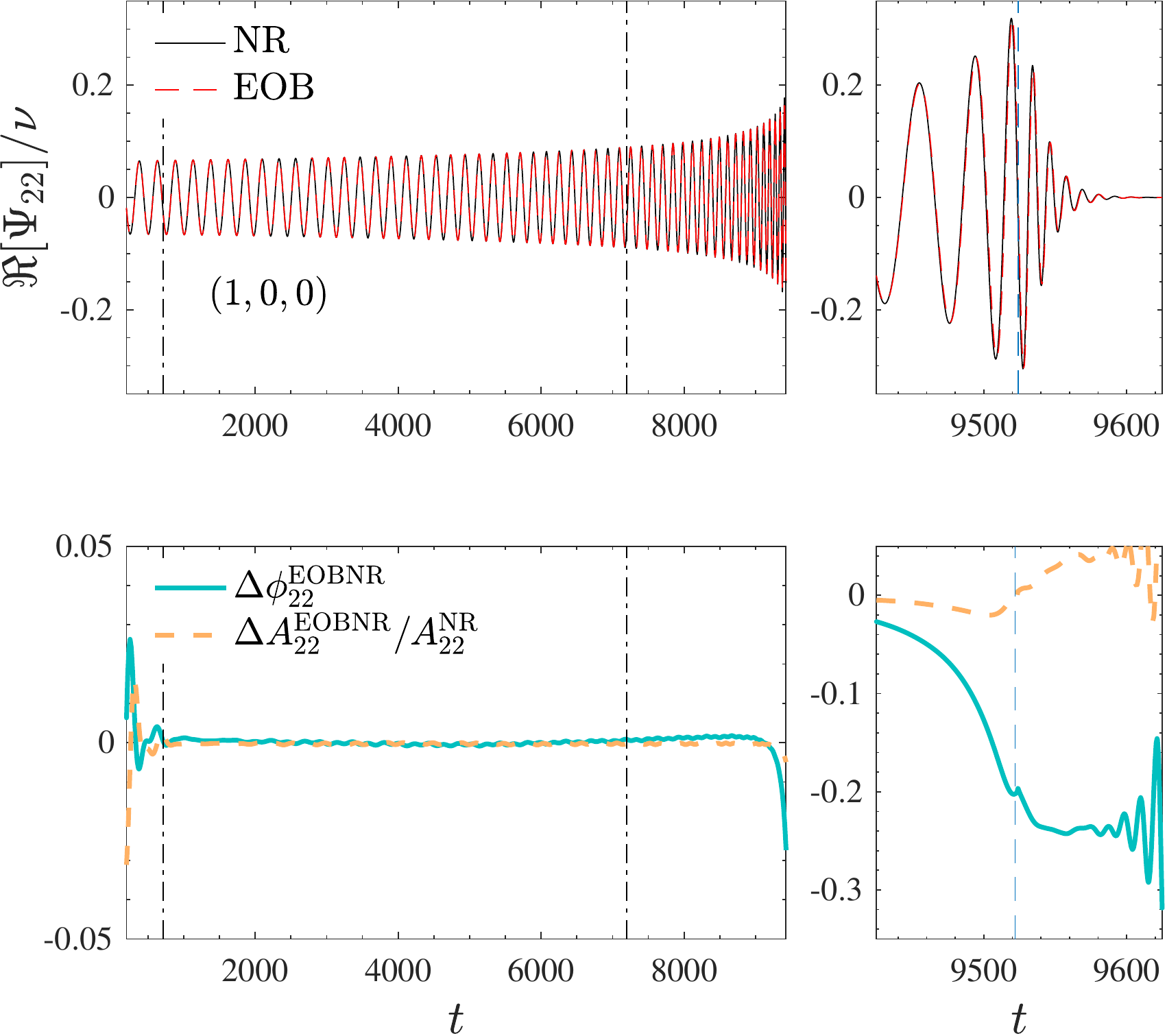}
	\includegraphics[width=0.32\textwidth]{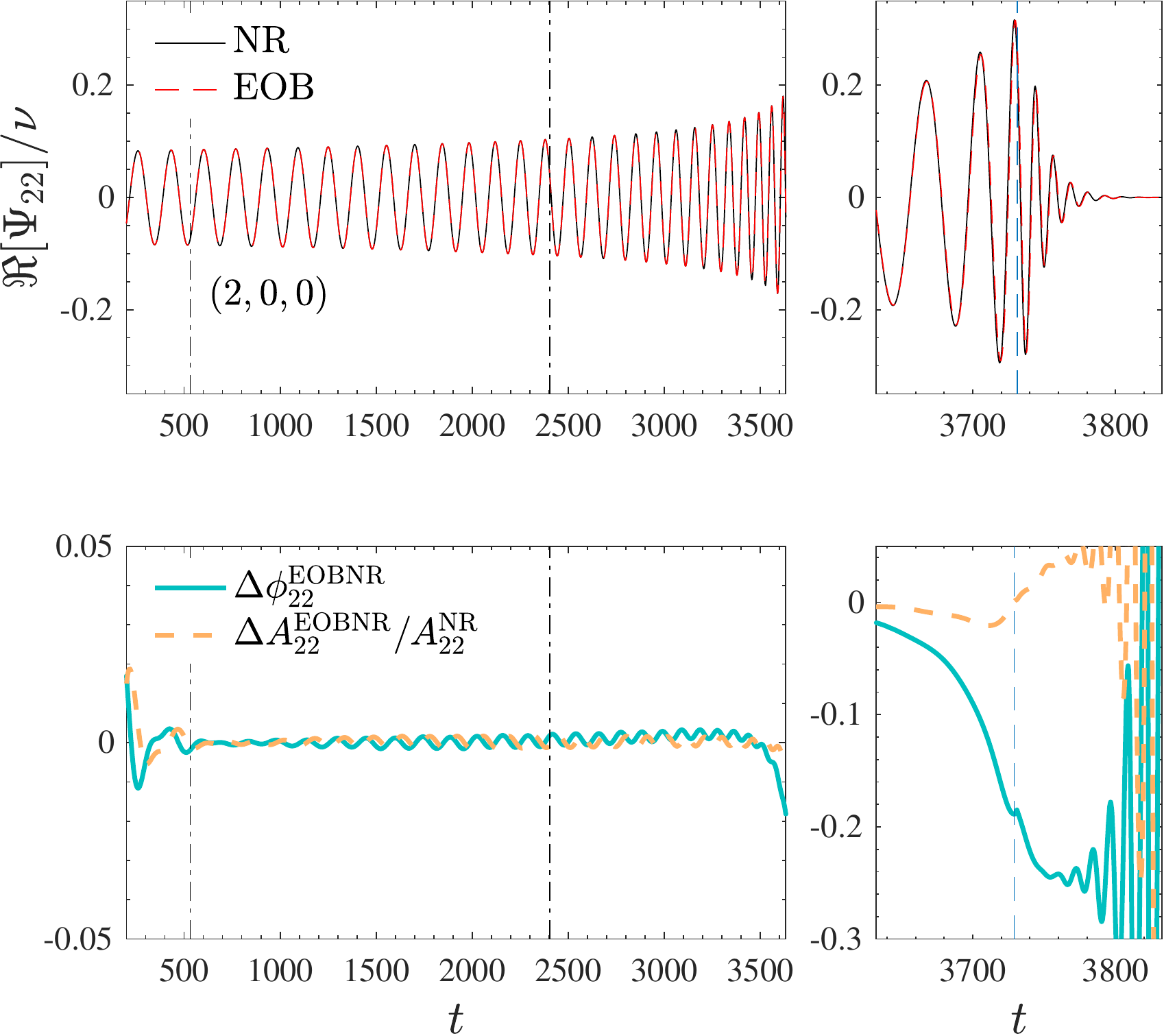}
	\includegraphics[width=0.32\textwidth]{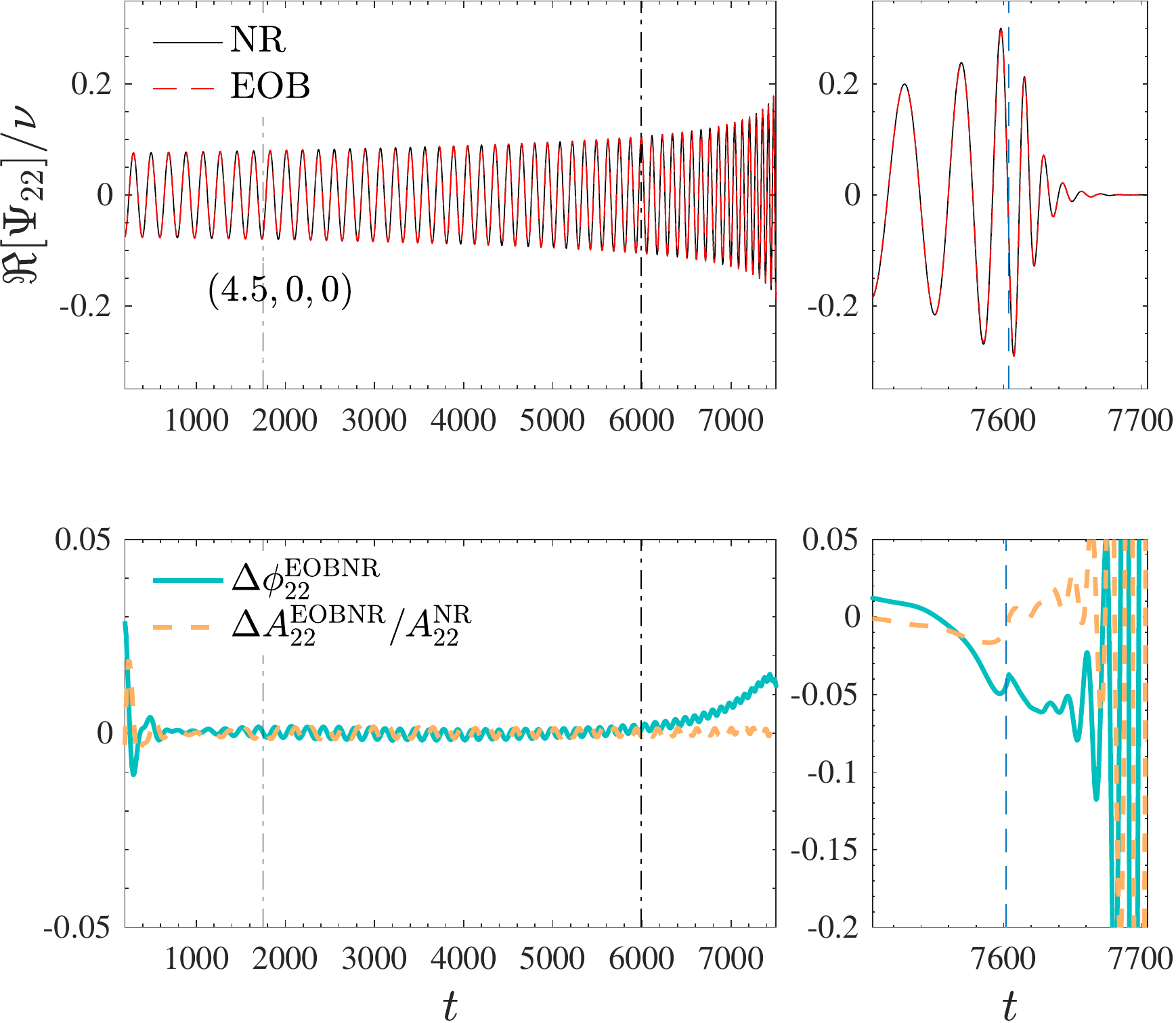}\\
	\includegraphics[width=0.32\textwidth]{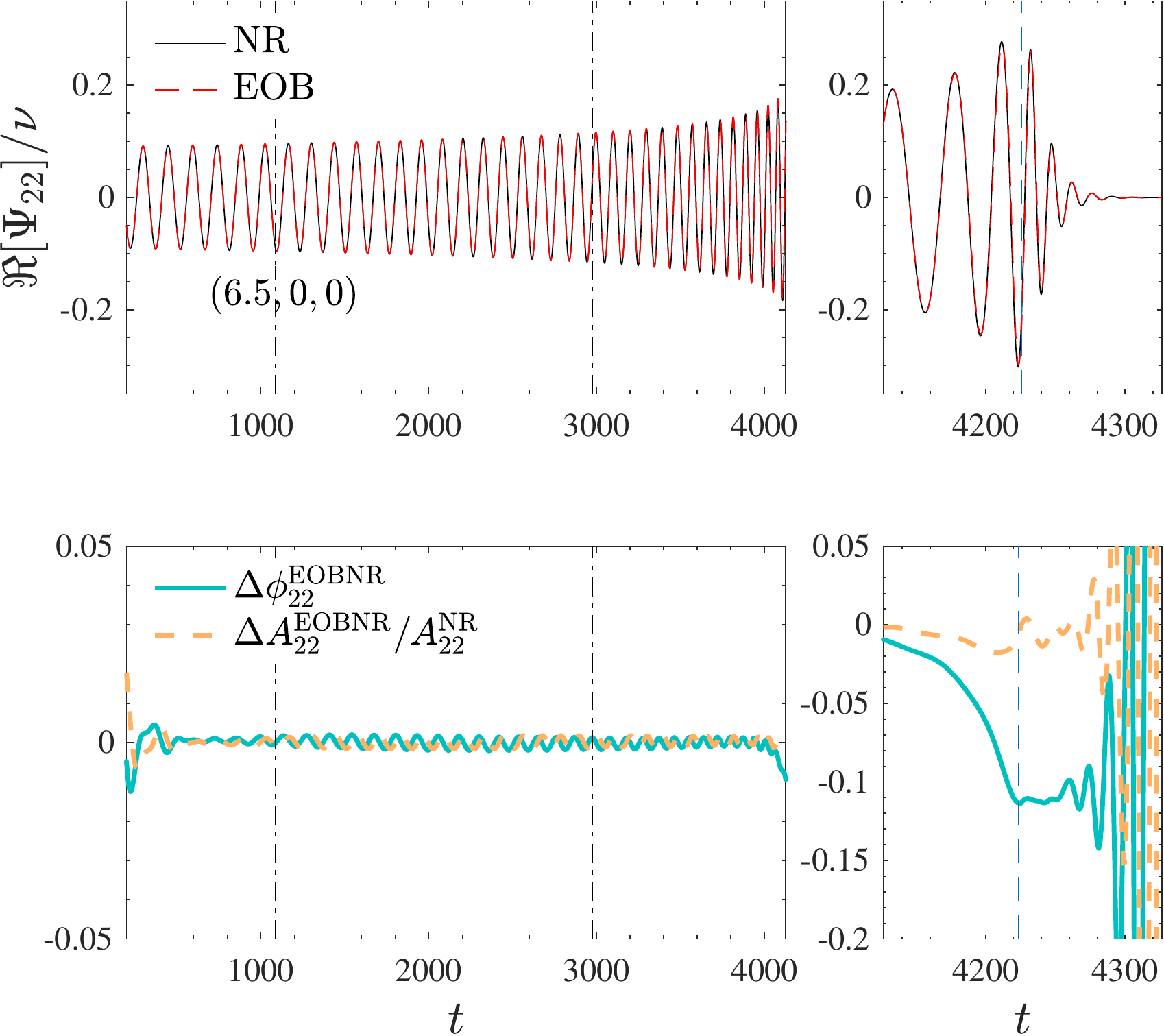}
	\includegraphics[width=0.32\textwidth]{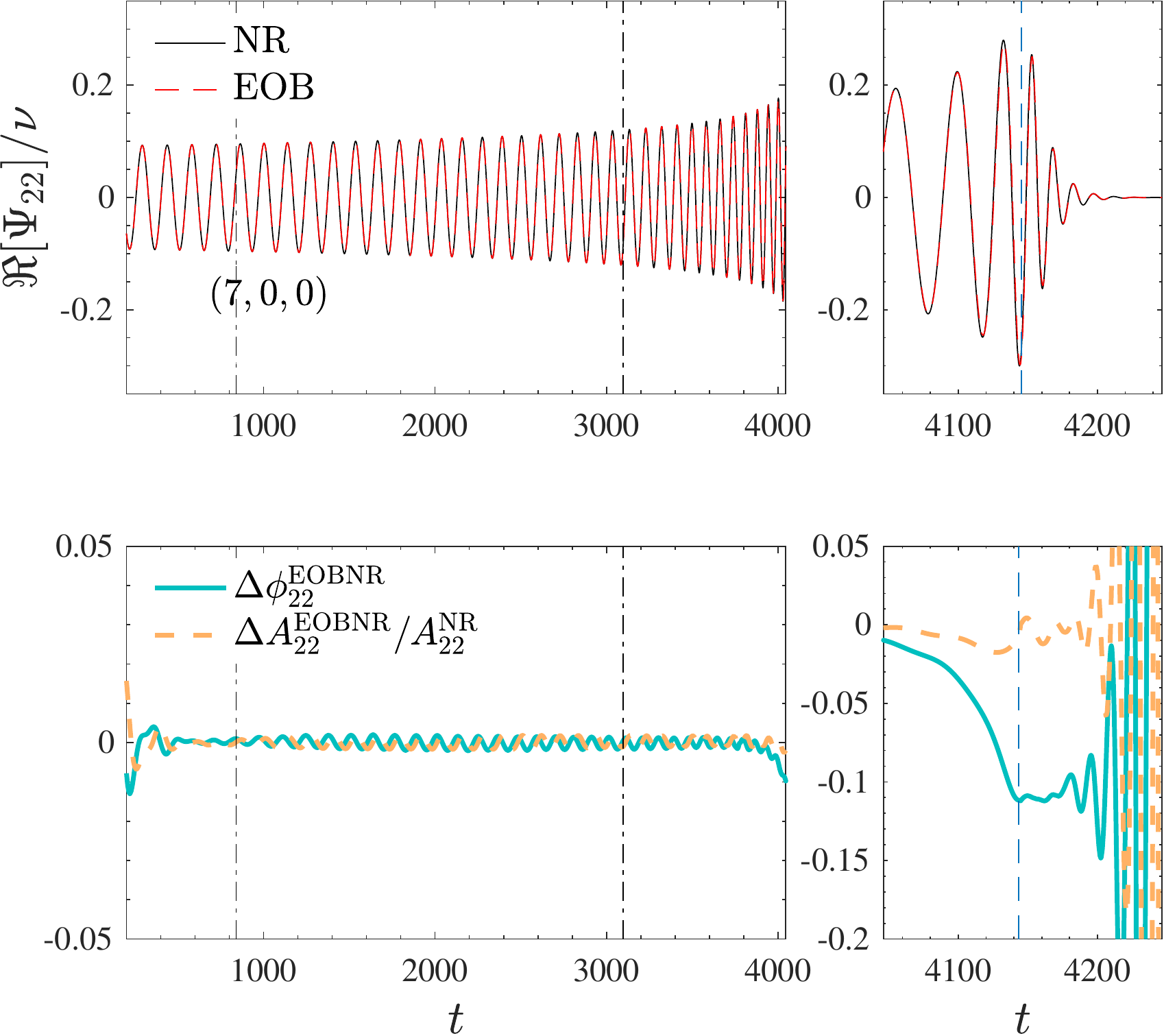}
	\includegraphics[width=0.32\textwidth]{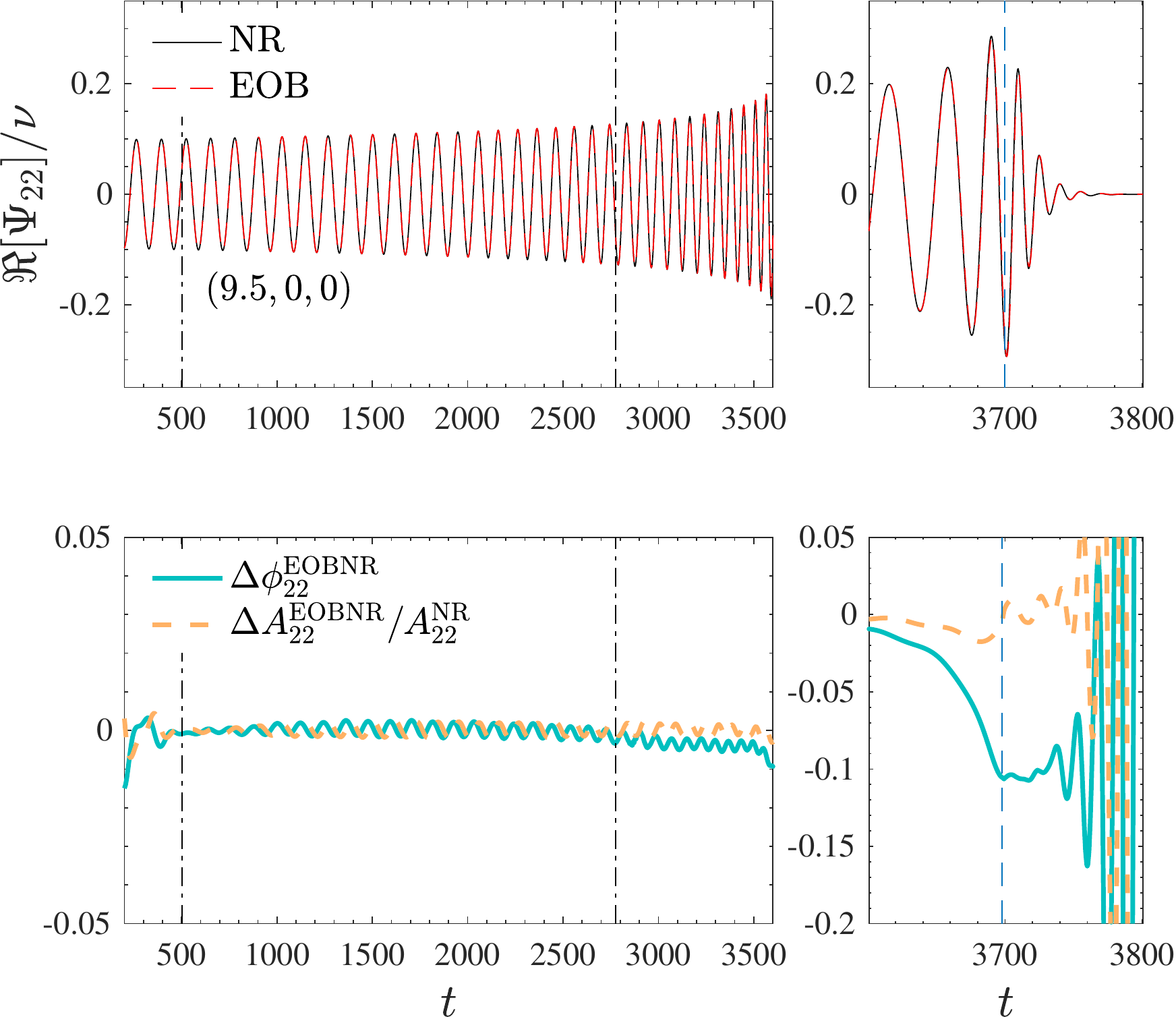}
	\caption{\label{fig:raw_phasing}{\tt D3Q3\_NQC} model: EOB/NR phasing agreement obtained with the new fit for 
	$a_6^c$ for the mass ratios $q=\{1, 2,4.5,6.5,7,9.5\}$  of Table~\ref{tab:a6c}. Differently from previous work, 
	$a_6^c$ is tuned so that $\Delta\phi_{22}^{\rm EOBNR}$ decreases monotonically through 
	merger and ringdown. Note how visible are the modulations  due to residual NR eccentricity as
	the mass ratio is increased.}.
	\center
\end{figure*}

\begin{figure}[t]
	\center
	\includegraphics[width=0.45\textwidth]{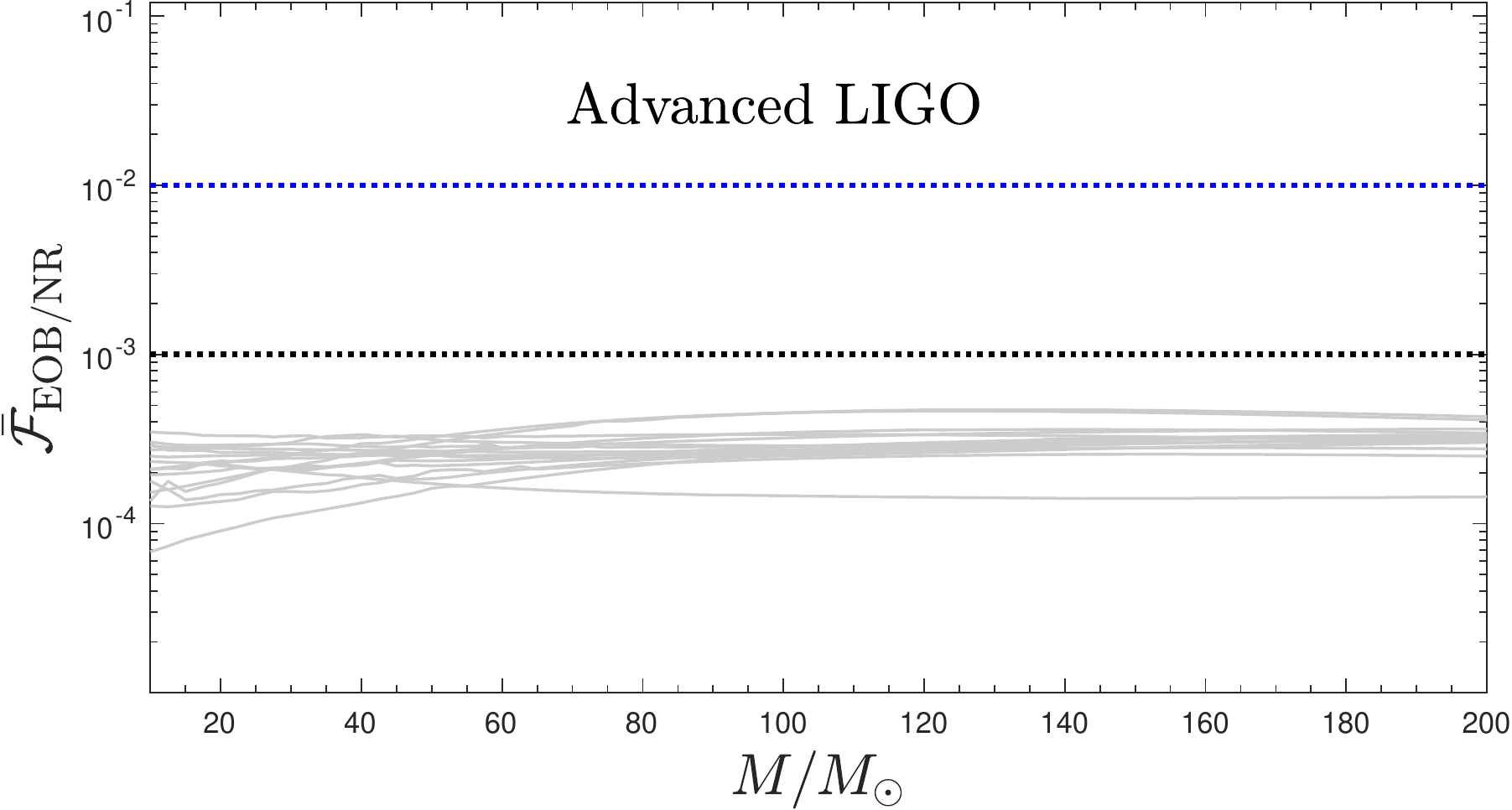}\\
	\vspace{5 mm}	
	\includegraphics[width=0.45\textwidth]{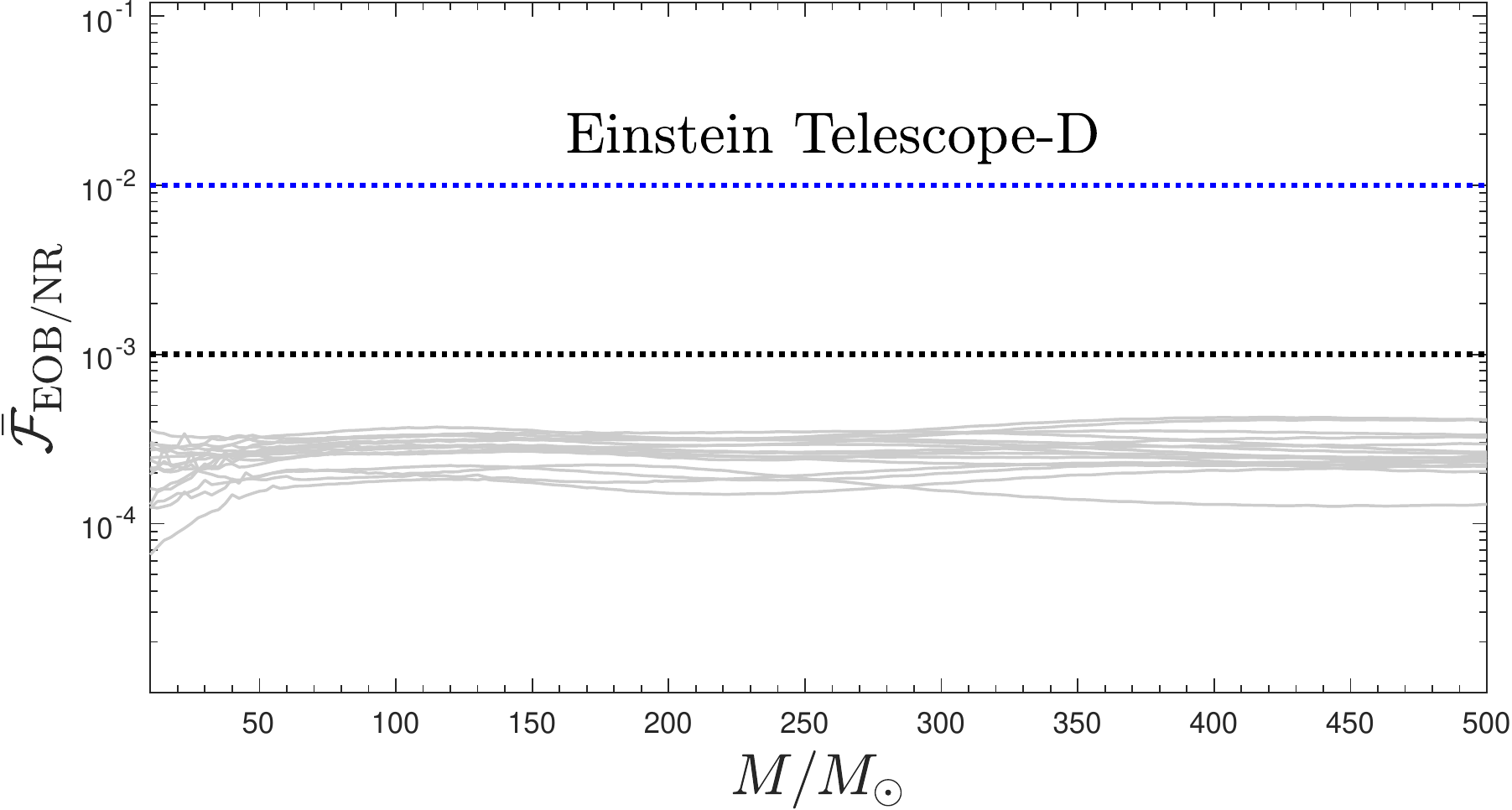}\\	
	\vspace{5 mm}	
	\includegraphics[width=0.45\textwidth]{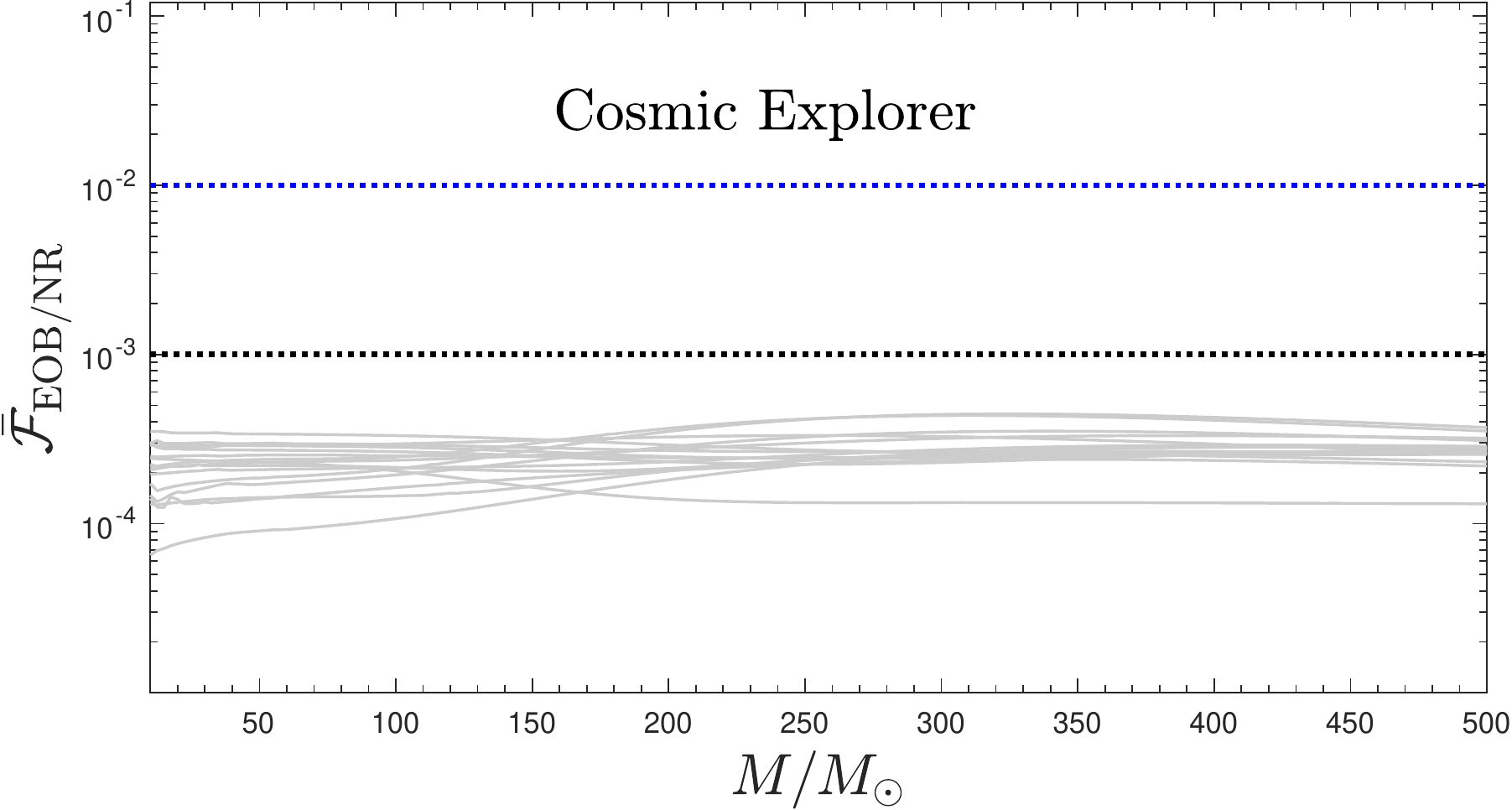}
	\caption{\label{fig:new_barF}EOB/NR unfaithfulness over all SXS nonspinning datasets
	listed of Table~\ref{tab:a6c}. The top panel is obtained using the Advanced LIGO  PSD,
	the central panel uses the ET-D configuration for the Einstein Telescope 
	(see also Fig.~11 of Ref.~\cite{Albertini:2021tbt}), and the bottom panel uses 
	the expected PSD for Cosmic Explorer. The dotted blue and black lines mark
	respectively the $3\%$ and $1\%$ thresholds.
	}
	\center
\end{figure}

\section{Improving the conservative nonspinning sector of {\tt TEOBResumS-GIOTTO} }
\label{new:a6c}
The nonspinning Hamiltonian of {\tt TEOBResumS-GIOTTO} (only  \TEOBResumS{} in the following for simplicity) 
depends on a single NR-informed function $a_6^c(\nu)$, which plays the role of an effective 5PN correction 
in the EOB radial potential $A(r)$~\cite{Damour:2009kr}, where $r\equiv R/M$ is the dimensionless relative 
separation between the two bodies.
The original function $a_6^c(\nu)$ employed in \TEOBResumS{} dates back to Ref.~\cite{Nagar:2019wds},
and, for simplicity, was never changed since (see in particular Table~I of~\cite{Nagar:2019wds} 
and related simulations of the Simulating eXtreme Spacetimes (SXS) catalog~\cite{SXS:catalog,Buchman:2012dw,Chu:2009md,Hemberger:2013hsa,Scheel:2014ina,
Blackman:2015pia,Lovelace:2011nu,Lovelace:2010ne,Mroue:2013xna,Lovelace:2014twa,Kumar:2015tha,Lovelace:2016uwp,Chu:2015kft,Varma:2018mmi} used to determine it). This expression 
of $a_6^c(\nu)$ was rather conservative, allowing for EOB/NR phase differences 
of the same order of (or larger than) the nominal NR phase uncertainty at merger. 
The latter is typically defined, for a given simulation, as the phase difference between the highest 
and second highest resolutions available in the SXS catalog.
This uncertainty estimate, however, is quite conservative on average
\footnote{Note in passing that this is a property of the SpEC code. 
Though finite-difference codes give results that are in general less accurate during
the inspiral, at least one can extrapolate to infinite resolution and get a sense of
the uncertainties due to finite resolution. See for example~\cite{Nagar:2022icd}.}, and may not
properly reflect systematic effects affecting the NR waveform (see e.g. Ref~\cite{Nagar:2020pcj} for some discussions).
Therefore, in this work we {\it assume} that the uncertainty on the highest resolution available 
is  substantially negligible. For each dataset, we tune $a_6^c$ so to have an
accumulated EOB/NR phase difference that is as small as possible at NR merger
(say, $\lesssim 0.05$~rad) and moreover (mostly) decreases  {\it monotonically}.
This qualitative feature of the phase difference is {\it crucial} to obtain EOB/NR
unfaithfulness of a few parts in $\sim 10^{-4}$, as we will see 
below~\footnote{In previous EOB models, see for example Ref.~\cite{Nagar:2015xqa} 
or~\cite{Nagar:2017jdw} we NR-informed the EOB model aiming at $10^{-2}$ 
EOB/NR unfaithfulness only. This is easily reachable also in case the phase difference
is not monotonic but oscillates around merger, see for example Fig.~1 of~\cite{Nagar:2015xqa}.}.
Here we determine $a_6^c$ point-wise using the SXS datasets listed in Table~\ref{tab:a6c}.
The raw values can be accurately fitted versus $\nu$ with the usual rational function
\begin{align}
\label{eq:a6c_new}
a_6^{c}&= n_0\dfrac{1+n_1\nu + n_2\nu^2 + n_3\nu^3}{1+d_1\nu} \ ,
\end{align}
with the coefficients listed in Table~\ref{tab:models} under {\tt D3Q3\_NQC}.

Figure~\ref{fig:raw_phasing}  shows a sample of EOB/NR time-domain phasings obtained
using this expression of $a_6^c(\nu)$. Following our usual habits, the figure depicts the Regge-Wheeler-Zerilli
normalized quadrupolar waveform $\Psi_{22}$ that is connected to the (otherwise commonly used)
strain multipoles $h_\lm$ as $\Psi_{\lm}\equiv h_\lm/\sqrt{(\ell+2)(\ell+1)\ell(\ell-1)}$ where 
\be
h_+ - i h_\times= \dfrac{1}{{\cal D}_L}\sum_{\ell =2}^{\infty}\sum_{m=-\ell}^{\ell} h_\lm\,{}_{-2} Y_\lm, 
\ee
with ${\cal D}_L$ is the luminosity distance and ${}_{-2} Y_\lm$ are the $s=-2$ spin-weighted spherical harmonics.
The waveform $\Psi_{22}$ is then decomposed in amplitude and phase as $\Psi_{22}=A_{22}e^{-i\phi_{22}}$.
Following standard procedures, the EOB and NR
waveforms are aligned by applying an arbitrary time and phase shift, obtained following the 
standard procedure delineated in e.g. Ref.~\cite{Damour:2012ky}.
Note that, to maximize accuracy, this comparison is obtained with 5 iterations on the
NQC amplitude parameters $(a_1^{22},a_{2}^{22})$, that correct
the $\ell=m=2$ waveform, in order to determine them self-consistently,  and not using 
the fits of Ref.~\cite{Riemenschneider:2021ppj} that were obtained  with a different 
expression for $a_6^c(\nu)$. To ease the discussion here we refer the reader to Sec.~II
of Ref.~\cite{Riemenschneider:2021ppj} for the definition and implementation of NQC
corrections in \TEOBResumS{}.

As additional evaluation of the quality of the EOB waveform, we compute the EOB/NR 
unfaithfulness. Given two waveforms $(h_1,h_2)$, the unfaithfulness 
is a function of the total mass $M$ of the binary and it is defined as
\be
\label{eq:barF}
\bar{\cal F}(M) \equiv 1-{\cal F}=1 -\max_{t_0,\phi_0}\dfrac{\langle h_1,h_2\rangle}{||h_1||||h_2||},
\ee
where $(t_0,\phi_0)$ are the initial time and phase. We used $||h||\equiv \sqrt{\langle h,h\rangle}$,
and the inner product between two waveforms is defined as 
$\langle h_1,h_2\rangle\equiv 4\Re \int_{f_{\rm min}^{\rm NR}(M)}^\infty \tilde{h}_1(f)\tilde{h}_2^*(f)/S_n(f)\, df$,
where $\tilde{h}(f)$ denotes the Fourier transform of $h(t)$, $S_n(f)$ is the detector power spectral density (PSD),
and $f_{\rm min}^{\rm NR}(M)=\hat{f}^{\rm NR}_{\rm min}/M$ is the initial frequency of the
NR waveform at highest resolution, i.e. the frequency measured after the junk-radiation
initial transient.
For $S_n$, in our comparisons we use either the zero-detuned, high-power noise spectral density of 
Advanced LIGO~\cite{aLIGODesign_PSD} or the predicted sensitivity of 
Einstein Telescope~\cite{Hild:2009ns, Hild:2010id} and Cosmic Explorer~\cite{Evans:2021gyd}.
Waveforms are tapered in the time-domain to reduce high-frequency 
oscillations in the corresponding Fourier transforms. 
The computation is done over a sample of nonspinning SXS simulations with mass ratio 
ranging from $1$ to $9.5$, and the EOB/NR unfaithfulness is then 
denoted as $\bar{\cal F}_{\rm EOB/NR}$. 
\begin{figure}[t]
	\center
	\includegraphics[width=0.45\textwidth]{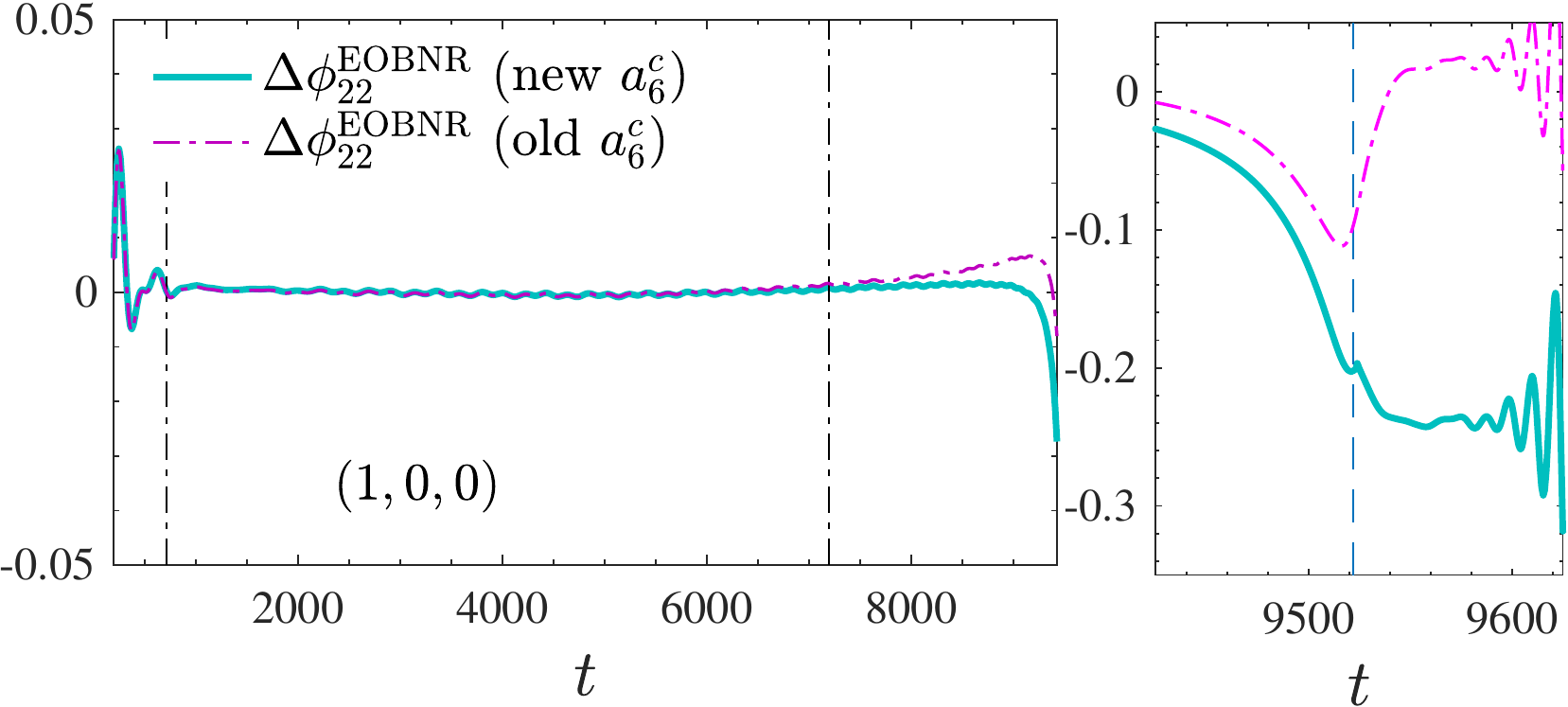}\\
	\vspace{5 mm}	
	\includegraphics[width=0.45\textwidth]{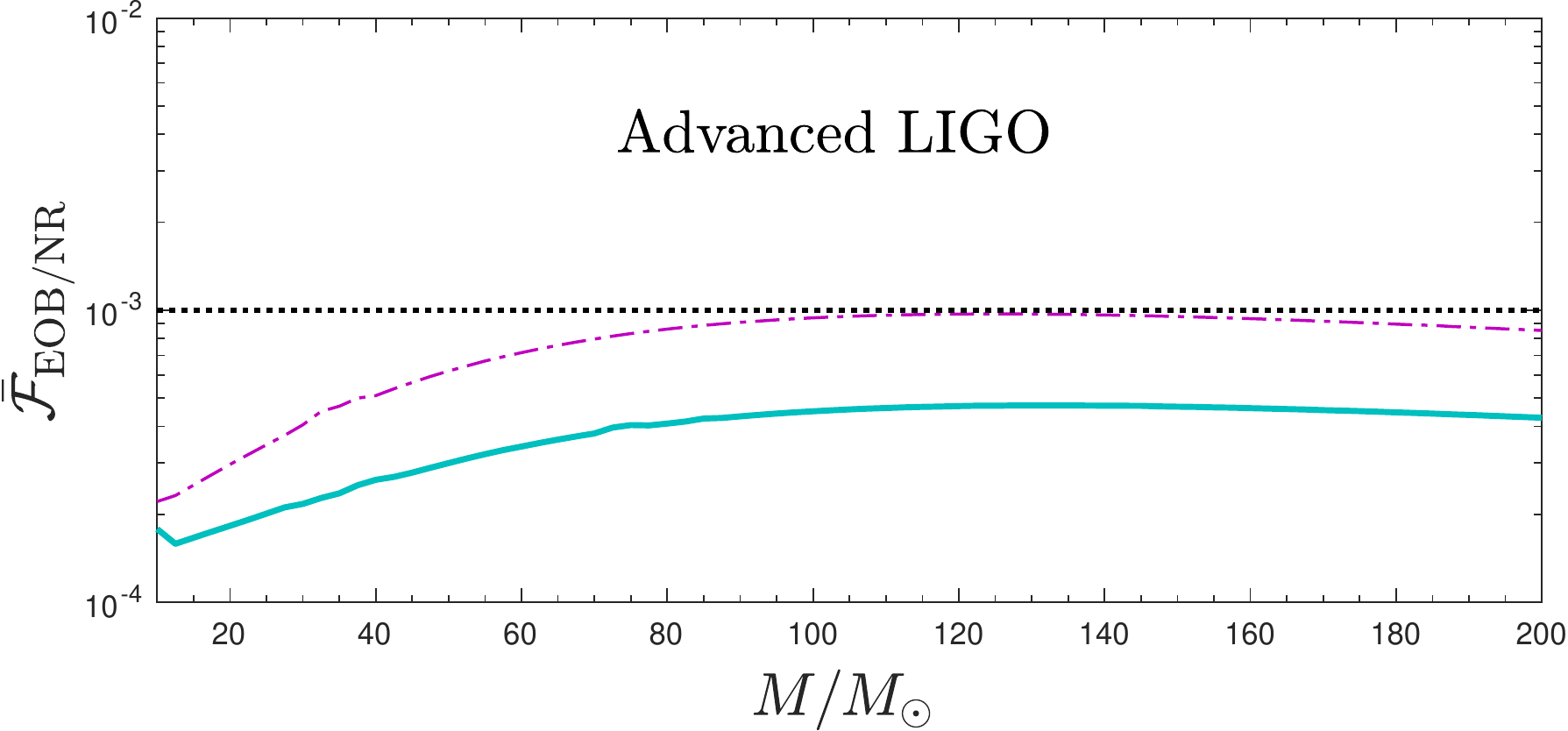}
	\caption{\label{fig:test_a6c}Effect of small changes in $a_6^c$ for $q=1$. 
	The old value is $a_6^c\simeq-41.16$, the standard \TEOBResumS{} one, 
	while the new one is $a_6^c\simeq-35.93$. The figure highlights the relation between
	the monotonically decreasing $\Delta\phi_{22}^{\rm EOBNR}$ and the unfaithfulness.
	A value of $a_6^c$ that is suboptimal corresponds to approximately a loss of $5\times 10^{-4}$
	in unfaithfulness.}
	\center
\end{figure}

The result of this computation is displayed in Fig.~\ref{fig:new_barF}. For aLIGO, 
the largest values graze the $5\times 10^{-4}$ level, correspond to $q=1$ and $q=1.5$ and
are somehow outliers with respect to the other configurations, where one easily gets to $2\times 10^{-4}$.
Inspecting the time-domain phasings of Fig.~\ref{fig:raw_phasing} one understands that this
is the effect of lowering the phase difference at merger from $\Delta\phi_{22}^{\rm EOBNR}\equiv \phi_{22}^{\rm EOB}-\phi_{22}^{\rm NR}\sim 0.2$~rad
to $\Delta\phi_{22}^{\rm EOBNR}\sim 0.1$ rad (see e.g. the $q=9.5$ case). Changing to ET-D and
CE the worst values of $\bar{\cal F}_{\rm EOB/NR}$ are largely unchanged, with all 
datasets clustering around $10^{-4}$. It thus seems that obtaining a model that is {\it more} 
NR-faithful (i.e. at the $10^{-5}$ level or below) mostly amounts at {\it further reducing} the 
phase difference around merger.
By contrast, an imperfect control of $\Delta\phi_{22}^{\rm EOBNR}$ around merger yields
an important reduction of the EOB/NR agreement. 

The reader should now be reminded that we stated that $a_6^c$ was carefully 
chosen such that the phase difference $\Delta\phi^{\rm EOBNR}_{22}$
is small and decreases monotonically, mentioning this as an essential 
qualitative feature to be controlled in the NR-information process of $a_6^c$.
Our statement can be understood by inspecting Fig.~\ref{fig:test_a6c}. 
The top panel of the figure compares two EOB/NR 
phase differences for $q=1$: one obtained with the standard value of $a_6^c$ of \TEOBResumS{}
(that is $a_6^c\simeq-41.16$) and the other with the improved value obtained above 
$a_6^c\simeq-35.93$. We see that in the former case the phase difference, 
that is not monotonic around merger, yields $\bar{\mathcal{F}}_{\rm EOB/NR}$ 
two times larger than in the optimized case.

In conclusion, this understanding allows to put on solid basis what is additionally required to further 
improve the model: one needs to {\it flatten} the phase difference further through merger-ringdown. 
This can be achieved by experimenting with the analytic content of the model, notably the 
nonradial pieces of the dynamics or augmenting the importance of NR data through
merger and ringdown. Some of these effects will be investigated in the next section.
Let us however mention, in passing, that an important improvement may come 
from a different determination of the NR-informed NQC correction 
to the phase. The procedure currently implemented in the model is still the one outlined 
in Refs.~\cite{Nagar:2019wds} (see Sec.~IIID therein) and might need to be 
revised or updated, e.g. with and additional NQC phase parameter determined
by imposing continuity between the EOB and NR third derivative of the frequency.
To do this properly, dedicated studies in the test-particle limit are currently 
ongoing~\cite{Albanesi:2023bgi}, and we do not pursue this investigation further 
here but limit ourselves to mention it.

\section{Analytic systematics}
\label{sec:systematics}

In this Section we attempt a preliminary investigation of what we call {\it analytic systematics}.
By this term we address features of the EOB waveforms that depend on specific choices of 
analytic elements entering the model. 
A rather naive way of thinking about analytic systematics is 
within post-Newtonian (PN) theory, assuming that they arise from the lack of higher PN orders and studying the impact of such missing terms on the waveform generation~\cite{Owen:2023mid}.
This approach, while useful to test the sensitivity of parameter estimation to some "small" modifications of a model, is rather misleading because of the well-known asymptotic, non-converging nature of PN series in strong field, exactly where such high order terms are expected to become most relevant.
Indeed, in place of simple PN expansions, it is desirable to use resummed analytical expressions~\cite{Damour:1997ub}, even for the description of inspiral waveforms.
In light of this fact, the EOB approach -- with all its different avatars -- represents the most natural formalism to \textit{quantitatively} study these kind of systematic effects, given its robustness and reliability also in the strong field regime.

There are many elements of EOB-based models that can
be affected by systematics. To quote a few: (i) the PN truncation of 
the EOB potentials $(A,D,Q)$, their resummed expressions and 
their NR-completion (preliminary explored in~Refs.\cite{Rettegno:2019tzh,Nagar:2020xsk,Nagar:2021xnh,Nagar:2022icd,Bonino:2022hkj}); 
(ii) the PN truncation and resummation of the angular momentum flux, 
that impacts the radiation reaction and becomes more and more important 
for long inspirals and large mass 
ratios~\cite{Nagar:2016ayt,Messina:2018ghh,Nagar:2019wrt,Albertini:2022rfe,Albertini:2022dmc};
(iii) the way NR-information is incorporated in the model.
Below we focus on two analytic systematics: (i) the importance of the time where the
inspiral-to-plunge EOB waveform is attached to an analytic description of the ringdown 
and (ii) the importance of terms beyond 3PN in the EOB potentials $(D,Q)$, 
and their link to NQC parameters.

\subsection{The role of the matching point}
\begin{table*}[t]
 \caption{\label{tab:models} Table listing the main features of the various, nonspinning, quasi-circular models developed throughout this paper.}
   \begin{center}
     \begin{ruledtabular}
\begin{tabular}{c c c c | c c c c c c} 
Model  & $D$ & $Q$ & NQC iteration & \multicolumn{6}{c}{$a^6_c=n_0(1+n_1\nu+n_2\nu^2+n_3\nu^3+n_4\nu^4)/(1+d_1\nu)$ }\\
       &     &     &                & $n_0$ & $n_1$ & $n_2$ & $n_3$ & $n_4$ & $d_1$\\
\hline
{\tt D3Q3\_NQC} & $3$PN, $P^0_3[D]$ & $3$PN & \checkmark         &  $46.5524$ & $-24.2516$ & $120.9594$ &  $-167.2242$ & -- & $-3.3998$ \\
{\tt D5Q5\_NQC} & $5$PN, $P^3_2[D]$ & $5$PN & \checkmark         & $104.6595$ & $-23.2539$ & $113.8091$ &  $-261.8068$ & -- & $3.6511$ \\
{\tt D5Q5}      & $5$PN, $P^3_2[D]$ & $5$PN & \text{\sffamily X} & $331.1899$ & $-27.9217$ & $268.4658$ & $-1138.1009$ & $1784.4727$ & $0.063909$\\
{\tt D3Q3}      & $3$PN, $P^0_3[D]$ & $3$PN & \text{\sffamily X} &   $41.803$ & $-24.5764$ & $251.4175$ &  $-926.6667$ & $1080.9227$ & $0.71904$\\
{\tt D5Q3}      & $5$PN, $P^3_2[D]$ & $3$PN & \text{\sffamily X} & $-42.0938$ & $-28.8863$ & $332.9101$ & $-1494.0405$ & $2275.5579$ & $-2.3295$
\end{tabular}
\end{ruledtabular}
\end{center}
\end{table*}
So far, we have seen that it is relatively easy to improve the performance of the state-of-the-art 
\TEOBResumS{} model simply by introducing slight changes in the way the single EOB-flexibility function 
$a_6^c(\nu)$ is determined. This yields EOB/NR unfaithfulness values that are at most $\sim 5\times 10^{-4}$,
with a gain of a factor two in the worst cases with respect to the previous model. 
This value is still larger than, though compatible with, the nominal NR uncertainty and reflects a phase difference
at merger that is of order $\sim 0.2$~rad. As such, one is expecting that the EOB/NR agreement
can be improved further. In this respect, we remind the reader that we are NR-informing a {\it single} free
parametric functions, while the {\rm SEOBNR} lineage of models~\cite{Bohe:2016gbl,Cotesta:2018fcv,Pompili:2023tna} 
uses {\it two} for the same setup (nonspinning BBHs) to obtain comparable results (see~\cite{Pompili:2023tna}). 
More specifically, the {\tt SEOBNR} family NR-calibrates: (i) the analogous of $a_6^c$, 
a function called $K$ (or even $a_6^c$ in the latest {\tt SEOBNRv5} model~\cite{Pompili:2023tna}), 
and (ii) the temporal location of the peak of the $\ell=m=2$ waveform. 
Within \TEOBResumS{} the analogous of this second parameter is called $\Delta t_{\rm NQC}$, 
that is defined as follows. Calling $\Omega_{\rm orb}$ the orbital frequency\footnote{In the presence of spin, this is replaced by 
the {\it pure} orbital frequency, $\Omega_{\rm orb}=\de_{p_\varphi}\hat{H}_{\rm orb}$, i.e.
the orbital frequency without the contribution coming from the spin-orbit Hamiltonian, see~\cite{Damour:2014sva}.
The use of this function was inspired by features in the test-mass 
limit~\cite{Damour:2014sva,Harms:2014dqa} that indicate that the peak of the quadrupolar waveform
is always close to the peak of $\Omega_{\rm orb}$~\cite{Harms:2014dqa} for values of the spin
compatible to those of the final black hole that can be generated from BBH coalescences.
As such, it is the pivotal element of the \TEOBResumS{} construction, since it offers a natural, and simple,
anchor point to define the merger time. Note in addition that $\Omega_{\rm orb}$
always has a maximum, differently from the complete orbital frequency $\Omega$, whose structure 
depends on subtle interplays between the (necessarily approximated)
descriptions of the orbital and spin-orbit sector of the Hamiltonian.} and $t_{\Omega^{\rm peak}_{\rm orb}}$ the time when it peaks, the interval 
$\Delta t_{\rm NQC}$ identifies on the EOB time axis the time $t_{\rm NQC}$, 
where we compute the NQC corrections
\begin{figure}[t]
	\center
	\includegraphics[width=0.45\textwidth]{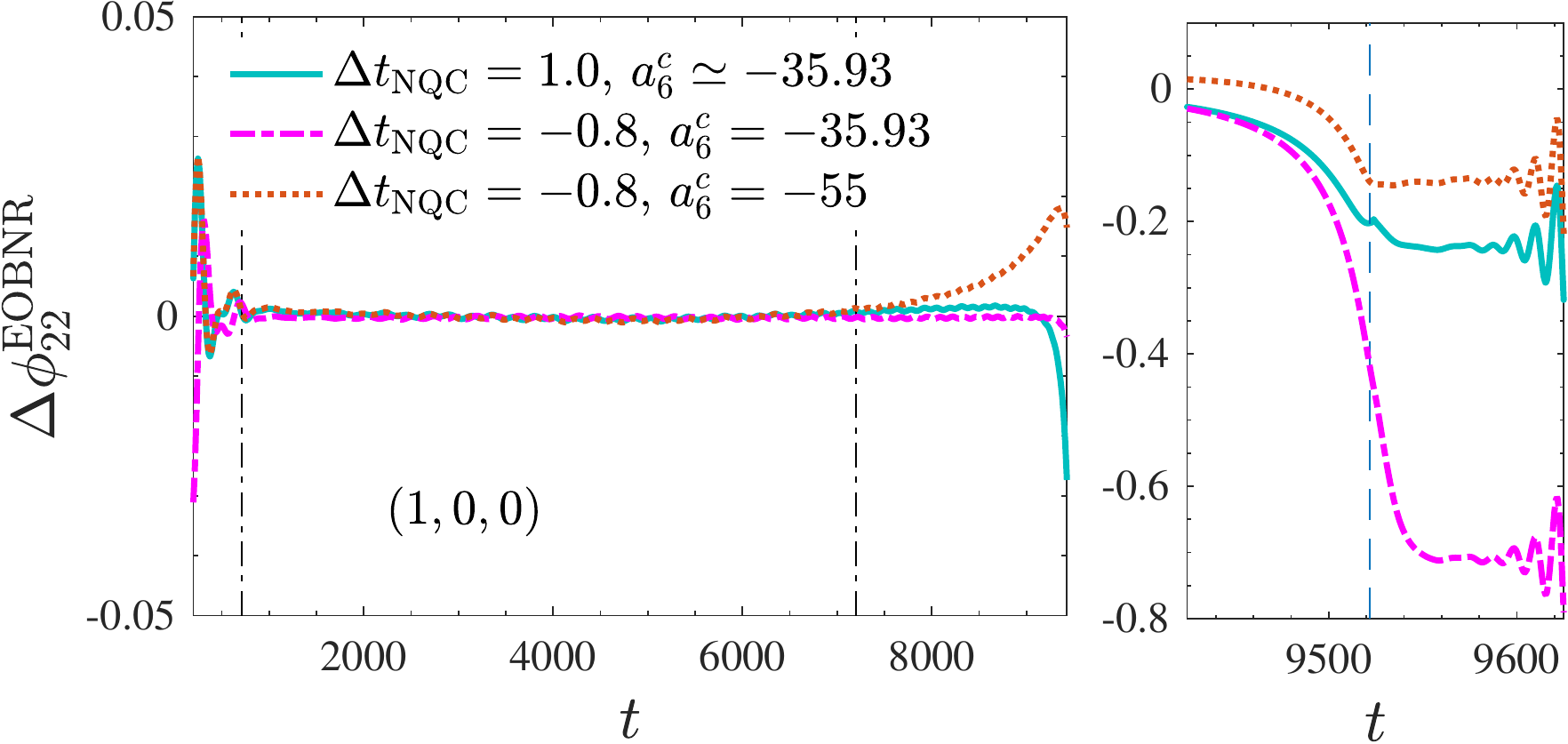}	
	\includegraphics[width=0.45\textwidth]{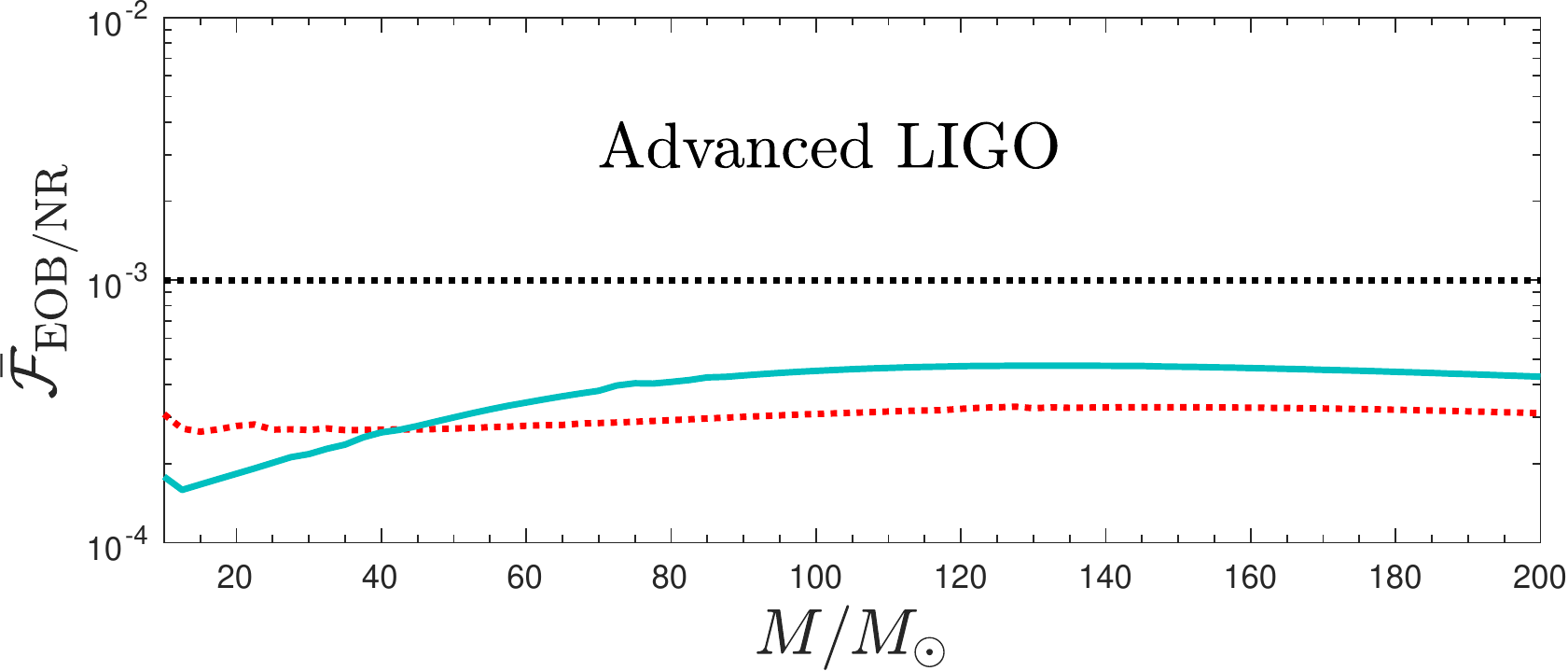}	
	\caption{\label{fig:tuningDeltaTnqc}Equal-mass case: effect of tuning, at the same 
	time, $a_6^c$ and $\Delta t_{\rm NQC}$. An improvement around merger ringdown
	seems to be balanced by a slight worsening during the late inspiral. This is apparent
	also looking at the EOB/NR unfaithfulness.}
	\center
\end{figure}
\be
t_{\rm NQC}^{\rm EOB}=t_{\Omega_{\rm orb}}^{{\rm peak}}-\Delta t_{\rm NQC} \ ,
\ee
by imposing there that the EOB waveform (amplitude, phase and their time derivatives) coincides 
with the NR one, the latter evaluated $2M$ after the NR merger.
See in particular Refs.~\cite{Damour:2014sva,Nagar:2018zoe} for the technical details of the 
procedure. Following Ref.~\cite{Nagar:2018zoe}, we have that
\be
\label{eq:Dt_nqc}
\Delta t_{\rm NQC} = 1 \ .
\ee
This choice is motivated by the fact that, in the test-mass limit, this separation is $\lesssim 1$.
More precisely, for the case of a test-mass plunging on a nonspinning black hole, one finds that
$\Delta t_{\rm NQC}= 0.56$ (see Table~3 of\footnote{Actually, the value is 0.38 or 0.56 depending
whether it is calculated using {\tt Teukode} or by solving directly the Regge-Wheeler-Zerilli equations
with an hyperboloidal layer~\cite{Bernuzzi:2010xj,Bernuzzi:2011aj}.} Ref.~\cite{Harms:2014dqa}).
Analogously, for a test-particle plunging on a {\it spinning} black hole,  we have 
that $\Delta t_{\rm NQC}\lesssim 1$ and it grows as the BH spin increases. 
See in particular Table~A3 of Ref.~\cite{Harms:2014dqa}. 
On the basis of this test-mass knowledge, the condition of Eq.~\eqref{eq:Dt_nqc} was considered 
as a good, and simple, compromise and we did not attempt to inform {\it also} this parameter using 
NR simulations, although a priori one is expecting it to depend on {\it both} the mass ratio and the spins. 
It is thus interesting to explore to which extent an additional tuning  of $\Delta t_{\rm NQC}$ 
can impact the model performance. To do so, we focus on the $q=1$ case, that is the one showing 
the largest disagreement with the NR waveform. 
The top panel of Fig.~\ref{fig:tuningDeltaTnqc} compares three phase differences 
$\Delta\phi_{22}^{\rm EOBNR}$: (i) the standard one, with $\Delta t_{\rm NQC}=1$
and $a_6^c\simeq 35.93$; (ii) the one obtained using  $\Delta t_{\rm NQC}=-0.8$
but keeping $a_6^c$ unchanged (magenta, dash-dotted line); (iii) the one with
$\Delta t_{\rm NQC}=-0.8$ and $a_6^c=-55$, i.e. after an additional tuning of this parameter.
Interestingly, one gains about a factor two around merger, although some +0.02~rad 
are now lost during the late inspiral. 
The bottom panel of the figure quantifies this information in terms of EOB/NR unfaithfulness:
to a gain of about $2\times 10^{-4}$ for high masses corresponds to a loss of $\sim 1\times 10^{-4}$
for low masses.
This analysis evidences that the tuning of $\Delta t_{\rm NQC}$ could actually be helpful in
improving the EOB/NR phase agreement around merger. However, the slight loss during the 
late inspiral seems to suggest it is not  the best way of proceeding and alternative routes should
probably be explored. Moreover, it requires an additional complication (tuning two parameters 
with NR data instead of one) that should possibly be faced only {\it after} other analytic elements 
are carefully considered and evaluated. It is certainly not impossible to NR-tune two parameters 
at the same time, and indeed this is currently done for EOB models of the {\tt SEOBNR} family~\cite{Bohe:2016gbl,Cotesta:2018fcv,Pompili:2023tna}. 
However, since \TEOBResumS{} allows for a simple and efficient iterative tuning of $a_6^c$ only that
yields highly accurate waveforms, we believe that this complication should be currently avoided.  
Possibly, it should be implemented {\it only after} a new analysis of  the structure  of the merger in 
the test-mass limit and its connection to the comparable-mass case is performed. In this respect, 
a seminal study was performed in Ref.~\cite{Damour:2012ky}, see Sec.~IVB therein. An update
of that analysis with the current, more accurate, SXS waveforms is in order and will be considered
in future work.

\subsection{$D$ and $Q$ up to 5PN accuracy.}
\label{sec:DandQ5pn}
\begin{figure}[t]
	\center
	\includegraphics[width=0.22\textwidth]{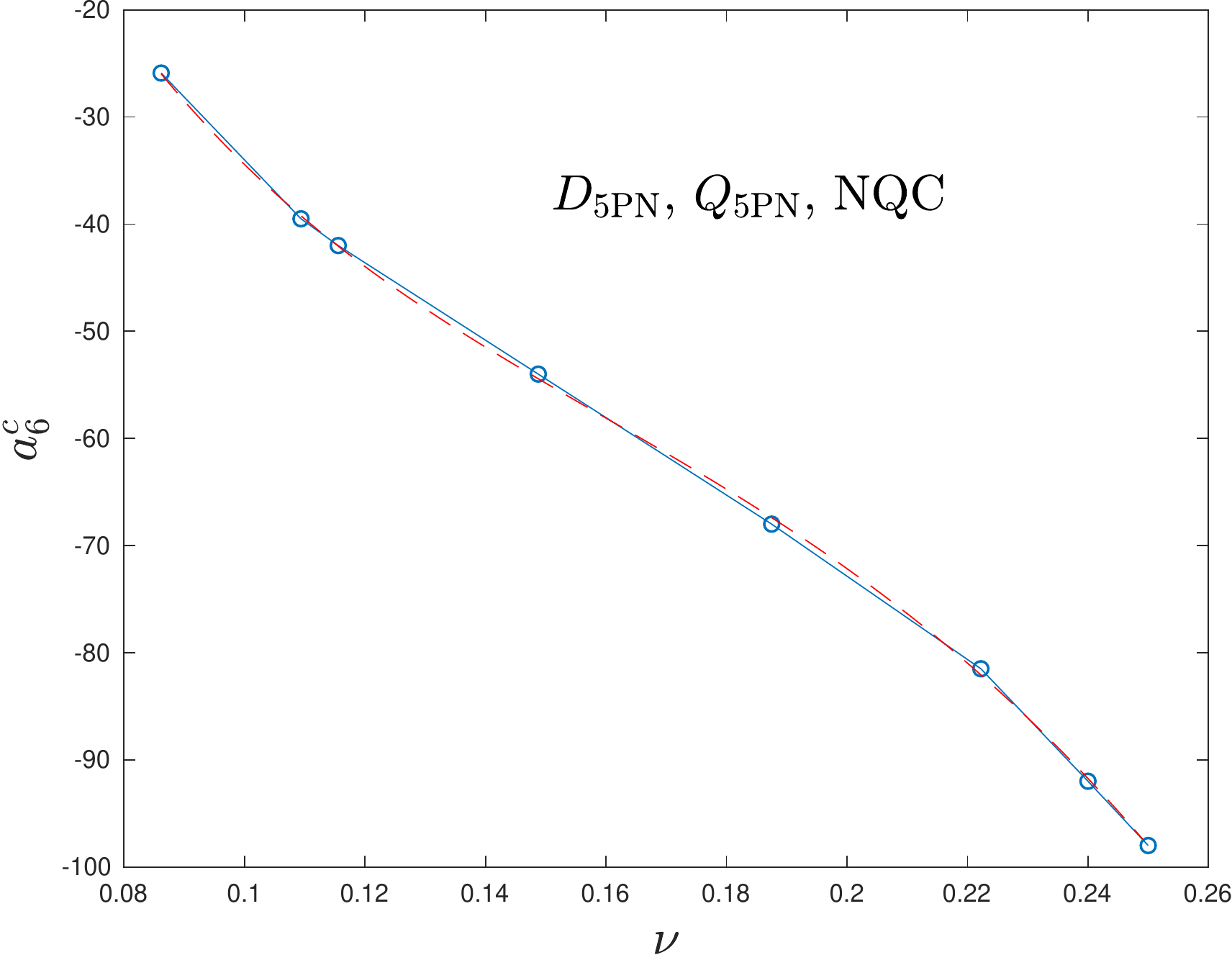}
	\includegraphics[width=0.22\textwidth]{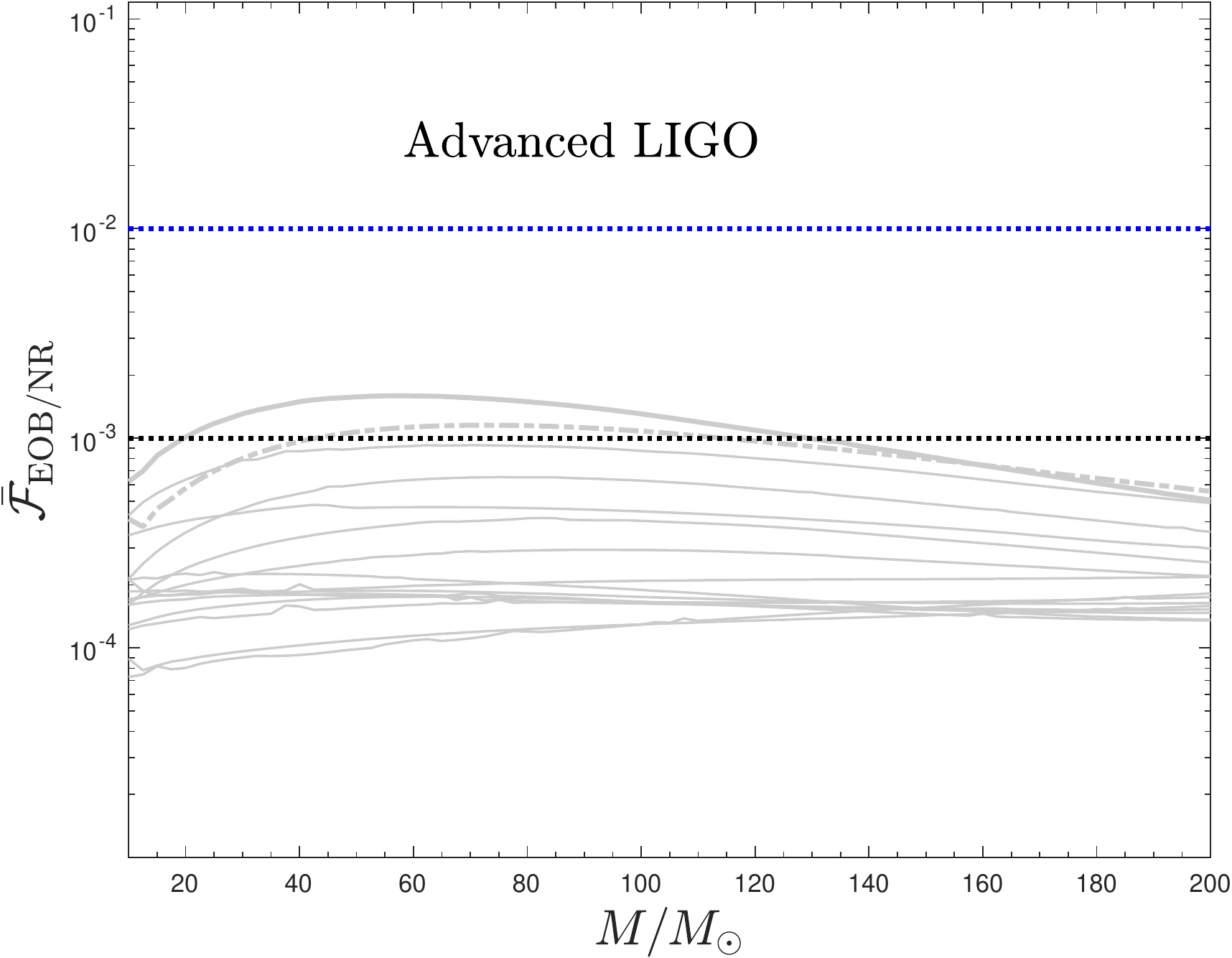}\\
	\includegraphics[width=0.22\textwidth]{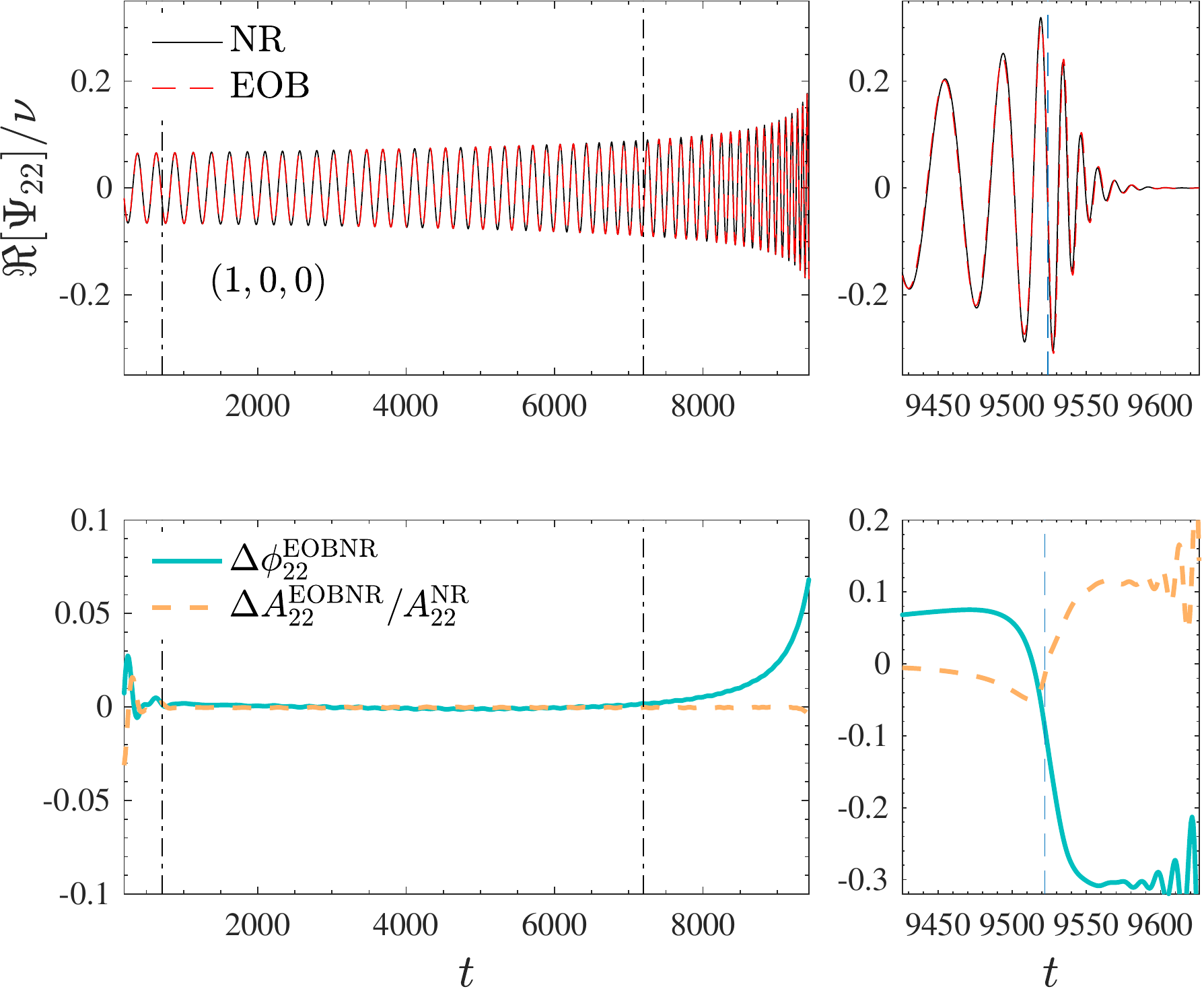}
	\includegraphics[width=0.22\textwidth]{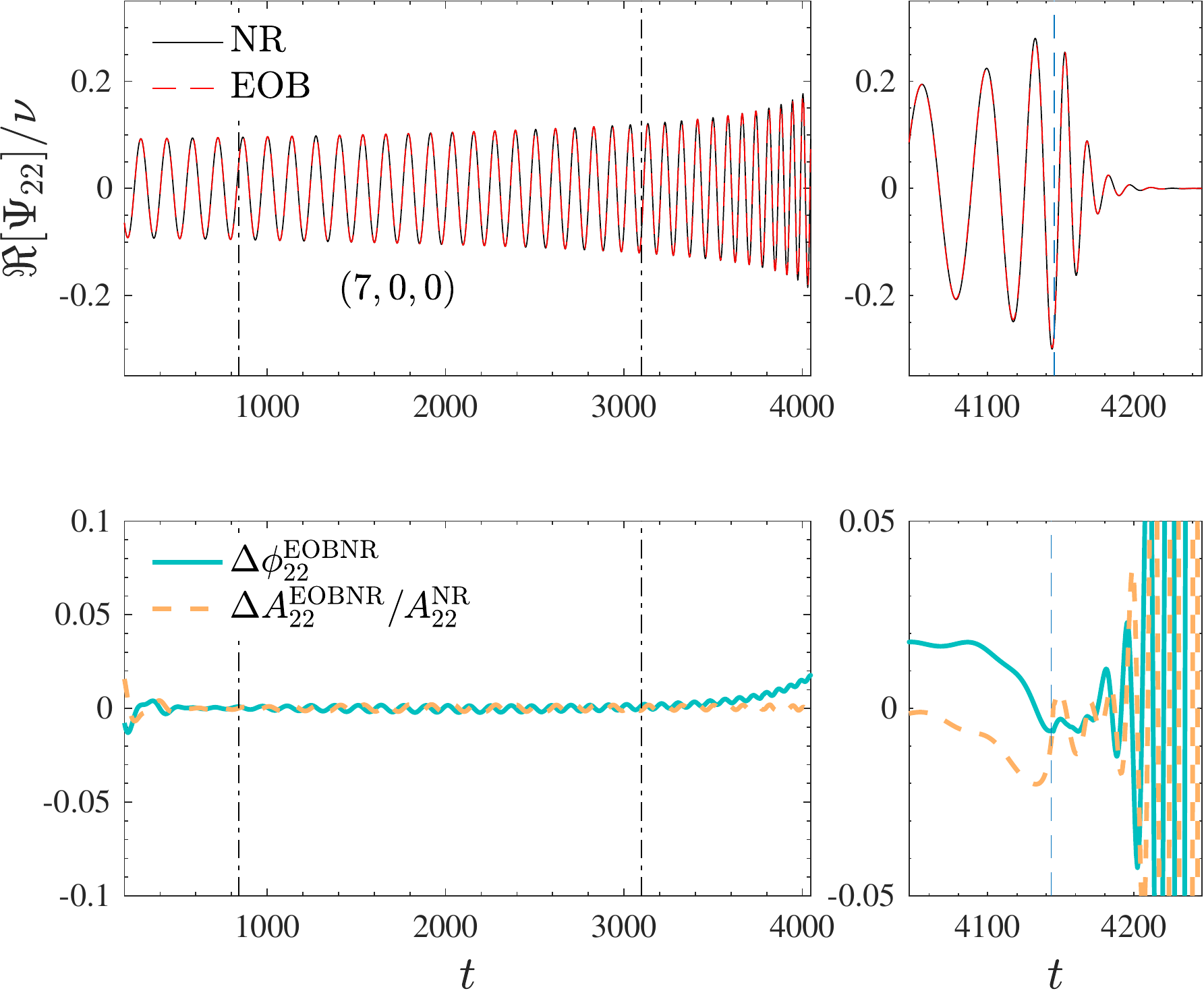}
	\caption{\label{fig:phase_P32D}{\tt D5Q5\_NQC} model: EOB/NR phasing analogous to Fig.~\ref{fig:raw_phasing},
	though using the complete, analytically known, 5PN information in $(D,Q)$, with the $D$ function 
	resummed using a $(3,2)$ Pad\'e approximant. Note that the phase difference accumulated already 
	{\it before} merger for $q=1$ is much larger than what occurs for {\tt D3Q3\_NQC}, Fig.~\ref{fig:raw_phasing}.}.
	\center
\end{figure}

Let us now analyze in detail the other obvious source of analytic systematics,
namely the PN accuracy of the EOB potentials.  The impact of currently known analytical 
information at high PN order (notably, up to 6PN) was partially explored in Refs.~\cite{Nagar:2020xsk} 
and~\cite{Nagar:2021xnh}. In particular, in the context of constructing a spin-aligned 
EOB model for generic (non-quasi-circular) planar orbits, Ref.~\cite{Nagar:2021xnh}
already explored the effect of the currently known analytical information at 5PN in
$(A,D,Q)$. To do so, in order to avoid the occurrence of spurious poles, the resummation 
of the $A$ function was performed using a $(3,3)$ Pad\'e approximant; similarly,
$D$ was resummed using a $(3,2)$ approximant, while the $Q$ function was kept
in PN-expanded form. In addition, Ref.~\cite{Nagar:2021xnh} also compared
various Pad\'e-based resummations for $D$, concluding that the $(3,2)$ is the best
compromise at this PN order. Since the focus of~\cite{Nagar:2021xnh} was on
eccentric orbits (and scattering), the performance of the 5PN-accurate $(D,Q)$ potentials was
not explicitly spelled in the context of quasi-circular binaries, i.e. using the standard
quasi-circular radiation reaction of \TEOBResumS{}.

The aim of this section is to fill this gap in our knowledge by exploring how
changes in $(D,Q)$ affect the quasi-circular model discussed above.
We do so either by (i) following our standard approach of iterating on the
NQC amplitude parameters $(a_1,a_2)$  or (ii) removing the iteration, similar to
the procedure followed for the eccentric model~\cite{Nagar:2021xnh}.
We will explore different PN truncations of $(D,Q)$ and in each case we
will determine a new $a_6^c(\nu)$ function, compute EOB/NR time-domain
phasing and unfaithfulness.
The properties of the nonspinning models we are going to compare and contrast
are listed in Table~\ref{tab:models}. Let us now analyze them one by one.

\subsubsection{{\tt D5Q5\_NQC}: 5PN and iteration on NQC parameters}
Let us start by considering a simple modification of \TEOBResumS{}, 
dubbed {\tt D5Q5\_NQC}, where we replace
$D$ and $Q$ at 3PN with their 5PN counterparts. The analytic expressions are 
precisely those of Eqs.~(3) and (5) of Ref.~\cite{Nagar:2021xnh}. In particular, for simplicity
we use only the {\it local} part of $Q$ and omit its nonlocal contributions.
Similarly, in $D$ we set to zero the analytically unknown coefficient $d_5^{\nu^2}$. 
The $D$ function is then resummed with a (3,2) Pad\'e approximant.
For each mass ratio considered, we determine the best $a_6^c$ value inspecting
the EOB/NR phasing and requiring, as in the previous case, that the phase difference
possibly decreases monotonically through merger and ringdown. The corresponding 
values of $a_6^c$,  shown in the top-left panel of Fig.~\ref{fig:phase_P32D}, can be easily 
fitted with the same rational function given by Eq.~\eqref{eq:a6c_new} above. 
The fitting coefficients can be read from Table~\ref{tab:models}.
The corresponding EOB/NR time-domain phasings for $q=(1,2,4.5,6.5,9.5)$, obtained with the 
fitted $a_6^c$ function, are shown in Fig.~\ref{fig:phase_P32D}. From visual inspection, it is evident 
that the EOB/NR phasing agreement is less good than the one of Fig.~\ref{fig:raw_phasing}, obtained 
using $D_{\rm 3PN}$ and $Q_{\rm 3PN}$.
Note in particular that there is a nonnegligible positive phase difference that accumulates {\it already} 
during the late inspiral up to merger and that cannot be absorbed by tuning $a_6^c$. 
This effect is more evident when $q$ is small: for $q=1$ one reaches $\Delta\phi^{\rm EOBNR}_{22}\sim 0.1$ 
at merger time, which then changes sign to reach $-0.3$ during the ringdown. 
The EOB performance degrades further for $q=2$, although -- for larger mass ratios -- 
$\Delta\phi^{\rm EOBNR}_{22}$ is compatible with the corresponding cases of Fig.~\ref{fig:raw_phasing}.
The sign change of $\Delta\phi^{\rm EOBNR}_{22}$ around merger time
negatively impacts on the unfaithfulness calculation, as illustrated in Fig.~\ref{fig:phase_P32D}.
The curves are generally more spread than in Fig.~\ref{fig:new_barF}, with a few configurations
(those with mass ratios $q=(1,1.5)$) above the $0.1\%$ level.

This exercise demonstrates that it is possible to construct an EOB model,
within a given EOB paradigm, that incorporates currently available 5PN information in the
$(D,Q)$ potentials, that is {\it less} NR faithful than an analogous model that 
only incorporates 3PN information in the same functions. This looks somehow counterintuitive.
A commonly accepted statement within the gravitational wave modeling community 
is that the quality of the waveform model {\it improves} by increasing the order of
the PN information used to construct it. As a consequence, high-PN results are usually
considered an essential element to obtain highly faithful waveform models for current 
and future GW detectors.
Our simple exercise demonstrates that this simple statement {\it may not always be} true, 
at least for the considered EOB potentials, as it depends on other features of the model.
In particular, in an attempt to frame this statement in a wider context, in the next section 
we will explore what happens when the well-established practice of iterating on the 
amplitude NQC parameters $(a_1,a_2)$ is removed.
\begin{figure}[t]
	\center
	\includegraphics[width=0.45\textwidth]{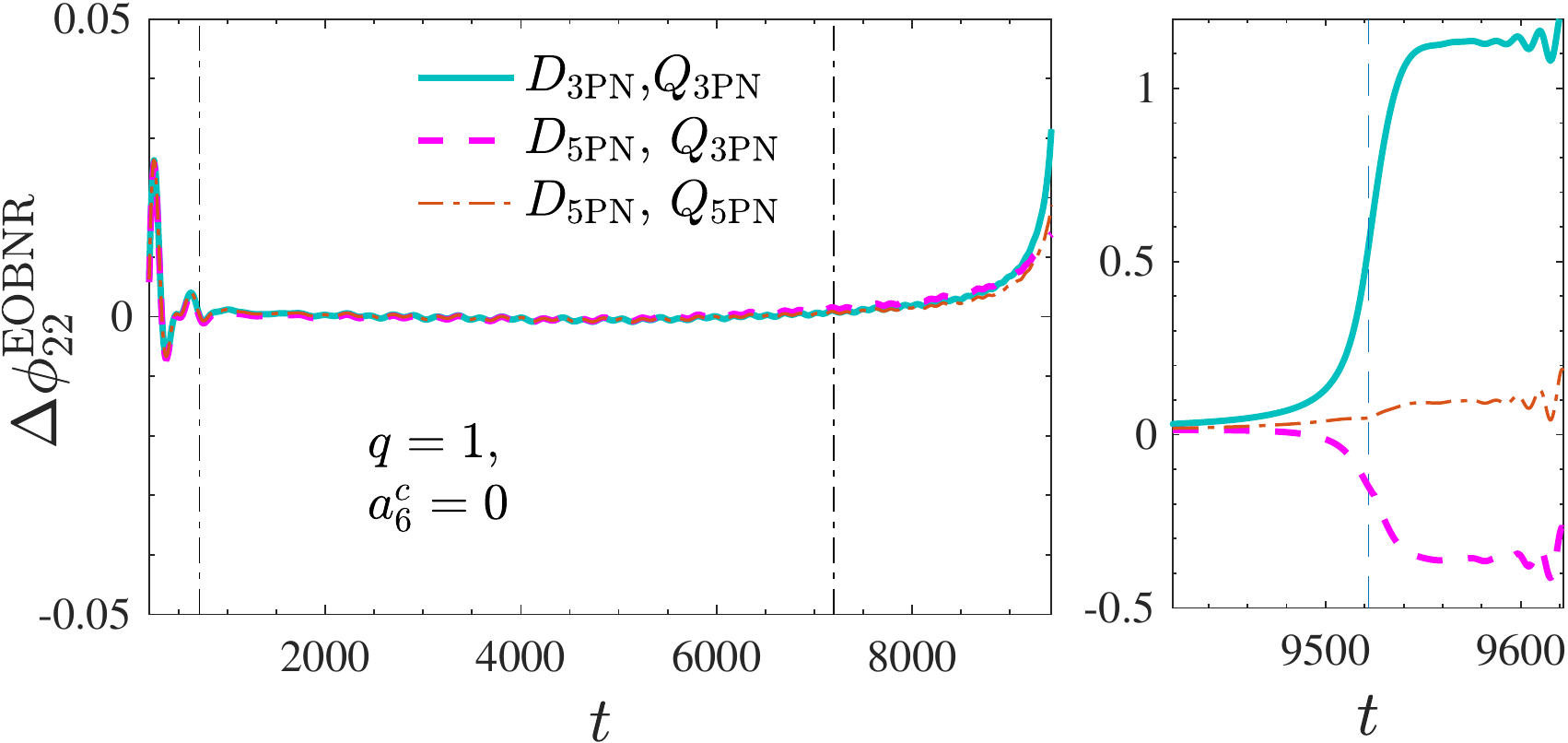}	
	\caption{\label{fig:noNQC_PN_impact}EOB/NR phasing comparison for $q=1$. Here we 
	fix $a_6^c=0$ and do not iterate on the NQC parameters $(a_1,a_2)$. Increasing the amount
	of PN information in the $D$ and $Q$ functions seem to go in the good direction and improve
	the EOB/NR phasing agreement during the plunge up to merger. See text for discussion.}
	\center
\end{figure}
%
\begin{figure}[t]
	\center
	\includegraphics[width=0.45\textwidth]{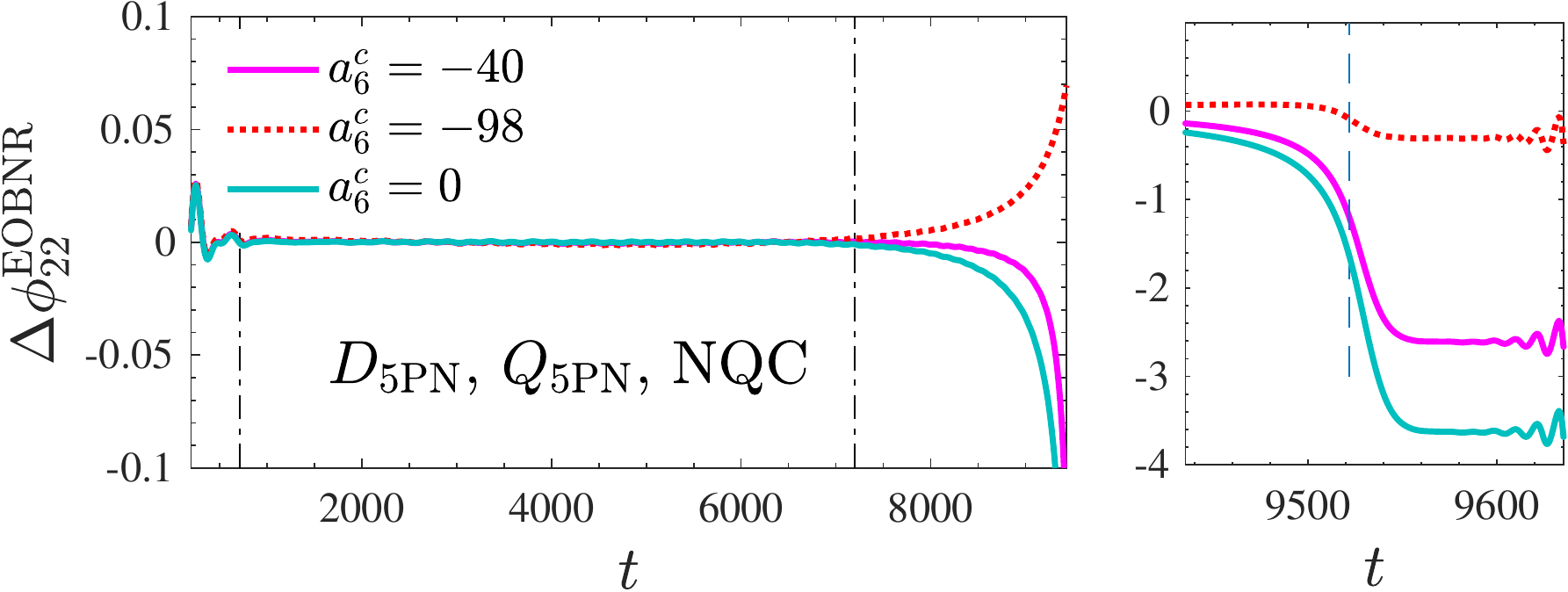}	
	\caption{\label{fig:NQC_impact}Complement to Fig.~\ref{fig:noNQC_PN_impact}. The iteration on the NQC parameters
	is equivalent to a {\it repulsive} effect in the potential, i.e. the merger gets strongly delayed. Note that the $a_6^c=0$ line
	here is how the red dashed curve of Fig.~\ref{fig:noNQC_PN_impact} is modified due to the action of the iterated NQC.
	This feature can be {\it partly compensated} by tuning $a_6^c$, although the price to pay to have an acceptable phasing at 
	merger is a progressive worsening of the phasing during the late inspiral (see line for $a_6^c=-98$, that is the value 
	corresponding to {\tt D5Q5\_NQC}). This curve here corresponds to $\Delta\phi_{22}^{\rm EOBNR}$ in the bottom
	left panel of Fig.~\ref{fig:phase_P32D}.}
	\center
\end{figure}
\begin{figure}[t]
	\center
	\includegraphics[width=0.45\textwidth]{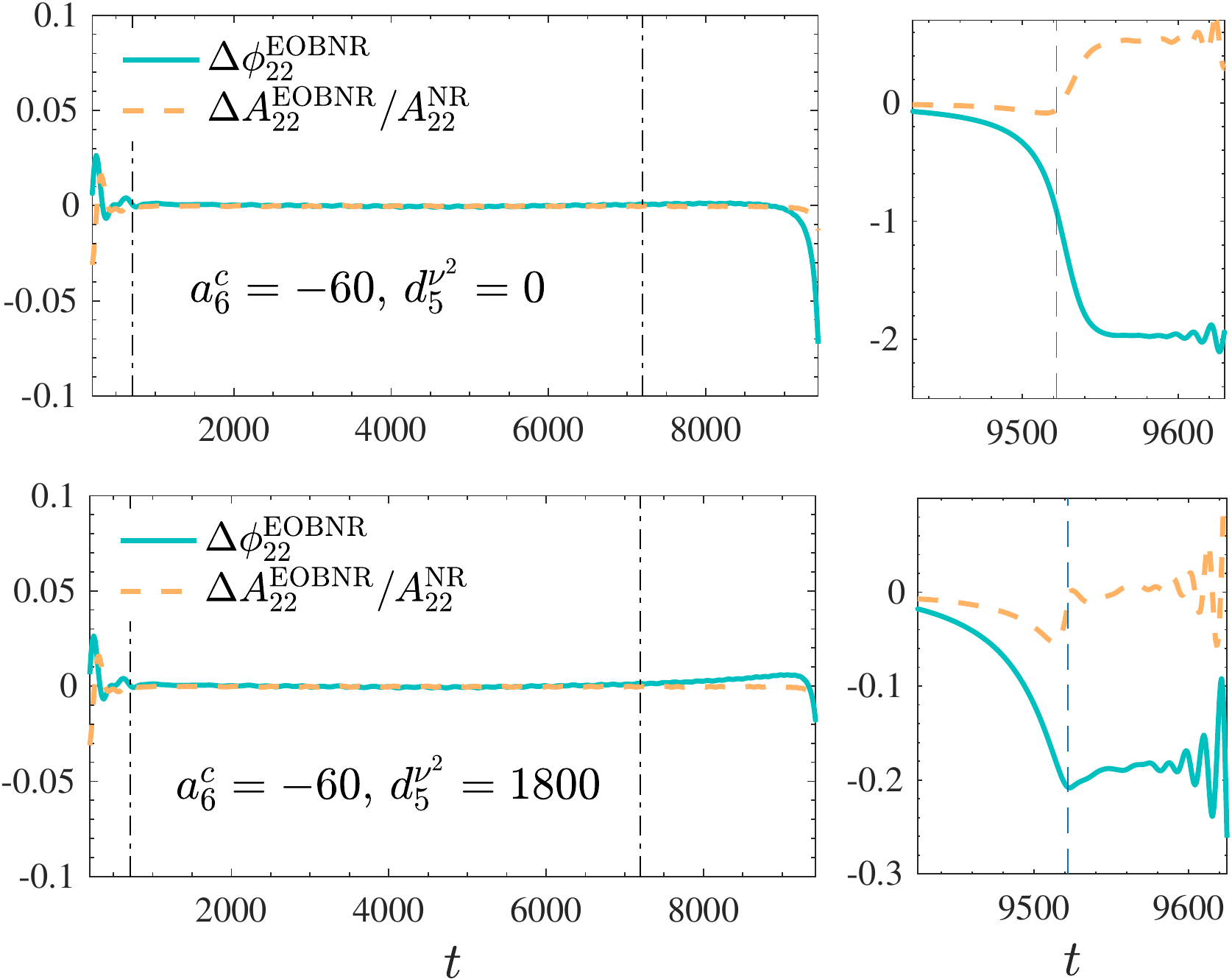}	
	\caption{\label{fig:D_tuning}Tuning in progression first $a_6^c$ and then 
	$d_5^{\nu^2}$ keeping the iteration on NQC parameters. The performance, 
	for the $d_5^{\nu^2}$-tuned case, is visually comparable to {\tt D3Q3\_NQC}
	of Fig.~\ref{fig:raw_phasing}.}
	\center
\end{figure}
\begin{figure*}[t]
	\center
        \includegraphics[width=0.31\textwidth]{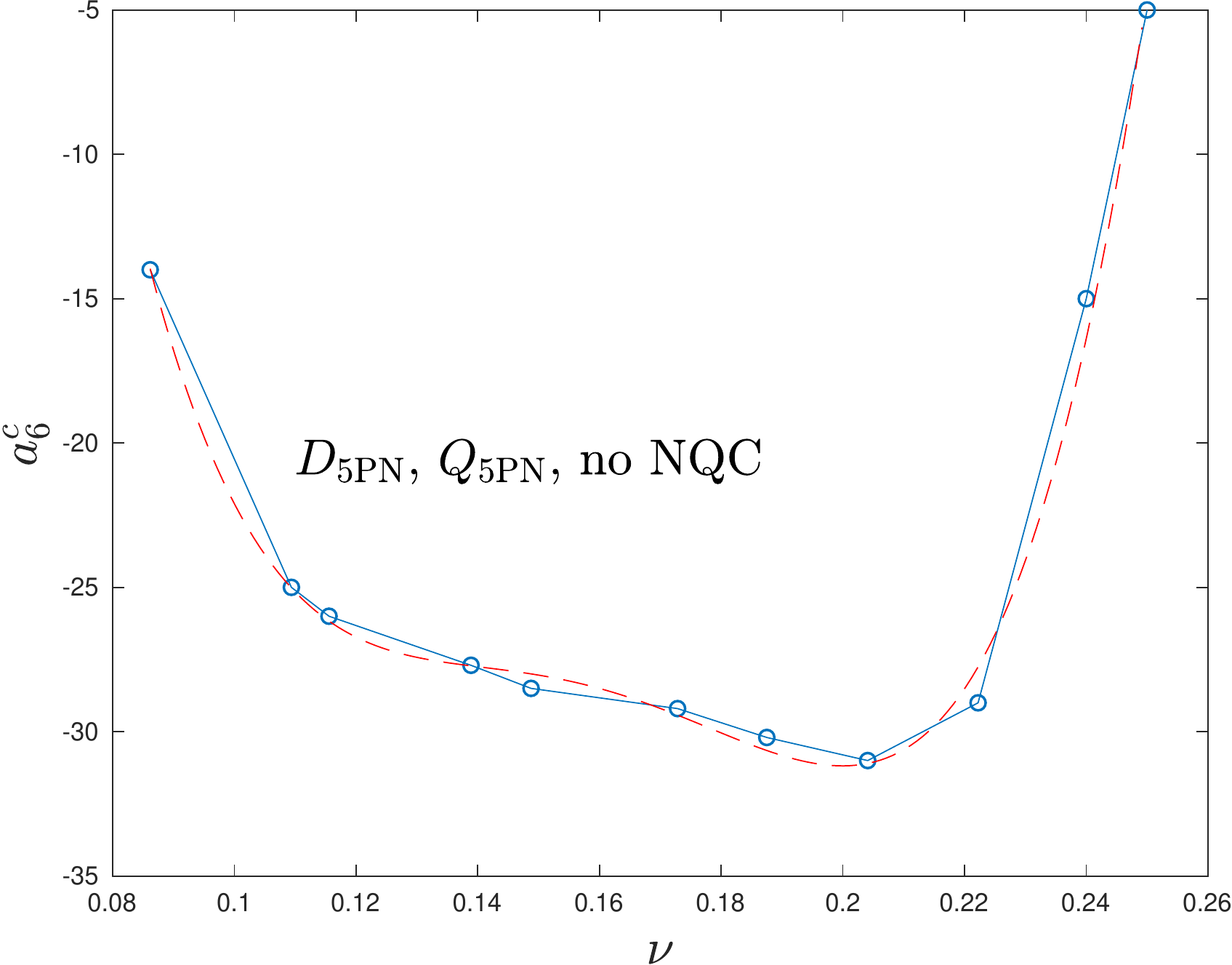}	
	\includegraphics[width=0.31\textwidth]{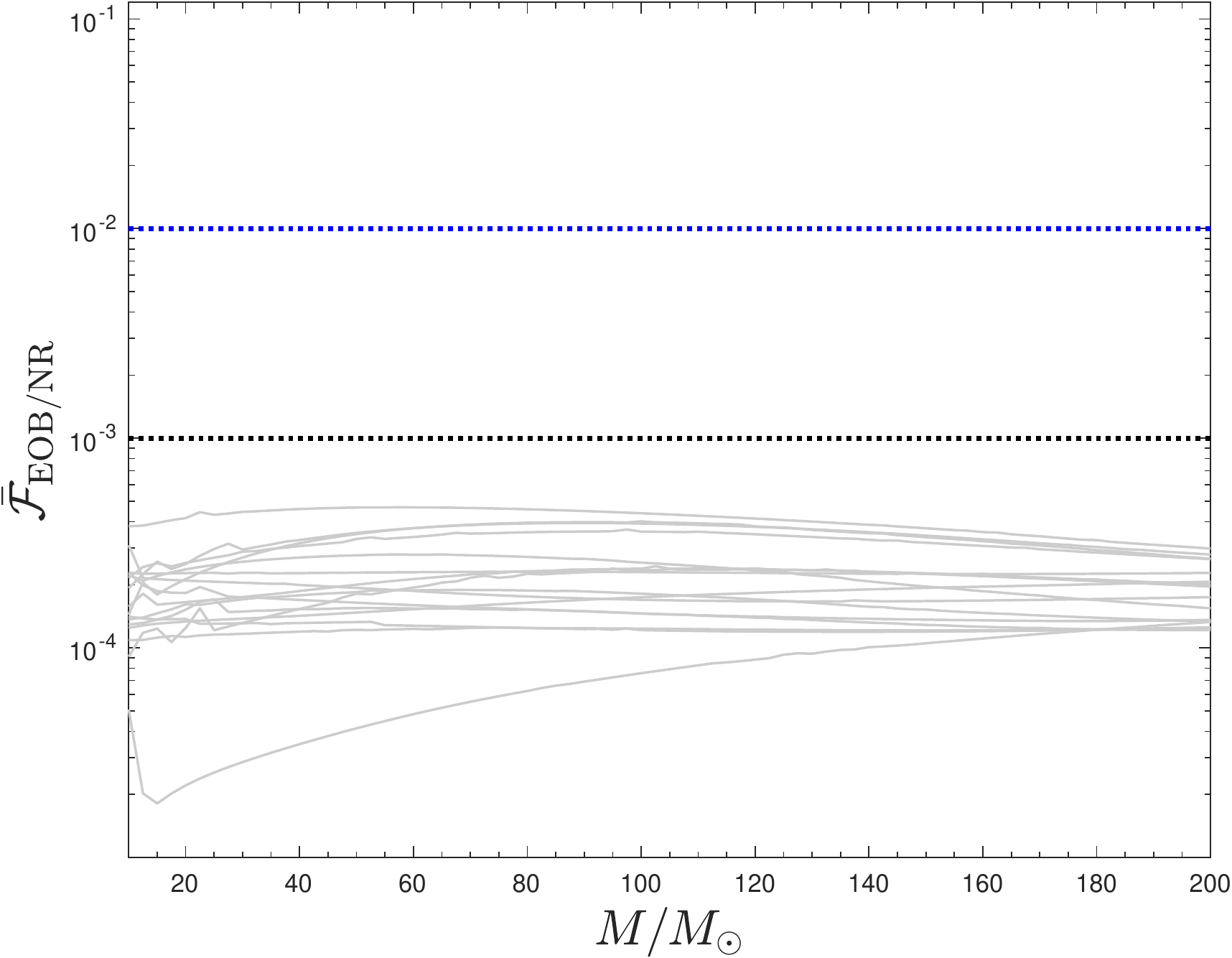}	
	\includegraphics[width=0.31\textwidth]{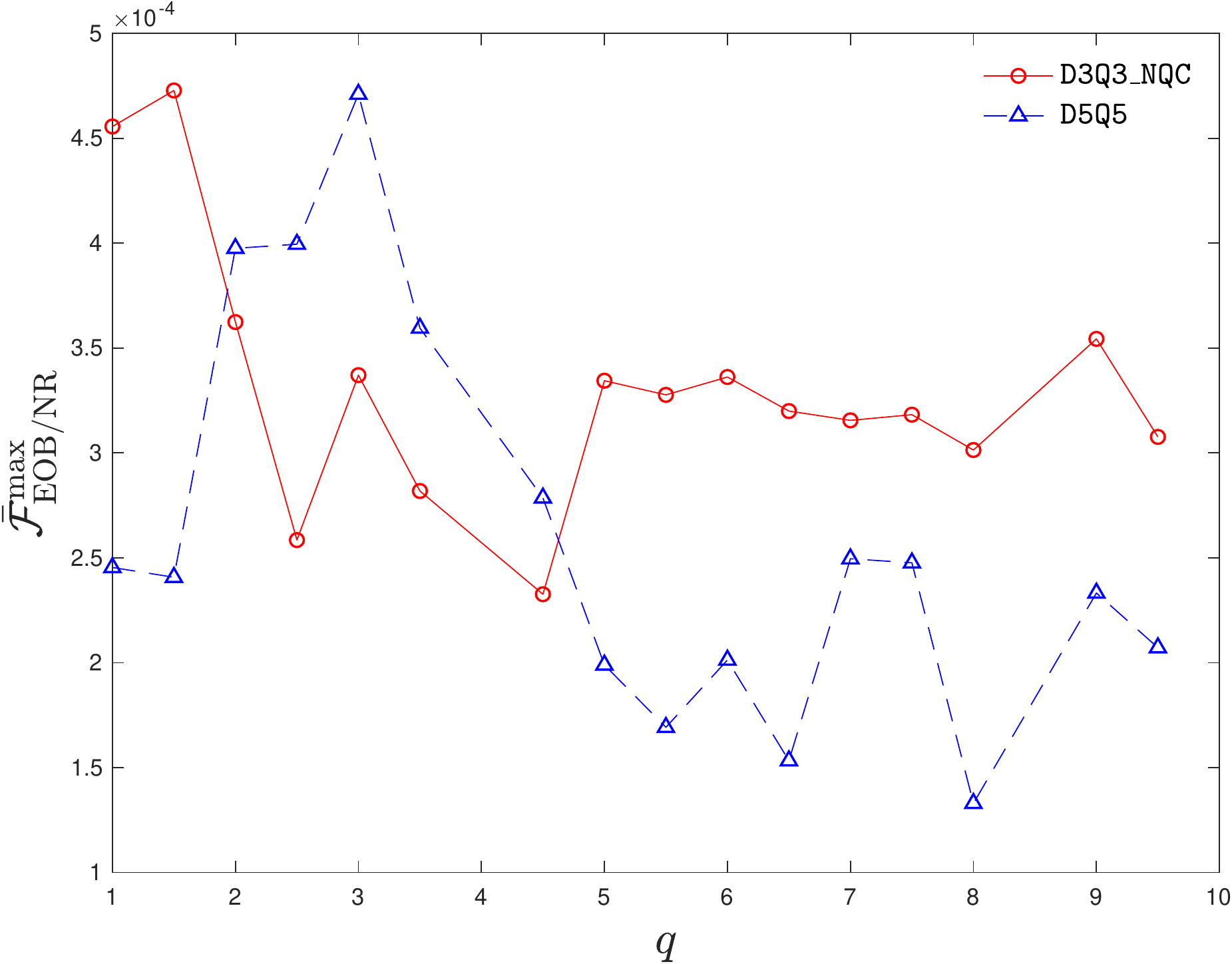} 
	\vspace{5mm}
	\includegraphics[width=0.31\textwidth]{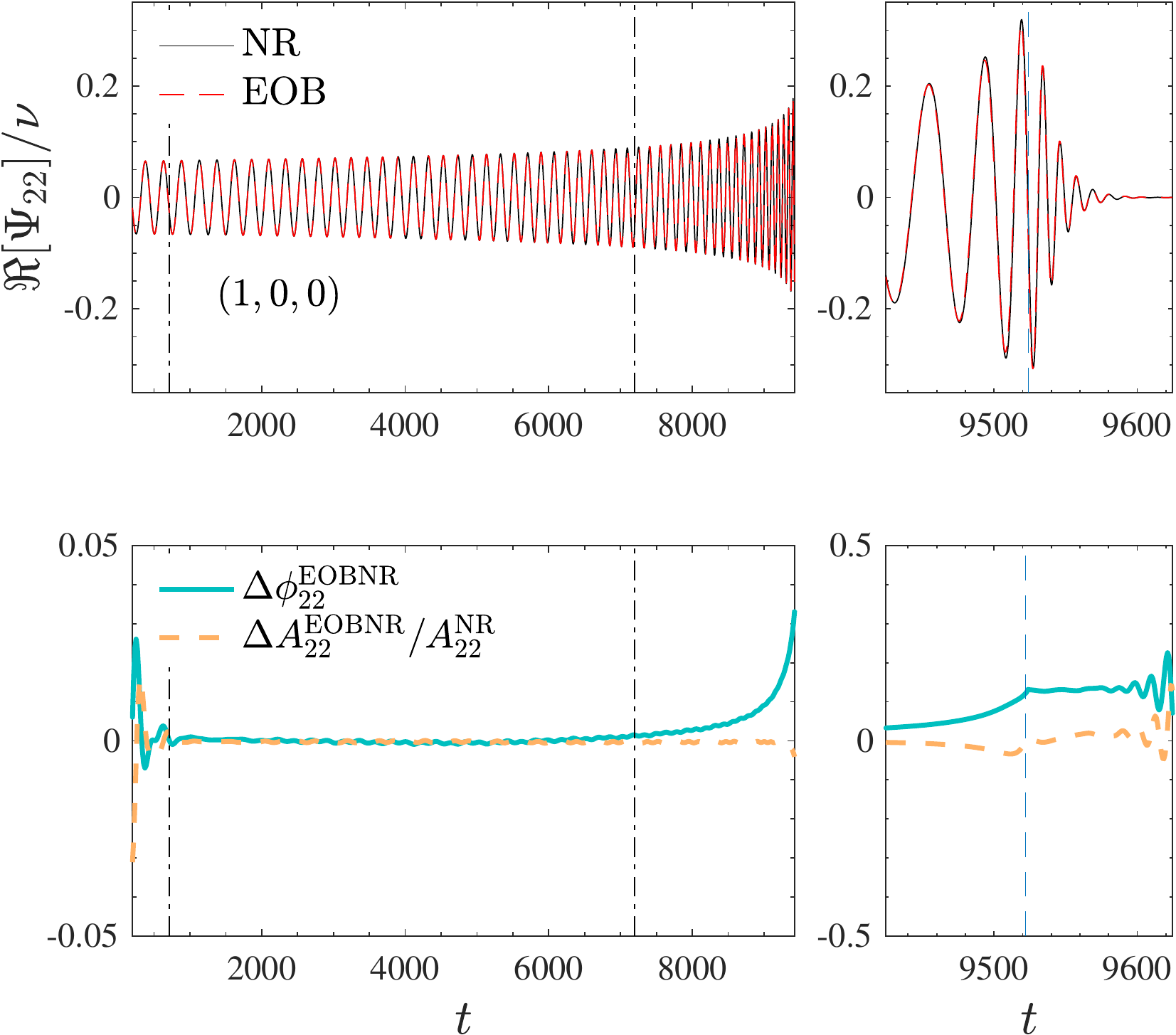}	
	\includegraphics[width=0.31\textwidth]{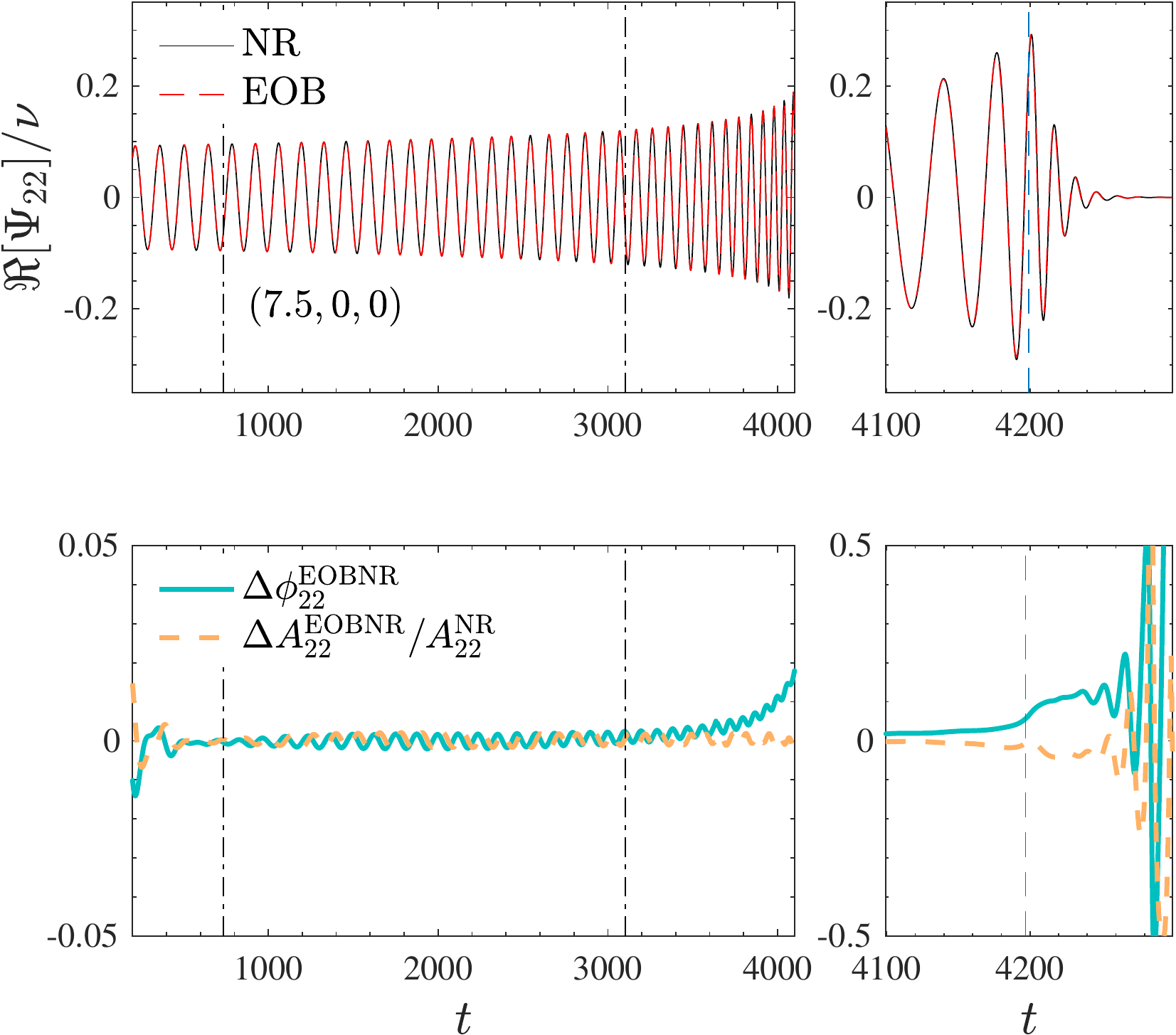}	
	\includegraphics[width=0.31\textwidth]{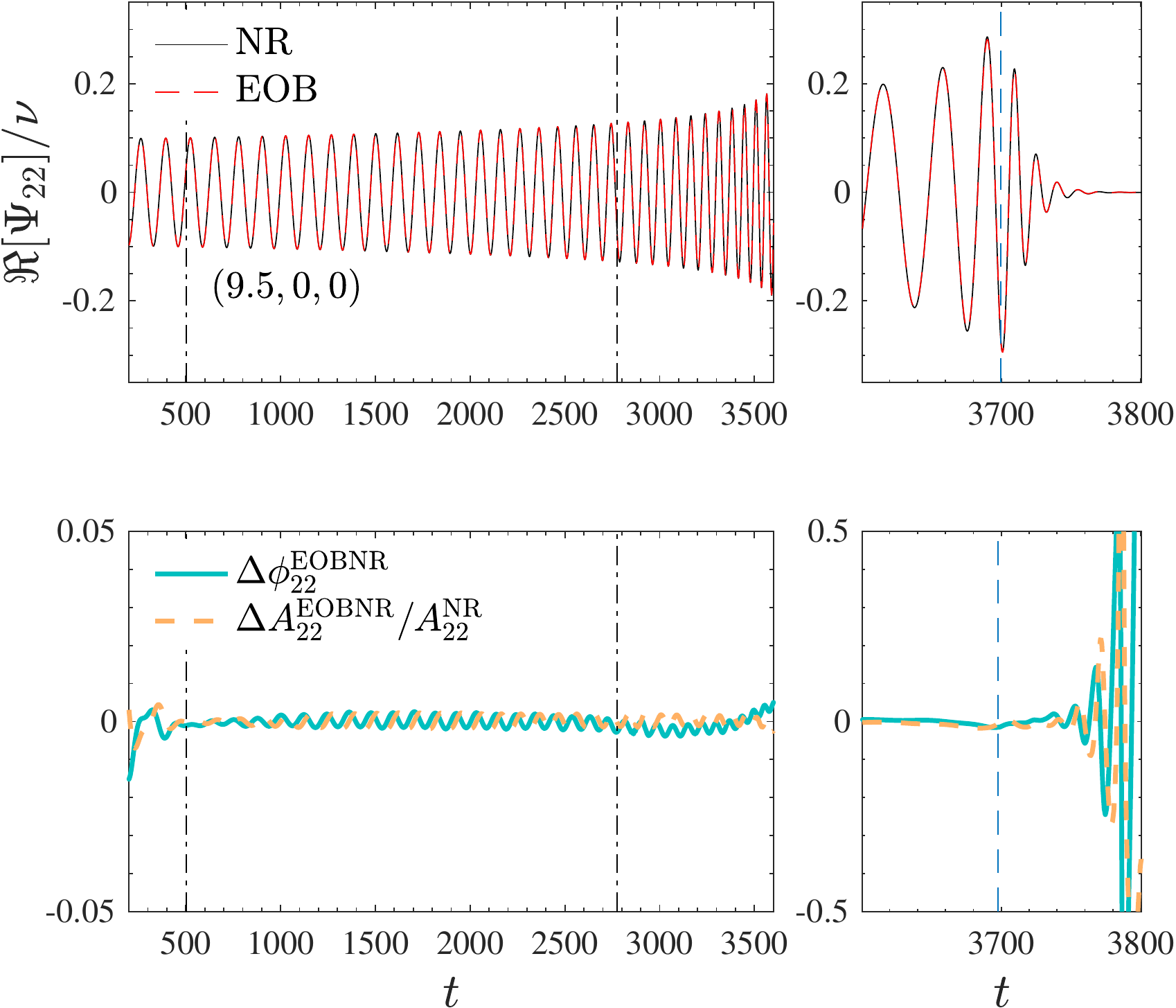}
	\caption{\label{fig:model_noNQC}{\tt D5Q5}: performance of the model with $(D_{\rm 5PN},Q_{\rm 5PN})$ and
	no iterations on the NQC amplitude parameters $(a_1,a_2)$, see Table~\ref{tab:models}. Top row: the left panel shows 
	the NR-informed $a_6^c$, the middle $\bar{\cal F}_{\rm EOB/NR}(M)$, and right one compares the corresponding 
	$\bar{\cal F}_{\rm EOB/NR}^{\rm max}$ with that of {\tt D3Q3\_NQC}. A few illustrative time-domain phasings 
	are also reported in the bottom row.}
	\center
\end{figure*}

\subsubsection{{\tt D5Q5}: 5PN and no iteration on NQC parameters}
It is well known that NQC corrections are an essential element of EOB models. 
They were introduced long ago to correctly match the structure of the EOB waveform 
with the numerical one around merger~\cite{Damour:2007xr,Damour:2007yf,Damour:2007vq}. 
In particular, Ref.~\cite{Damour:2009kr} introduced the {\it iteration} on the NQC amplitude 
parameters, in order to accomplish consistency between the waveform and the flux. 
This proved important in several cases so to get a high level of consistency between 
analytical and numerical data (see e.g.~Ref.~\cite{Albertini:2021tbt} and references therein).
However, as we will better understand at the end of this section, the iteration on $(a_1,a_2)$
introduces an additional coupling between the conservative (i.e. the Hamiltonian) and the 
nonconservative (i.e. the radiation reaction) parts of the dynamics, so that the actual effect
of each part is somehow hidden by this nonlinear, though extremely effective, procedure.
To comply with the guiding philosophy of this paper, i.e. understanding the effect of each
single theoretical element of the model, let us now {\it drop} the NQC iteration in order
to get a clear understanding of the strong-field action of $(D,Q)$.
To start with, we focus on an equal-mass configuration and fix $a_6^c=0$ for simplicity.
In Fig.~\ref{fig:noNQC_PN_impact} we compare the EOB/NR phase difference 
$\Delta\phi^{\rm EOBNR}_{22}\equiv \phi^{\rm EOB}_{22}-\phi^{\rm NR}_{22}$
for three different choices of $(D,Q)$: (i) the standard one of \TEOBResumS{} $(D_{\rm 3PN},Q_{\rm 3PN})$; 
(ii) the case $(D_{\rm 5PN},Q_{\rm 3PN})$ and (iii) $(D_{\rm 5PN},Q_{\rm 5PN})$.
In terms of the dynamics, the figure shows that the plunge driven by the 3PN functions
is {\it faster} than the NR one, so that a nonnegligible, {\it positive}, phase difference 
builds up to merger (vertical dashed line in the right panel of Fig.~\ref{fig:noNQC_PN_impact}).
By contrast, the $D_{\rm 5PN}$ function, resummed with $(3,2)$ Pad\'e approximant,
acts in the opposite direction, i.e. by {\it delaying the plunge} with respect to NR, so
that $\Delta\phi^{\rm EOBNR}_{22}\sim -0.2$ rad at merger. When we consider also 
5PN (local) terms in $Q$, the phase difference results almost flat throughout merger 
and ringdown.
This finding seems to indicate that the iterative procedure for obtaining self-consistent 
NQC parameters (either in the waveform and in the flux), as introduced long 
ago~\cite{Damour:2009kr}, conflicts with high-order PN corrections in the $D$ and $Q$ 
potentials. To gain a deeper understanding on what is going on for {\tt D5Q5\_NQC},
in Fig.~\ref{fig:NQC_impact} we switch on the NQC iteration keeping $a_6^c=0$ first
for consistency with Fig.~\ref{fig:noNQC_PN_impact}. One sees immediately that the
NQC iteration is equivalent to a large repulsive effect in the dynamics, with the EOB
merger delayed with respect to the NR one (vertical dashed line in the right panel of
the figure), so that $\Delta\phi_{22}^{\rm EOBNR}$ is large and {\it negative}. 
The tuning of $a_6^c$ allows one to partly {\it compensate} this effect, as the cases
$a_6^c=-40$ and $a_6^c=-98$ in Fig.~\ref{fig:noNQC_PN_impact} illustrate. 
However, the price to pay to have an acceptable phasing at merger
(such to yield $\bar{\cal F}^{\rm max}_{\rm EOB/NR}\lesssim 10^{-3}$) is a progressive
worsening of the phasing during the late inspiral, as already pointed out in the above 
discussion on {\tt D5Q5\_NQC}.  
This suggests that, at least within the current analytic context, we are actually 
{\it overfitting} $a_6^c$ (that belongs to the quasi-circular regime) 
to fix some missing physical effects that belong to the genuine noncircular regime. 
The drawback of this overfitting is the loss of performance during the late inspiral.
To overcome this difficulty, one could attempt to tune the noncircular potentials to further improve 
the behavior during plunge and merger {\it without} worsening the inspiral.
The tuning on the noncircular dynamics could be implemented, for example, by
tuning the currently unknown\footnote{See however Ref.~\cite{Blumlein:2021txe} for
a recent complete calculation of the 5PN dynamics.} residual 5PN coefficient 
$d_5^{\nu^2}$ in $D$. A successful attempt in this direction is shown in 
Fig.~\ref{fig:D_tuning}. In the top-panel we tune iteratively $a_6^c$ with the aim of reducing
the phase difference at merger as much as possible keeping $\Delta\phi_{22}^{\rm EOBNR}$
negative and monotonically decreasing. The bottom panel shows that a rather large
value of $d_5^{\nu^2}$ can actually reduce $\Delta\phi_{22}^{\rm EOBNR}$ 
during late plunge, merger and ringdown. Note that the phasing here, though still improvable,
is substantially comparable to the top-left panel of Fig.~\ref{fig:raw_phasing}.
This exercise thus shows that {\it even} working with $(D_{\rm 5PN},Q_{\rm 5PN})$ it is
possible to use NQC corrections in self-consistent way and obtain a highly accurate
model, but one should NR-inform both the circular and noncircular part of the dynamics.

Due to the level of additional complication of tuning two dynamical parameters at the 
same time, and given the already good performance of the model without NQC
iterations for $q=1$ and $a_6^c=0$ seen above, it is then interesting to investigate the 
performance attainable using $(D_{\rm 5PN},Q_{\rm 5PN})$ for nonspinning binaries {\it without} 
iterating on the NQC parameters $(a_1,a_2)$. Now and below we will refer to this model as {\tt D5Q5}.
We thus perform a new determination of $a_6^c$ versus $\nu$ using the usual procedure,
as shown in the top left panel Fig.~\ref{fig:model_noNQC}.
The $\nu$-dependence is more complicated than the previous cases, especially because the 
curve becomes rather steep as $\nu\to 0.25$. This entails that one would need more NR simulations 
in the range $0.2\lesssim \nu \lesssim 0.25$ to correctly determine the shape of the curve.
For consistency with our previous result, we add a few more datasets, but not many, and we
limit ourselves to those where the initial values of the dimensionless spins are below $10^{-4}$
to avoid contamination from the small spins. 
An accurate representation of the points is given by a $(4,1)$ rational function in $\nu$, with
the coefficients that can be read off Table~\ref{tab:models}. 
This gives a representation of the current data that is good enough, although certainly suboptimal to
what attainable using more NR simulations. The EOB/NR unfaithfulness for the usual nonspinning 
configurations is shown in the right panel of Fig.~\ref{fig:model_noNQC}. At first sight, the performance 
is globally compatible to the standard result of Fig.~\ref{fig:new_barF}, so that it seems that no special
gain is found. However, looking at $\bar{\cal F}^{\rm max}_{\rm EOB/NR}$ (top right panel of Fig.~\ref{fig:model_noNQC})
one discovers that {\tt D5Q5} performs {\it better} than {\tt D3Q3\_NQC} either for small or large values of $q$.
This is further highlighted by the plots in the bottom row of Fig.~\ref{fig:model_noNQC}, where the phasing
improvement for $q=9.5$ with respect to {\tt D3Q3\_NQC} is evident.

\subsubsection{{\tt D3Q3} and {\tt D5Q3}: varying D with no NQC}
\label{sec:3PN_nonqc}
\begin{figure}[t]
	\center
	\includegraphics[width=0.22\textwidth]{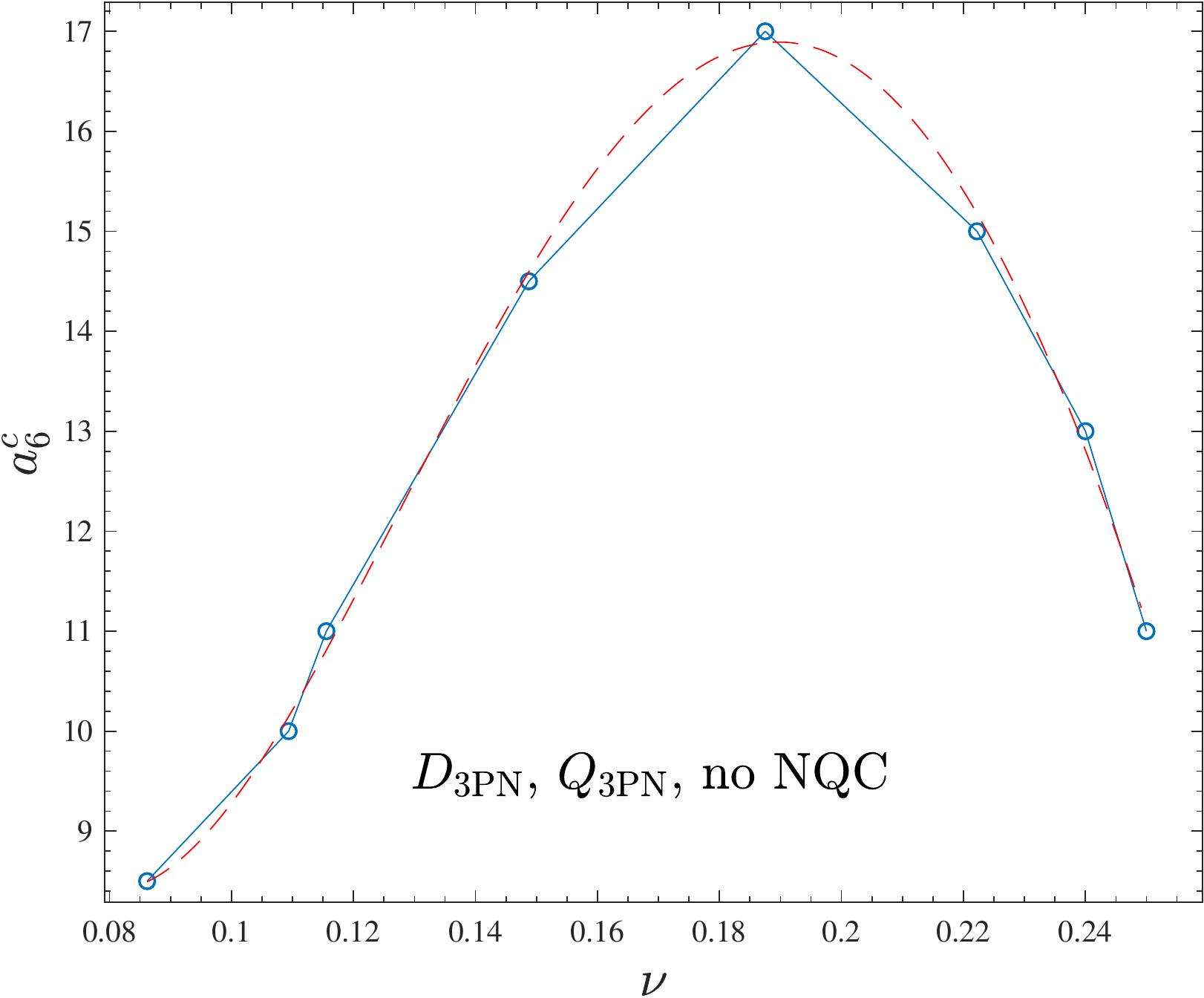}
	\includegraphics[width=0.22\textwidth]{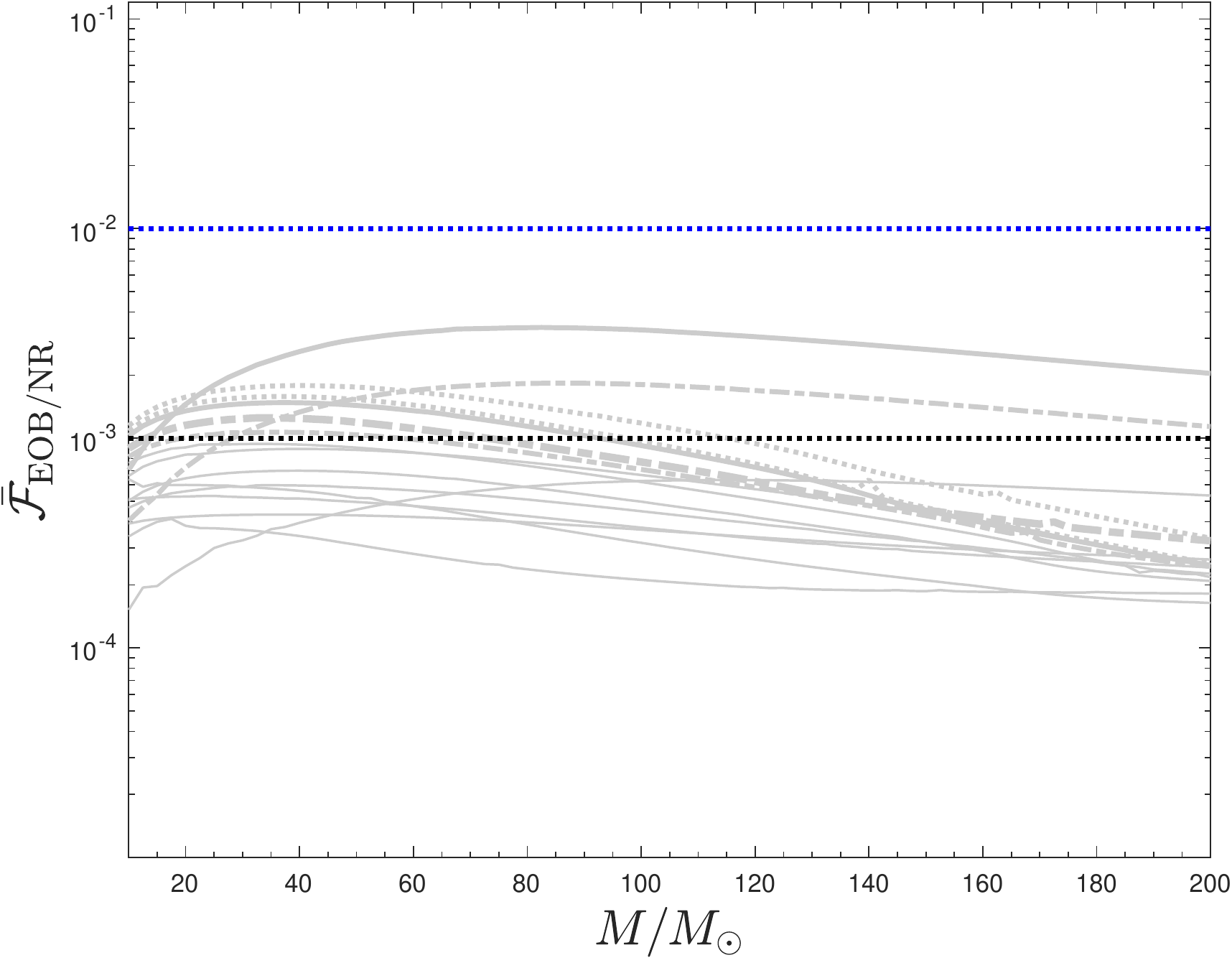}    \\   
          \includegraphics[width=0.22\textwidth]{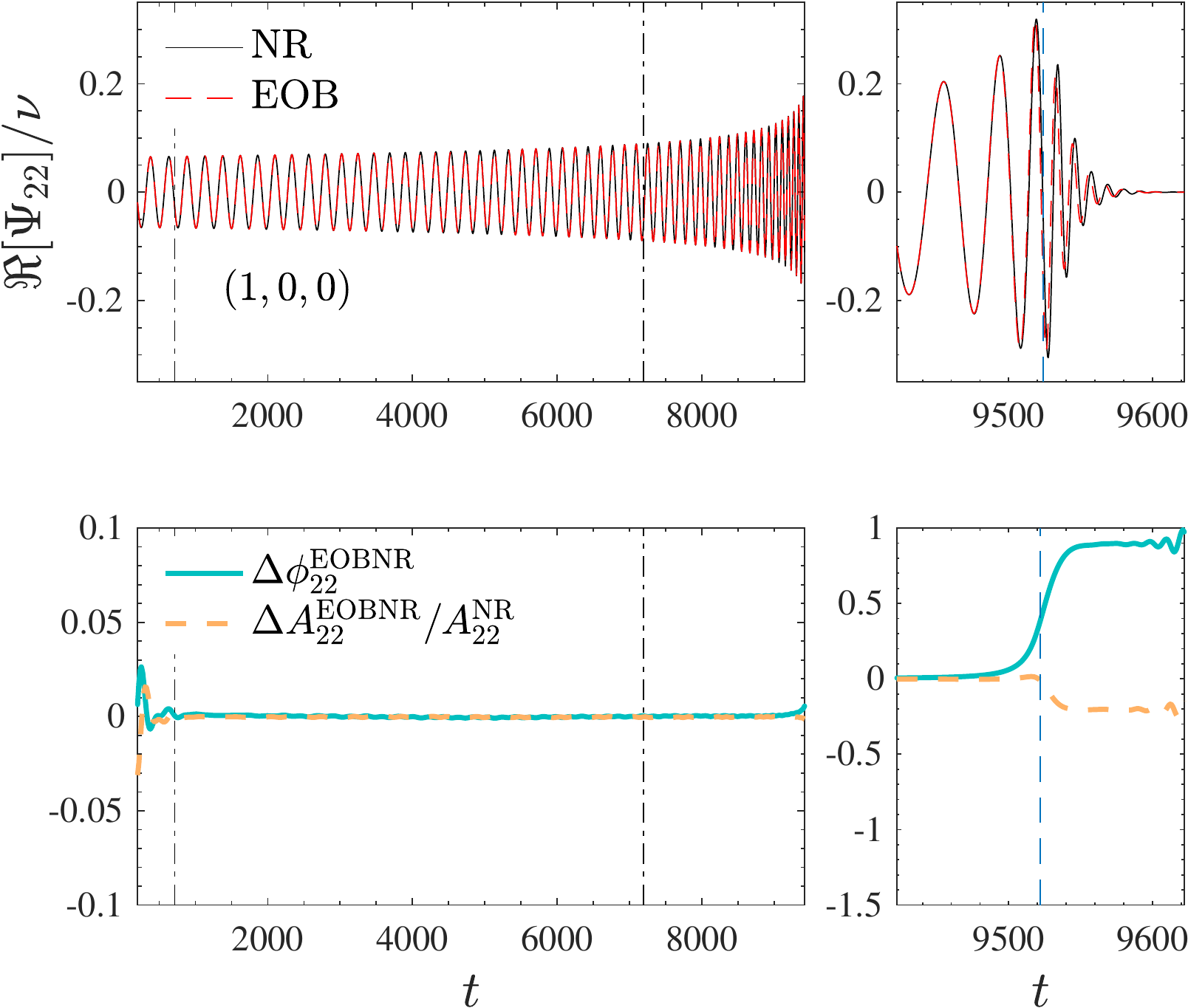}
          \includegraphics[width=0.22\textwidth]{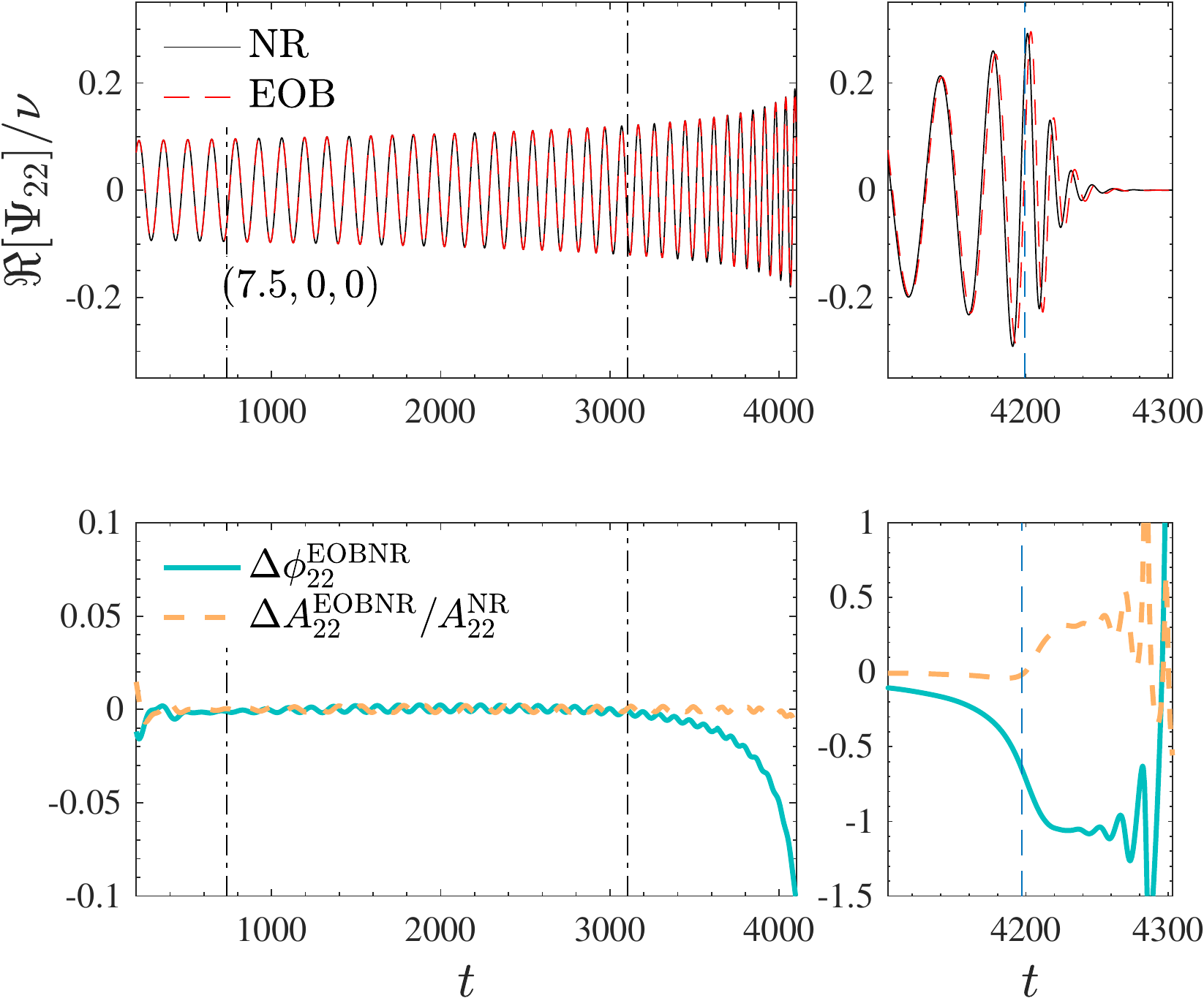}
         	\caption{\label{fig:3PN_nonqc}{\tt D3Q3}: performance of the model with $(D_{\rm 3PN},Q_{\rm 3PN})$ 
	           and no NQC iteration. From left to right, top bottom: the behavior of $a_6^c(\nu)$,
	           $\bar{\cal F}_{\rm EOB/NR}$ and two illustrative time-domain phasings. Note that
	           the phase difference is positive and is accumulated only during the last orbit.
	           Consistently with Fig.~\ref{fig:NQC_impact}, and the effective repulsive effect due to the NQC iteration
	           that here is missing, the EOB waveform is shorter than the NR one, with the plunge occurring earlier on.}
\end{figure}
To complete the findings of the previous section, let us also explore the performance of \TEOBResumS{} 
with $Q$ fixed at 3PN order, {\it without} the iteration on NQC parameters and varying $D$.
To start, we fix $(D_{\rm 3PN},Q_{\rm 3PN})$ with $D$ resummed with a $(0,3)$ Pad\'e approximant,
while $A$ is resummed with the $(1,5)$ one and $a_6^c$ is informed in the usual way 
using NR simulations. This model is simply dubbed {\tt D3Q3}, see Table~\ref{tab:models}.
We find that the absence of the NQC correction in the radiation
reaction (and thus the absence of the related repulsive effect discussed above) 
reduces the flexibility and accuracy of the model. As in the main text, 
a good fit of $a_6^c$ is given by a $(4,1)$ rational function (see Table~\ref{tab:models}).
The fitted points and the fitting functions are shown in the top left panel of  
Fig.~\ref{fig:3PN_nonqc}. The corresponding EOB/NR unfaithfulness, with
the usual nonspinning datasets used above, on the Advanced LIGO PSD 
is shown in the top-right panel of Fig.~\ref{fig:3PN_nonqc}, while the 
bottom panels show two time-domain phasings for two illustrative
configurations. Interestingly, the largest EOB/NR phase
differences (up to $\sim1$~rad) build up only during the last part of the plunge up to merger.
It is useful to compare these plots with the corresponding panels in Fig.~\ref{fig:raw_phasing},
that are obtained using the iterated NQC corrections in the flux. The phasing plots indicate 
how the NQC flux corrections, with their intrinsic repulsive character in strong field, are crucial to improve 
the dynamics {\it especially} through the late plunge up to merger, thus playing an essential role 
to achieve EOB/NR values of $\bar{\cal F}_{\rm EOB/NR}$ {\it below} $10^{-3}$ found for ${\tt D3Q3\_NQC}$.
In addition, thanks to the NQC iteration, the NR-informed $a_6^c$, that is a priori expected to
only affect the circular part of the dynamics, in practice propagates its action {\it also} on the 
noncircular part of the dynamics, allowing for a fine tuning of the plunge-merger dynamics that 
is otherwise impossible.
\begin{figure}[t]
	\center
	\includegraphics[width=0.22\textwidth]{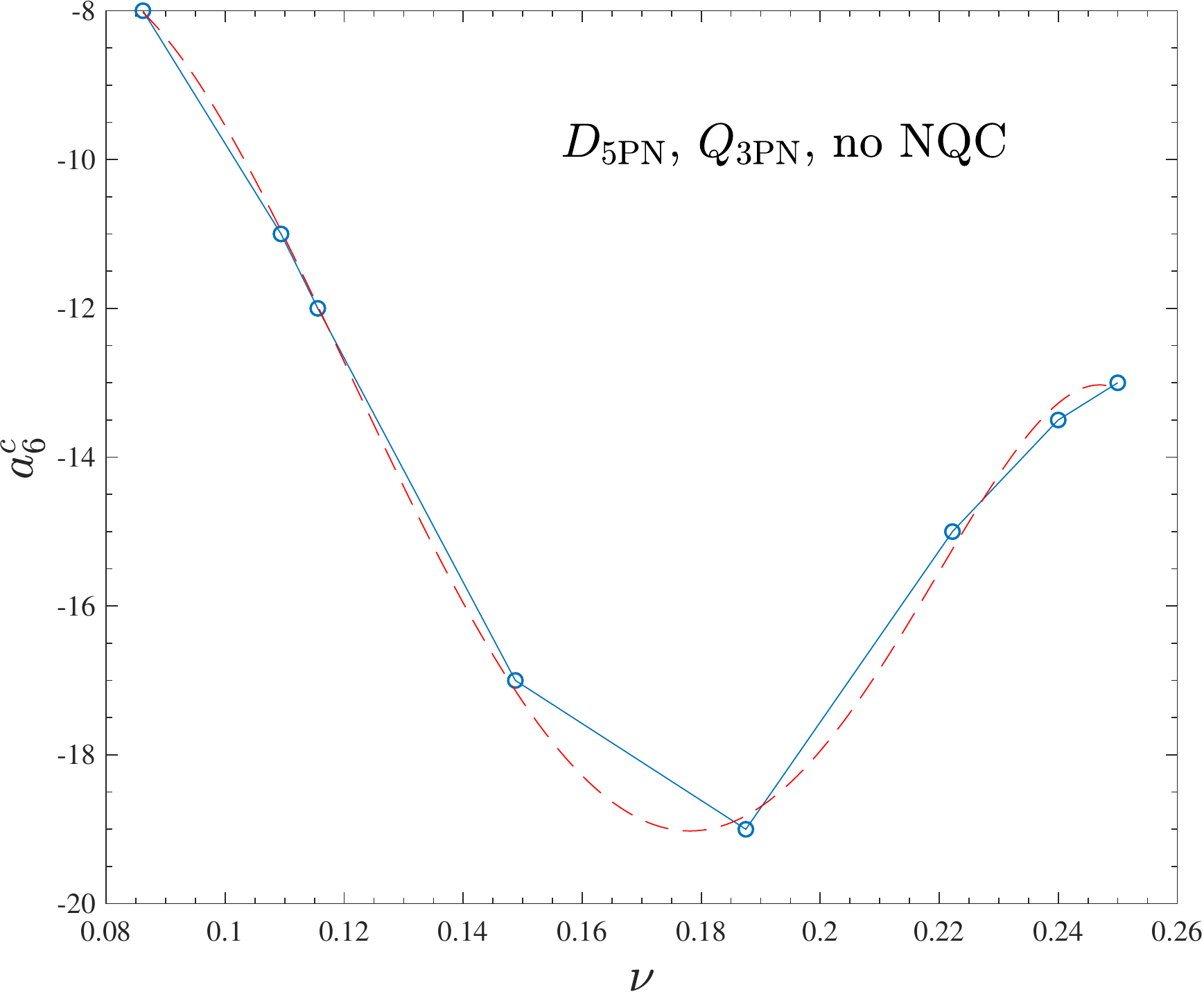}
	\includegraphics[width=0.22\textwidth]{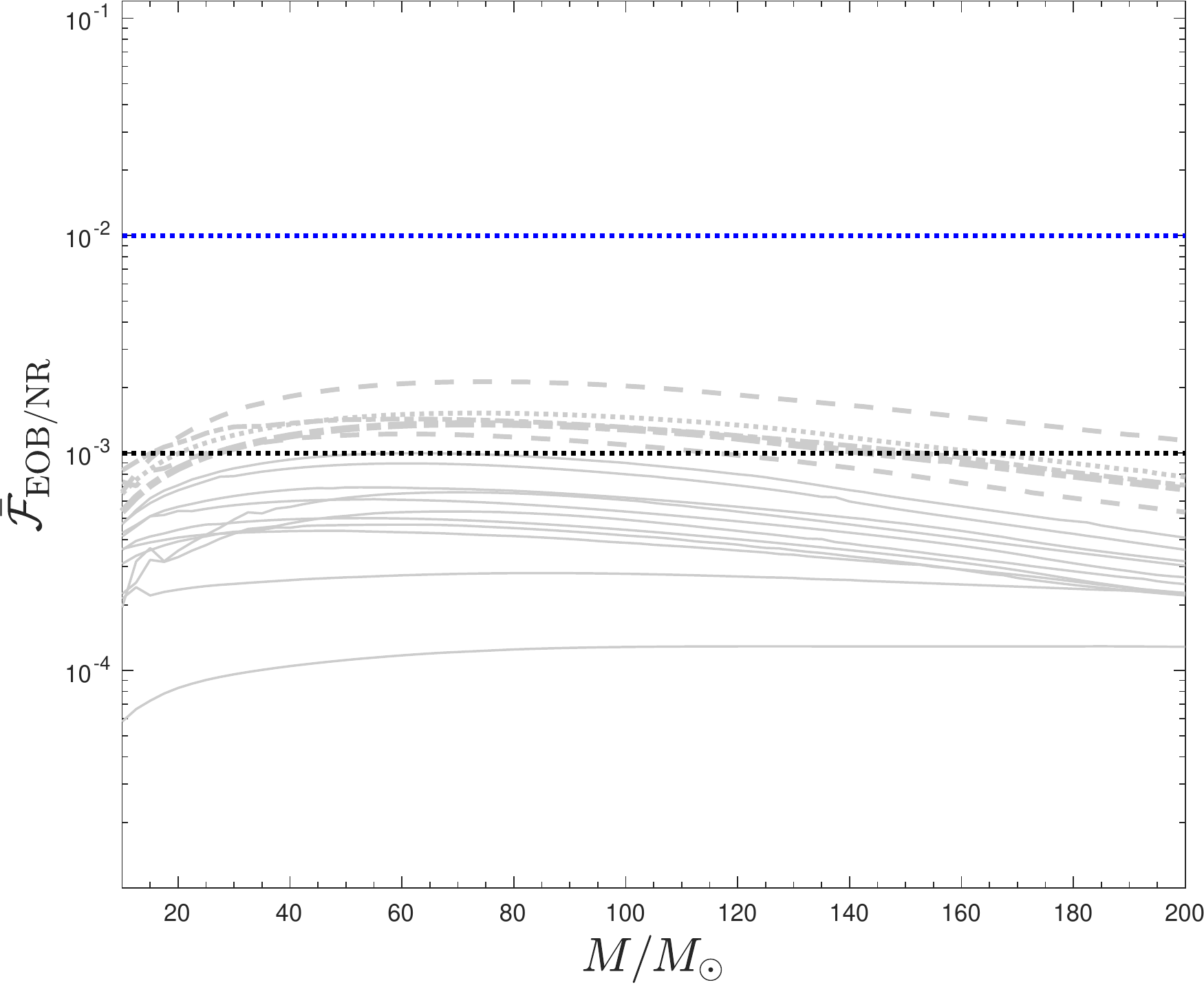}    \\   
          \includegraphics[width=0.22\textwidth]{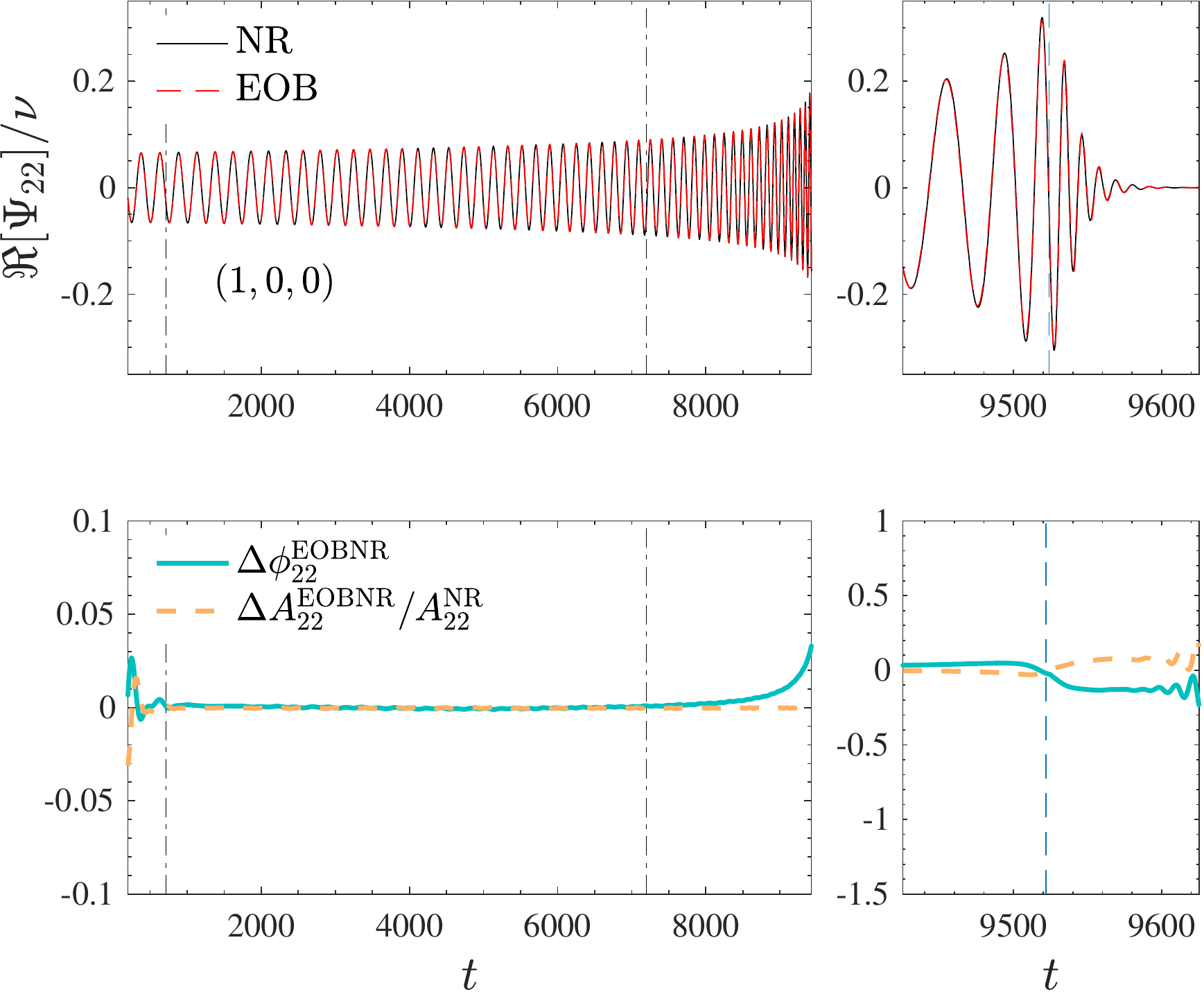}
          \includegraphics[width=0.22\textwidth]{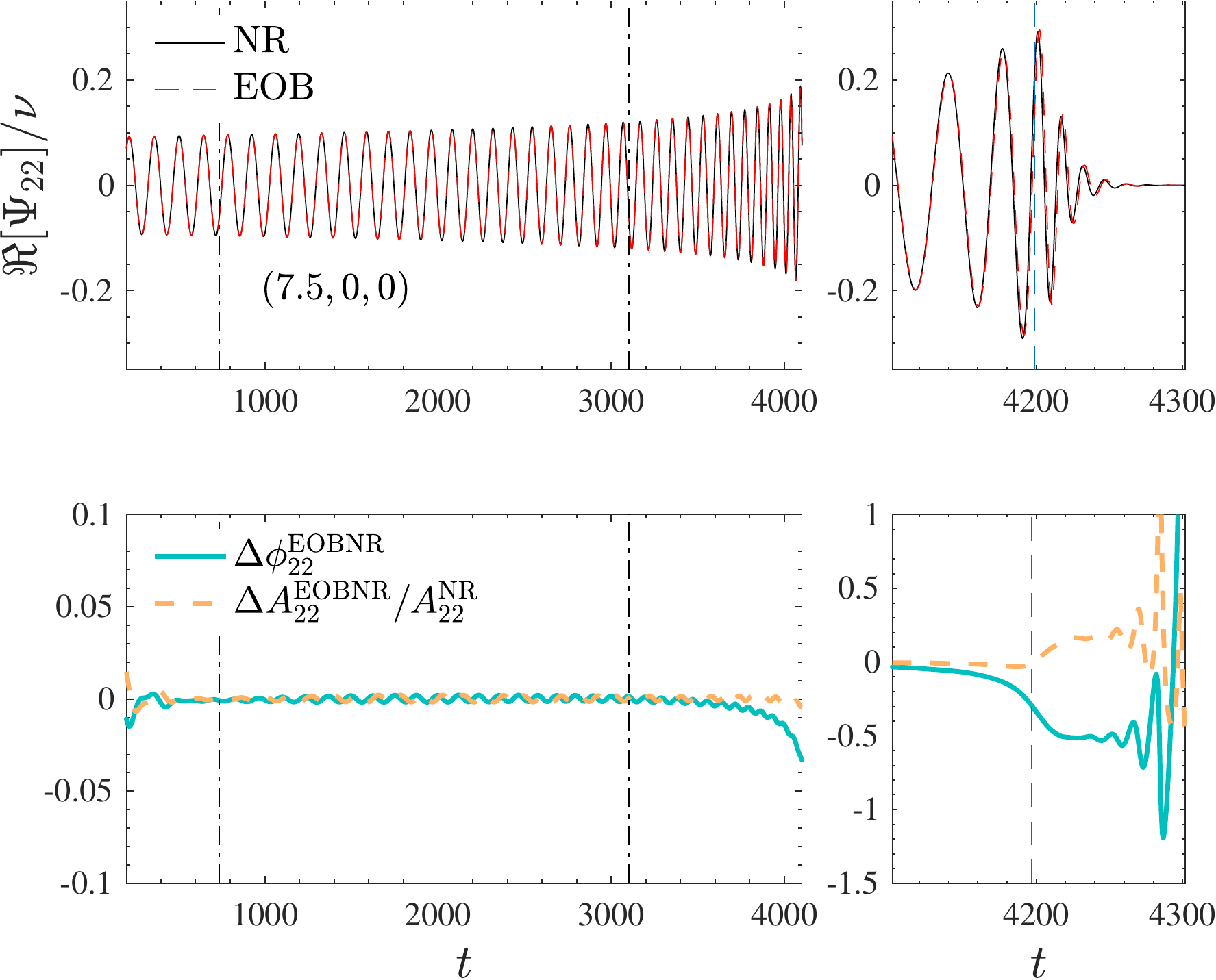}
         	\caption{\label{fig:5PN3PN_nonqc}{\tt D5Q3}: model with $(D_{\rm 5PN},Q_{\rm 3PN})$ 
	           and no NQC iteration. From left to right, top bottom: the behavior of $a_6^c(\nu)$,
	           $\bar{\cal F}_{\rm EOB/NR}$ and two illustrative time-domain phasings. Note 
	           the improved performance with respect to the {\tt D3Q3} model shown in 
	           Fig.~\ref{fig:3PN_nonqc}.}
\end{figure}

To further support the importance of a the description of the radial part of the dynamics,
Fig.~\ref{fig:5PN3PN_nonqc} shows the performance of the model {\tt D5Q3}, obtained with 
$(D_{\rm 5PN},Q_{\rm 3PN})$. By comparing Fig.~\ref{fig:3PN_nonqc} and 
Fig.~\ref{fig:5PN3PN_nonqc}, it is apparent the improvement brought by moving from 
$D_{\rm 3PN}$ to $D_{\rm 5PN}$, although the global performance is also influenced 
by the new determination of $a_6^c$.

\section{Eccentricity and scattering}
\label{sec:nc}
\begin{figure}[t]
	\center	
	\includegraphics[width=0.22\textwidth]{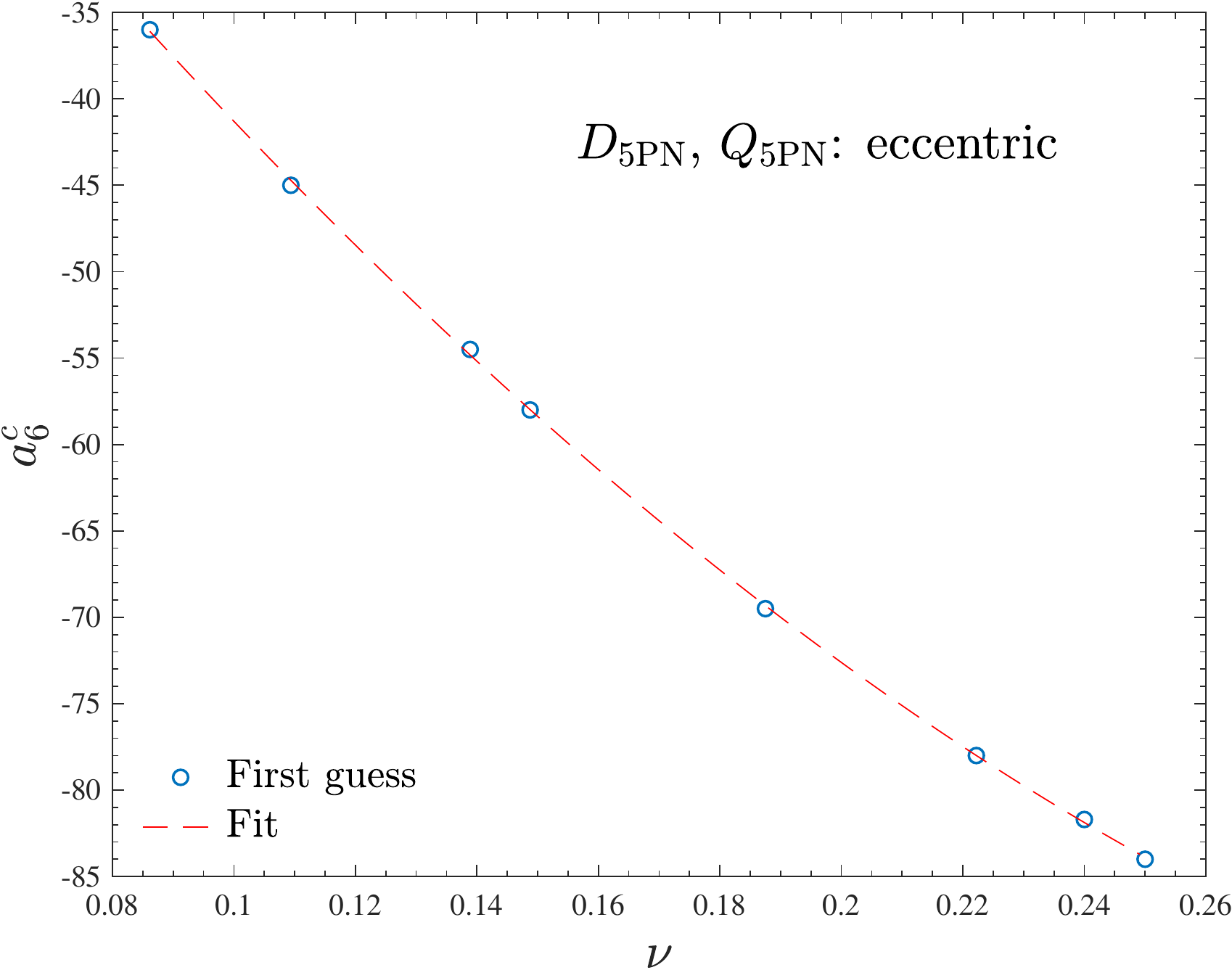} 
         \includegraphics[width=0.22\textwidth]{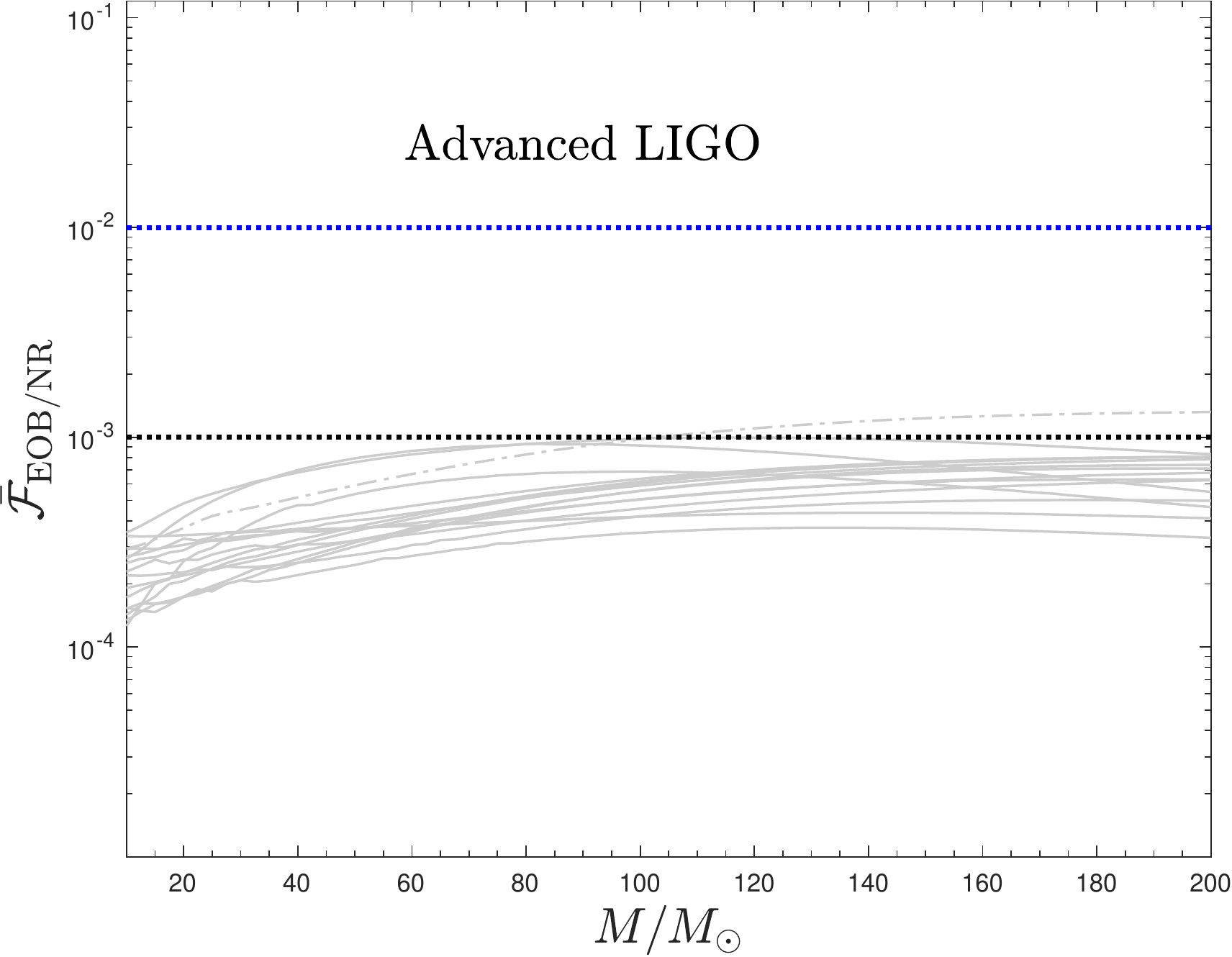}  \\
         \includegraphics[width=0.22\textwidth]{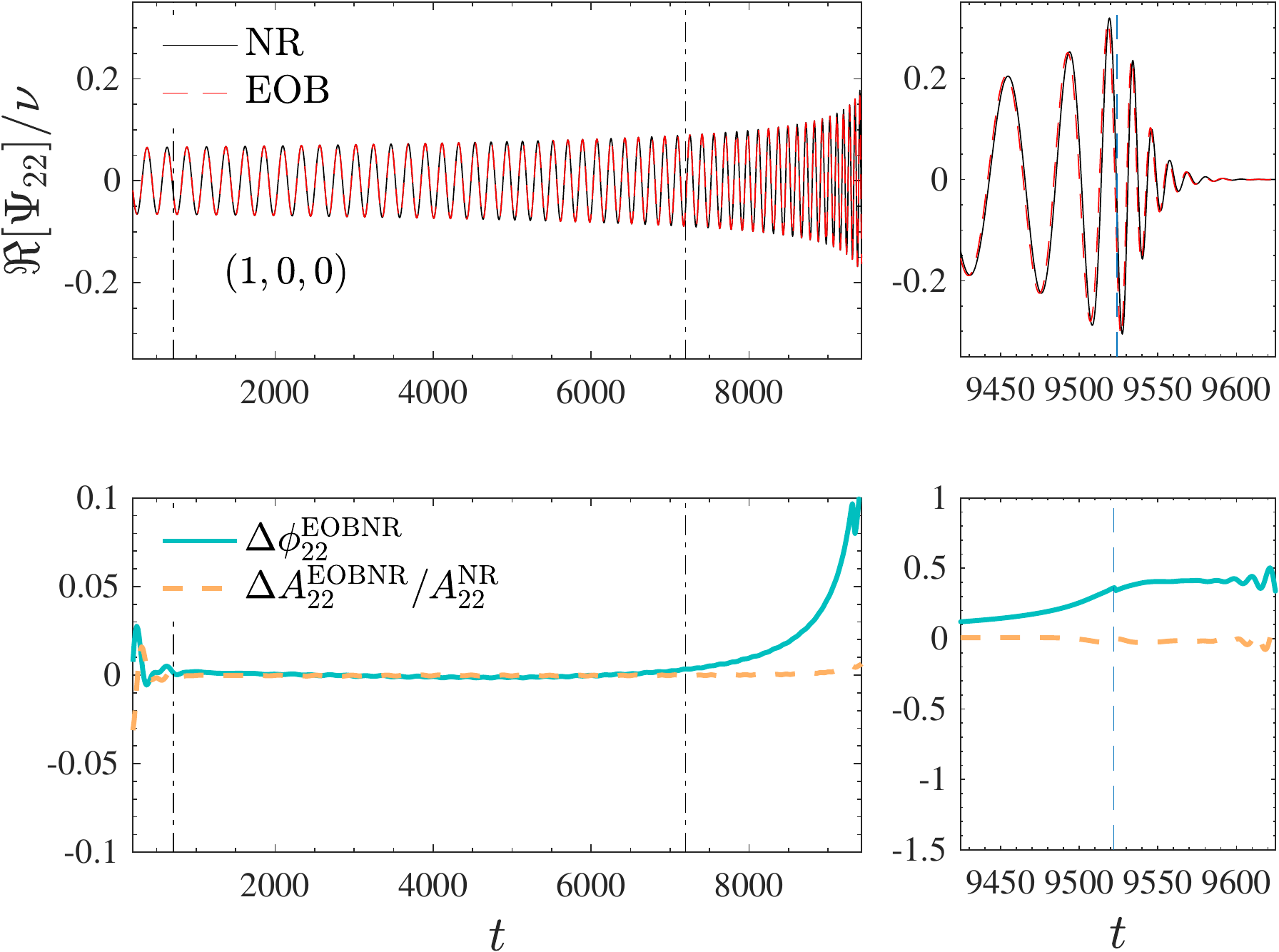} 
         \includegraphics[width=0.22\textwidth]{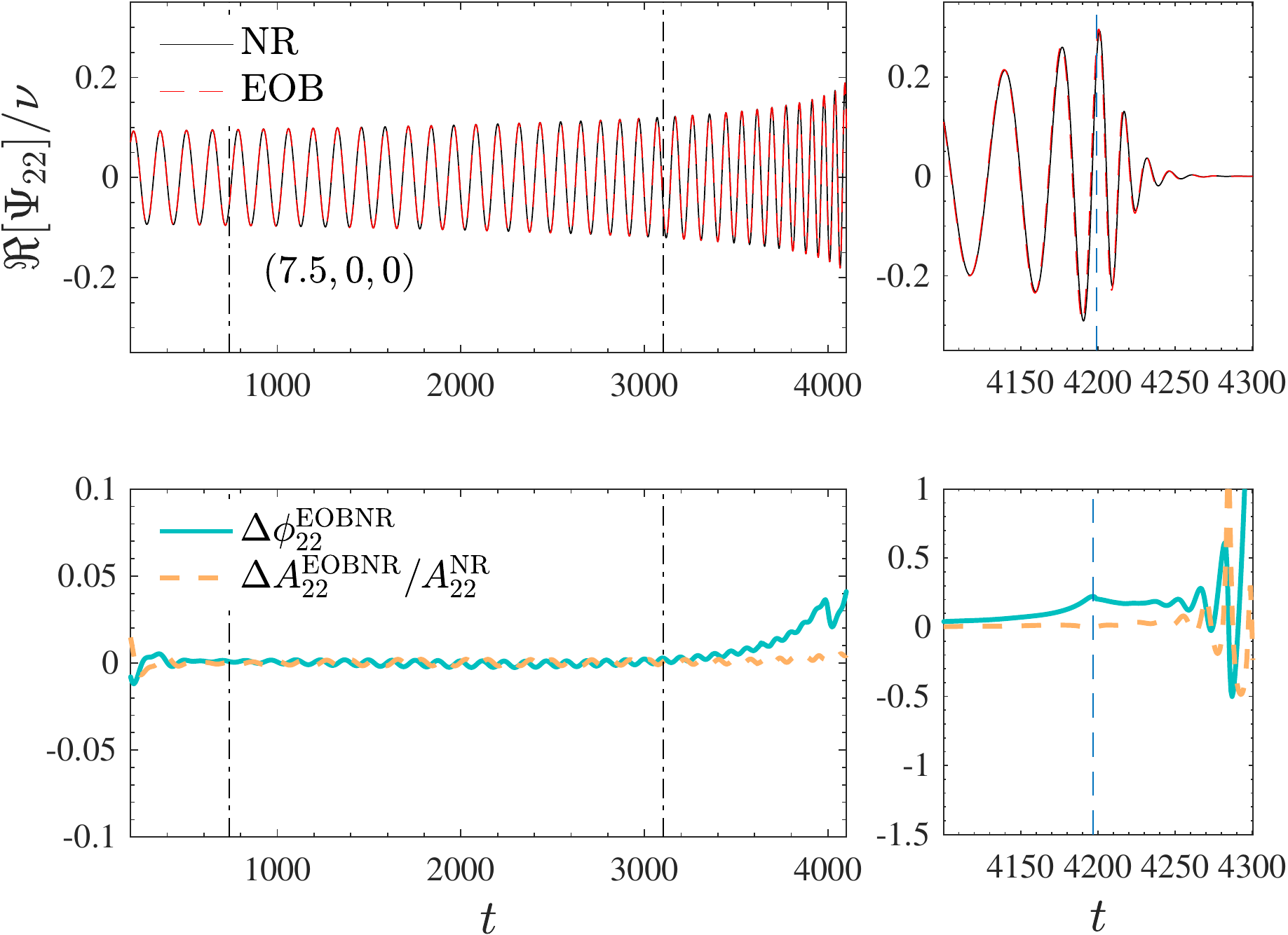}
         	\caption{\label{fig:Fecc}Eccentricity: quasi-circular limit of the eccentric nonspinning model.
	         The EOB/NR performance is comparable to the improved \TEOBResumS{} quasi-circular 
	         model (with NQC iterations), and slightly better than the eccentric model of Ref.~\cite{Nagar:2021xnh}
	         for $q\approx 1$.}
\end{figure}

\begin{figure}[t]
	\center	
         \includegraphics[width=0.45\textwidth]{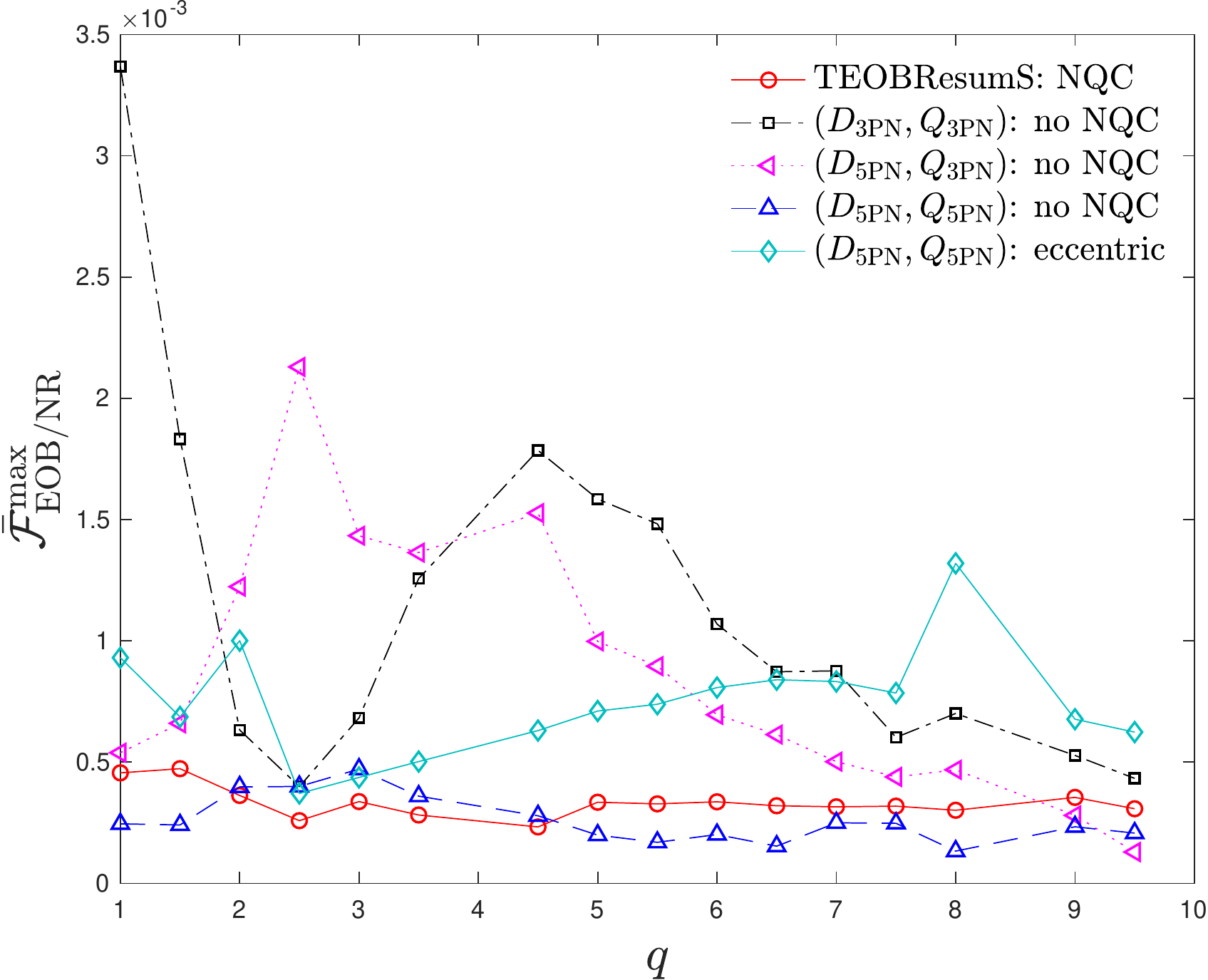}  
         	\caption{\label{fig:maxF_vs_PN} The values of $\bar{\cal F }_{\rm EOB/NR}^{\rm max}$ 
	         for the quasi-circular limit of the generic, noncircular (eccentric) model discussed in this
		 section compared with the corresponding values obtained with quasi-circular models 
		 discussed in Sec.~\ref{sec:systematics}.}
\end{figure}
\begin{figure}[t]
	\center	
	\includegraphics[width=0.23\textwidth]{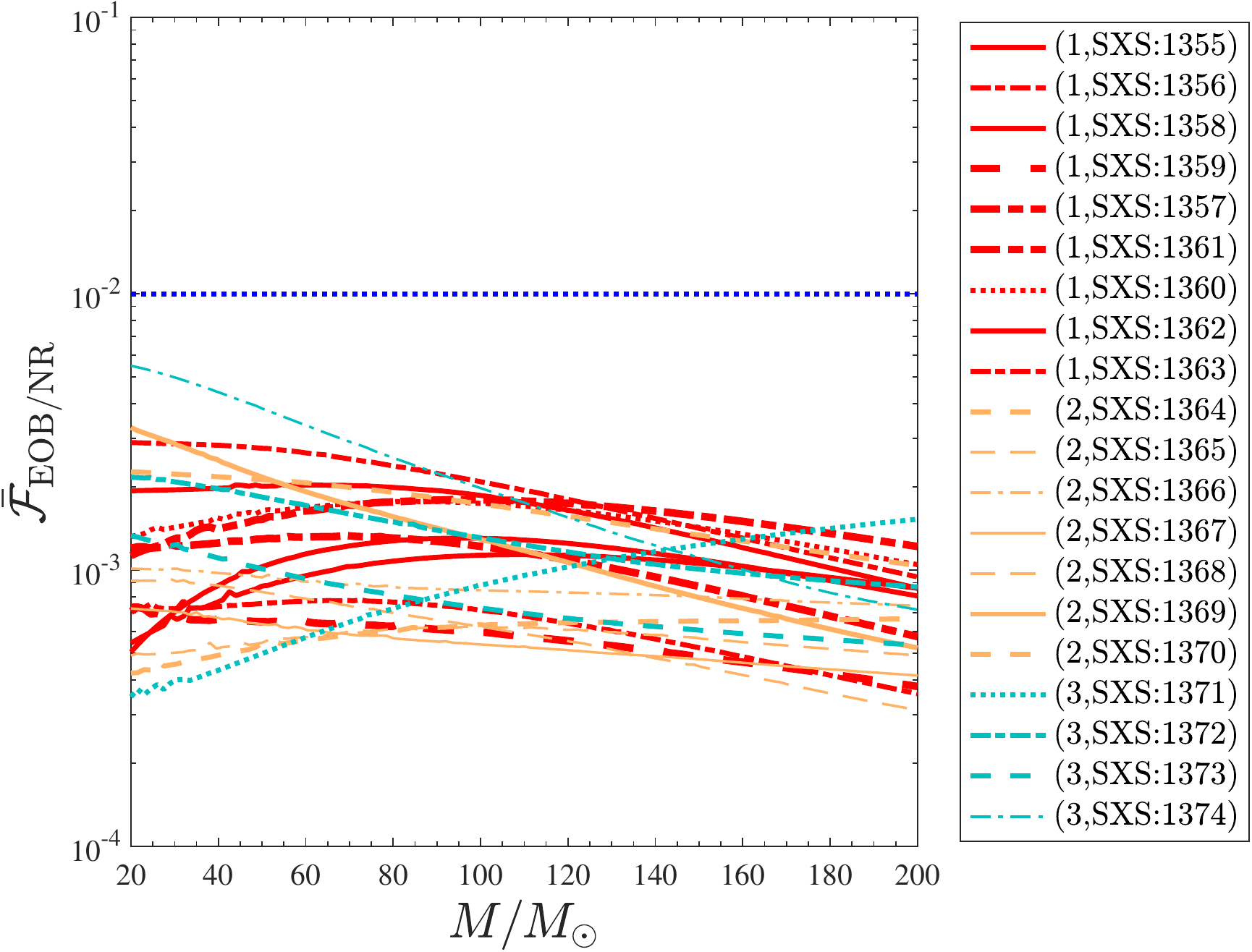} 
         \includegraphics[width=0.22\textwidth]{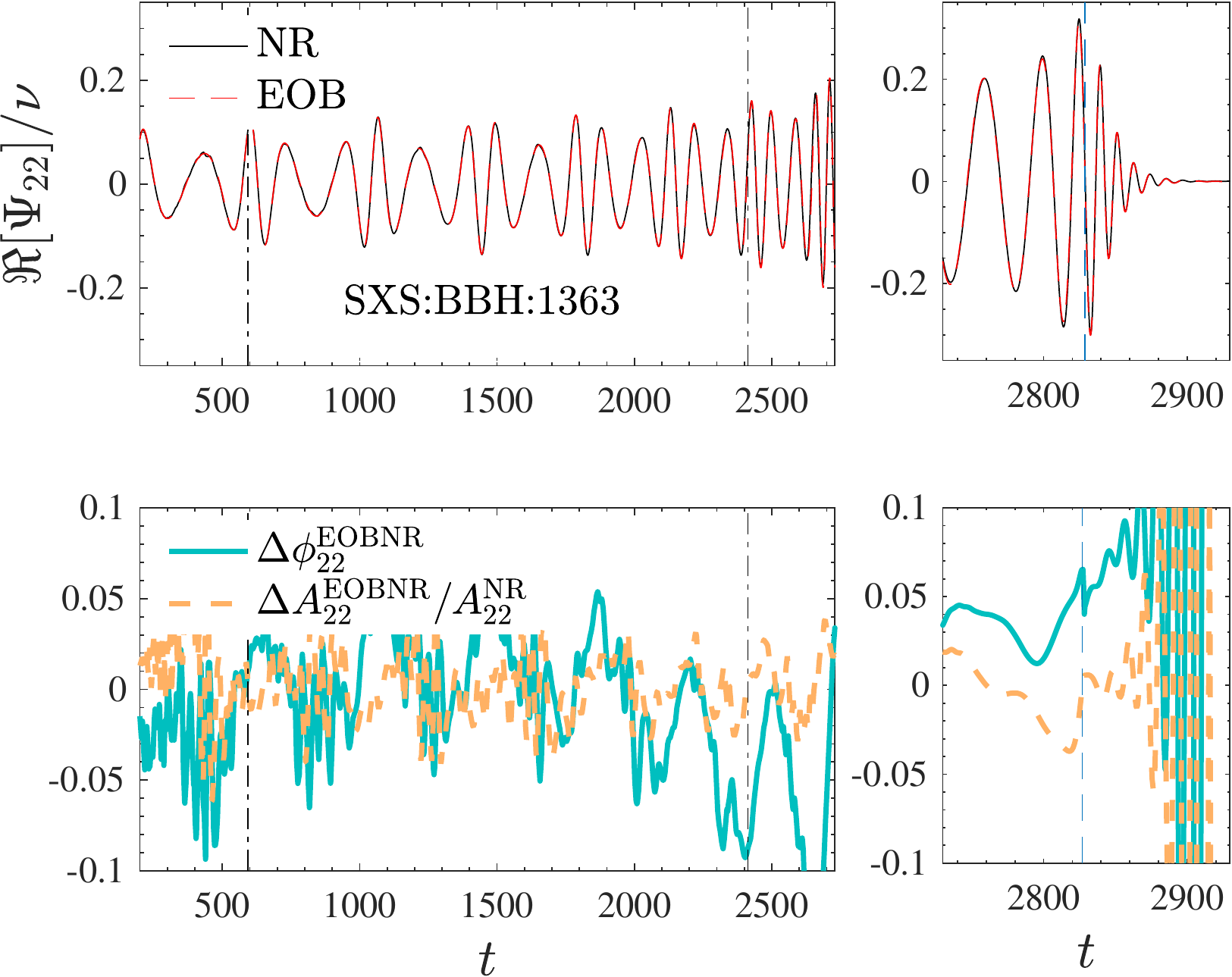}  \\
         \vspace{2.6mm}
         \includegraphics[width=0.22\textwidth]{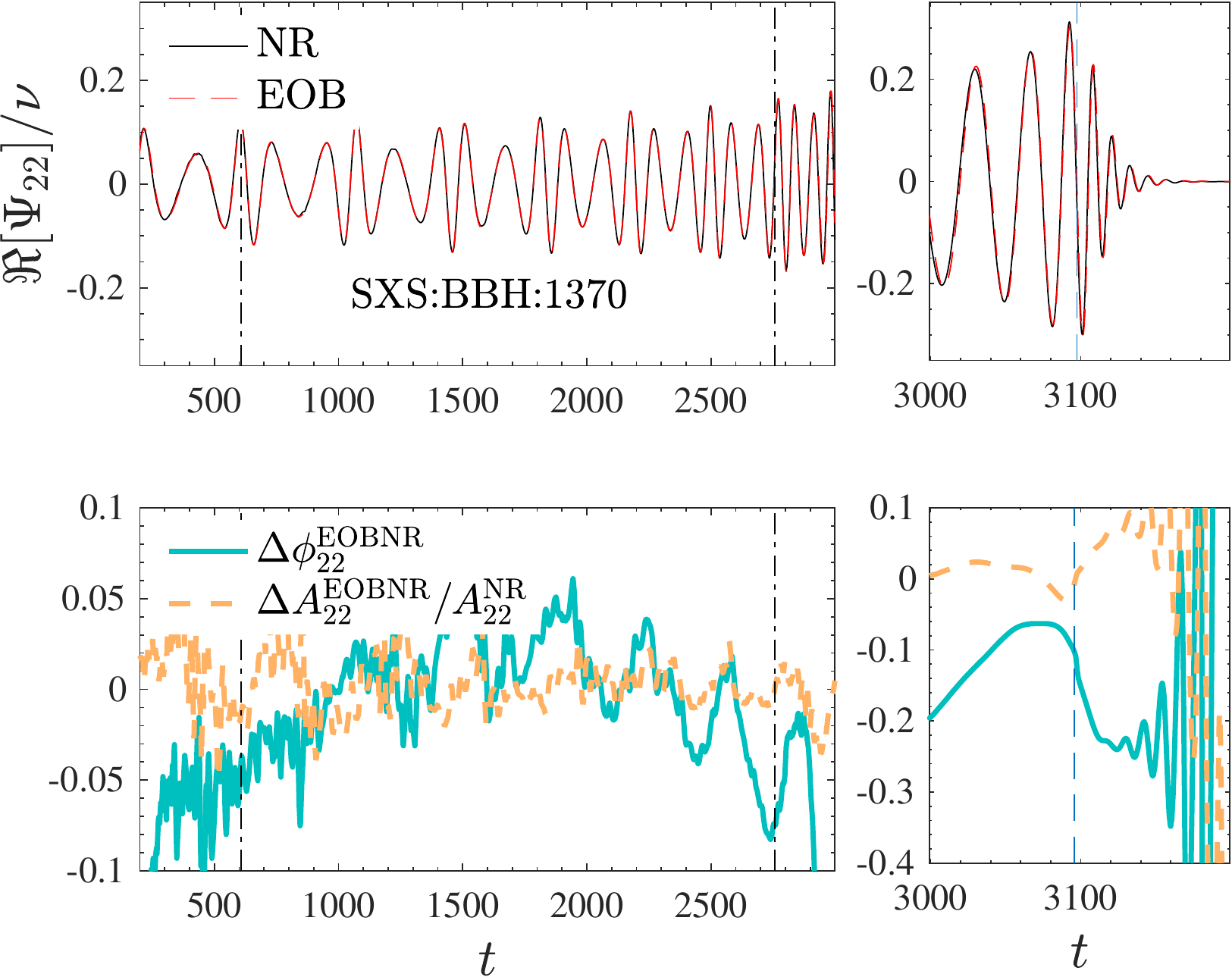} 
         \includegraphics[width=0.22\textwidth]{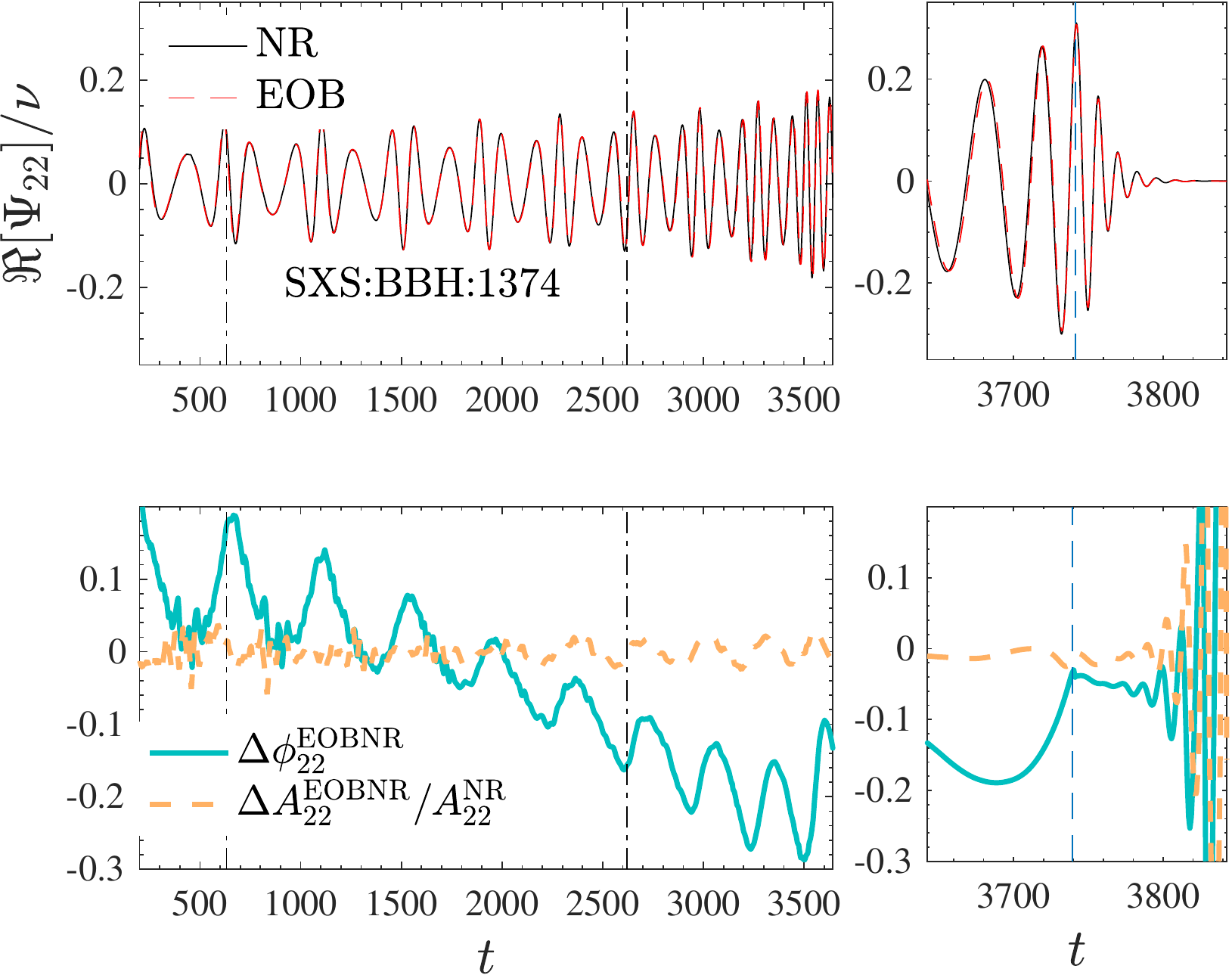}
         	\caption{\label{fig:sxs_ecc}Eccentricity: EOB/NR comparison for nonspinning configurations
	        with different initial eccentricities. Top left panel: EOB/NR unfaithfulness. 
	        Other panels: time-domain phasing plots for the most eccentric configurations.}
\end{figure}

In Ref.~\cite{Nagar:2021xnh} we developed an eccentric waveform model relying on 
the same  $(D_{\rm 5PN},Q_{\rm 5PN})$ functions discussed above, but with an $A$ 
function resummed with a $(3,3)$ Pad\'e approximant. In addition, the model of 
Ref.~\cite{Nagar:2021xnh} was including explicit noncircular effects in radiation reaction 
(at leading Newtonian order) and in the waveform
(including up to 2PN corrections~\cite{Placidi:2021rkh,Albanesi:2022ywx,Albanesi:2022xge}),
both in factorized (and possibly resummed) form. In particular, the state-of-the-art
\TEOBResumS{} eccentric model now incorporates the new waveform introduced
in Ref.~\cite{Albanesi:2022xge} that relies on the direct differentiation of the 
quadrupole moment without explicit replacement of the 2PN-accurate equations of
motion. As discussed in Ref.~\cite{Albanesi:2022xge} this yields improved accuracy
and robustness with respect to previous approaches~\cite{Placidi:2021rkh,Albanesi:2022ywx}.
Given our detailed analysis of the circular model of previous section, it is then interesting
to compare the performance of the quasi-circular limit of the eccentric model.
To do so, we do not precisely use the model of Ref.~\cite{Nagar:2021xnh,Albanesi:2022ywx},
but a version that is improved in three directions: (i) post-adiabatic initial conditions, valid
in the eccentric case, that continuously reduce to the post-adiabatic conditions in the circular
limit. This is important to avoid systematics in going to the circular limit of \TEOBResumS{}
that were recently pointed out~\cite{Knee:2022hth}; (ii) a new expression of the
radial force ${\cal F}_{r_*}$ where the quasi-circular part is explicitly factorized, as proposed 
in Ref.~\cite{Bini:2012ji}; (iii) we find unnecessary to use NR-informed NQC corrections to 
the waveform phase. Among these features, let us only note that the change of the radiation 
reaction is motivated by the fact that the ${\cal F}_{r_*}$ used in Ref.~\cite{Nagar:2021xnh} is
the main reason why the model is less NR-faithful  (see Fig.~8 of Ref.~\cite{Nagar:2021xnh}) 
than the native quasi-circular model for $q\sim 1$.
As we will see below, this feature disappears once the following, 
circular-factorized, radial force is used
\be
{\cal F}_{r_*} =-\dfrac{5}{3}\dfrac{p_{r_*}}{p_\varphi}{\cal F}_\varphi\hat{f}_{p_{r_*}} \ ,
\ee
where $\hat{f}_{p_{r_*}}$ is a quadratic function in $u$ that reads
\begin{align}
\hat{f}_{p_{r_*}} &= 1+\left(\dfrac{5317}{1680} - \dfrac{227}{140}\nu\right)u  \nonumber\\
                         &+\left(\dfrac{1296935}{1016064} - \dfrac{274793}{70560}\nu + \dfrac{753}{560}\nu^2\right)u^2 \ ,
\end{align}
that is then resummed using a $(0,2)$ Pad\'e approximant. 
The corresponding best $a_6^c$ values are fitted by the function
\be
a_6^c=175.5440\nu^3 + 487.6862\nu^2 -471.7141\nu + 0.8178 \ ,
\ee
that is shown in the top-left panel of Fig.~\ref{fig:Fecc}. Note that, although the
values of $a_6^c$ are slightly different from the best ones determined in 
Ref.~\cite{Nagar:2021xnh} the qualitative shape of the function remains unchanged.
The top-right panel of Fig.~\ref{fig:Fecc} shows the EOB/NR unfaithfulness, while
the bottom panel the time-domain phasings for two illustrative configurations.
We see that the performance of the quasi-circular limit of the model is more than
acceptable, though about one order of magnitude less good than the native quasi-circular
model with the same analytic choices for $(A,D,Q)$ (see top-middle panel of Fig.~\ref{fig:model_noNQC}
as well as Fig.~\ref{fig:maxF_vs_PN}, where we compare $\bar{\cal F}^{\rm max}_{\rm EOB/NR}$ 
for all noncircular nonspinning models considered so far).
Note that $\bar{\cal F}^{\rm max}_{\rm EOB/NR}\sim 0.1\%$ (also when $q=1$)
which shows an improvement by a factor 2 with respect to Ref.~\cite{Nagar:2021xnh} 
(see Fig.~8 therein) due to the different choice of ${\cal F}_{r_*}$.
When moving to eccentric configurations (on bound orbits) we perform the same kind 
of comparisons with the available SXS datasets discussed in previous 
works~\cite{Nagar:2021gss,Nagar:2021xnh}. The various modifications we have introduced
to the model call for small modifications to the initial EOB eccentricity and frequency 
at periastron $(e_{\omega_a}^{\rm EOB},\omega_a^{\rm EOB})$.
Once this is taken into account, we compute the EOB/NR unfaithfulness,
that is shown in the top-left panel of Fig.~\ref{fig:sxs_ecc}. On this diagnostics,
the performance is substantially equivalent to previous versions of the model.
On the contrary, the time-domain phasings, shown for the most eccentric configurations
of the set, evidence some improvement with respect to previous versions of the
model, especially through merger and ringdown.
\begin{table*}[t]
 \caption{\label{tab:chi_scattering}Comparison between EOB and NR scattering angle. From left to right the columns report:
 the ordering number; the EOB impact parameter $r_{\rm min}$; the NR and EOB radiated energies, 
 $(\Delta E^{\rm NR}/M,\Delta E^{\rm EOB}/M)$; the NR and EOB radiated angular momentum, 
 $(\Delta J^{\rm NR}/M^2,\Delta J^{\rm EOB}/M^2)$; the NR and EOB scattering angles $(\chi^{\rm NR},\chi^{\rm EOB})$ and
 their fractional difference  $\hat{\Delta}\chi^{\rm NREOB}\equiv |\chi^{\rm NR}-\chi^{\rm EOB}|/\chi^{\rm NR}$.}
   \begin{center}
     \begin{ruledtabular}
\begin{tabular}{c c c c c c c c c } 
$\#$  & $r_{\rm min}$ & $\Delta E^{\rm NR}/M$ & $\Delta E^{\rm EOB}/M$ & $\Delta J^{\rm NR}/M^2$ & $\Delta J^{\rm EOB}/M^2$ & $\chi^{\rm NR}$ [deg] & $\chi^{\rm EOB}$[deg] & $\hat{\Delta}\chi^{\rm NREOB}[\%]$ \\
\hline
1 & 3.43    & 0.01946(17)       & 0.018003     &   0.17007(89)  &  0.166250    &   305.8(2.6)  &  315.6022 & 3.20  \\
2 & 3.76     & 0.01407(10)       & 0.012124  &  0.1380(14)     & 0.124784  & 253.0(1.4)   &  258.2949   &   2.09  \\
3 & 4.06    & 0.010734(75)     & 0.008743    &   0.1164(14)    & 0.098849   &  222.9(1.7)   & 225.0784   & 0.98   \\
4 & 4.86    & 0.005644(38)     & 0.004152    & 0.076920(80)  &  0.058812  &  172.0(1.4)   & 171.5614  &  0.25 \\
5 & 5.35     & 0.003995(27)     &0.002842     & 0.06163(53)     & 0.045379  &152.0(1.3)   & 151.2741     &  0.48 \\
6 & 6.50     & 0.001980(13)    & 0.001370   & 0.04022(53)    & 0.027736    &  120.7(1.5)  & 119.9820  &   0.59\\
7 & 7.60     & 0.0011337(90)  &  0.000789  &  0.029533(53) &  0.019227    &  101.6(1.7)  &  101.0896 & 0.50  \\
 8 & 8.68    & 0.007108(77)    & 0.000505   &  0.02325(47)    &  0.014333 & 88.3(1.8)    &  87.9789   &  0.36 \\
 9 &  9.73   &  0.0004753(75) &  0.000347  &  0.01914(76)    &  0.011213  &78.4(1.8)    & 78.1769   &  0.28 \\
10 & 10.79 & 0.0003338(77)  & 0.000251   &   0.0162(11)     &  0.009081  &  70.7(1.9)    &  70.4980     & 0.28  \\
\hline
11 & 3.03  & 0.0281(11) &    0.0291   & 0.2220(64)&0.2366  &  307.1316  &  337.6476  &     9.94 \\
12 & 3.91 & 0.01194(27)  &    0.0101   &0.1252(10) &0.1098  &  225.5430  &  229.8746  &     1.92 \\
13 & 4.41 &  0.00793(34)  & 0.0062  & 0.09456(70) &0.0780  &  207.0259  &  207.4499  &      0.20 \\
14 & 4.99 &  0.004925(30) &    0.0038   & 0.069504(39)&0.0560  &  195.9340  &  194.5736  &     0.69 \\
15 & 6.68 &  0.001625(16) &    0.0013   &0.034511(71) &0.0280   & 201.9091   & 200.1535    &    0.87       
  \end{tabular}
 \end{ruledtabular}
 \end{center}
 \end{table*}
 %
\begin{figure}[t]
	\center
	\includegraphics[width=0.235\textwidth]{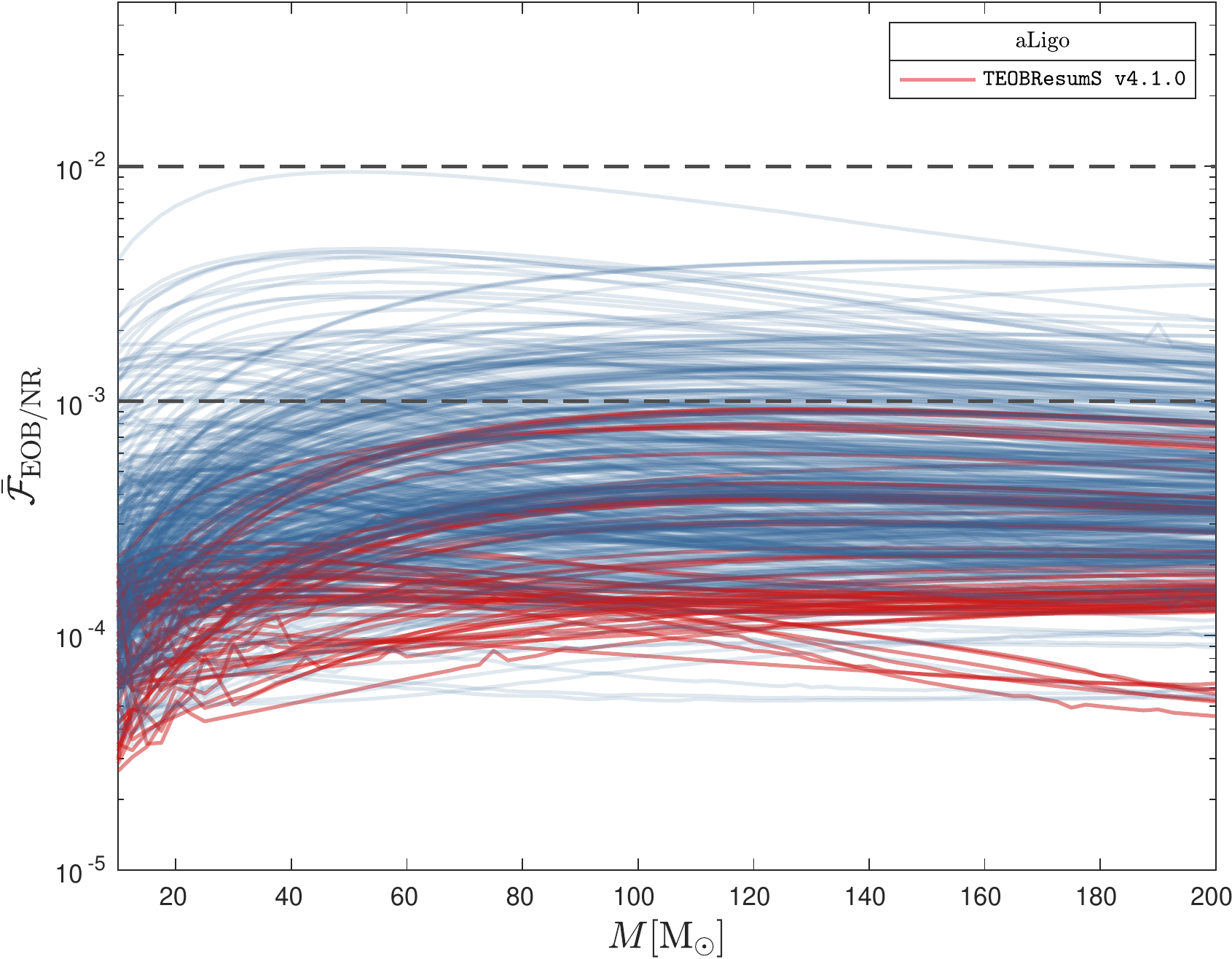}
	\includegraphics[width=0.235\textwidth]{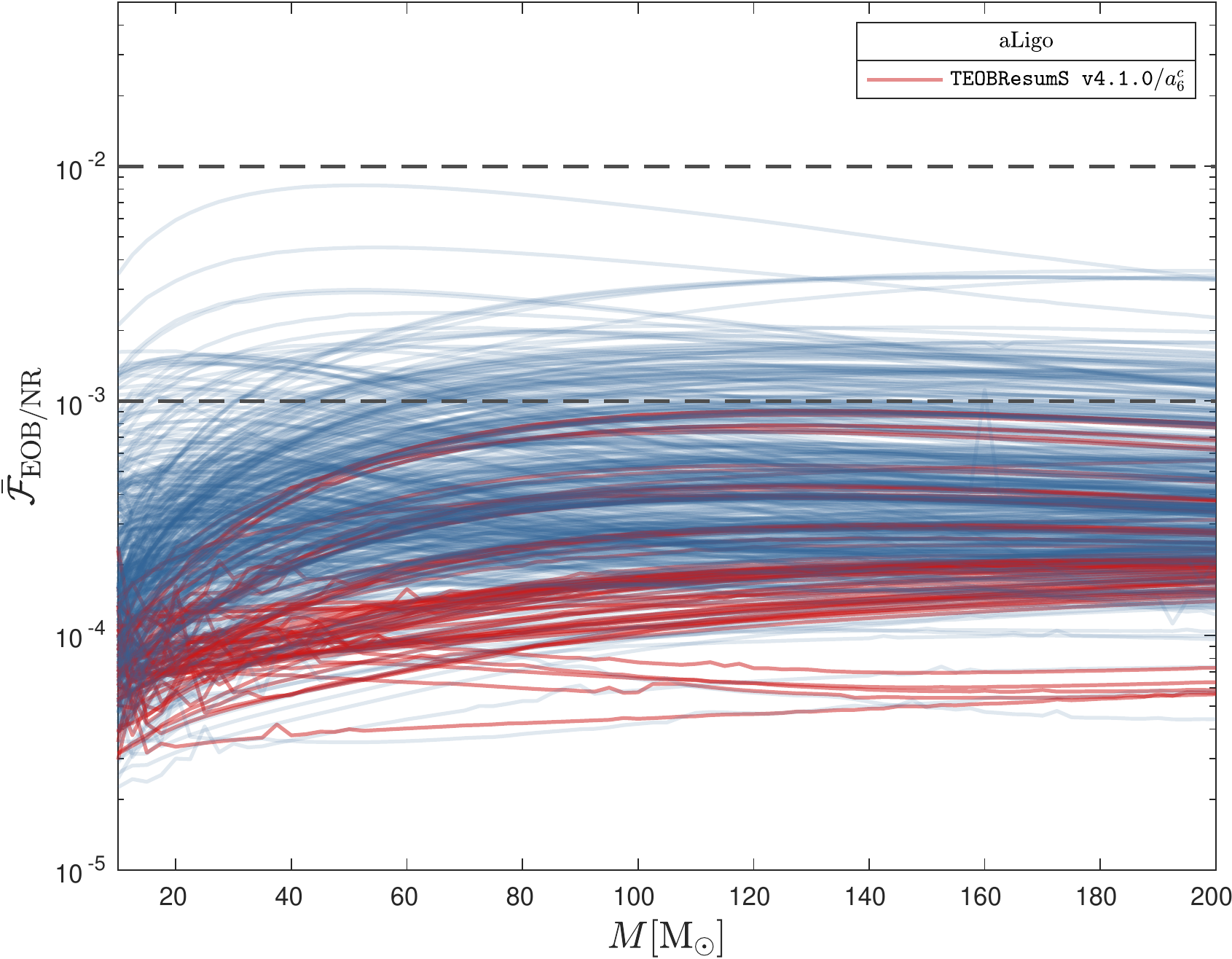}
	\caption{\label{fig:all_spin}EOB/NR unfaithfulness comparisons. Left panel: standard \TEOBResumS{}.
	Right panel: improved model with the {\tt D3Q3\_NQC} $a_6^c(\nu)$ function of Table~\ref{tab:models} 
	but with the standard, NR-informed, 
	N$^3$LO spin-orbit effective parameter $c_3$. The red curves correspond to all 
	reliable nonspinning datasets. In this second case, the only outlier above $0.5\%$ 
	is $(1.5,+0.95,+0.95)$, {\tt SXS:BBH:1146}.}
	\center
\end{figure}
%
\begin{table*}[t]
 \caption{\label{tab:models_vs_c3}Coefficients for the fit of $c_3$ for models that share the same new determination of $a_6^c$ but change mostly because of $c_3$.
 More precisely \TEOBResumSvfourtwo{} uses the $c_3$ function of Ref.~\cite{Albertini:2021tbt}; \TEOBResumSvfourthree{} shares the same $c^=_3$ but implements
 a new $c^\neq_3$ informed with NR datasets not used in Ref.~\cite{Albertini:2021tbt} and all listed in Table; \TEOBResumSvfourthreeone{} implements an improved 
 representation of $c^=_3$ informed from the dataset of Table~\ref{tab:c3_v4p3p1} that also yields a correspondingly updated $c^{\neq}_3$. As a last step \TEOBResumSvfourthreetwo{}
 implements a further modification of $c^{\neq}_3$ related to having changed the first-guess value for the SXS:BBH:1432 dataset from $c_3^{\rm first-guess}=25$ 
 to $c_3^{\rm first-guess}=21$. Furthermore, this the model also implements a different analytical representation of the NR NQC point used to 
 determine (iteratively) the NQC corrections that is essential to remove the few outliers with unfaithfulness slightly above $0.2\%$ always present for 
 the other models.}
   \begin{center}
     \begin{ruledtabular}
\begin{tabular}{c |c c c c c c | c c c c c c} 
Model   & \multicolumn{11}{c}{\hspace{-28mm}$c_3^{=}\equiv p_0\left(1 + n_1\tilde{a}_0 + n_2\tilde{a}_0^2 + n_3\tilde{a}_0^3 + n_4\tilde{a}_0^4\right)/\left(1 + d_1\tilde{a}_0\right)$}\\
            &  \multicolumn{11}{c}{$c_3^{\neq}\equiv \left(p_1\tilde{a}_0 + p_2\tilde{a}_0^2  + p_3\tilde{a}_0^3\right)\sqrt{1-4\nu}+ p_4\tilde{a}_0\nu \sqrt{1-4\nu} + \left(p_5\tilde{a}_{12}+ p_6\tilde{a}_{12}^2\right)\nu^2$}\\
            \hline
       & $p_0$ & $n_1$ & $n_2$   &  $n_3$   &     $n_4$  & $d_1$  & $p_1$ & $p_2$ & $p_3$ & $p_4$ & $p_5$ & $p_6$\\
\hline
\TEOBResumSvfourtwo{}         &  $43.873$ &    $-1.849$            &  $1.0112$            &   $-0.0864$                &    $-0.0384$             & $-0.888$                   & 26.553& $-8.6584$&  0 &  $-84.7473$ &  24.0418 & 0 \\
\TEOBResumSvfourthree{}       &  $43.873$ &    $-1.849$            &  $1.0112$            &   $-0.0864$                &    $-0.0384$             & $-0.888$    & $16.6957$ & $2.0250$  & $-6.6009$  & $-53.1461$  &  $34.0979$ & $-101.0037$ \\
\TEOBResumSvfourthreeone{} & 42.195 & $-2.0107$ & 1.258  & $-0.1210$ & $-0.1063$ &   $-0.9665$  & $20.9956$ & $1.5806$ &  $-10.428$ & $-61.1980$  & $37.1134$ & $-37.6681$ \\
\TEOBResumSvfourthreetwo{}  & 42.195 & $-2.0107$ & 1.258  & $-0.1210$ & $-0.1063$ &   $-0.9665$  &   $18.8003$  & $0.6175$ &  $-10.398$ &  $-47.1696$ & $33.4449$ & $-32.5157$   \\
\end{tabular}
\end{ruledtabular}
\end{center}
\end{table*}
\begin{table}[t]
 \caption{\label{tab:outliers}SXS datasets of Fig.~\ref{fig:max_global} considered outliers, 
 with $\bar{{\cal F}}_{\rm EOB/NR}^{\rm max}>0.2\%$. Note that datasets SXS:BBH:0258 and SXS:BBH:2132
 represent the same configuration.}
   \begin{center}
     \begin{ruledtabular}
\begin{tabular}{c c c c c} 
$\#$ & ID & $(q,\chi_1,\chi_2)$ &$\bar{{\cal F}}_{\rm EOB/NR}^{\rm max} [\%]$ \\
\hline
\hline
1 & BBH:0552 & $(1.750100,+0.799926,-0.399972)$ & 0.2126 \\
2 & BBH:1466 & $(1.896882,+0.698849,-0.799666)$ & 0.2182 \\
3 & BBH:0258 & $(1.999666,+0.871258,-0.849486)$ & 0.3899 \\
4 & BBH:2132 & $(1.999854,+0.871263,-0.849645)$ & 0.3920 \\
5 & BBH:1453 & $(2.352106,+0.800164,-0.784292)$ & 0.3684 \\
6 & BBH:0292 & $(2.999266,+0.731359,-0.849301)$ & 0.3887 \\
7 & BBH:1452 & $(3.641386,+0.800138,-0.426550)$ & 0.4252 \\
8 & BBH:1428 & $(5.516491,-0.800166,-0.699341)$ & 0.2201 \\
9 & BBH:1440 & $(5.638278,+0.769754,+0.306334)$ & 0.2094 \\
10 & BBH:1437 & $(6.037524,+0.799933,+0.147520)$ & 0.3061 \\
11 & BBH:1419 & $(7.997128,-0.799957,-0.798880)$ & 0.2364 \\
\end{tabular}
 \end{ruledtabular}
 \end{center}
 \end{table}
As a last investigation, we obtain the scattering angle for the  $q=1$ {\it nonspinning} configurations whose corresponding
NR values are computed in Refs.~\cite{Damour:2014afa,Hopper:2022rwo}. Also here one finds
a small, though nonnegligible, improvement for most of the configurations, see Table~\ref{tab:chi_scattering}.
This is particularly evident for the configurations simulated in Ref.~\cite{Hopper:2022rwo}, that generally 
span a stronger field regime than those of Ref.~\cite{Damour:2014afa}. In particular, configurations $\# 11$
in Table~\ref{tab:chi_scattering} has a EOB/NR fractional difference $~10\%$, which means a $\sim 2.6~\%$
improvement with respect to Ref.~\cite{Hopper:2022rwo} (see Table~II therein) that was relying on the 
EOB model of Ref.~\cite{Nagar:2021xnh}. The table also lists the values of the NR and EOB radiated energy
and angular momentum as well as the closest EOB separation reached, indicated as $r_{\rm min}$.

\section{Spinning configurations: improving the \TEOBResumS{} model}
\label{teob:best}

\subsection{From \TEOBResumSvfour{} to \TEOBResumSvfourtwo{}}

Now that we have understood the importance of the modelization of the noncircular part 
of the dynamics in the nonspinning case, let us move to considering spins in \TEOBResumS{}.
In particular, we want to evaluate how the quality of the {\tt D3Q3\_NQC} expression of 
$a_6^c(\nu)$ propagates on the spin sector of the model. To do so efficiently, 
we mostly use the  $C$-implementation of \TEOBResumS{} complemented, for simplicity, 
by the fits for the NQC parameters\footnote{We have verified that the differences with the iterated model
are practically negligible.} entering the radiation reaction presented in~\cite{Riemenschneider:2021ppj}. 
Figure~\ref{fig:all_spin} shows the EOB/NR unfaithfulness all over the full SXS catalog obtained
with the N$^3$LO $c_3$ parameter of \TEOBResumS{} but with different choices for $a_6^c$.
The left panel of Fig.~\ref{fig:all_spin} is  $\bar{\cal F}_{\rm EOB/NR}$ with the standard 
$a_6^c(\nu)$ of \TEOBResumS{}, while the right panel the same quantity obtained with 
the {\tt D3Q3\_NQC}  $a_6^c$.
%
\begin{figure*}[t]
	\center
	\includegraphics[width=0.31\textwidth]{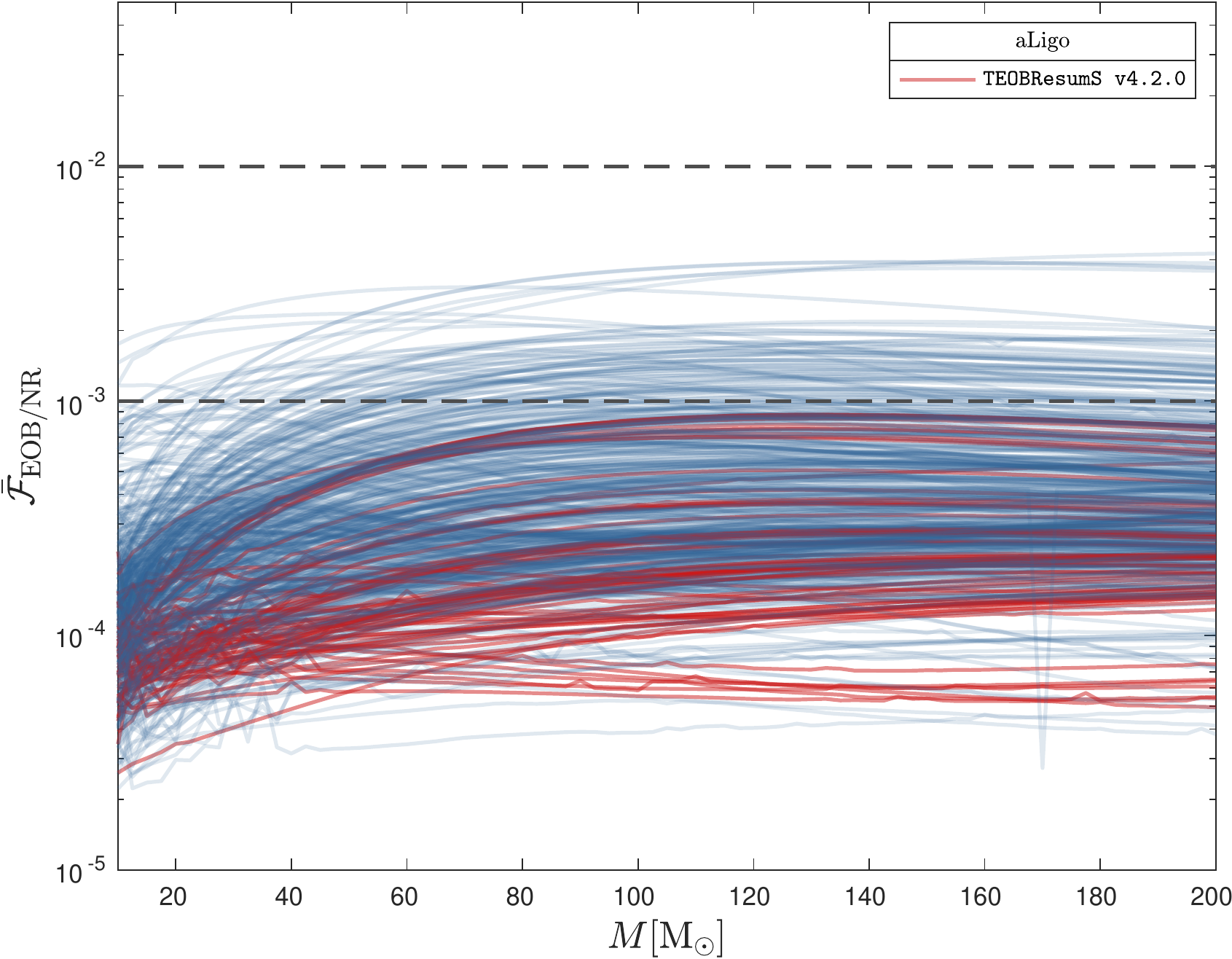} 
         \includegraphics[width=0.33\textwidth]{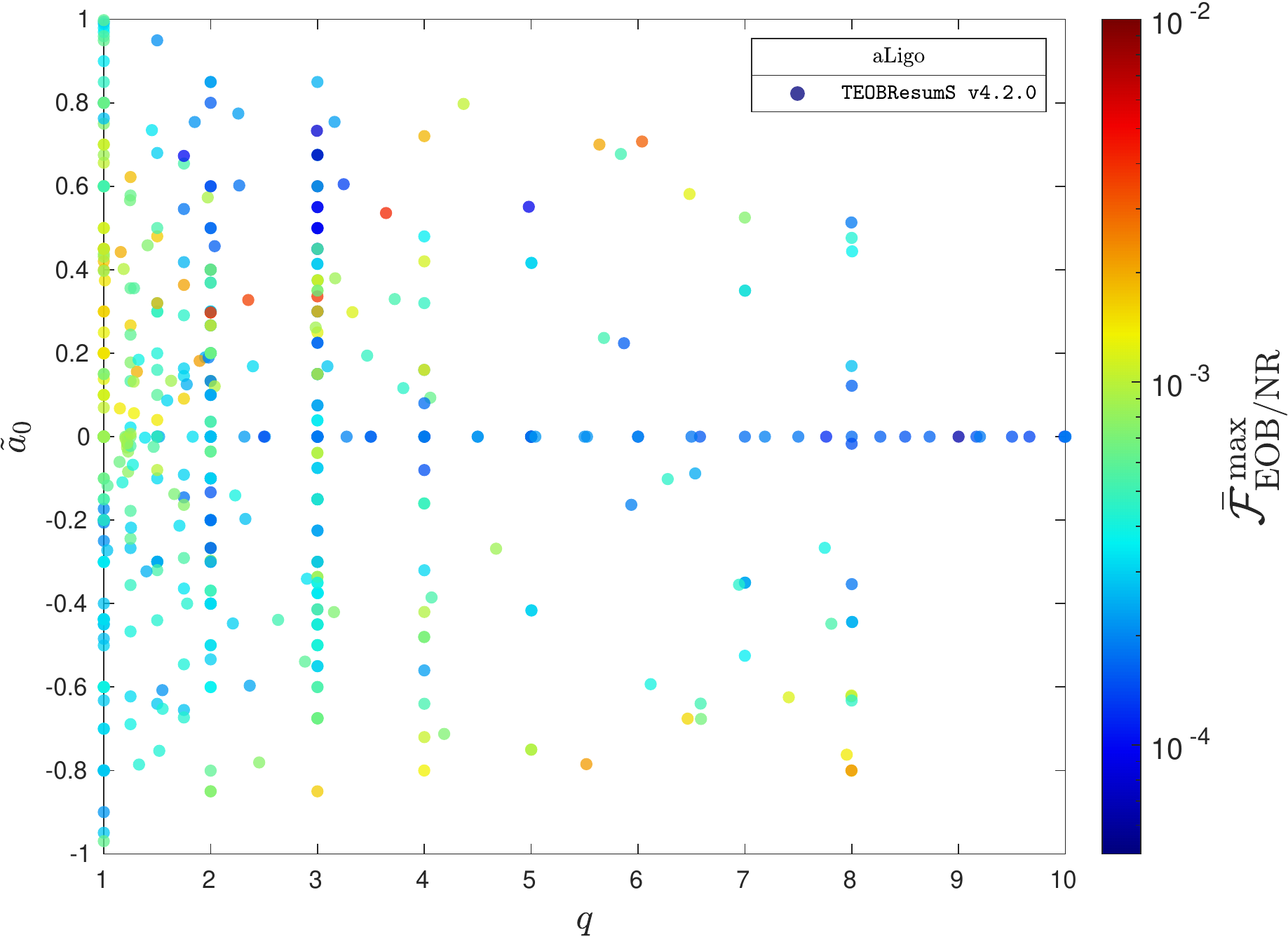}
         \includegraphics[width=0.33\textwidth]{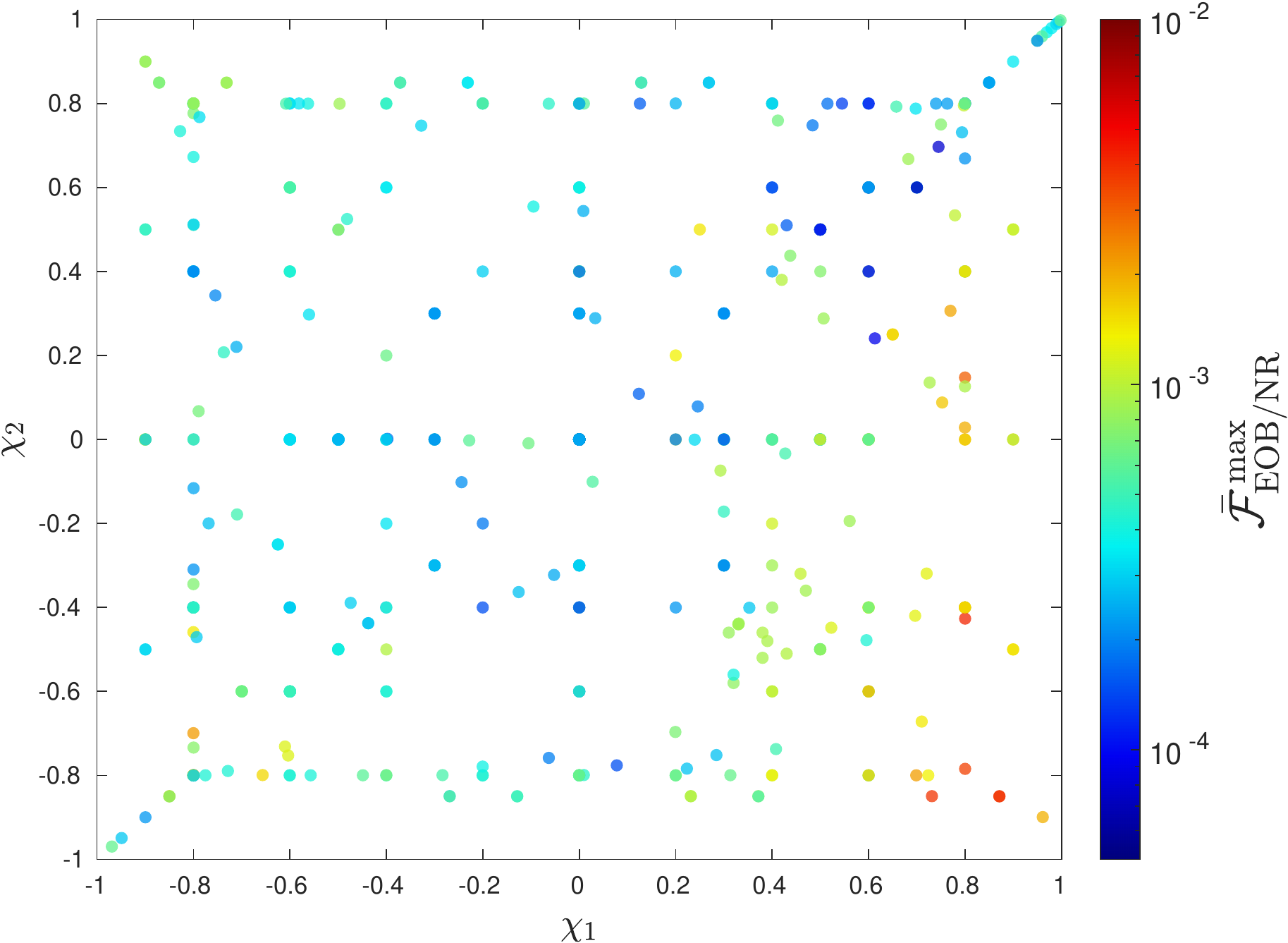}
	\caption{\label{fig:max_global}Performance of \TEOBResumSvfourtwo{}: the EOB/NR unfaithfulness
	over 534 datasets of the SXS catalog. The plot is obtained using the public implementation of the C \TEOB{} 
	code were the  native expression of $a_6^c$ is replaced by the {\tt D3Q3\_NQC} one 
	of Table~\ref{tab:models} for $a_6^c$ while $c_3$ is given by the first row of Table~\ref{tab:models_vs_c3}.
        Left panel: the EOB/NR unfaithfulness $\bar{\cal F}_{\rm EOB/NR}$ versus the total mass of the binary.
	Middle: $\bar{\cal F}_{\rm EOB/NR}^{\rm max}$ highlighting the dependence on $q$ and $\tilde{a}_0$. 
	Right panel: $\bar{\cal F}_{\rm EOB/NR}^{\rm max}$  versus the dimensionless spins $(\chi_1,\chi_2)$.
	One finds 11 configurations with $\bar{\cal F}_{\rm EOB/NR}^{\rm max}>0.2\%$, that are listed in 
	Table~\ref{tab:outliers} for convenience.
	}
\end{figure*}
\begin{figure*}[t]
	\center
	\includegraphics[width=0.31\textwidth]{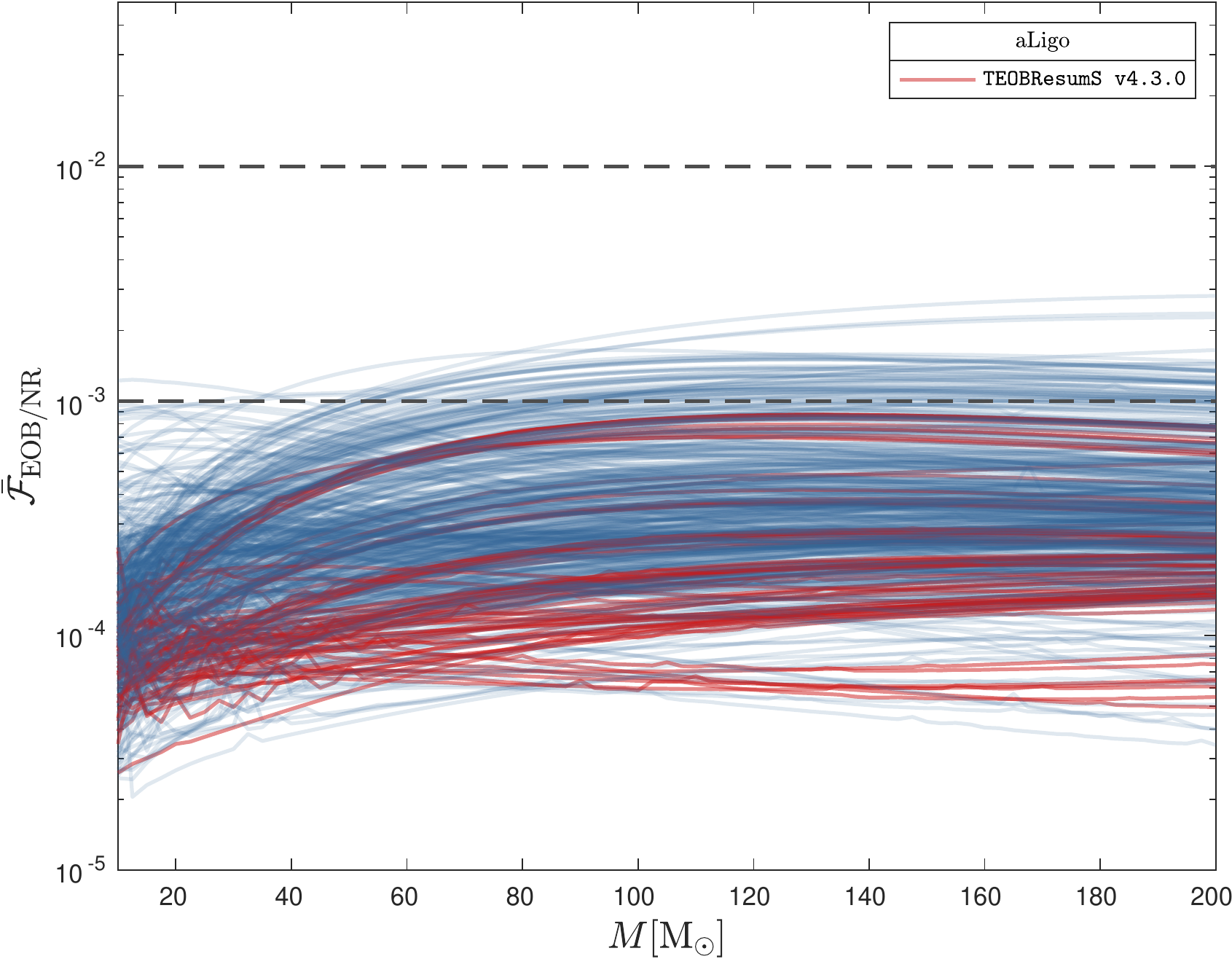}       
	\includegraphics[width=0.33\textwidth]{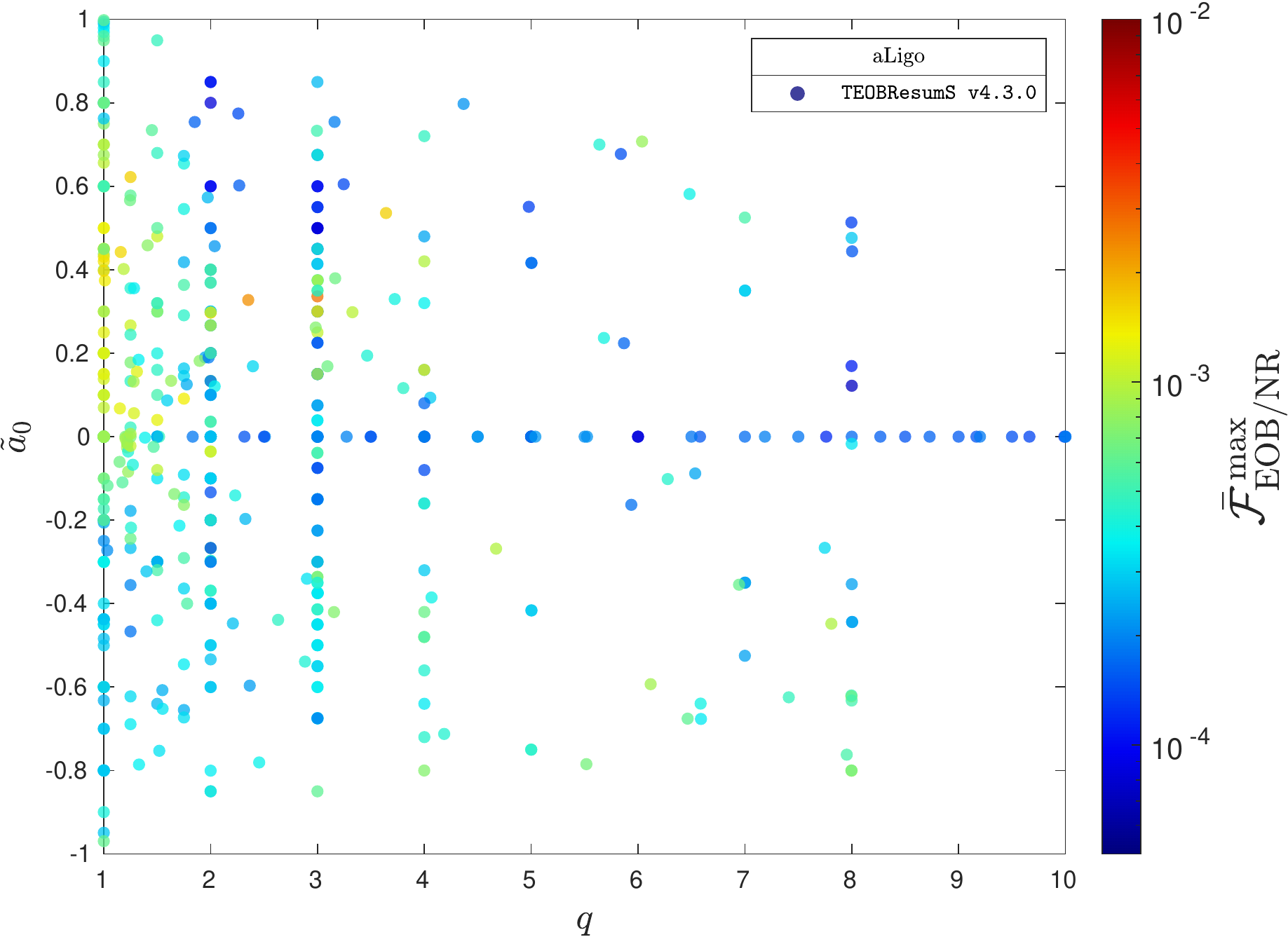}       
         \includegraphics[width=0.33\textwidth]{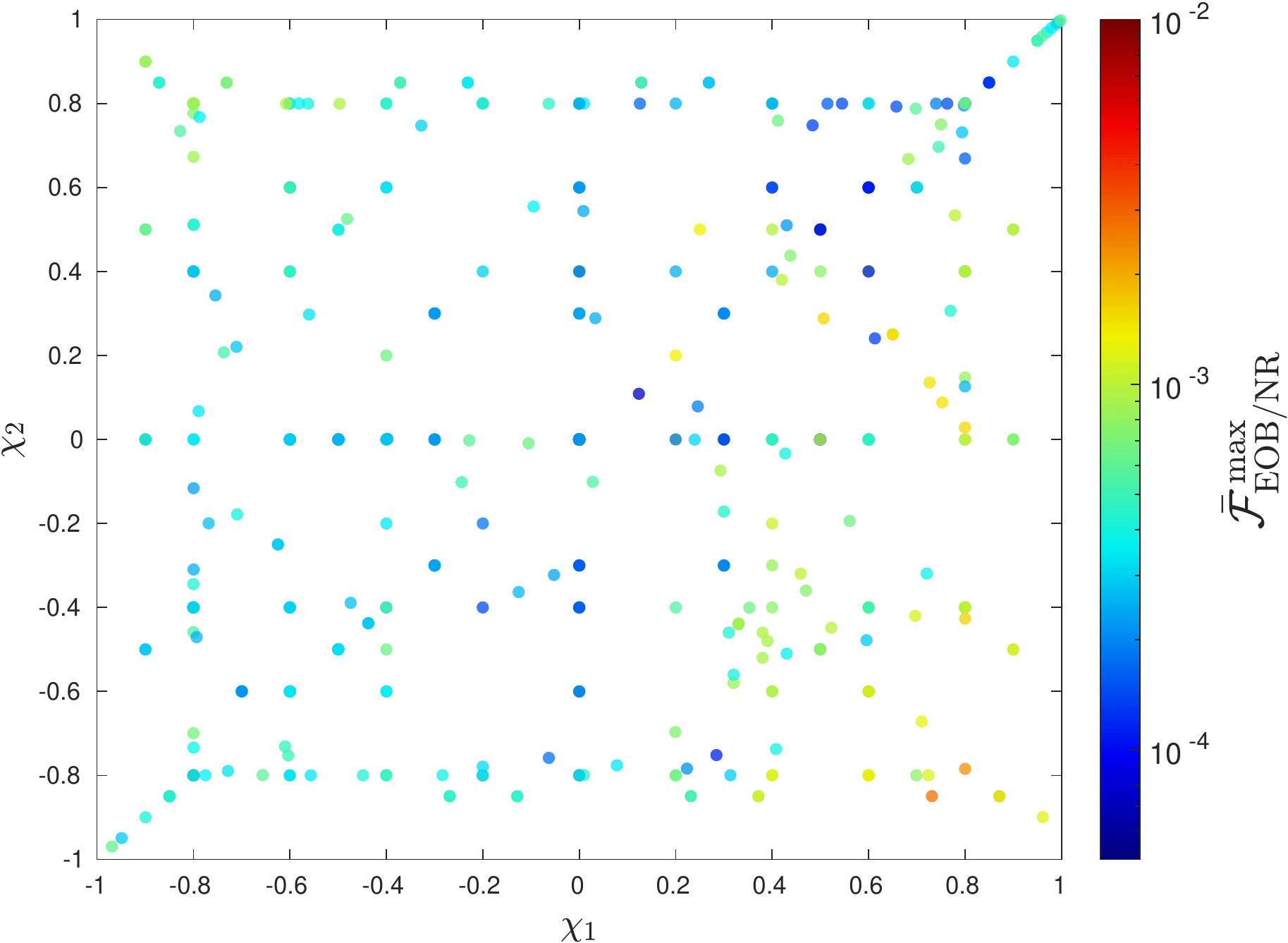}
	\caption{\label{fig:max_global_v4p30}Performance of \TEOBResumSvfourthree{} looking at the EOB/NR unfaithfulness
	over 534 datasets of the SXS catalog. The model implements a new fit for the unequal-mass part of the function
	of $c_3$. Left panel: the EOB/NR unfaithfulness versus the total mass of the binary.
	Right panel: $\bar{\cal F}_{\rm EOB/NR}^{\rm max}$ highlighting the dependence on $q$ and $\tilde{a}_0$. 
	Only two configurations are left with $\bar{\cal F}_{\rm EOB/NR}^{\rm max}>0.2\%$ that are listed in Table~\ref{tab:outliers_v4p3p0}.	
	}
\end{figure*}
%
 \begin{table}[t]
   \caption{\label{tab:c3_v4p3p0}First-guess values for $c_3$ determined modifying the 
   unequal-mass part. This gives the \TEOBResumSvfourthree{} model whose performance
   is illustrated in Fig.~\ref{fig:max_global_v4p30}.}
   \begin{center}
 \begin{ruledtabular}
   \begin{tabular}{lllc|cc}
     $\#$ & ID & $(q,\chi_1,\chi_2)$ & $\tilde{a}_0$ &$c_3^{\rm first\;guess}$ & \\
     \hline
15 & BBH:0004 & $(   1, -0.   50,  0.    0)$ & $-0.25$ & 55.5 &  \\ 
16 & BBH:0005 & $(   1, +0.   50,  0.    0)$ & $+0.25$ & 35 & \\ 
17 & BBH:2105 & $(   1, +0.   90,  0.    0)$ & $+0.45$ & 27.7 &  \\ 
18 & BBH:2106 & $(   1, +0.   90, +0.   50)$ & $+0.70$ & 19.1 &  \\ 
19 & BBH:0016 & $( 1.5, -0.   50,  0.    0)$ & $-0.30$ & 56.2 &  \\ 
20 & BBH:1146 & $( 1.5, +0.   95, +0.   95)$ & $+0.95$ & 14.35 &  \\ 
21 & BBH:0552$^*$ & $(1.75,+0.80,-0.40)$& $+0.36$ & 29 & \\
22 & BBH:1466$^*$ & $(1.90,+0.70,-0.80)$& $+0.18$ & 33 &  \\
23 & BBH:2129 & $(   2, +0.   60,  0.    0)$ & $+0.40$ &  29.5 &  \\ 
24 & BBH:0258$^*$ & $(2,+0.87,-0.85)$& $+0.296$ & 32 &  \\
25 & BBH:2130 & $(   2, +0.   60, +0.   60)$ & $+0.60$ & 23 &  \\ 
26 & BBH:2131$^\dag$ & $(   2, +0.   85, +0.   85)$ & $+0.85$  & 15.8 &  \\ 
27 & BBH:1453$^*$ & $(2.352,+0.80,-0.78)$& $+0.328$ & 29 &  \\
28 & BBH:2139 & $(   3, -0.   50, -0.   50)$ & $-0.50$ & 65.3 &  \\ 
29 & BBH:0036$^\dag$ & $(   3, -0.   50,  0.    0)$ & $-0.38$ & 61 &  \\ 
30 & BBH:0174 & $(   3, +0.   50,  0.    0)$ & $+0.37$ & 28.5 &  \\ 
31 & BBH:2158 & $(   3, +0.   50, +0.   50)$ & $+0.50$ & 27.1 &  \\ 
32 & BBH:2163 & $(   3, +0.   60, +0.   60)$ & $+0.60$ & 24.3 &  \\ 
33 & BBH:0293$^\dag$ & $(   3, +0.   85, +0.   85)$ & $+0.85$ & 16.0 &  \\ 
34 & BBH:0292$^*$ & $(3,+0.73,-0.85)$& $+0.335$ & 30.6&  \\
35 & BBH:1447 & $(3.16, +0.7398, +0.   80)$ & $+0.75$ & 19.2 &  \\ 
36 & BBH:1452$^*$ & $(3.641,+0.80,-0.43)$& $+0.534$ & 25.6&  \\
37 & BBH:2014 & $(   4, +0.   80, +0.   40)$ & $+0.72$ & 21.5 &  \\ 
38 & BBH:1434 & $(4.37, +0.7977, +0.7959)$ & $+0.80$ & 19.8 &  \\ 
39 & BBH:0111 & $(   5, -0.   50,  0.    0)$ & $-0.42$ & 54 & \\ 
40 & BBH:0110$^\dag$ & $(   5, +0.   50,  0.    0)$ & $+0.42$ & 29.5 &  \\ 
41 & BBH:1428$^*$ & $(5.516,-0.80,-0.70)$ & $-0.784$ & 80&  \\
42 & BBH:1440$^*$ & $(5.64,+0.77,+0.31)$& $+0.70$ & 21.5  \\
43 & BBH:1432 & $(5.84, +0.6577, +0. 793)$ & $+0.68$ & 25 & \\ 
44 & BBH:1437$^*$ & $(6.038,+0.80,+0.15)$& $+0.7076$ & 21.5&  \\
45 & BBH:1375$^\dag$ & $(   8, -0.   90,  0.    0)$ & $-0.80$ & 70 &  \\ 
46 & BBH:1419$^*$ & $(8,-0.80,-0.80)$& $-0.80$ & 81.5 &  \\
47 & BBH:0114$^\dag$ & $(   8, -0.   50,  0.    0)$ & $-0.44$ & 61 &  \\ 
48 & BBH:0065$^\dag$& $(   8, +0.   50,  0.    0)$ & $+0.44$ & 26.5 &  \\ 
49 & BBH:1426 & $(   8, +0.4838, +0.7484)$ & $+0.51$ & 30.3 &  \\ 
 \end{tabular}
 \end{ruledtabular}
 \end{center}
 \end{table}
It is interesting to note that the improvement in the orbital sector are by themselves sufficient to reduce
the number of outliers above the  $2\times 10^{-3}$ level. Note however that little improvement is obtained,
for example, for the $(1.5,0.95,0.95)$ configuration, although it remains the only outlier above $0.5\%$.
In any case, the global lowering of the EOB/NR unfaithfulness is non negligible already up to $M=M_\odot\sim 100$, 
i.e. for the total mass range covered by most of the events detected so far by the LVK collaboration~\cite{LIGOScientific:2021djp}.
Evidently, the EOB/NR performance is expected to improve further by a new determination of $c_3$
that is either more consistent with the current choice of $a_6^c$ or uses more NR data to better determine
its dependence on the spins and mass ratio. In this respect, Ref.~\cite{Albertini:2021tbt} introduced a new 
version of \TEOBResumS{} that differs from the standard one because of: (i) more (NR-informed) NQC corrections 
in the flux, so to achieve a closer EOB/NR flux consistency during the plunge phase up to merger 
and (ii) a different expression for $c_3$ obtained using a carefully chosen set of SXS NR simulations, 
see Table~II in Ref.~\cite{Albertini:2021tbt}. As a start, we can simply replace the standard 
\TEOBResumS{} $c_3$ with the one given in  Eq.~(22) of Ref.~\cite{Albertini:2021tbt} and explore 
whether the EOB/NR unfaithfulness is reduced. In general, the analytic expression of $c_3$ is made 
by two terms, one determined using only equal-mass and equal-spin configurations, $c_{3}^{=}$, 
and another one corresponding to all other combinations of mass ratios and spins, $c_{3}^{\neq}$.
Globally, the $c_3$ function reads
\begin{align}
  \label{eq:c3fit}
c_3(\nu,\tilde{a}_0,\tilde{a}_{12})= c^{=}_{3}+c_3^{\neq} \ ,
\end{align}
where
\begin{align}
c_3^{=}&\equiv p_0\dfrac{1 + n_1\tilde{a}_0 + n_2\tilde{a}_0^2 + n_3\tilde{a}_0^3 + n_4\tilde{a}_0^4}{1 + d_1\tilde{a}_0}\\
c_3^{\neq}&\equiv \left(p_1\tilde{a}_0 + p_2\tilde{a}_0^2  + p_3\tilde{a}_0^3\right)\sqrt{1-4\nu} \nonumber\\
                 &+ p_4\tilde{a}_0\nu \sqrt{1-4\nu} + \left(p_5\tilde{a}_{12}+ p_6\tilde{a}_{12}^2\right)\nu^2 \ ,
\end{align}
and $\tilde{a}_i\equiv X_i\chi_i$, $\tilde{a}_0=\tilde{a}_1+\tilde{a}_2$ and $\tilde{a}_{12}\equiv \tilde{a}_1-\tilde{a}_2$,
while $(p_0,n_i,d_1,p_i)$ are fitting coefficients obtained by fitting the point-wise $c_3^{\rm first-guess}$ values obtained for
a (possibly limited) number of NR configurations. The coefficients for the fit are found in the first row of Table~\ref{tab:models_vs_c3}, 
with the model dubbed \TEOBResumSvfourtwo{}. Note that, following~\cite{Albertini:2021tbt}, we exactly impose $p_3=p_6=0$.
This constraint will be eventually relaxed below. Figure~\ref{fig:max_global} illustrates that this expression of $c_3$ already 
brings a relevant improvement, so that \TEOBResumSvfourtwo{} is  closer to NR than its previous avatars. 
This is by itself remarkable considering
that this $c_3$ was determined for  a model with a different $a_6^c$ and radiation reaction. 
One is left with only 11 outliers with $\bar{{\cal F}}_{\rm EOB/NR}^{\rm max}>0.2\%$ (listed in Table~\ref{tab:outliers}) 
with the worst performance ($~0.42\%$) obtained for a $(3.64,+0.80,-0.43)$ configuration. From the table we see
that most of the largest values are in the range $2\lesssim q \lesssim 3$ and large values of the
individual (unequal) spins. This is a priori not surprising considering that the $c_3$ above was determined
using {\it only three} $q=2$  and six $q=3$ dataset, see Table~II of Ref.~\cite{Albertini:2021tbt},
mostly equal-spin ones. On the one hand, this is a proof of the robustness of the analytic
structure of \TEOBResumS{}, as it can somehow automatically accommodate for the lack of
additional NR-information. On the other hand, it makes us a priori confident that a different
determination of $c_3$ that relies on some more unequal-spin datasets should allow us
to additionally lower $\bar{\cal F}_{\rm EOB/NR}^{\rm max}$, possibly below $10^{-3}$ 
for all configurations.

\subsection{From \TEOBResumSvfourtwo{} to \TEOBResumSvfourthreetwo{}}
Let us then embark into the enterprise of improving \TEOBResumSvfourtwo{} further.
We anticipate that, by carefully understanding the origin of the EOB/NR (small) discrepancies,
we will eventually succeed in obtaining a model, dubbed \TEOBResumSvfourthreetwo{},
with $\bar{\cal F}_{\rm EOBNR}^{\rm max}\sim 0.1\%$ all over the public SXS catalog.
We note that this result will be achieved by {\it only} working on the NR-informed part of
the model, without incorporating additional analytical information.

To start with, the simplest way to proceed seems to compute a new $c_3$ function
by incorporating in the NR-informing data the 10 outlier configurations of Table~\ref{tab:outliers}.
Since these are all unequal-mass configurations, we will only update $c_3^{\neq}$.
In following this procedure, we realized that it is useful to also update some of the first-guess 
values used in Ref.~\cite{Albertini:2021tbt}. All datasets used to this aim are listed 
in Table~\ref{tab:c3_v4p3p0}. For the reader's convenience, we mark with an $*$ the 
{\it new} datasets (coming from Table~\ref{tab:outliers}) added to inform $c_3$, 
while with a $\dag$ the configurations that receive an updated value of $c_3^{\rm first-guess}$.
\begin{table}[t]
 \caption{\label{tab:outliers_v4p3p0}SXS Outliers with $\bar{{\cal F}}_{\rm EOB/NR}^{\rm max}>0.2\%$ 
 for \TEOBResumSvfourthree{}.}
   \begin{center}
     \begin{ruledtabular}
\begin{tabular}{c c c c c} 
$\#$ & ID & $(q,\chi_1,\chi_2)$ & $\bar{{\cal F}}_{\rm EOB/NR}^{\rm max} [\%]$ \\
\hline
\hline
1 & BBH:0258 & $(1.999666,+0.871258,-0.849486)$  &0.227161\\
2 & BBH:0292 & $(2.999266,+0.731359,-0.849301)$  &0.280737\\
3 & BBH:1453 & $(2.352106,+0.800164,-0.784292)$  &0.236153
\end{tabular}
 \end{ruledtabular}
 \end{center}
 \end{table}
%
 \begin{table}[t]
   \caption{\label{tab:c3_v4p3p1}First-guess values for new $c_3$ values for 
   equal-mass, equal-spin configurations. Together with the unequal-mass configurations
   of Table~\ref{tab:c3_v4p3p0} this gives the \TEOBResumSvfourthreeone{} model whose
   performance is evaluated in  Fig.~\ref{fig:max_global_v4p3p1}.}
   \begin{center}
 \begin{ruledtabular}
   \begin{tabular}{lllc|c}
     $\#$ & ID & $(q,\chi_1,\chi_2)$ & $\tilde{a}_0$ &$c_3^{\rm first\;guess}$ \\
     \hline
1 & BBH:1137 & $(   1, -0.   97, -0.   97)$ & $-0.97$ & 89.7   \\ 
2 & BBH:0156 & $(   1, -0.9498, -0.9498)$ & $-0.95$ & 88.5   \\ 
3 & BBH:0159 & $(   1, -0.90, -0.   90)$ & $-0.90$ & 84.5   \\ 
4 & BBH:2086 & $(   1, -0.80, -0.   80)$ & $-0.80$ & 82   \\ 
5 & BBH:2089 & $(   1, -0.60, -0.   60)$ & $-0.60$ & 71   \\ 
6 & BBH:2089 & $(   1, -0.20, -0.   20)$ & $-0.60$ & 52   \\
7 & BBH:0150 & $(   1, +0.   20, +0.   20)$ & $+0.20$ & 33   \\ 
8 & BBH:0170 & $(   1, +0.   4365, +0.   4365)$ & $+0.20$ & 33   \\ 
9 & BBH:2102 & $(   1, +0.   60, +0.   60)$ & $+0.60$ & 21.0   \\ 
10 & BBH:2104 & $(   1, +0.   80, +0.   80)$ & $+0.80$ & 15.9   \\ 
11 & BBH:0153 & $(   1, +0.   85, +0.   85)$ & $+0.85$ & 15.05   \\ 
12 & BBH:0160 & $(   1, +0.   90, +0.   90)$ & $+0.90$ & 14.7   \\ 
13 & BBH:0157 & $(   1, +0.   95, +0.   95)$ & $+0.95$ & 14.3   \\ 
14 & BBH:0177 & $(   1, +0.   99, +0.   99)$ & $+0.99$ & 14.2   
\end{tabular}
 \end{ruledtabular}
 \end{center}
 \end{table}
%
\begin{table}[t]
 \caption{\label{tab:outliers_v4p3p1}SXS outliers, with $\bar{{\cal F}}_{\rm EOB/NR}^{\rm max}>0.2\%$,
 for  \TEOBResumSvfourthreeone{}, see of Fig.~\ref{fig:max_global_v4p3p1}. Some configurations are
 the same of Table~\ref{tab:outliers}.}
   \begin{center}
     \begin{ruledtabular}
\begin{tabular}{c c c c c} 
$\#$ & ID & $(q,\chi_1,\chi_2)$ &$\bar{{\cal F}}_{\rm EOB/NR}^{\rm max} [\%]$ \\
\hline
\hline
1 & BBH:2132 &$(1.999854,+0.871263,-0.849645)$   & 0.258972 \\
2 & BBH:0258 &$(1.999666,+0.871258,-0.849486)$  & 0.257633 \\
3 & BBH:1453 &$(2.352106,+0.800164,-0.784292)$   & 0.247826 \\
4 & BBH:0292 &$(2.999266,+0.731359,-0.849301)$  & 0.273188 \\
5 & BBH:1452 &$(3.641386,+0.800138,-0.426550)$  & 0.294905 
\end{tabular}
 \end{ruledtabular}
 \end{center}
 \end{table}
The parameters of this new model, dubbed \TEOBResumSvfourthree{}, are listed in the second row of
Table~\ref{tab:models_vs_c3}. The corresponding accuracy checks, versus NR, are exhibited in 
Fig.~\ref{fig:max_global_v4p30}. The number of outliers above $2\%$ is now reduced, with only three remaining 
configurations, listed in Table~\ref{tab:outliers_v4p3p0}. 
However, from Fig.~\ref{fig:max_global_v4p30} we see that the performance is visibly lower for equal-mass
binaries, and in particular in the range $0\lesssim \tilde{a}_0\lesssim 0.4$. This calls for an improved determination
of $c_3^=$ in that spin range. To do so, we list in Table~\ref{tab:c3_v4p3p1} the equal-mass, equal-spin configurations
with the corresponding first-guess values. The table includes two datasets more than in Ref.~\cite{Albertini:2021tbt} 
to constrain better the function for small, positive, spins. The corresponding new coefficients for {\it both} $c_3^=$ 
and $c_3^{\neq}$ are listed in the third row of Table~\ref{tab:models_vs_c3}, with the model now 
dubbed \TEOBResumSvfourthreeone{}.
\begin{figure*}[t]
	\center
	\includegraphics[width=0.31\textwidth]{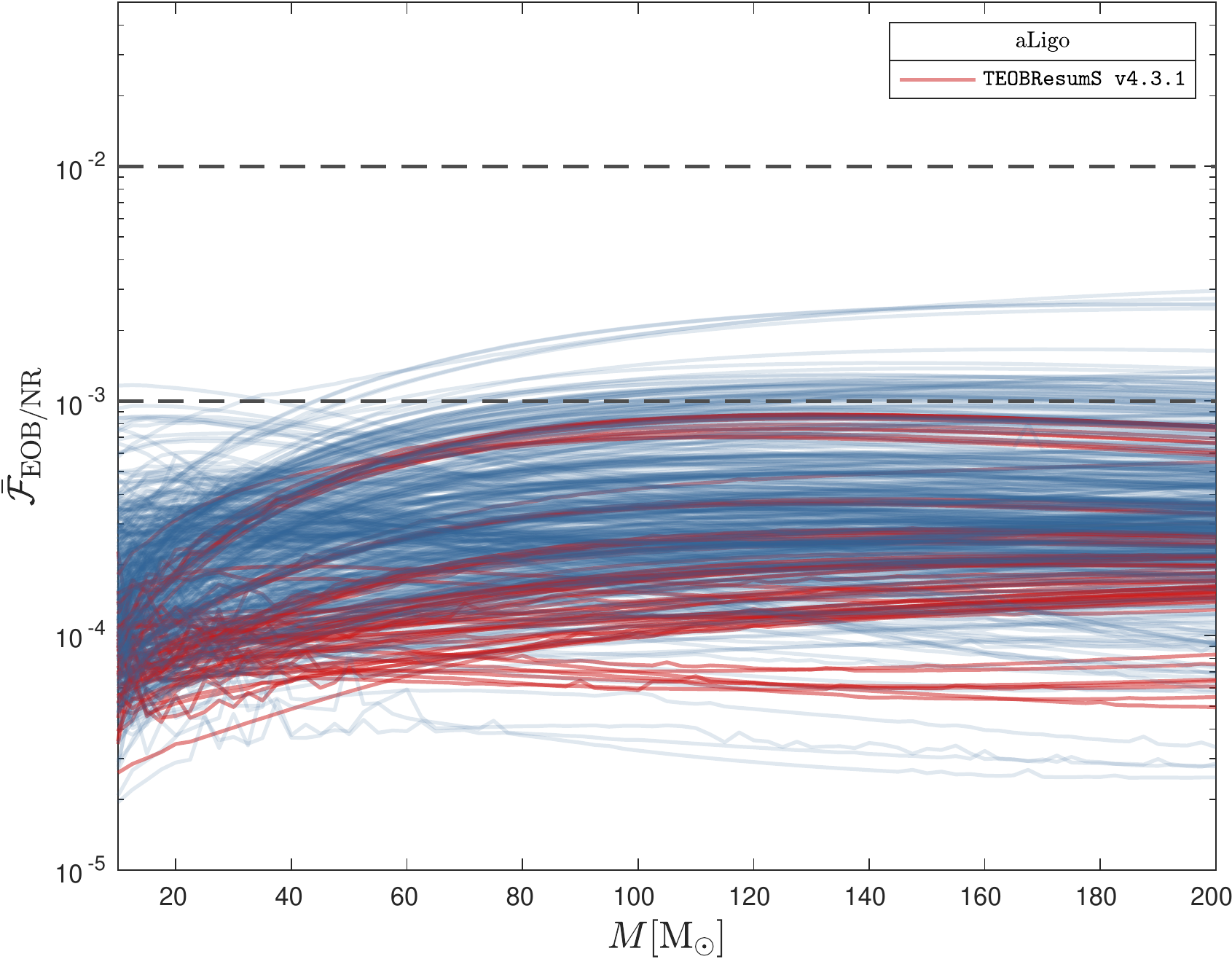}
	\includegraphics[width=0.33\textwidth]{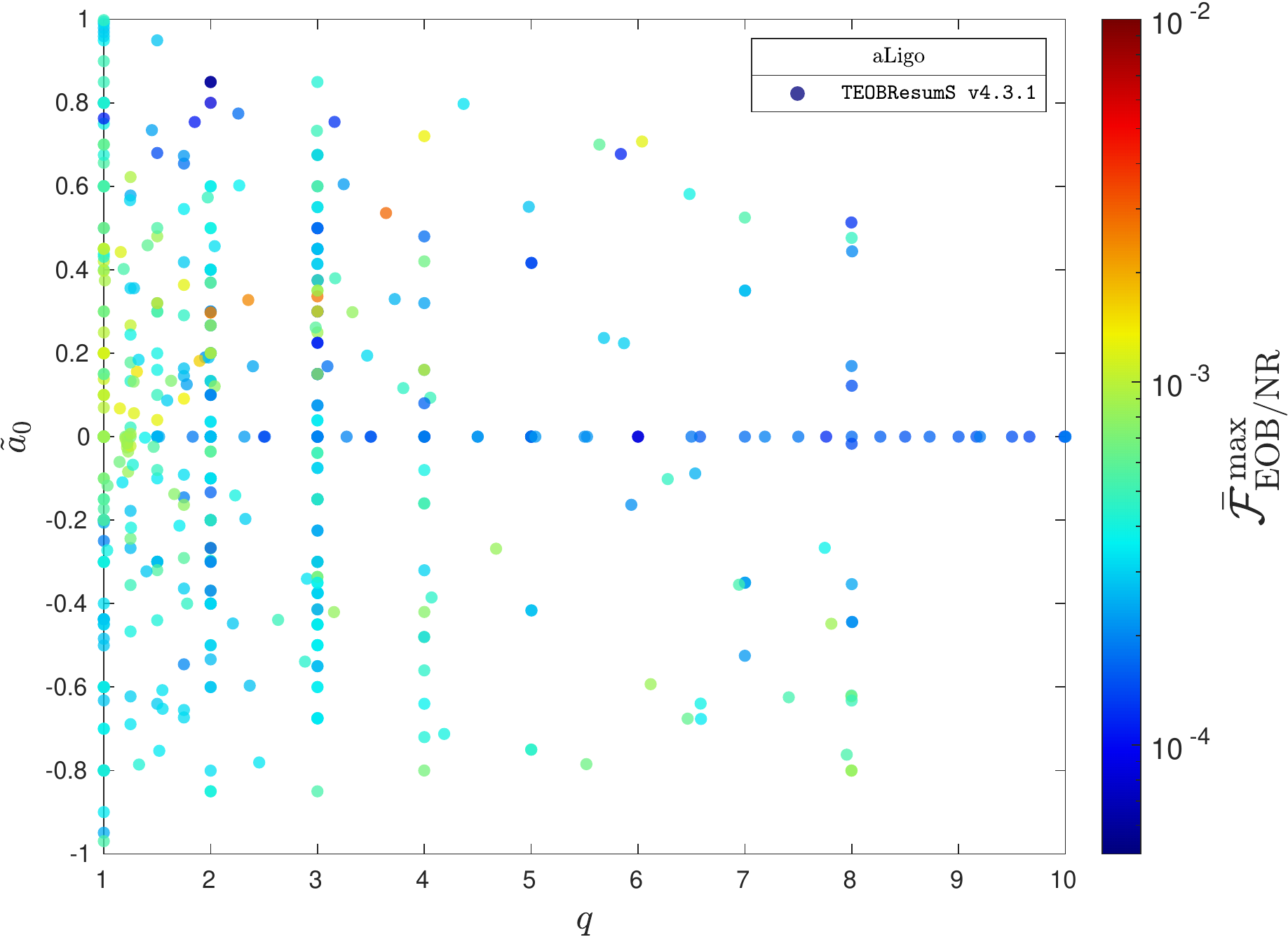}            
         \includegraphics[width=0.33\textwidth]{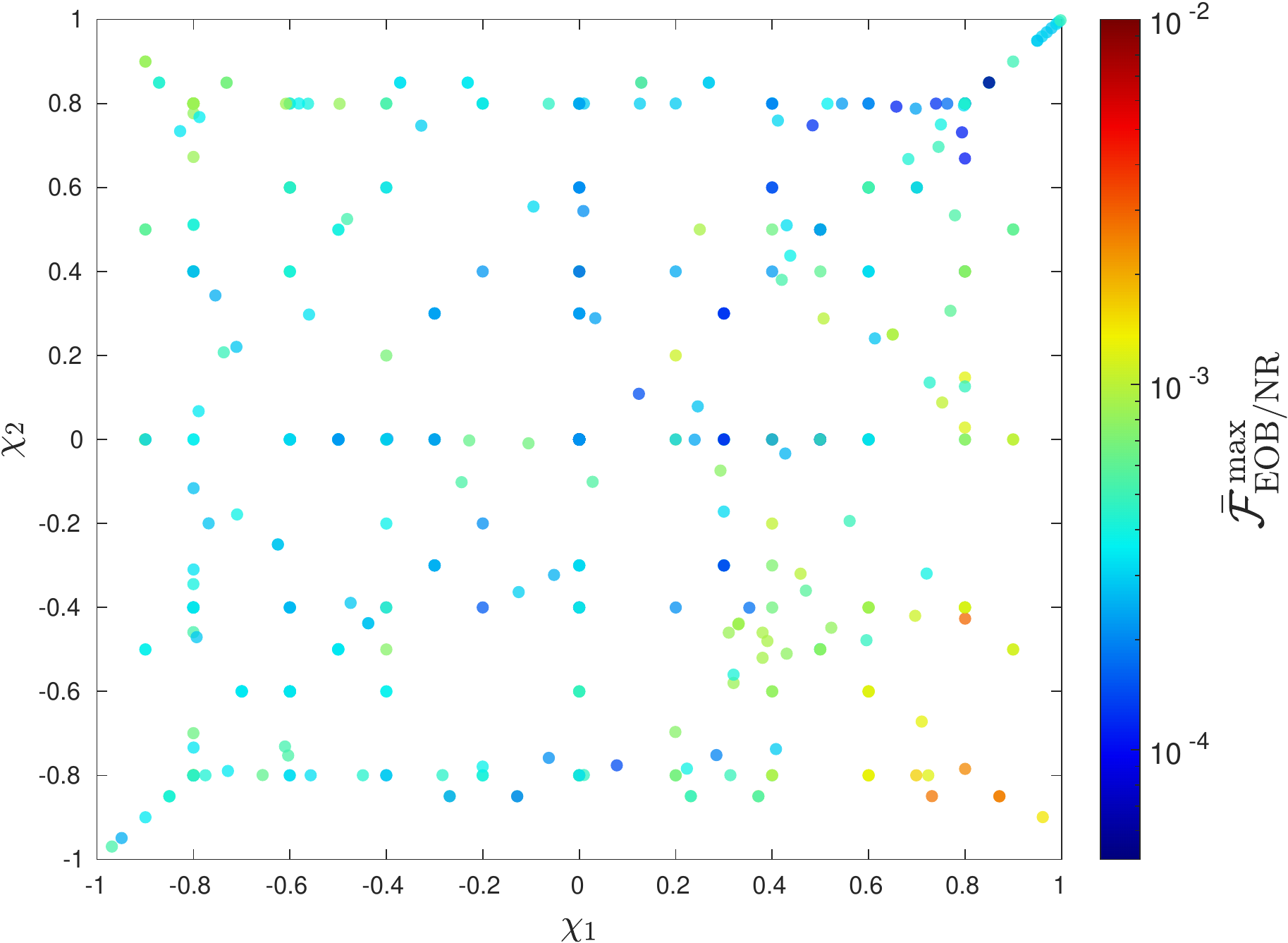}
	\caption{\label{fig:max_global_v4p3p1}Performance of \TEOBResumSvfourthreeone{} looking at the EOB/NR unfaithfulness
	over 534 datasets of the SXS catalog. Despite the global improvement, there are still four outliers above the 0.2\% level}
\end{figure*}
\begin{figure*}[t]
	\center
	\includegraphics[width=0.31\textwidth]{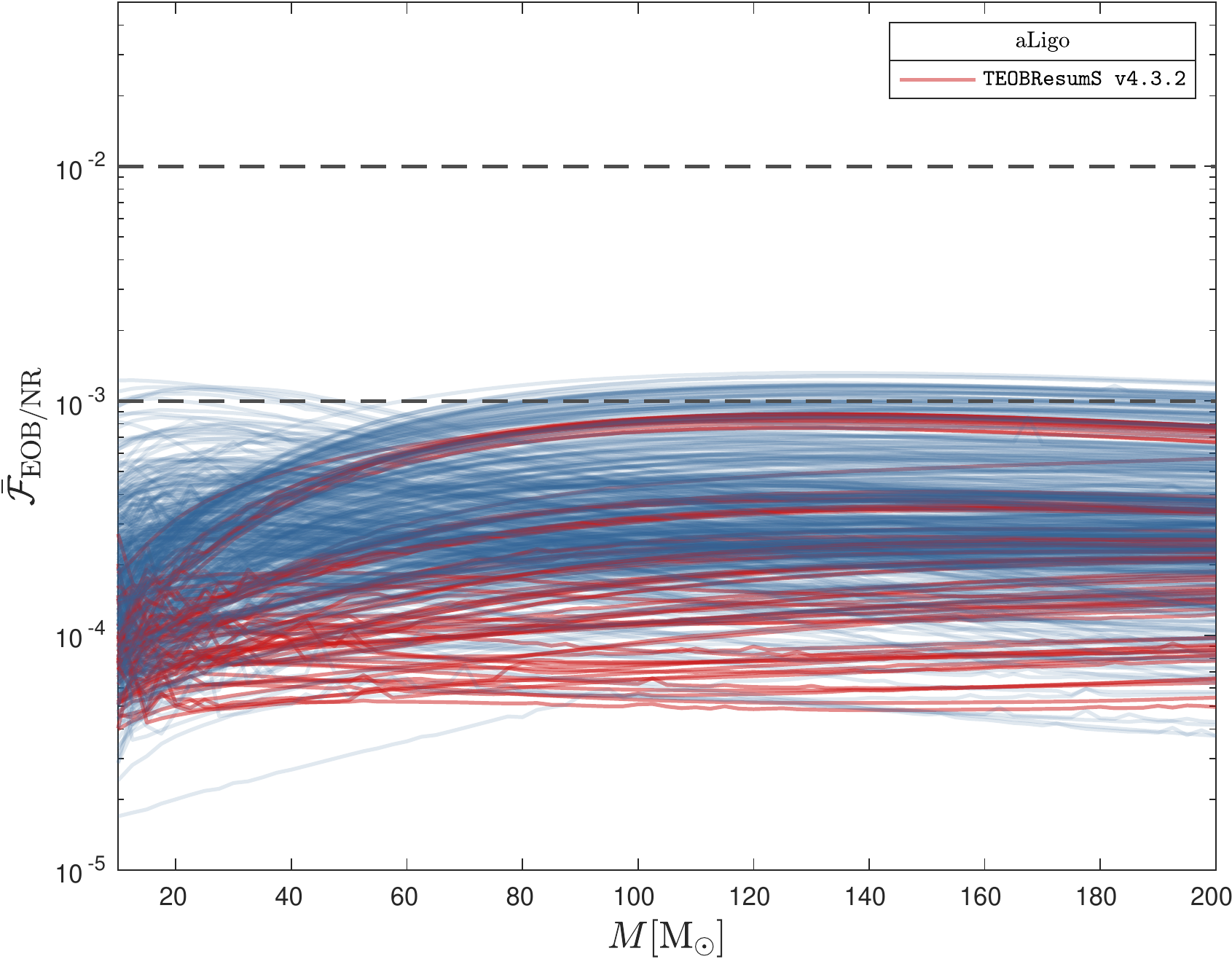}
	\includegraphics[width=0.33\textwidth]{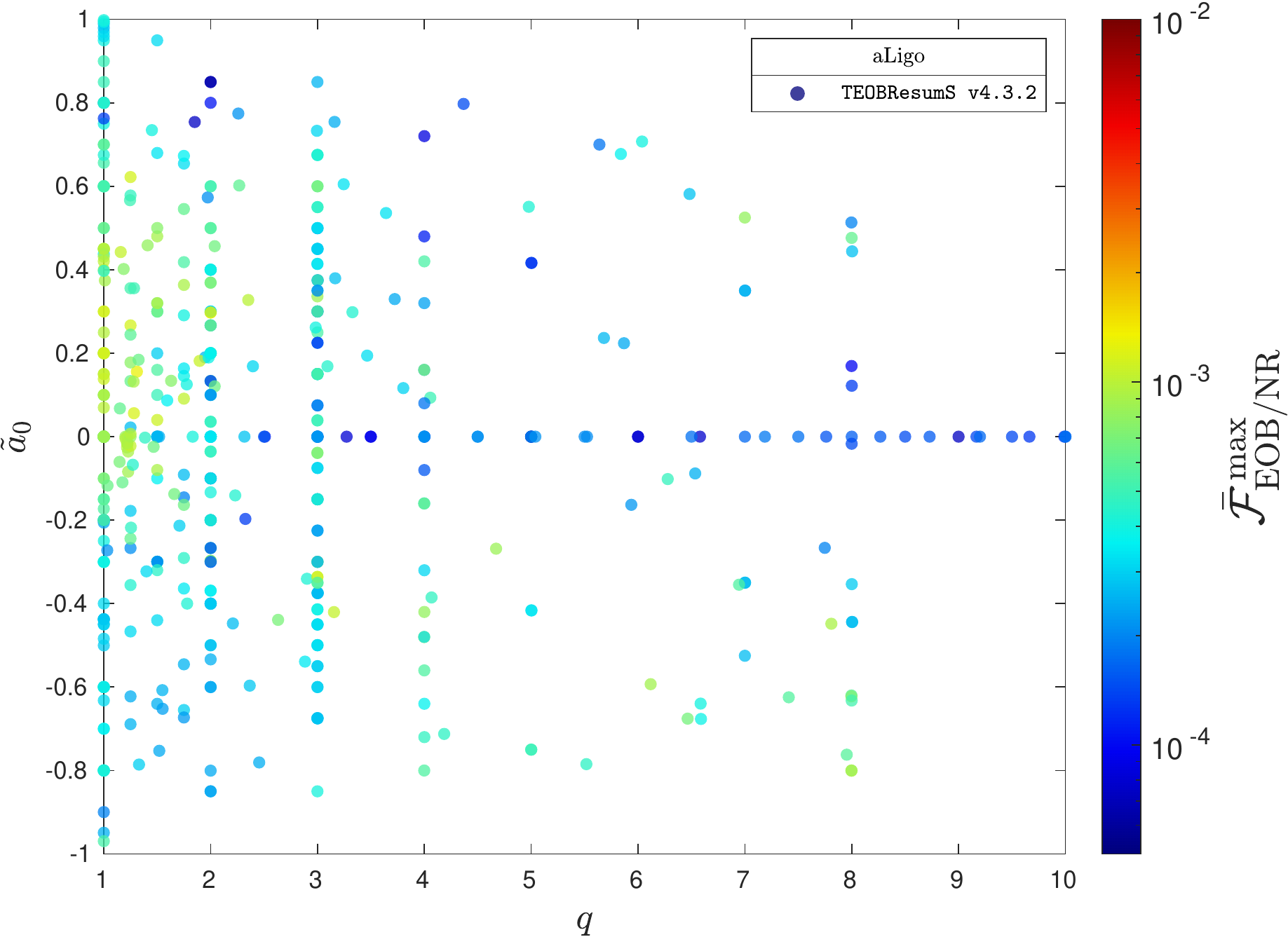}       
         \includegraphics[width=0.33\textwidth]{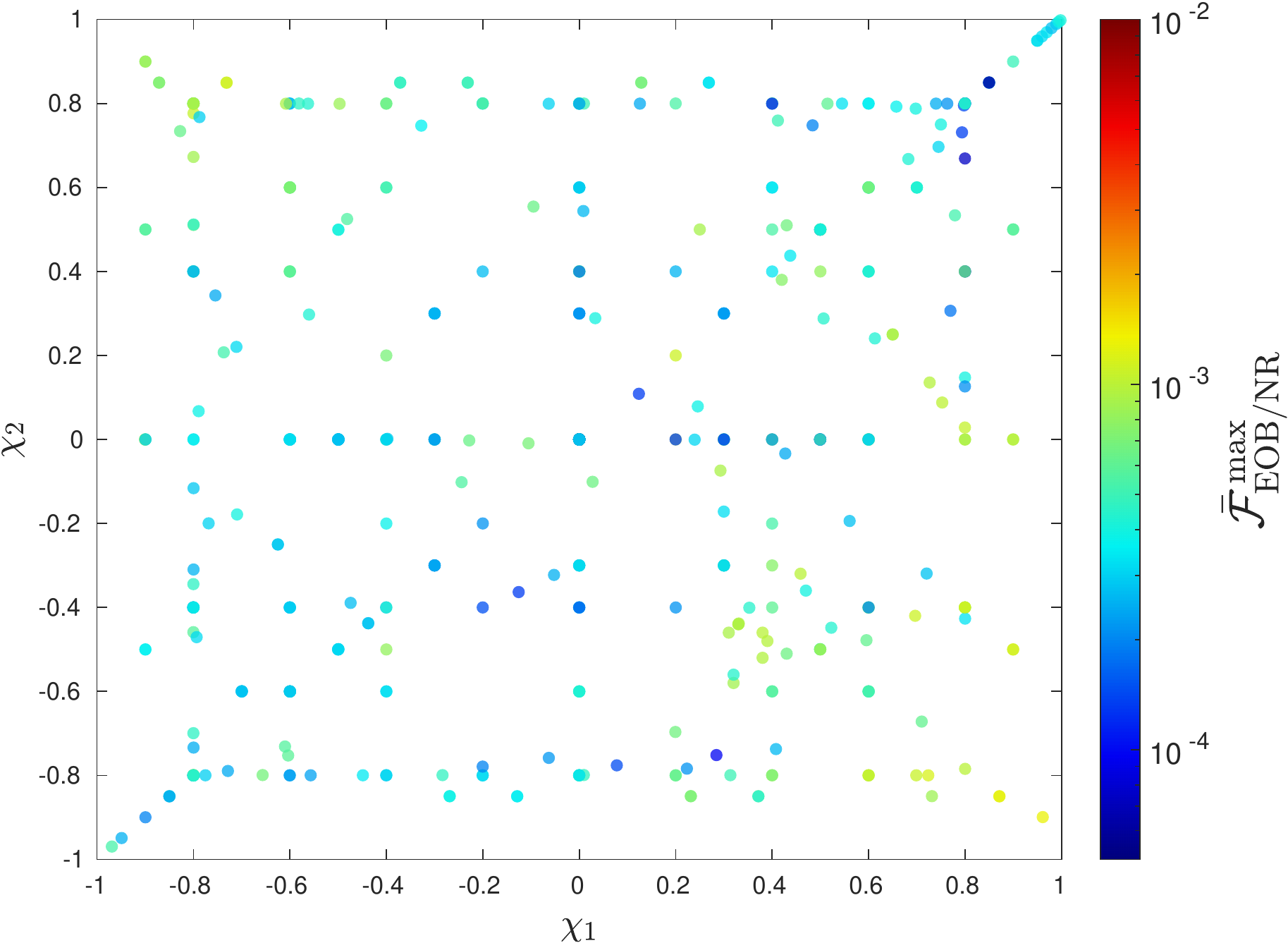}    \\
         \includegraphics[width=0.33\textwidth]{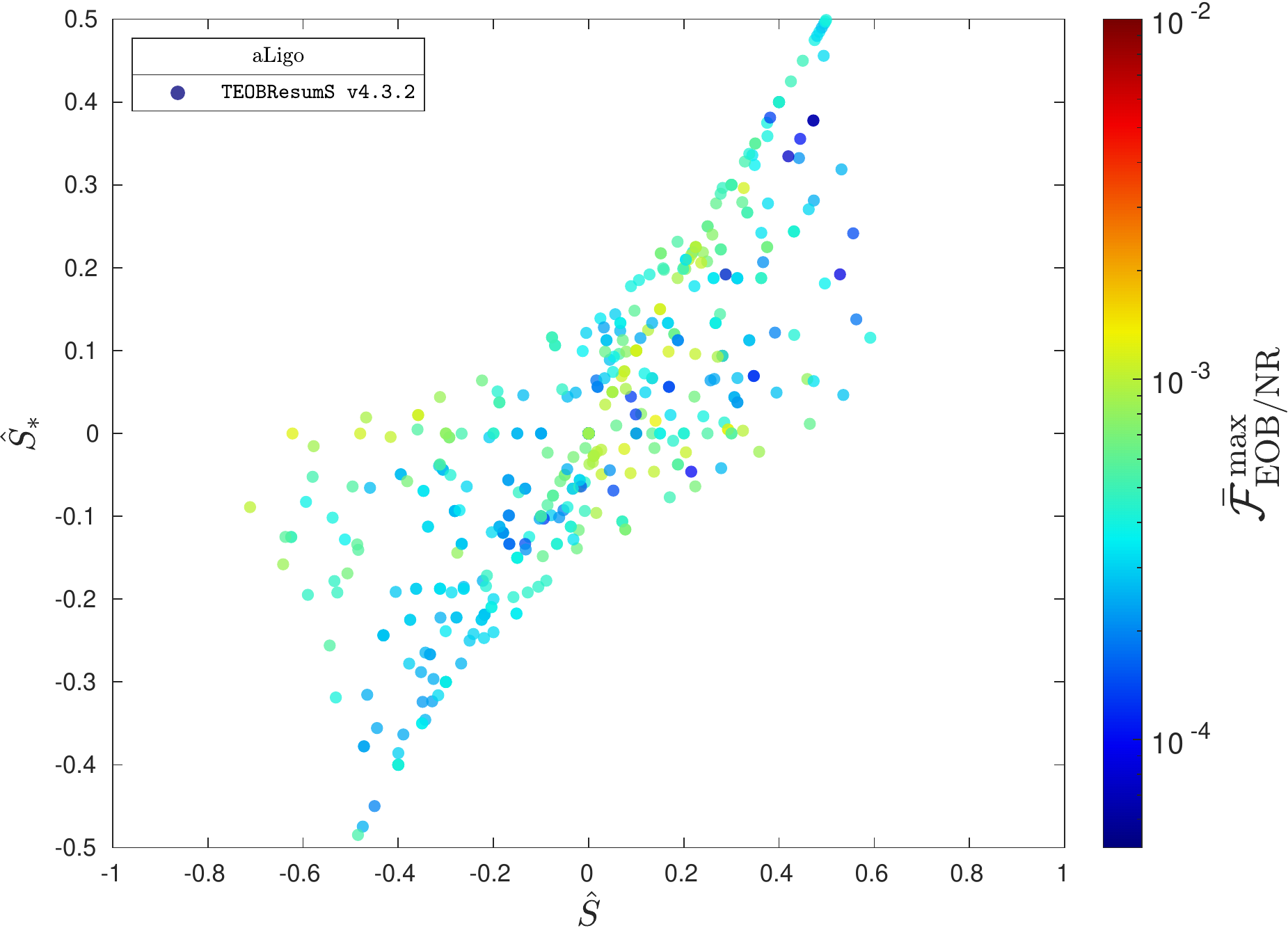}   
	  \includegraphics[width=0.32\textwidth]{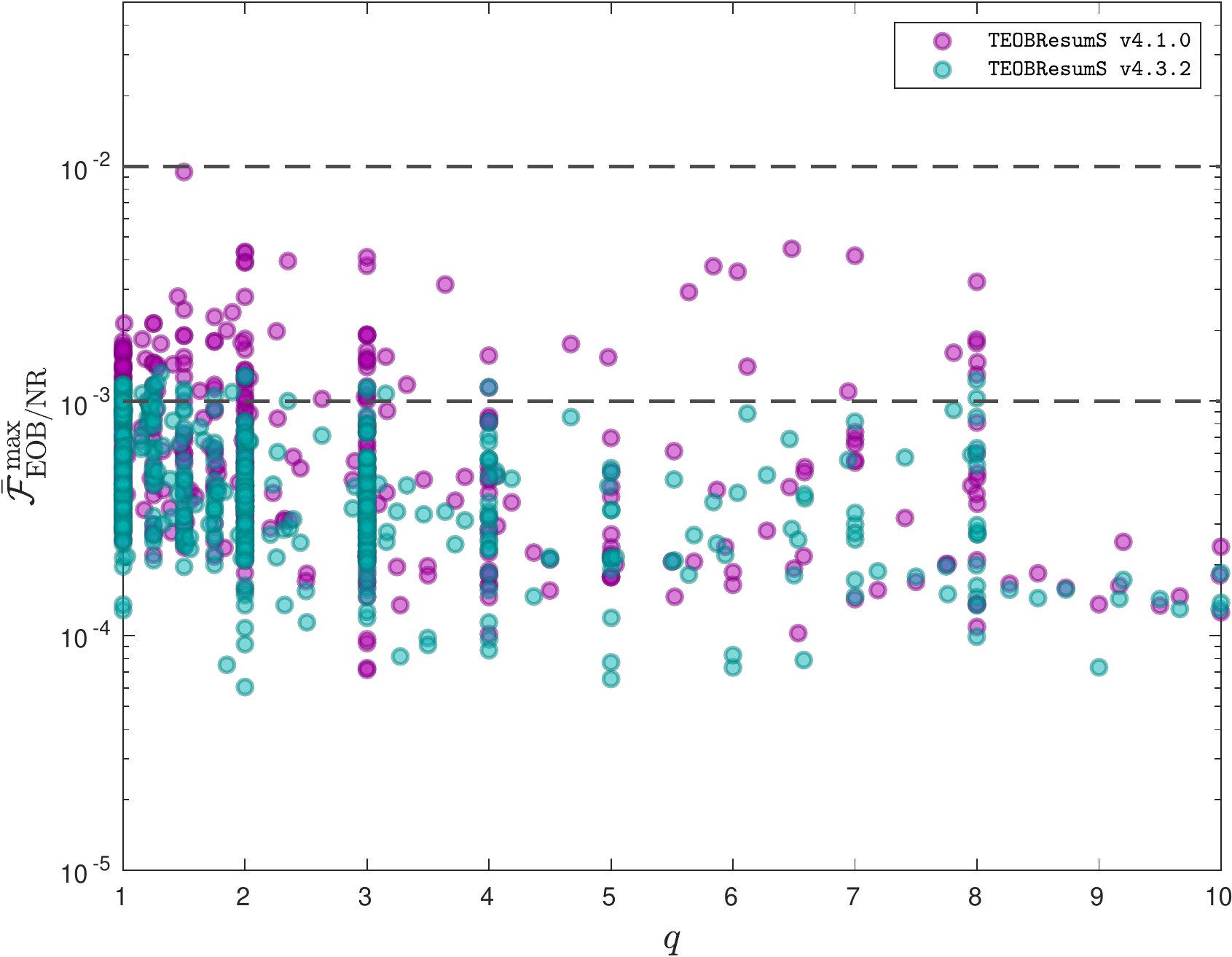}	      	   	  
         \includegraphics[width=0.31\textwidth]{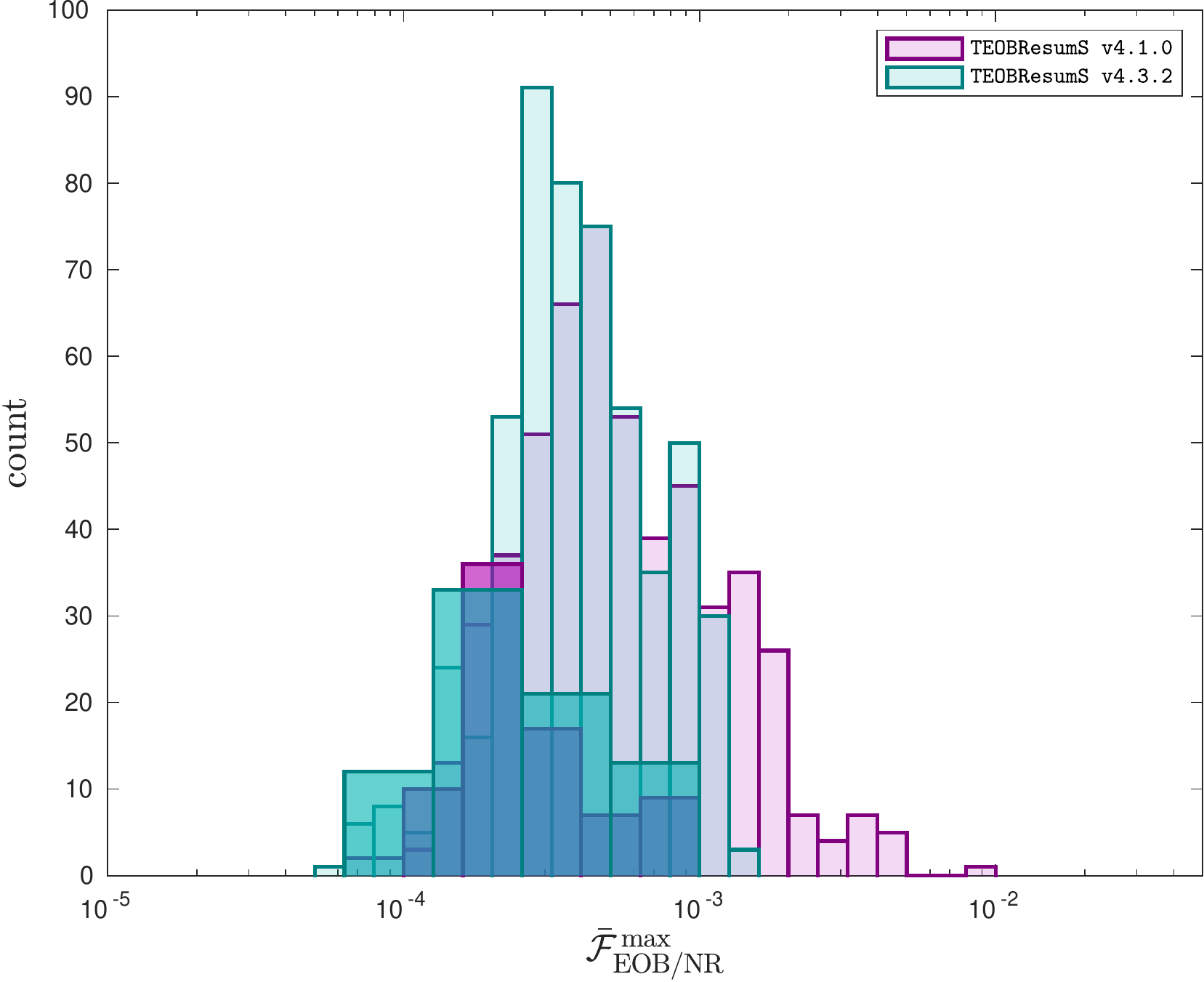}	        
	\caption{\label{fig:max_global_v4p3p2}Final result: the \TEOBResumSvfourthreetwo{} model and its performance looking at 
	the EOB/NR unfaithfulness over 534 datasets of the SXS catalog. The differences with respect the original  \TEOBResumSvfourthreeone{}
	model are: (i) the improved analytical representation of $a_6^c$; (ii) the improved analytical representation of $c_3$; 
	(iii) the improved calculation of the NQC corrections based on a more accurate analytical representation of the NR NQC determination
	point that removes the residual outliers with $0.3\lesssim\tilde{a}_0\lesssim 0.6$ found with \TEOBResumSvfourthreeone{}. 
	Top left, the EOB/NR unfaithfulness $\bar{\cal F}_{\rm EOB/NR}$. Top, middle and top right: $\bar{\cal F}_{\rm EOB/NR}^{\rm max}$ 
	versus $(\tilde{a}_0,q)$ and $(\chi_1,\chi_2)$. Bottom, left: $\bar{\cal F}_{\rm EOB/NR}^{\rm max}$  versus $(\hat{S},\hat{S}_*)$. 
	Bottom middle and right: comparing $\bar{\cal F}^{\rm max}_{\rm EOB/NR}$  of  \TEOBResumSvfour{}  
	and  \TEOBResumSvfourthreetwo{} . The solid histograms refer to nonspinning configurations.}
\end{figure*}
The performance is evaluated in Fig.~\ref{fig:max_global_v4p3p1}. Globally, the number of dataset below
$0.1\%$ unfaithfulness is slightly larger than before. However, we find now {\it four} outliers above $0.2\%$,
see Table~\ref{tab:outliers_v4p3p1}. Notably, some are the same as Table~\ref{tab:outliers}. These suggest
that, whatever is incorrect in the EOB model in this special corner of the parameter space, it should not be
related to the determination of $c_3$. We thus performed a more careful investigation of the waveform properties 
for a particular configuration, the SXS:BBH:0258 one. We concluded that the waveform inaccuracies effectively 
come from the determination of the NQC corrections to the waveform and {\it not} from the choice of $c_3$.
Let us discuss this in some detail. We remind the reader that the determination of the NQC parameters 
$(a^{22}_1,a^{22}_2,b^{22}_1,b^{22}_2)$ needs some representation of the NR amplitude and frequency
(and first derivatives) $(A_{22},\dot{A}_{22},\omega_{22},\dot{\omega}_{22})$ across the parameter space.
More precisely, Ref.~\cite{Nagar:2020pcj} represented these quantities via NR-informed fits, in the form discussed
in Appendix~C.5. By computing the waveform using the {\it exact} NR values of $(A_{22},\dot{A}_{22},\omega_{22},\dot{\omega}_{22})$
for SXS:BBH:0258 we concluded that these fits are not sufficiently accurate in this corner of the parameter space
and that are thus responsible of all outliers. Luckily, a way to overcome this problem was already found
and discussed in Ref.~\cite{Nagar:2020pcj}, i.e. computing the quantities useful for the NQC determination
directly from the NR-informed model for the postmerger discussed therein. If we do so, the inaccuracies are
largely reduced and the resulting waveform is much closer to the one that employs the exact values.
In doing so, we realized that the former $c_3$ fits need a bit of adjustment, that is obtained by simply changing
the first-guess value for SXS:BBH:1432 from $c_3^{\rm first-guess}=25$ to $c_3^{\rm first-guess}=21$. This
eventually yields new fitting coefficients for $c_3^{\neq}$ and a new model, dubbed \TEOBResumSvfourthreetwo{},
as detailed in the last row of Table~\ref{tab:models_vs_c3}.
\begin{figure}[t]
	\center
	\includegraphics[width=0.4\textwidth]{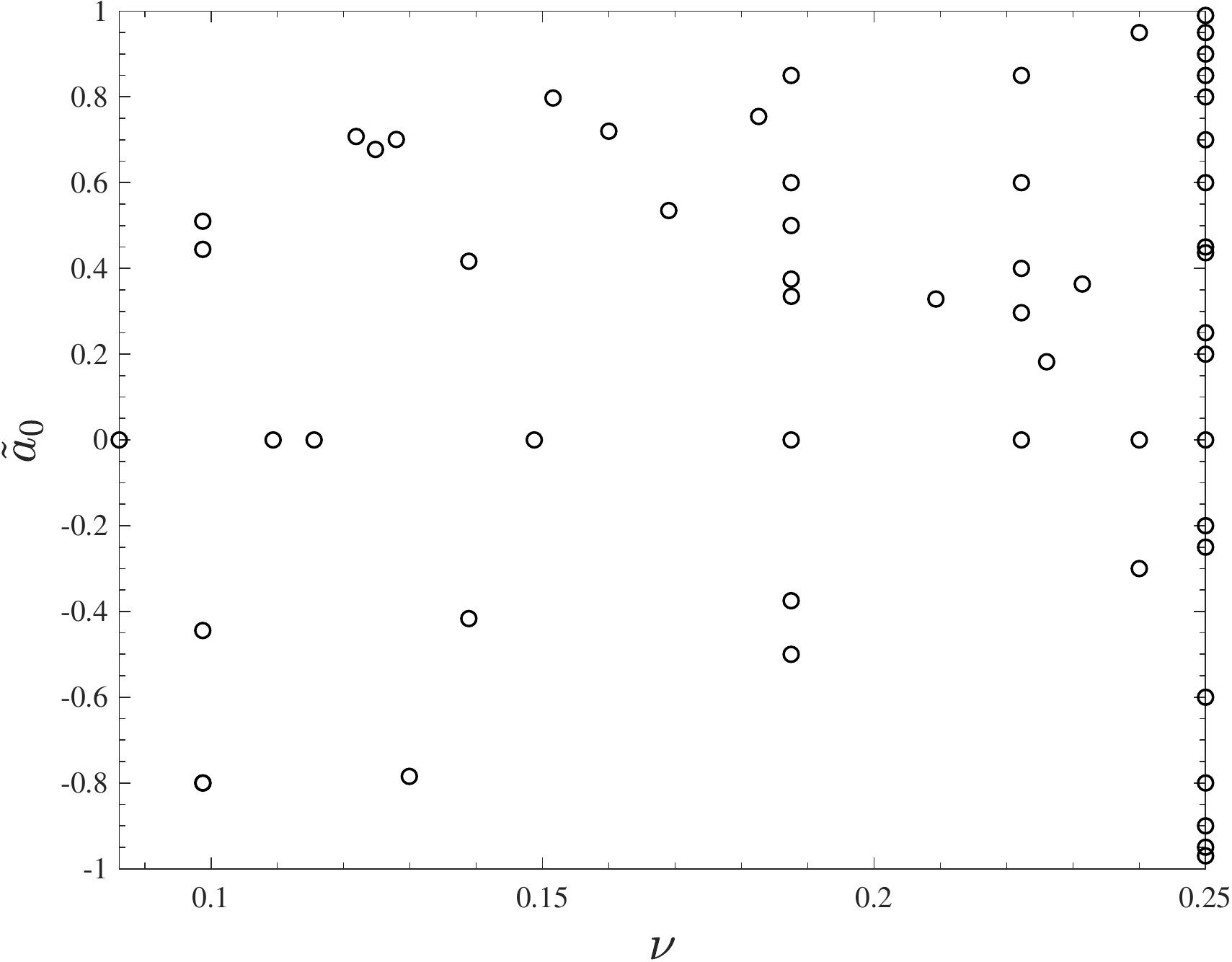}
	\caption{\label{fig:parspace}Location in the $(\nu,\tilde{a}_0)$ space of the 55 SXS simulations
	used to NR-inform the $(a_6^c,c_3)$ functions for \TEOBResumSvfourthreetwo{}.}
\end{figure}
The performance of the model versus the SXS catalog is evaluated in Fig.~\ref{fig:max_global_v4p3p2}.
The top row of the figure shows the same analyses as above. We see now that the (minor) changes implemented
in the model are such that $\bar{F}_{\rm EOB/NR}^{\rm max}\sim 0.1\%$ all over the catalog. 
A more careful look at the other plot is in order. Combining the middle and right panel of the first row of 
Fig.~\ref{fig:max_global_v4p3p2} one sees that the model is (relatively) less accurate for $1\leq q \leq 2$
and large spins. Possibly, this is related to the fact that, to determine $c_3$ we used just $\sim 10$ datasets 
in that mass ratio range, with spins that are not extremely high. The use of more datasets should be helpful
to further reduce $\bar{F}_{\rm EOB/NR}^{\rm max}$. Modulo this little island, it is worth stressing that the
model performs equally well either at {\it low} and at {\it large} mass ratios. This is a priori expected, also 
on the basis of the findings of Ref.~\cite{Nagar:2022icd}. It thus does not seem an issue, at least for \TEOBResumS{},
to produce accurate waveforms for large mass ratios. We will come back to discussing this topic below.
Let us additionally comment the bottom row of Fig.~\ref{fig:max_global_v4p3p2}.
To have a better handle on where the improvements are needed most, it is instructive to plot $\bar{\cal F}_{\rm EOB/NR}^{\rm max}$
versus $(\hat{S},\hat{S}_*)$ that are the actual spin variables entering the Hamiltonian. The plot indicates that
the criticalities mostly occur around the $\hat{S}=\hat{S}_*$ and $\hat{S}_*=0$ lines. This suggests that
a new determination of $c_3$ using NR data along these lines could easily improve the model further.
Finally, the middle and rightmost panels in the bottom row of Fig.~\ref{fig:max_global_v4p3p2} show
the improvement with respect to the original \TEOBResumSvfour{} model.  
For convenience, Fig.~\ref{fig:parspace} shows the location in the $(\nu,\tilde{a}_0)$ plane of
the 55 SXS simulations used to determine $(a_6^c,c_3)$.

\subsection{Higher modes: comparisons with NR surrogates}
\label{sec:HM}
\begin{figure*}[t]
	\center
	\includegraphics[width=0.45\textwidth]{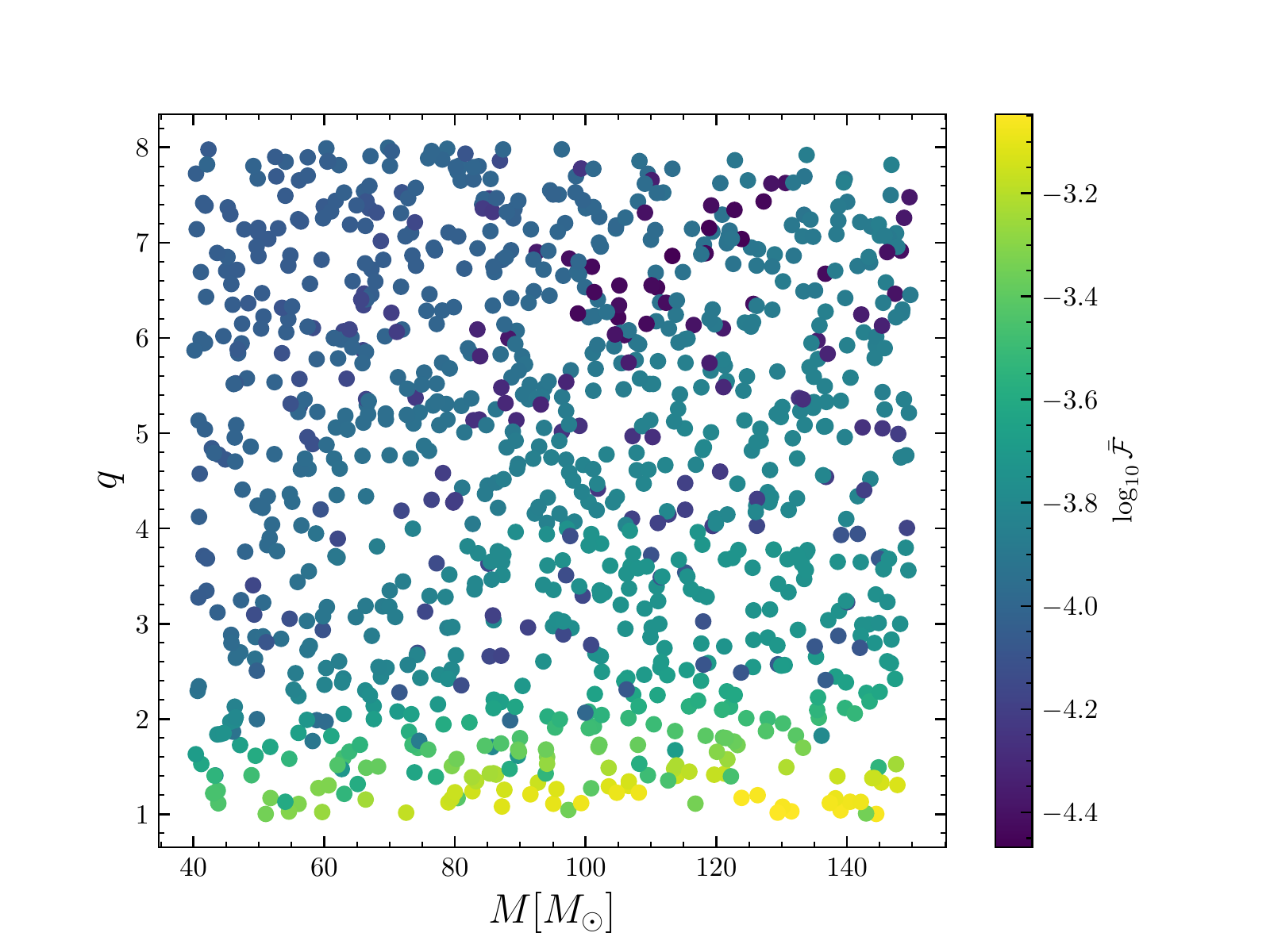}
	\includegraphics[width=0.45\textwidth]{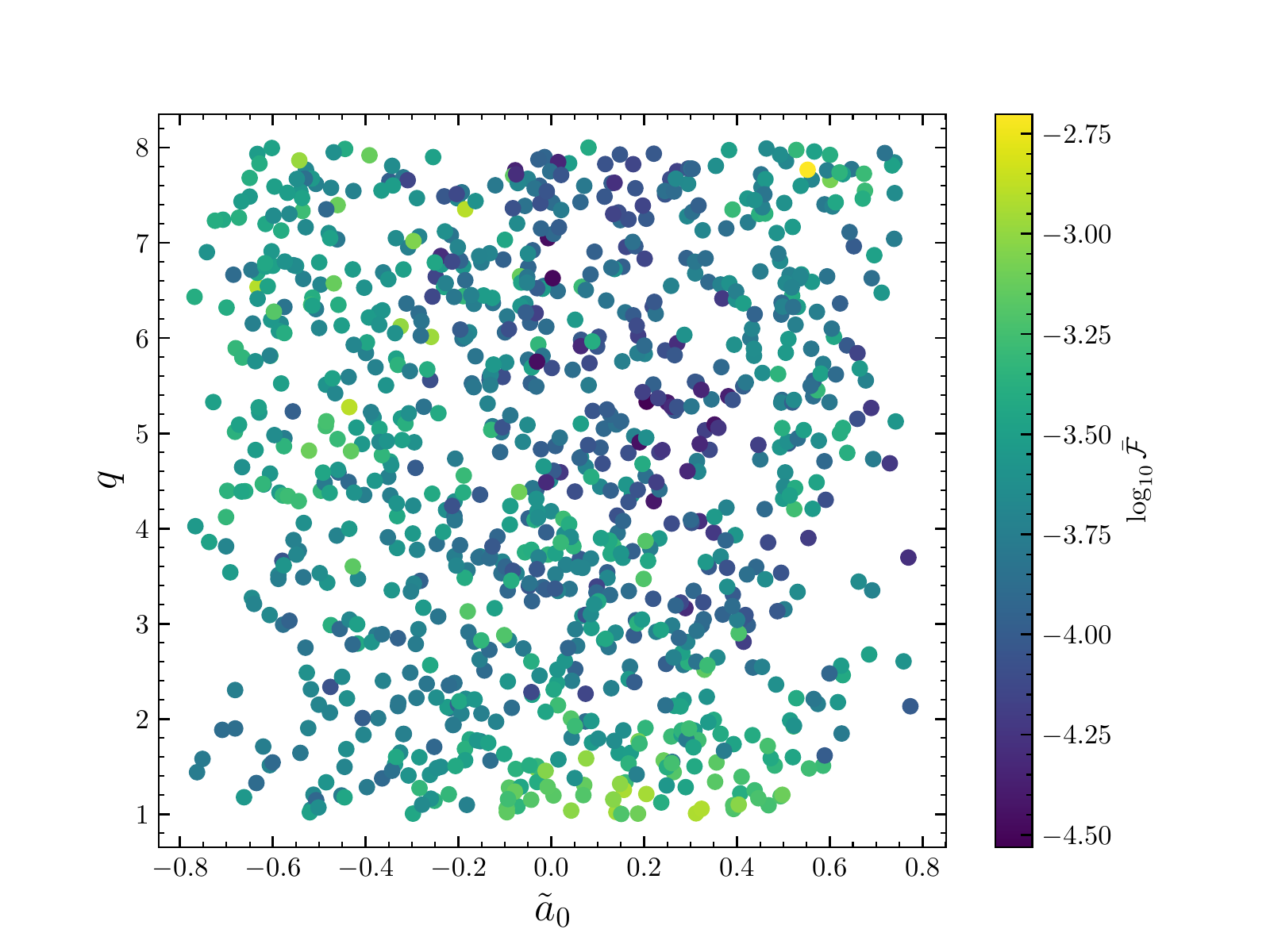}
	\caption{\label{fig:teob_surrogate_22}Left panel: nonspinning configurations. Right panel: aligned spin configurations.
	Performance of the \TEOBResumSvfourthreetwo{} model against the NR surrogate 
	{\tt NRHybSur3dq8}, including only the $(\ell,|m|) = (2,2)$ mode. We consider systems with mass ratio $q \in [1, 8]$, 
	total mass $M \in [40, 140] \Msun$ and dimensionless spins $|\chi_i|<0.8$ (right panel).  The maximal values of 
	$\bar{\cal F{}}_{\rm EOB/NR}$ correspond to high mass systems, for which the merger-ringdown gives the largest in-band 
	contribution, and systems with large positive effective spin $\tilde{a}_0$. Notably, however, the largest values of 
	unfaithfulness is at most $2 \times 10^{-3}$ for both scenarios.}
\end{figure*}
\begin{figure}
	\includegraphics[width=0.42\textwidth]{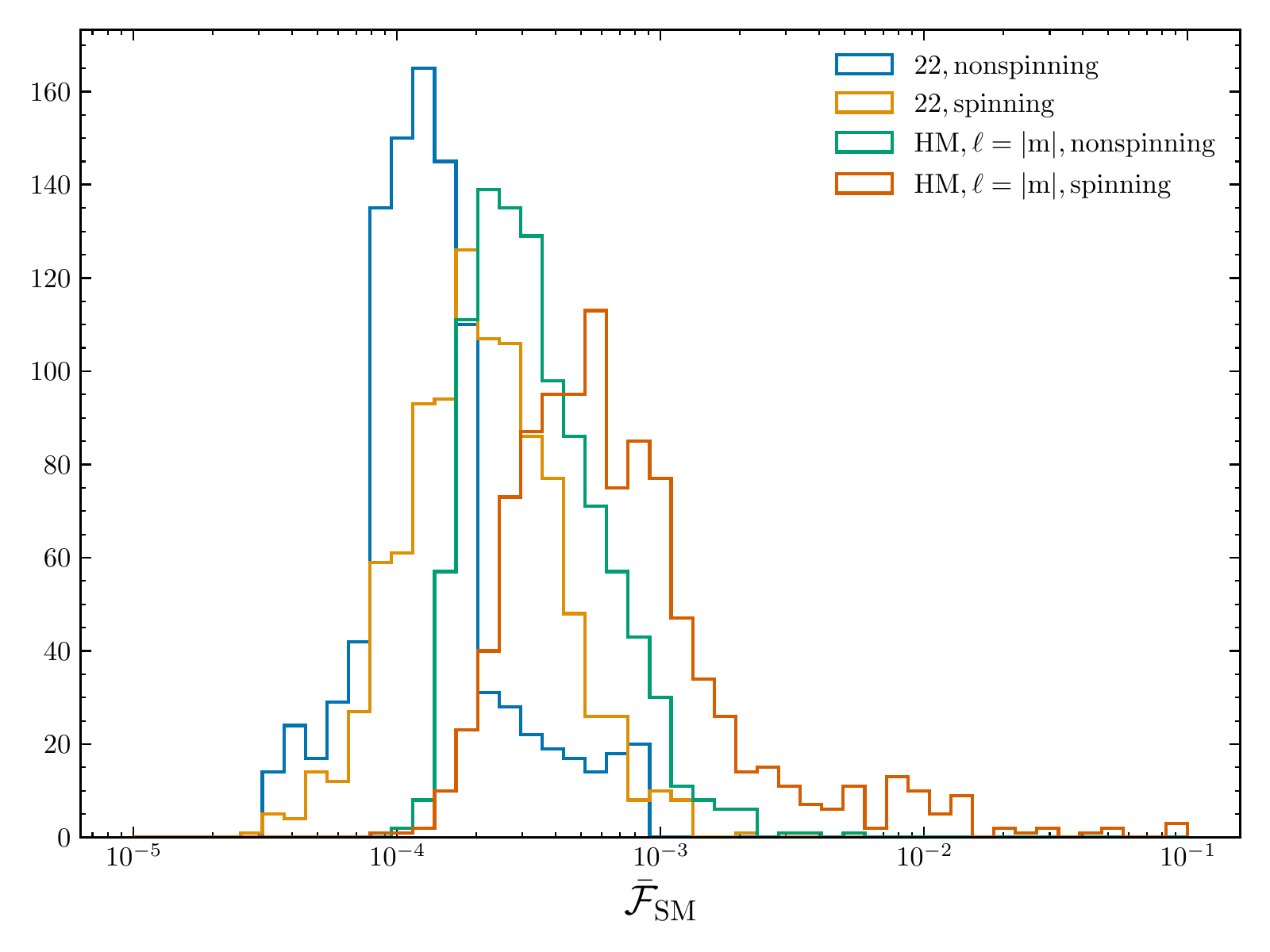}
	\caption{\label{fig:teob_surrogate_summary}Summary of all the mismatches computed in this section between \TEOB{} 
	and {\tt NRHybSur3dq8}. Even when including progressively more information (subdominant modes, spins), the performance of \TEOB{} is overall satisfactory, with 
	$\sim 97\%$ of mismatches below $1\%$ and $75\%$ below $0.1\%$.}
\end{figure}

\begin{figure}
	\includegraphics[width=0.4\textwidth]{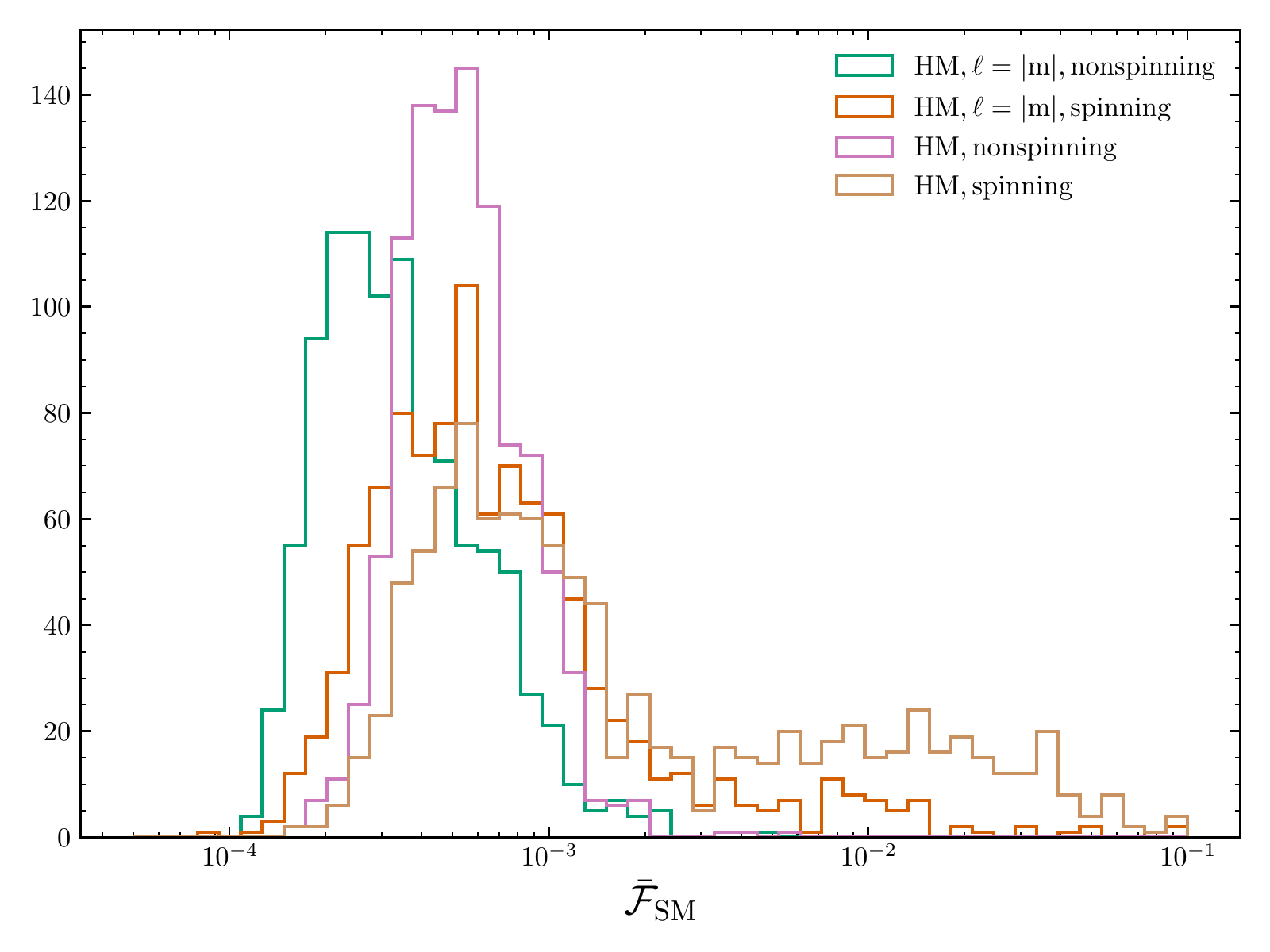}
	\includegraphics[width=0.4\textwidth]{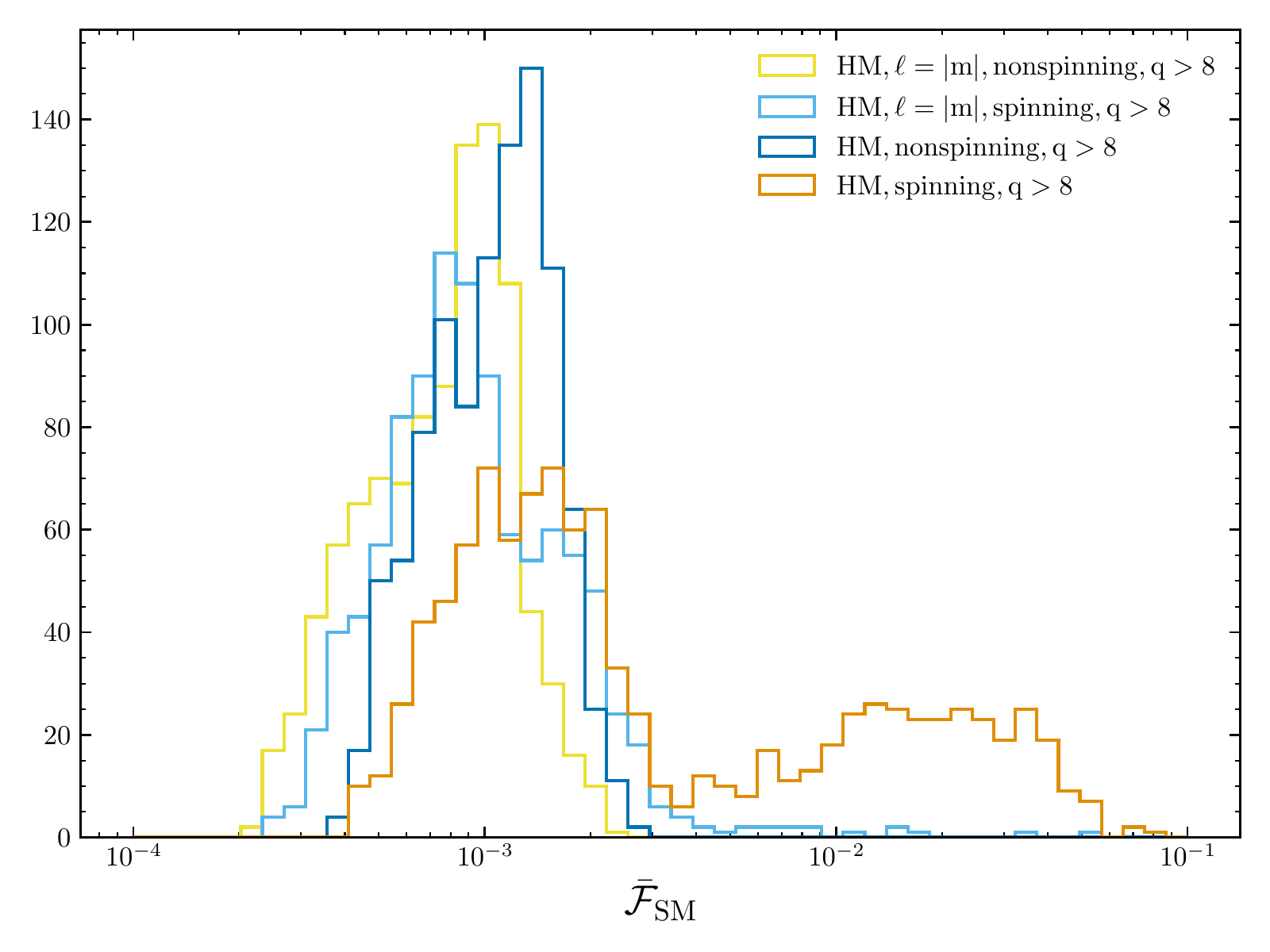}	
	\caption{\label{fig:teob_surrogate_summary_HM}Comparison between mismatches computed between \TEOBResumSvfourthreetwo{} 
	and {\tt NRHybSur3dq8} (top) or \nrsurqfifteen~(bottom) with $\ell = |m|$ modes only or with $(2,2)$, $(2,1)$, $(3,3)$, $(3,2)$, 
	$(4,4)$ and $(5,5)$ modes. The inclusion of the $(2,1)$ and $(3,2)$ modes shifts the mismatch distribution to higher values, 
	especially when considering spinning binaries. This is due to the performance of the $(2,1)$ and $(3,2)$ modes, that 
	are known to display unphysical behavior when spins are large and anti-aligned with the orbital angular 
	momentum~\cite{Nagar:2020pcj}.}
\end{figure}

\begin{figure}[t]
	\center
	\includegraphics[width=0.42\textwidth]{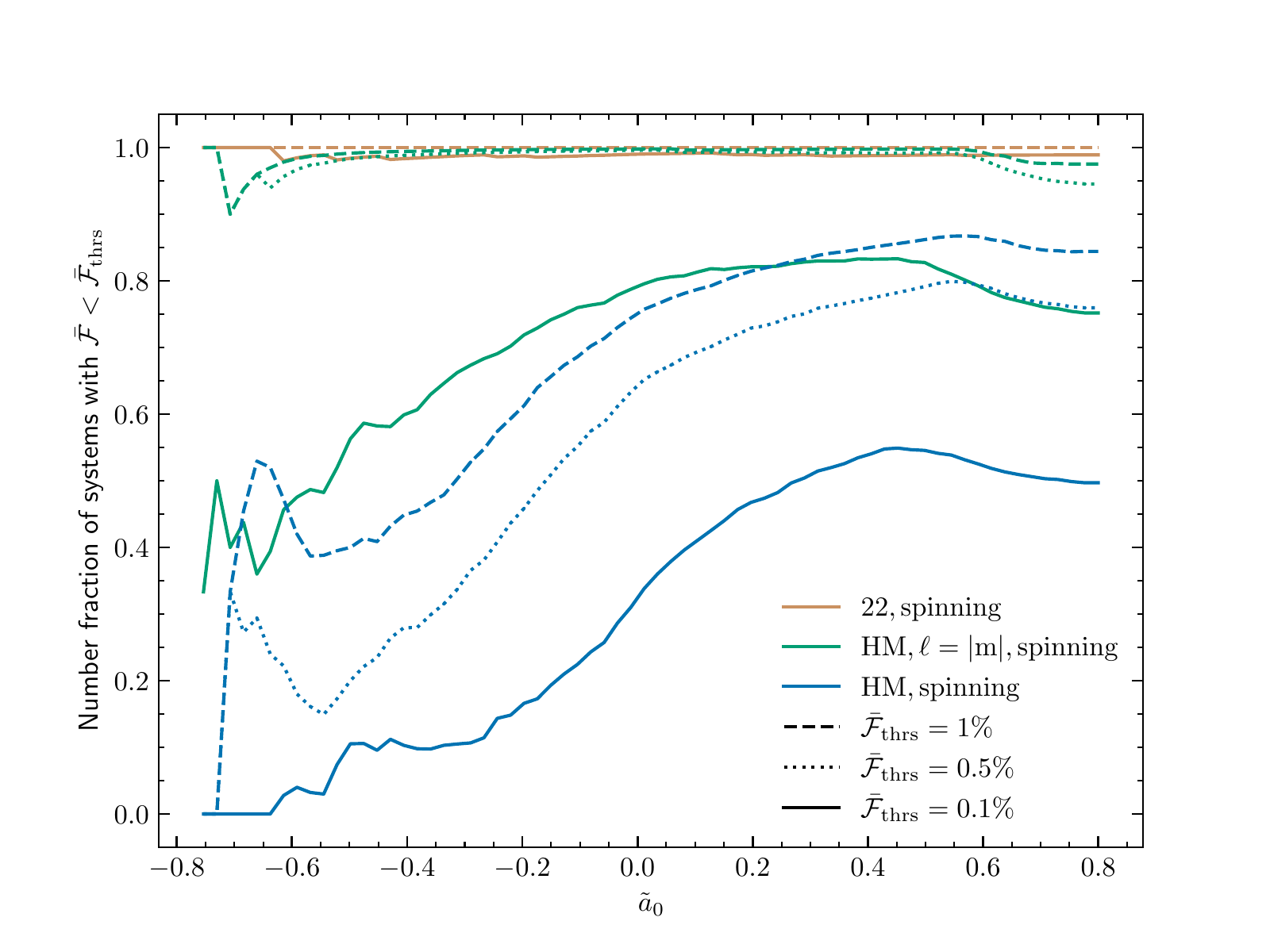}
	\caption{\label{fig:frac_HM}Summary of the   \TEOBResumSvfourthreetwo{}/{\tt NRHybSur3dq8} performance as higher
	modes are progressively included. It is shown the fraction of configurations with $\bar{\cal F}<{\bar{\cal F}}^{\rm thrs}$,
	where ${\bar{\cal F}}^{\rm thrs}$ can take different values indicated in the legend.
	The model performs less well as long as the effective spin $\tilde{a}_0$ becomes negative
	and large. This is mostly due to the (currently inefficient) modelization of modes $(2,1)$ and $(3,2)$ during the plunge up to
	merger in that corner of parameter space.}
\end{figure}
\begin{figure*}[t]
	\center
	\includegraphics[width=0.49\textwidth]{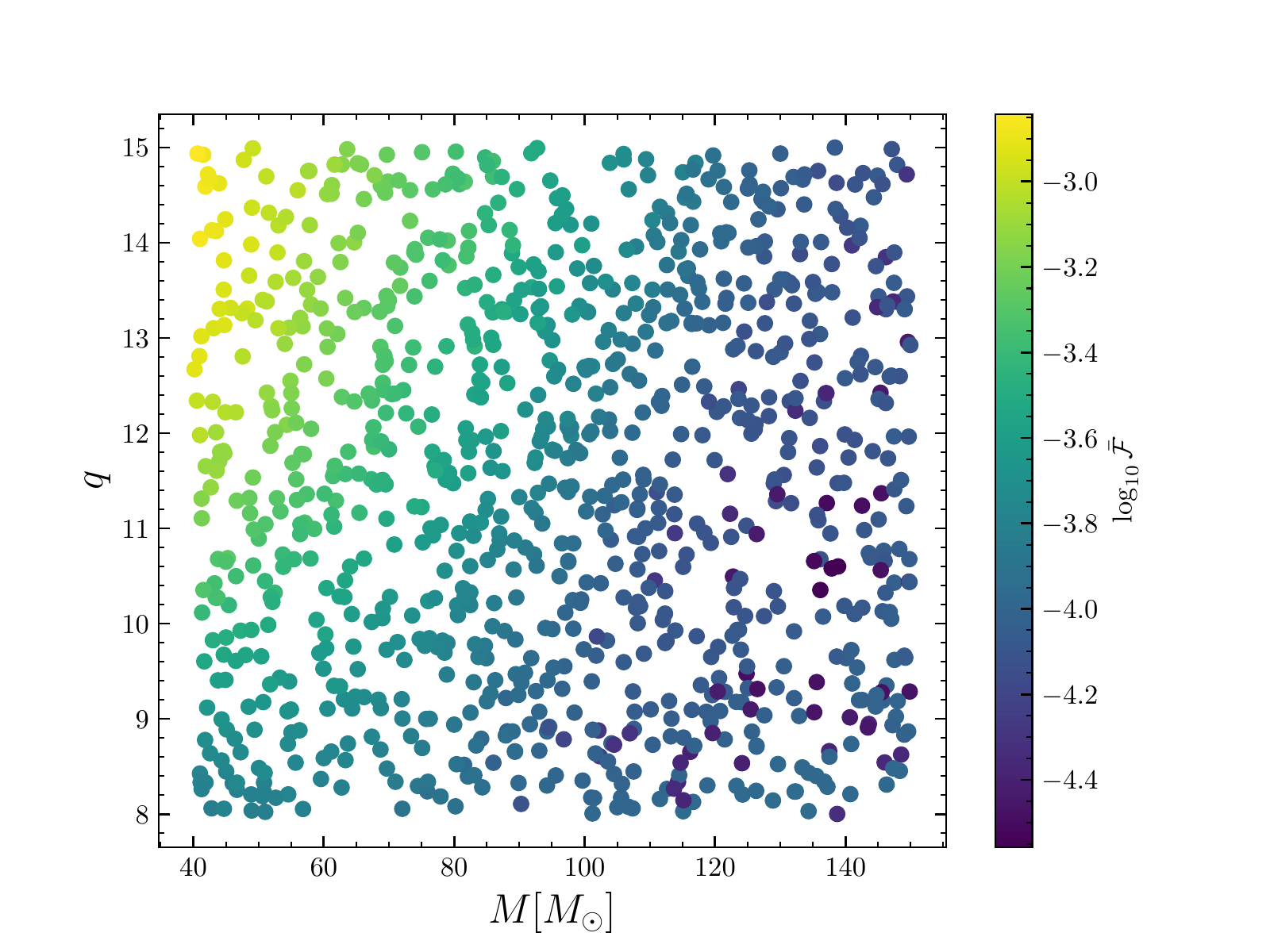}
	\includegraphics[width=0.49\textwidth]{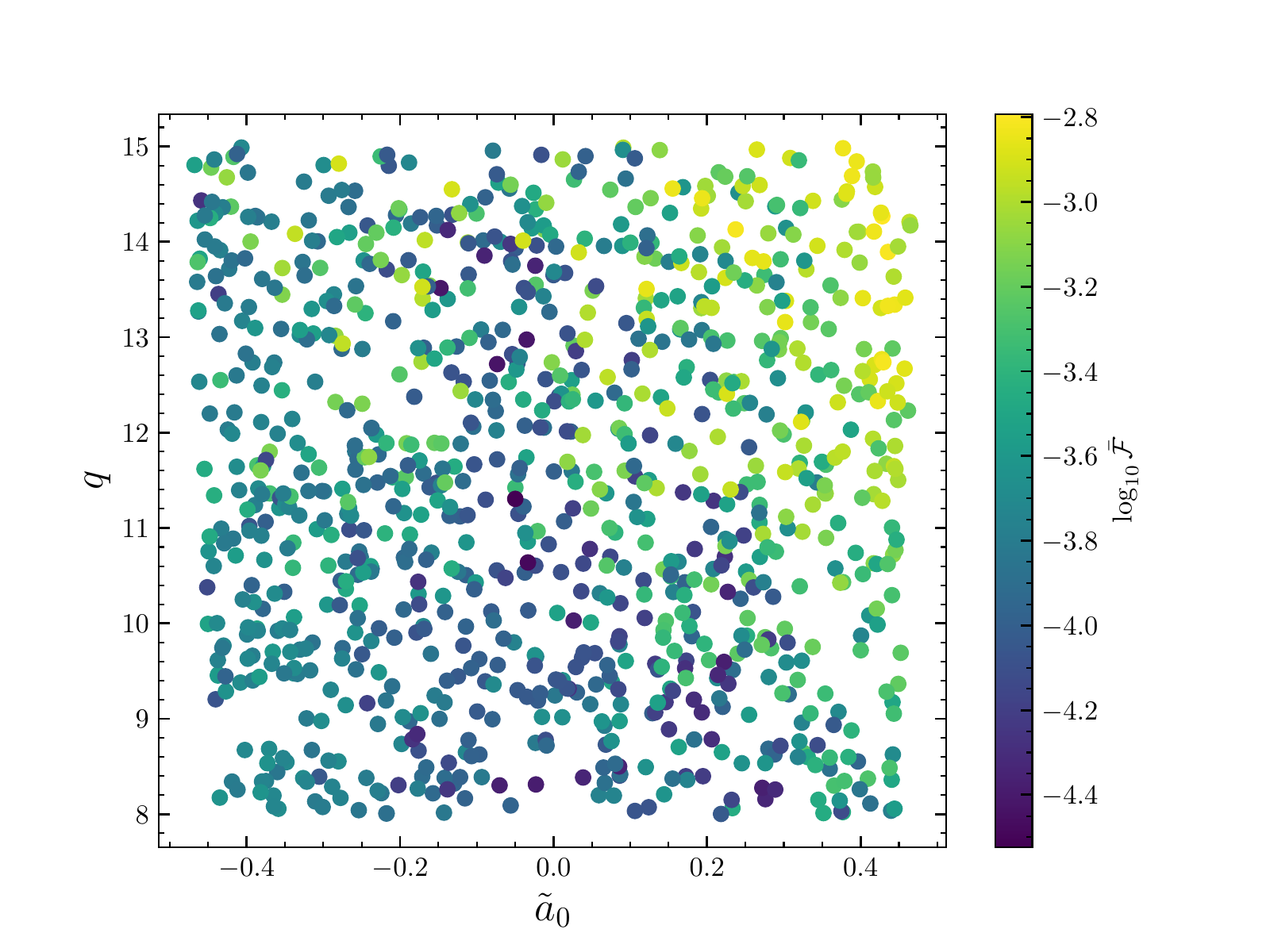}
	\caption{\label{fig:teob_surrogate_22_q15}Performance of \TEOBResumSvfourthreetwo{} against the \nrsurqfifteen~ NR surrogate,
	including only the $(\ell,|m|) = (2,2)$ mode. We consider systems with mass ratio $q \in [8, 15]$, 
	total mass $M \in [40, 140] \Msun$, zero spins (left panel) or aligned spins with dimensionless spin 
	magnitudes $|\chi_1|<0.5$ and $\chi_2 = 0$ (right panel). The largest values of unfaithfulness 
	($\bar{\mathcal{F}}\sim 10^{-3}$) are found for unequal mass systems with large, positive, spin.
	}
\end{figure*}

\begin{figure*}[t]
	\center
	\includegraphics[width=0.31\textwidth]{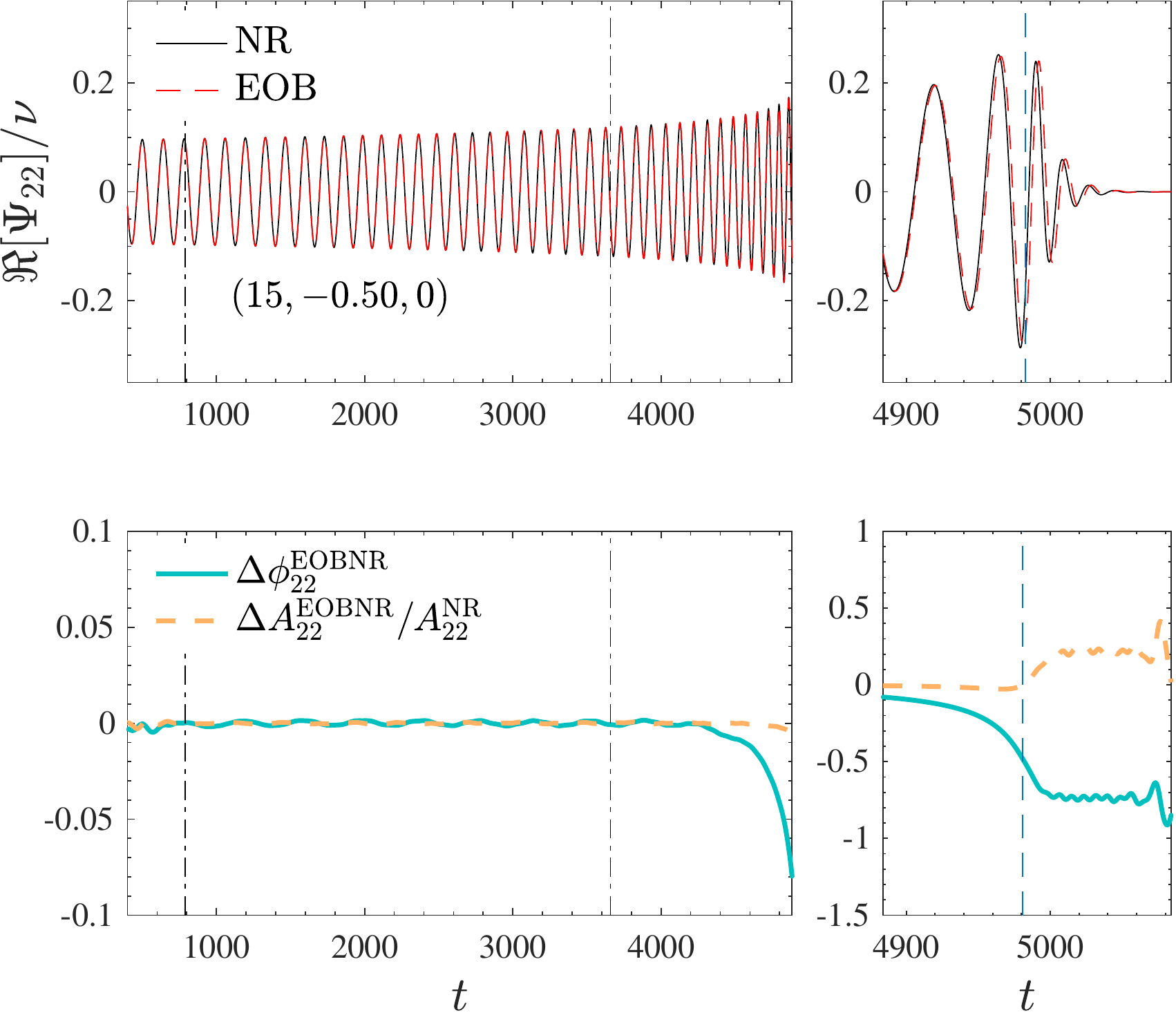}
	\includegraphics[width=0.31\textwidth]{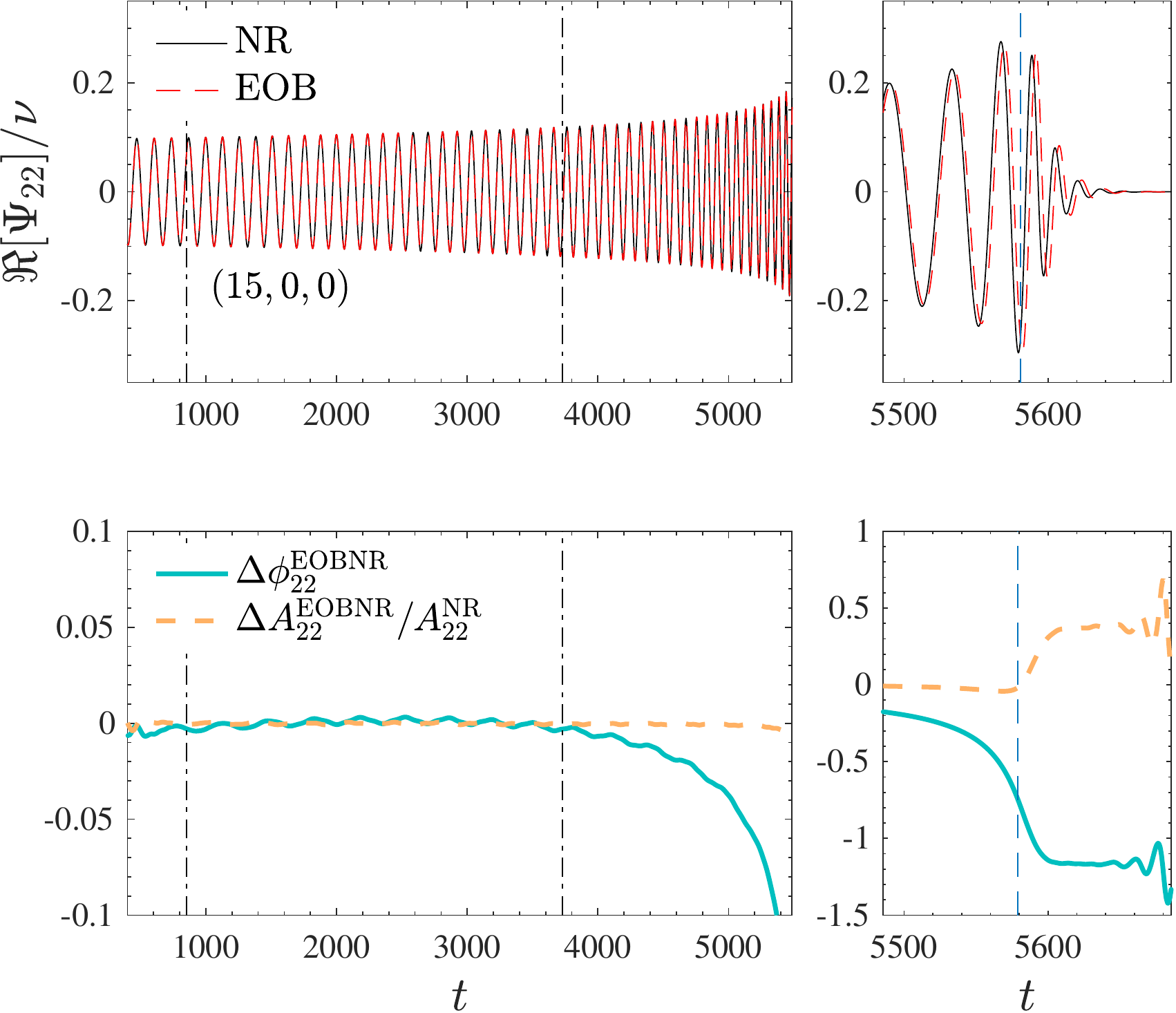}
	\includegraphics[width=0.31\textwidth]{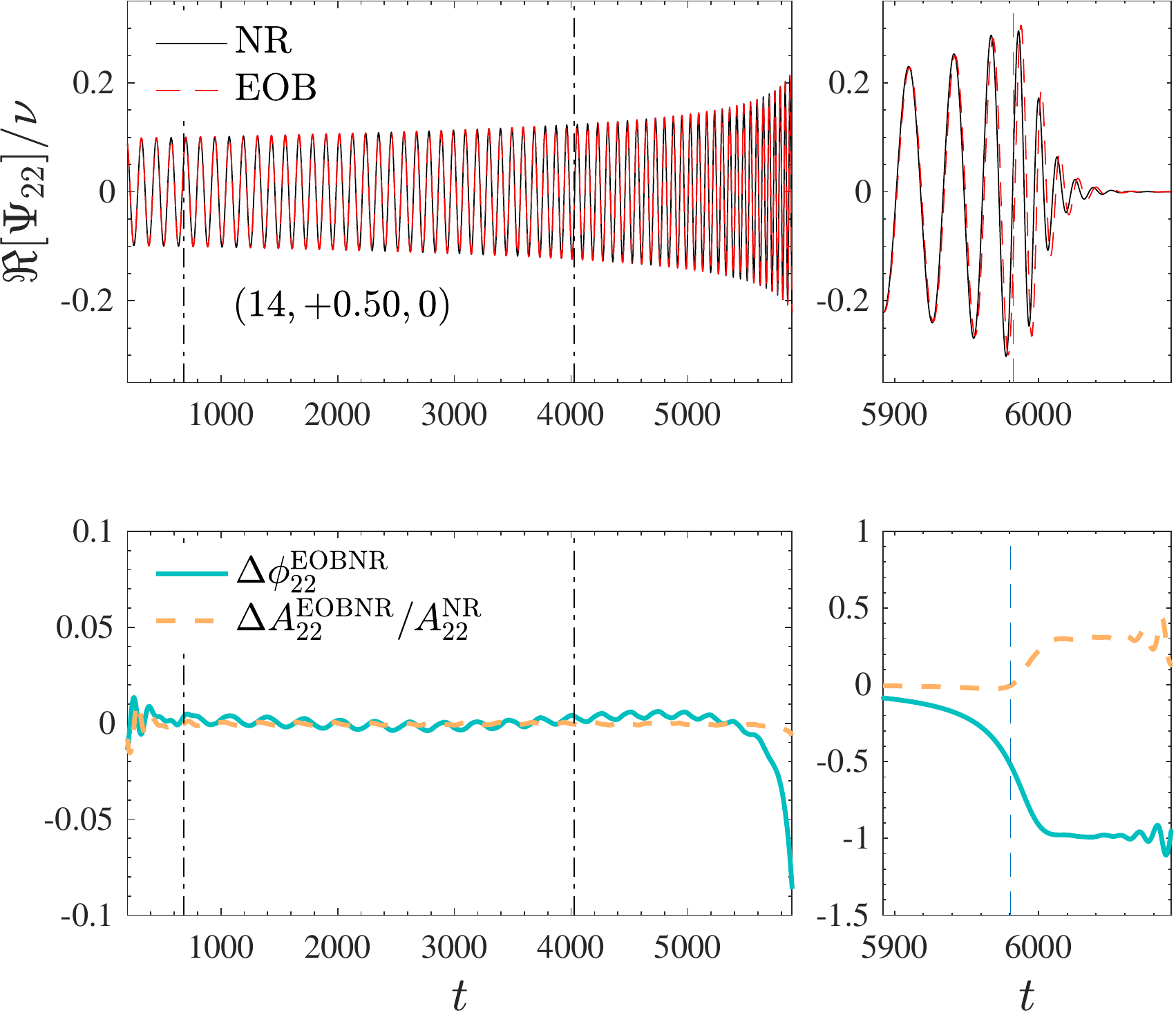}
	\caption{\label{fig:eobnr_q15}EOB/NR phasing comparisons with some of the SXS simulations 
	of Ref.~\cite{Yoo:2022erv} used to construct the \nrsurqfifteen NR surrogate. Note that the accumulated
	phase difference is substantially independent of the spin value. The origin of the EOB/NR discrepancies might be
	due to missing physics in the analytical description of the fluxes, as advocated in Ref.~\cite{Nagar:2022icd}.
	Despite the phase differences may look large, $\bar{\cal F}^{\rm max}_{\rm EOB/NR}\sim10^{-3}$ for 
	these configurations, see Fig.~\ref{fig:teob_surrogate_22_q15}.}
\end{figure*}
%

%
\begin{figure}
	\includegraphics[width=0.4\textwidth]{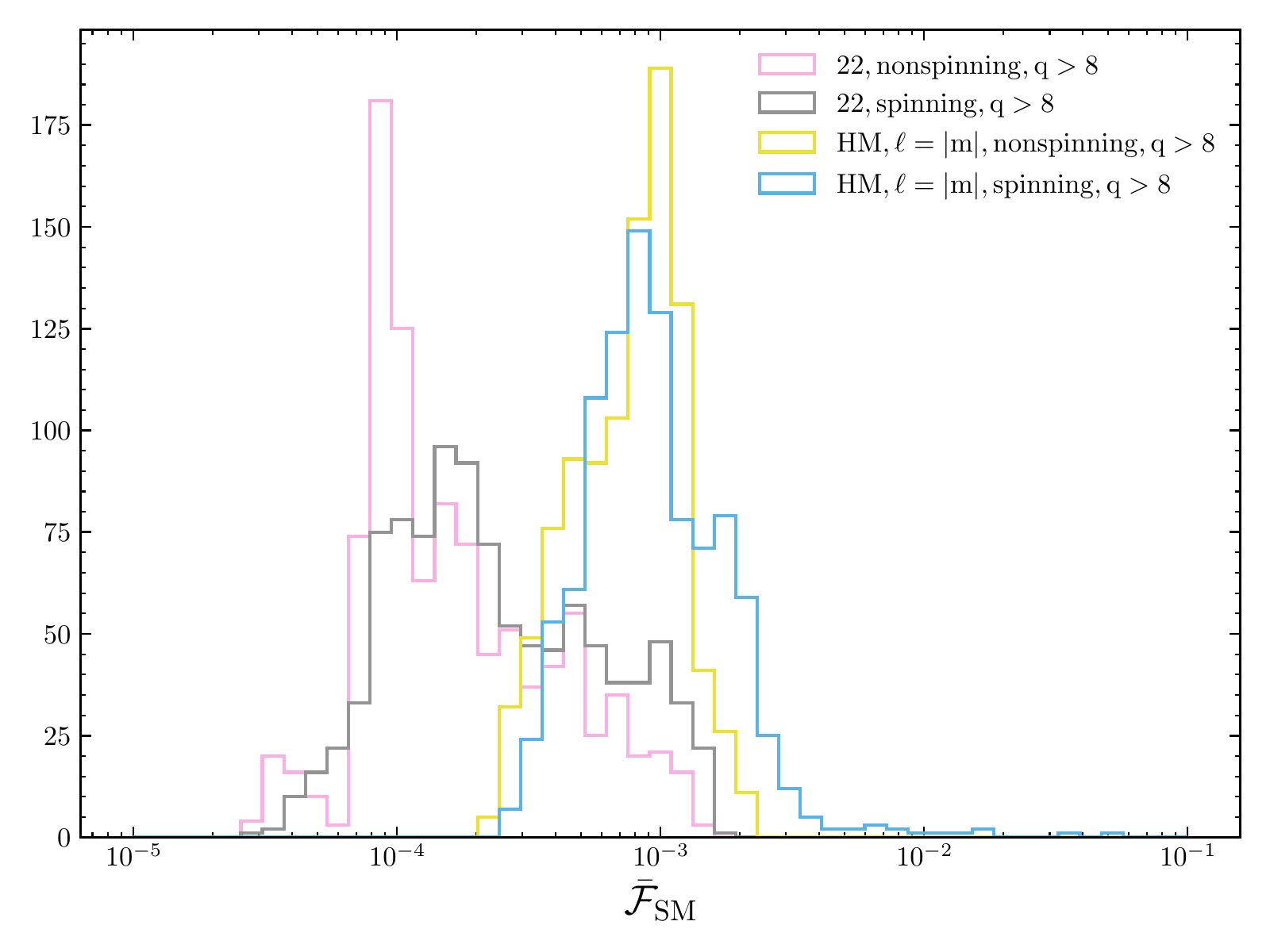}
	\caption{\label{fig:teob_surrogate_HM_q15}Summary of all the mismatches computed in this section 
	between \TEOBResumSvfourthreetwo{} and \nrsurqfifteen. When considering spinning systems including
	subdominant modes with $\ell=|m|$, we find $\sim 99.3\%$ of mismatches below $1\%$.}
\end{figure}

To have a more robust and informative handle on the performance of the new
model we conclude this section by comparing it with the NR surrogates \nrsurqeight~\cite{Varma:2018mmi}
and \nrsurqfifteen~\cite{Yoo:2022erv}
To do so, we use the $C$, public, implementation of \TEOBResumS{} upgraded with the
new $a_6^c$ fit and $c_3$ as described above (notably {\tt D3Q3\_NQC} of Table~\ref{tab:models} for $a_6^c$).
We compute the waveforms including either only the $(2,2)$ mode or the $(2,2)$, $(3,3)$, $(4,4)$ and $(5,5)$ 
subdominant modes\footnote{It is well known~\cite{Nagar:2020pcj} that the $(2,1)$ and $(3,2)$ modes are badly
modeled for large spins anti-aligned with the angular momentum due to a nonphysical behavior of the NQC corrections.
A way to overcome this problem has been found and is discussed in Ref.~\cite{Albanesi:2023bgi}
In addition, the $(3,2)$ mode is not modeled by \nrsurqfifteen, and will therefore not be included in our comparisons.}.
We compute mismatches from $20$ Hz and -- crucially -- to avoid noise in the FFT due to border effects generate the waveforms 
from $5 Hz$, and taper them at the beginning.
Notably, while mismatches computed with waveforms constructed from $(\ell, |m|) = (2,2)$ modes only are independent 
of the sky position and reference phase of the target waveform (\nrsurqeight or \nrsurqfifteen, in our case), when
subdominant modes are included in the waveform construction this simplification does not hold. 
The usual definition of mismatch depends on the extrinsic parameters of the source.
Following e.g. Ref.~\cite{Harry:2016ijz, Harry:2017weg, Garcia-Quiros:2020qpx, Pratten:2020ceb}, we define the template sky-maximized (SM)
unfaithfulness as
\begin{equation}
\bar{\mathcal{F}}_{\rm EOB/NR}^{\rm SM} = 1 - \max_{t_0^h, \varphi_0^h, \kappa^h} \frac{(s, h)}{\sqrt{(s,s)(h,h)}} \ ,
\end{equation}
where $s$ is the target waveform, $h$ is our template waveform, $\kappa^h$ is the template effective polarizability 
(which encodes information on the sky location of the binary~\cite{Harry:2016ijz, Harry:2017weg}) and we fix\footnote{We have found that the 
SNR-averaging $\mathcal{F}_{\rm SNR}$ over a grid of $\kappa^s$ and $\phi^s$ values typically gives very similar results 
to the non-averaged unfaithfulness. As such, for computational reasons, we choose to fix $\kappa^s$ and $\varphi^s$.} 
the target's reference phase $\varphi_0^s = 0$, effective polarizability $\kappa^s=0$ and inclination $\iota=\pi/3$ between 
the orbital angular momentum and the line of sight.

\subsubsection{Mass ratio $q \leq 8$}
The left panel of Fig.~\ref{fig:teob_surrogate_22} focuses on {\it nonspinning} systems and displays 
the (2,2)-only mismatches computed between \TEOBResumSvfourthreetwo{} and \nrsurqeight, 
using the Advanced LIGO noise curve in the frequency interval between $[20, 2048]$~Hz. 
We consider the NR surrogate in the mass ratio $q\in[1,8]$, total mass $M \in [40, 150] \Msun$ and 
dimensionless spins $|\chi_{1,2}|\leq 0.8$.
Mismatches always lie below the $8\times 10^{-4}$ threshold, with the largest values of $\bar{\cal F}^{\rm SM}_{\rm EOB/NR}$ 
observed for equal mass binaries, for which the effect of radiation reaction is larger causing fewer in-band cycles.
When we also explore the spins parameter space (right panel of Fig.~\ref{fig:teob_surrogate_22}), we find maximum values of $\bar{\cal F}^{\rm SM}_{\rm EOBNR}$ 
around $2 \times 10^{-3}$, corresponding to unequal mass systems with large spins.
In both cases, more than $99\%$ of the configurations we considered have mismatches below $10^{-3}$.

Further considering higher modes (HM), the scenario remains qualitatively similar, although mismatches degrade overall. 
Figure~\ref{fig:teob_surrogate_summary} summarizes our results, comparing the mismatches obtained 
with the various binaries and with different mode content.
When subdominant modes are included in the construction of the waveform, the median of mismatches 
distributions shifts to larger values, with $\sim 75\%$ of the total mismatches
below $0.1\%$. 
The impact of the $(3,2)$ and $(2,1)$ modes is gauged in Fig.~\ref{fig:teob_surrogate_summary_HM}. 
As expected, including such modes decreases the overall faithfulness of the model, especially
for spinning binaries. As previously discussed, this behavior can be traced back to the unphysical 
behavior of the NQC for the $(2,1)$ and $(3,2)$ modes for systems with large spins anti-aligned with
the orbital angular momentum. This produces an incorrect waveform behavior during the plunge
phase up to merger~\cite{Nagar:2020pcj}.
This result is further investigated in Fig.~\ref{fig:frac_HM}.
The plot shows, versus the effective spin $\tilde{a}_0$, the fraction of configurations with 
$\bar{\cal F}<{\bar{\cal F}}^{\rm thrs}$, where ${\bar{\cal F}}^{\rm thrs}$ can take the 
values indicated in the legend. The (negative) impact of the modes with $\ell\neq m$ is evident, while
the performance of the modes with $\ell=m$ is satisfactory.

\subsubsection{Mass ratio: $8 \leq q \leq 15$}
We now repeat the comparison for systems with $q$ up to $15$ and $|\chi_1|\leq 0.5$, $\chi_2 = 0$, comparing 
the NR surrogate  \nrsurqfifteen~of Ref.~\cite{Yoo:2022erv} with \TEOBResumSvfourthreetwo{},
employing the same range of frequencies and total mass used above, as well as the same detector PSD. 
We consider the NR surrogate in the mass ratio $q\in[8,15]$, and dimensionless spins $|\chi_{1,2}|\leq 0.5$, $\chi_2=0$, 
corresponding to the validity range of \nrsurqfifteen.
The left panel of Fig.~\ref{fig:teob_surrogate_22_q15} shows the (2,2)-only mismatches for nonspinning systems, 
while the right panel also explores the impact of spins. 
Over the set of studied configurations, $\sim 97\%$ ($\sim 92\%$) of the mismatches lie below the $10^{-3}$ threshold for nonspinning (spinning) systems.

To get a better understanding of the meaning of such values of the mismatches, let us produce a few phasing comparisons
with some of the original datasets used to compute the surrogate \nrsurqfifteen. Figure~\ref{fig:eobnr_q15} illustrates
EOB/NR phasings for $(15,-0.5,0)$, $(15,0,0)$ and $(14,+0.50,0)$, this latter belonging to the region of the parameter
space where the EOB/NR matches with the surrogate have the largest values (top-right panel of Fig.~\ref{fig:teob_surrogate_22_q15}).

It is a well-known fact that, when binaries are very asymmetric, the importance of subdominant modes increases.
As a consequence, the inclusion of HM for high mass ratio systems largely impacts the mismatches obtained.
Figure~\ref{fig:teob_surrogate_HM_q15} compares the mismatches obtained with $(2,|2|)$ modes with those computed using $\ell = m$ modes up to $(5,5)$
for both spinning and nonspinning configurations. When HM are employed, mismatches can increase by up to one order of magnitude, with $99.4\%$ of them
lying below $1\%$ and $\sim~59\%$ below $0.1\%$.

\subsection{Improving the higher modes: the importance of the $\ell=2$, $m=1$ mode.}
\label{sec:mode_h21}
Let us now explore in deeper detail the role of the $(2,1)$ mode and discuss a way to improve the mode further.
As mentioned above and in Ref.~\cite{Nagar:2020pcj} (see also Ref.~\cite{Nagar:2018zoe}) the problems in the
construction of the $(2,1)$ mode are related to the (incorrect) determination of the NQC corrections. The reason
for this is that, by construction, the NQC parameters are determined by imposing continuity between the EOB
and NR waveform at some time {\it after} the peak of each multipole, like the $(2,2)$ case. However, for the
procedure to work one needs that the NQC functions do not develop any unphysical behavior in the strong-field
region close to merger, like poles. In this respect, as already mentioned in Ref.~\cite{Nagar:2018zoe}, the origin 
of the unphysical features in the NQC basis for the $(2,1)$ mode is due to the fact that for anti-aligned spins 
the orbital frequency passes through zero and thus, entering at the denominator, introduces a singularity in the functions. 
Overcoming this difficulty is conceptually simple, as it is sufficient to determine NQC corrections at a moment where
the orbital frequency is sufficiently larger than zero so to avoid any pathological behavior. In practice, this means
taking as NQC-determination point any time {\it before} the peak of the $(2,1)$ mode.
The simplest choice (though not the only one) is to do it at the location of the peak of the $(2,2)$ mode 
on the EOB time axis, defined as $t^{\rm EOB}_{A_{22}^{\rm max}}$. In practice, this needs to extend the
ringdown model also before the peak of $(2,1)$, which implies that the NR-informed fits discussed in 
Ref.~\cite{Nagar:2020pcj} are not useful for this purpose and should be redone.
However, the approach of NR-informing the ringdown from $t^{\rm EOB}_{A_{22}^{\rm max}}$ for {\it all}
modes was followed in all  {\tt SEOBNR} waveform models that incorporated higher modes~\cite{Cotesta:2018fcv,Pompili:2023tna}.
In particular, Ref.~\cite{Pompili:2023tna} gives updated NR-informed fits that describe the ringdown
for several waveform modes (up to $\ell=5$) starting from $t^{\rm NR}_{A_{22}^{\rm max}}$ instead
of $t^{\rm NR}_{A_{\lm}^{\rm max}}$ as it is normally done for {\tt TEOBResumS}. 

%
\begin{figure}
	\includegraphics[width=0.4\textwidth]{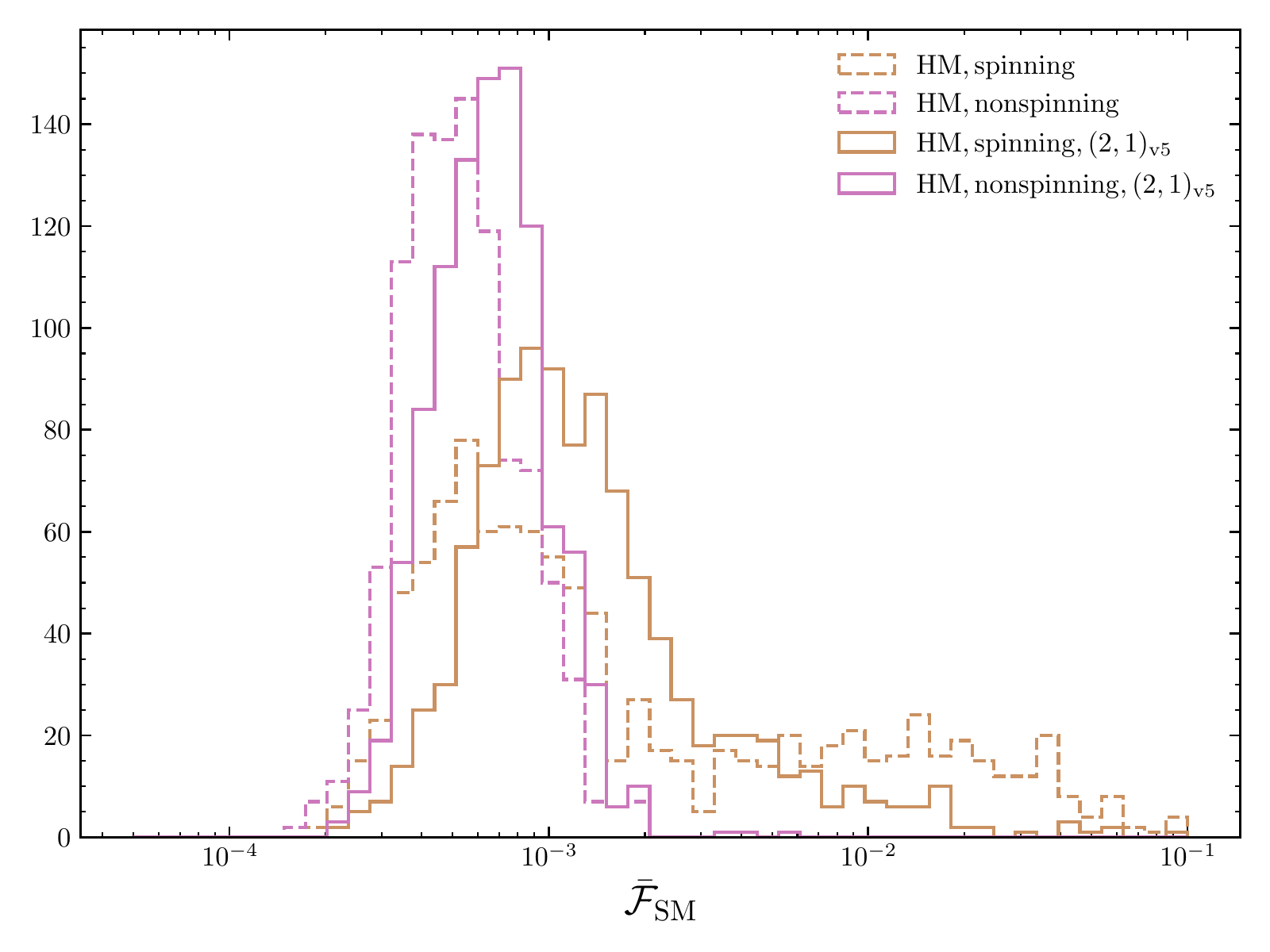}
	\includegraphics[width=0.4\textwidth]{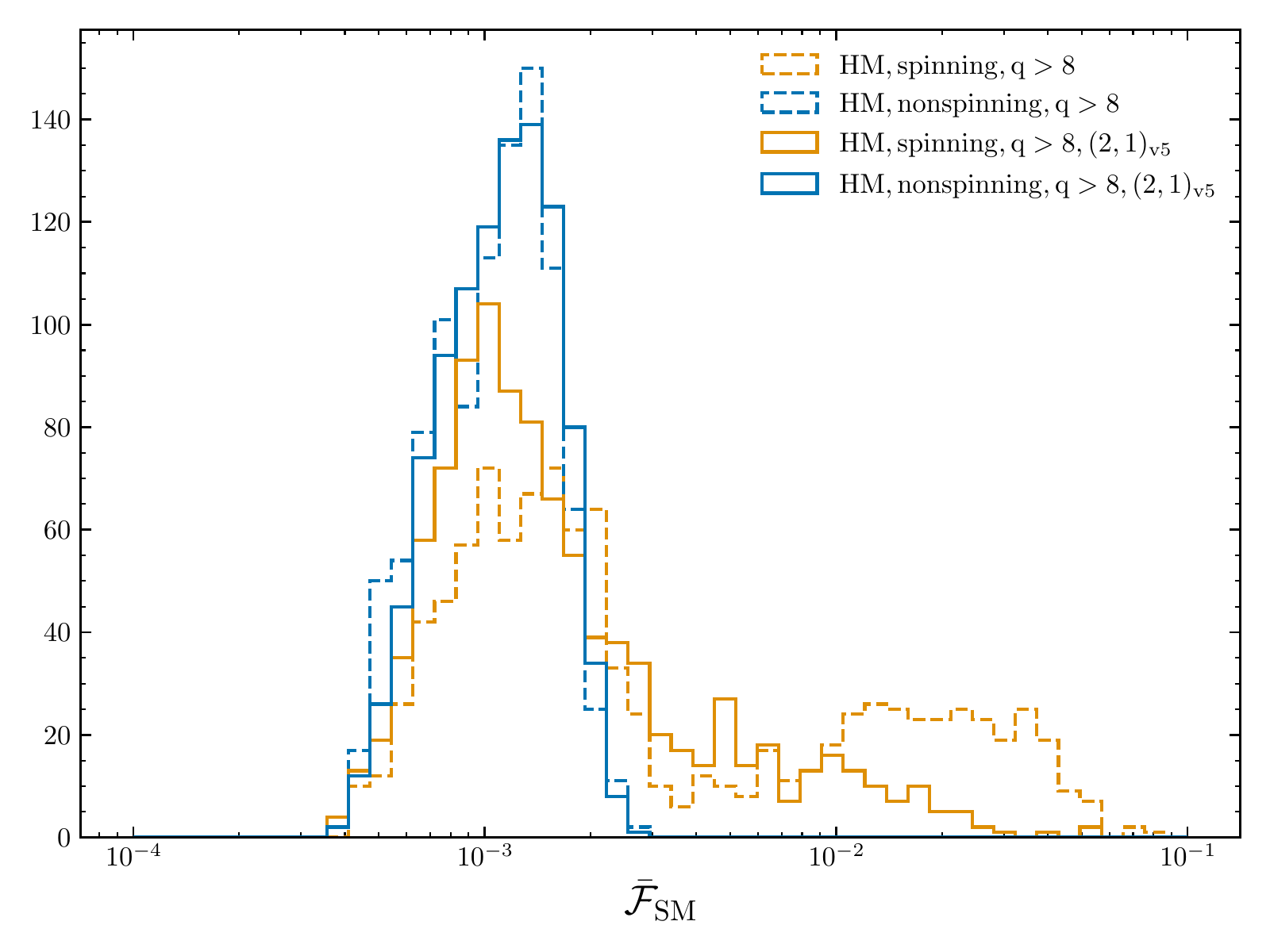}	
	\caption{\label{fig:teob_surrogate_summary_HM_21improved}Relevance of the $(2,1)$ mode. Unfaithfulness comparison between 
	\TEOBResumSvfourthreetwo{} and {\tt NRHybSur3dq8} (top panel) or \nrsurqfifteen~(bottom panel) with the $(2,2)$, $(2,1)$, $(3,3)$, $(3,2)$, 
	$(4,4)$ and $(5,5)$ modes, using either the previous prescription for the NQCs and merger/ringdown of the $(2,1)$ mode
	or the one from {\tt SEOBNRv5}~\cite{Pompili:2023tna}. The high mismatch tail of the distribution almost disappears 
	when employing this latter.}
\end{figure}
\begin{figure}[t]
	\includegraphics[width=0.5\textwidth]{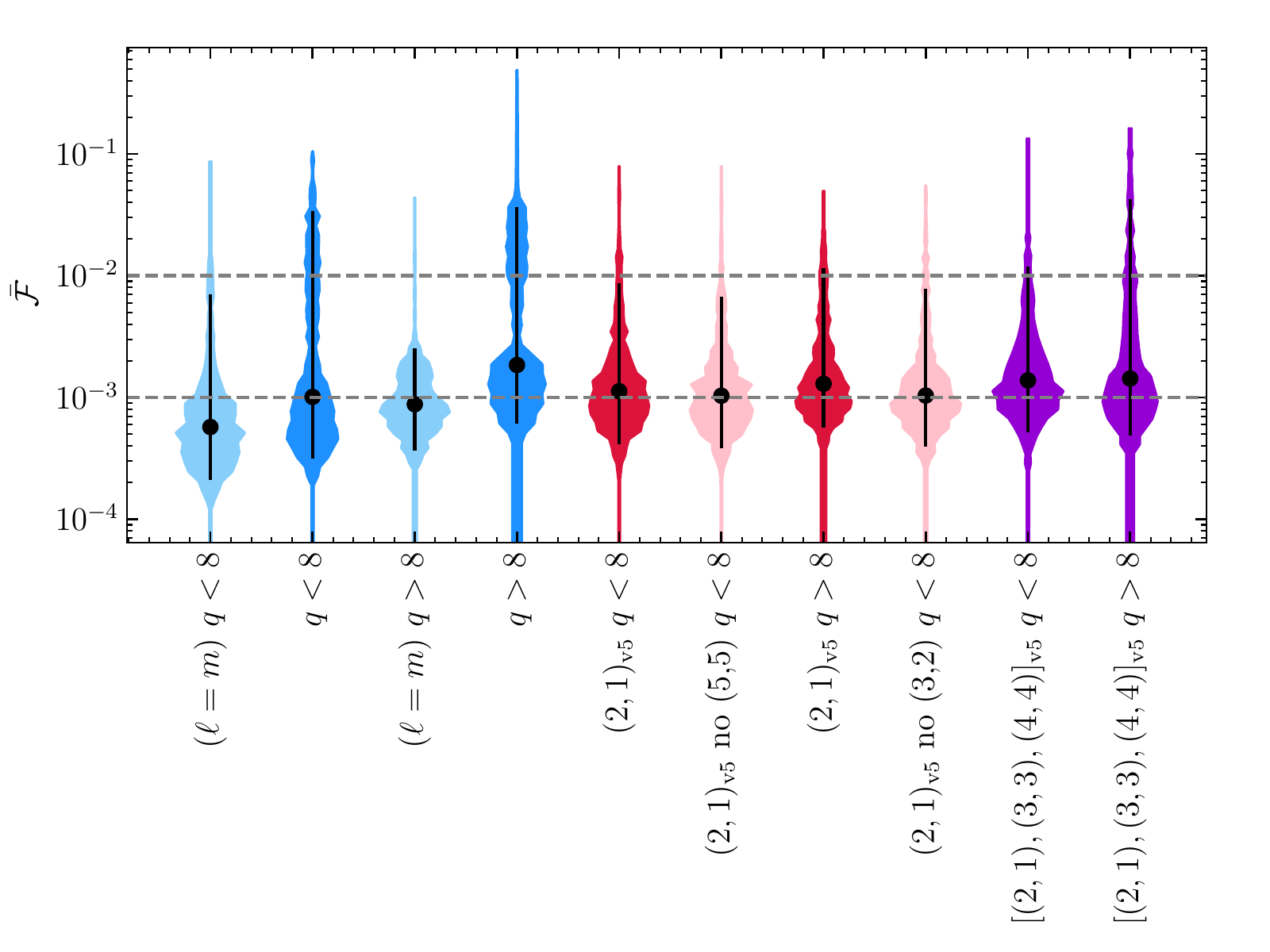}
	\caption{\label{fig:teob_surrogate_violin}Comparison between mismatches computed between \TEOBResumSvfourthreetwo{} 
	and {\tt NRHybSur3dq8} ($q<8$) or \nrsurqfifteen~($q>8$) with different mode content and various strategies for the computation
	of the $(2,1), (3,3)$ and $(4,4)$ modes.}
\end{figure}
\begin{figure}[t]
	\includegraphics[width=0.5\textwidth]{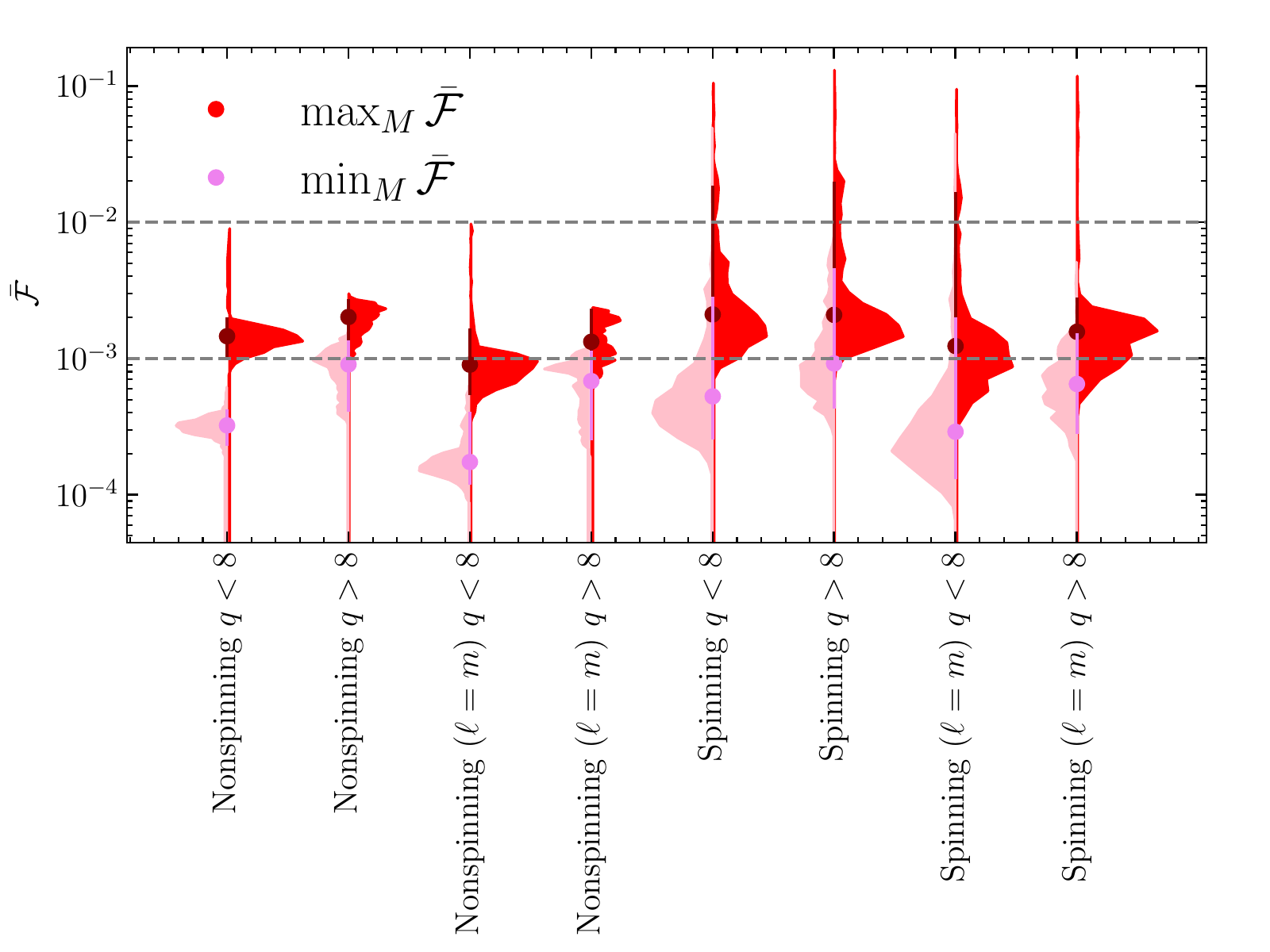}
	\caption{\label{fig:teob_surrogate_violin_M}Total mass maximized (red) or minimized (pink) mismatches computed between \TEOBResumSvfourthreetwo{} 
	and {\tt NRHybSur3dq8} ($q<8$) or \nrsurqfifteen~($q>8$). We consider here the same 1000 configurations of mass ratio (and spins) 
	employed within the rest of the paper, and vary the total mass 
	between $40$ and $200$ solar masses.}
\end{figure}
\begin{figure}[t]
	\includegraphics[width=0.5\textwidth]{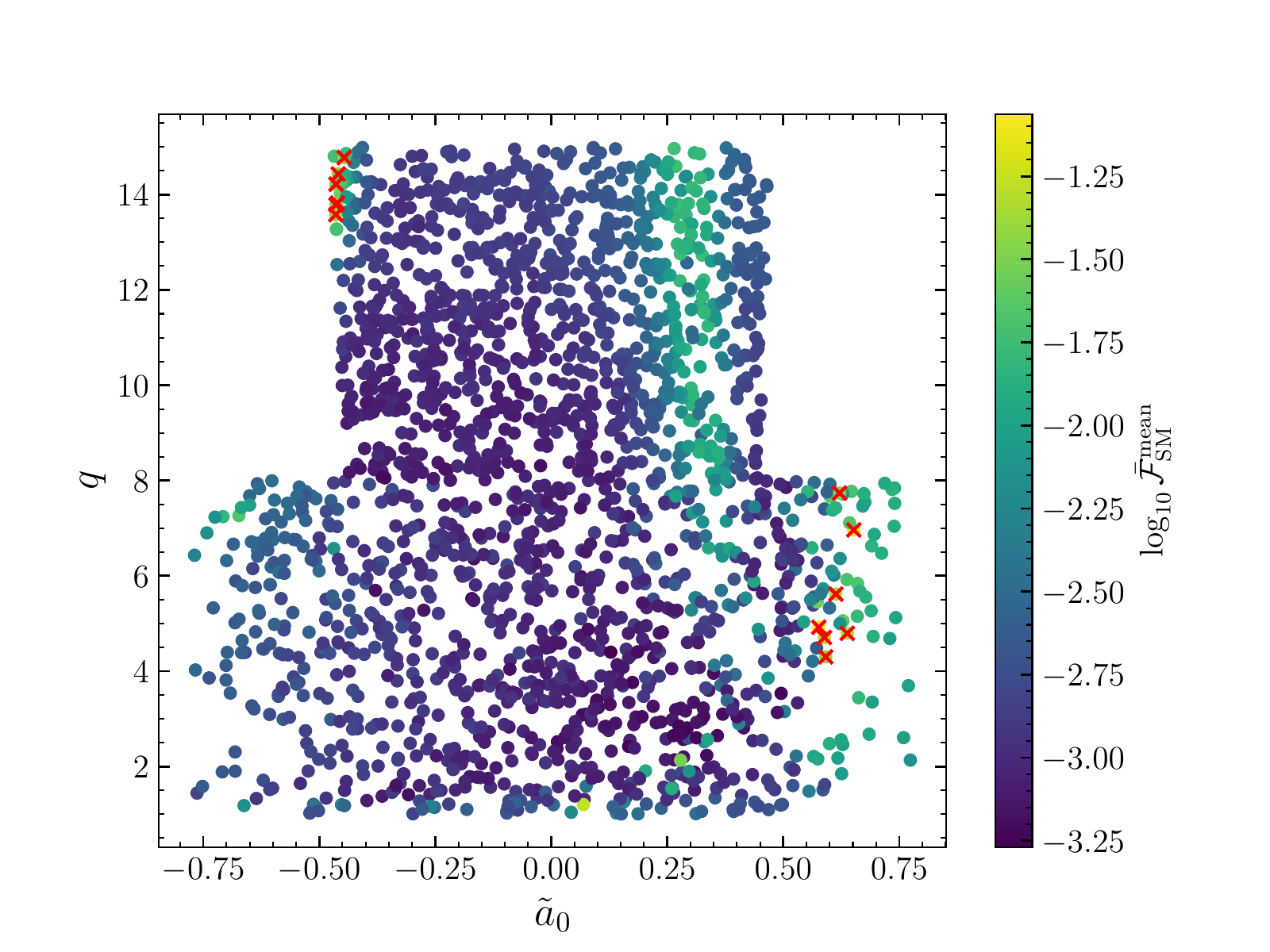}
	\caption{\label{fig:teob_surrogate_2d_M}Mass-averaged mismatches computed between \TEOBResumSvfourthreetwo{} 
	and {\tt NRHybSur3dq8} ($q<8$) or \nrsurqfifteen~($q>8$) for configurations with varying mass ratio and spins. Red crosses indicate
	systems for which the average unfaithfulness lies above the $3\%$ threshold. Such cases are typically characterized by large values of $q$ and
	large $|\tilde{a}_0|$.}
\end{figure}
Let us briefly review the general procedure for modeling the ringdown that is adopted by \TEOBResumS{} and {\tt SEOBNR},
that relies on the procedure of Ref.~\cite{Damour:2014yha}. The difference in the two model families are discussed below.
For each waveform mode, one fits the QNM-rescaled waveform
\be
\hbar_\lm = h^{\rm rng}_\lm e^{+\sigma_\lm \tau},
\ee 
where $h^{\rm rng}_\lm$ is the numerical ringdown waveform, $\sigma=\alpha_1 + \ii \omega_1$ is the fundamental complex QNM frequency and $\tau = t-\tmatch$. 
The QNM-rescaled wave is then written as $\hbar = A_\hbar e^{\ii \phi_\hbar}$,
where $h$ is always $\nu$-normalized and
\begin{align}
A_\hbar    & = c_1^A \tanh\left( c_2^A \tau + c_3^A \right) +c_4^A,  \label{eq:Abar} \\
\phi_\hbar & = \phi_0 - c_1^\phi \log\left( \frac{1+c_3^\phi e^{-c_2^\phi\tau}+ c_4^\phi e^{-2 c_2^\phi \tau}}{1 + c_3^\phi + c_4^\phi} \right) , \label{eq:Phbar}
\end{align}
where we adopt the notation\footnote{Note that in the {\tt SEOBNR} family, notably in 
Refs.~\cite{Bohe:2016gbl,Pompili:2023tna}, uses precisely the same functional form introduced in 
Ref.~\cite{Damour:2014yha} but the coefficients are named differently (and one omitted). 
More precisely, for the amplitude one has
$\left\lbrace 
c_1^A = c_{1,c}^\lm, \, 
c_2^A = c_{1,f}^\lm, \,
c_3^A = c_{2,f}^\lm, \,
c_4^A = c_{2,c}^\lm
\right \rbrace$, while for the phase
$\left\lbrace 
c_1^\phi = d_{1,c}^\lm, \,
c_2^\phi = d_{1,f}^\lm, \,
c_3^\phi = d_{2,f}^\lm,
c_4^\phi = 0\right\rbrace $.
Moreover, consider that there some different sign conventions; for the QNM frequencies we have 
$\alpha_1 = -\sigma^R_\lm$ and $\omega_1 = -\sigma^I_\lm$, and in our case the waveform frequency is
defined as positive, while the one used in Eq.~53 of Ref.~\cite{Pompili:2023tna} is negative.
Although we adopt the fits of Ref.~\cite{Pompili:2023tna} we prefer to stick to the notation of 
Ref.~\cite{Damour:2014yha} for consistency with previous work.} introduced in Ref.~\cite{Damour:2014yha}.
In passing, we also remind the reader that the fitting provided by Eq.~\eqref{eq:Abar} is not
suitable for large mass ratios, as discussed in Refs.~\cite{Albanesi:2021rby,Albanesi:2023bgi}.
To ensure continuity, some coefficients are constrained and are linked to NR quantities. Here, the two
EOB families adopt different strategies. In particular, \TEOBResumS{} adopts the strategy outlined in 
Ref.~\cite{Damour:2014yha,Nagar:2020pcj}, imposing the constraints at the peak of the ($\l,m$)-amplitude for each
mode and constraining $( c_1^A, c_2^A, c_4^A, c_1^\phi, c_2^\phi )$ in terms 
of QNM frequencies, $A^{\rm max}_{\lm}$ and $\omega_\lm^{\rm max}$. On the other hand, 
the {\tt SEOBNR} family imposes continuity conditions always at the peak of the quadrupolar amplitude
for each mode; this results in the constraints~\cite{Cotesta:2018fcv,Pompili:2023tna}
\begin{align} 
c_1^A = &\left( \dAmatch  + \alpha_1 \Amatch \right) \cosh^2 c_3^A/c_2^A, \\
c_4^A = & \Amatch - \nonumber \\
         & \left(\dAmatch + \alpha_1 \Amatch \right) 
         \cosh c_3^A  \sinh c_3^A /c_2^A \\
c_1^\phi = & (\omega_1 - \omgmatch) \frac{1+c_3^\phi}{c_2^\phi c_3^\phi}, \\
c_4^\phi = & 0,
\end{align}
that leave as free coefficients $( c_2^A, c_3^A, c_2^\phi, c_3^\phi)$.

In the following, we proceed implementing the {\tt SEOBNR} strategy for some of the higher modes. 
We obtain the fitted coefficients $( c_2^A, c_3^A, c_2^\phi, c_3^\phi)$
and the NR quantities $( \Amatch, \dAmatch, \omgmatch )$ 
from the global fits presented in Ref.~\cite{Pompili:2023tna} 
(see Appendix C and D therein). Although our main goal here is to fix the behavior
of the EOB (2,1) mode, we will also consider modes $(3,3)$ and $(4,4)$ 
to gain more precise insights on the relevance of tiny details in ringdown modeling.

Figure~\ref{fig:teob_surrogate_summary_HM_21improved} highlights the improvement
in the total EOB/NR unfaithfulness brought by replacing {\it only} the $(2,1)$ ringdown 
part with the one of Ref.~\cite{Pompili:2023tna}. The other modification implemented 
is that NQC corrections are now determined at $t^{\rm EOB}_{A_{22}^{\rm max}}$, though using
the same NQC basis that was used before. One appreciates how the number of configurations
with $\bar{\cal F}_{\rm SM}<0.01$ (solid lines) is now much smaller than before (dashed lines).
A better insight in the accuracy of the model is given by 
Figs.~\ref{fig:teob_surrogate_violin} and~\ref{fig:teob_surrogate_violin_M}.
The first figure shows that the correct modelization of the $(2,1)$ mode is 
essential. By contrast, the impact of changes in the $(3,3)$ and $(4,4)$ modelization 
are marginal and are in fact found to slightly worsen the global performance of the
model (see the violet violin plots in Fig.~\ref{fig:teob_surrogate_violin}).
Figure~30 reports the values of the unfaithfulness maximized and minimized as 
the total mass $M$ is varied between $40M_\odot$ and $200 M_\odot$. 
The largest values of unfaithfulness are typically obtained for heavy binaries ($M > 100 M_{\odot}$), indicating 
that the modelization of the  transition from late plunge to merger-ringdown needs further improvements.
More precisely, Fig.~\ref{fig:teob_surrogate_2d_M} offers an insight on the distribution of the 
total-mass-averaged mismatches in the two-dimensional parameter space of $q$ and $\tilde{a}_0$.
We find that the largest mismatches are obtained for systems with large mass ratio $q > 4$ and 
large spins $|\tilde{a}_0| > 0.5$. In these cases, the degradation of the performance of the model can 
be attributed to the modelization of the $(3,3)$ and $(4,4)$ modes close to merger.
Overall, $\sim 98\%$ ($99\%$) of the total-mass-maximized mismatches lie below 
the $3\%$ threshold when $q < 8$ ($q > 8$).

\subsubsection{Interference between co-rotating and counter-rotating QNMs for $(2,1)$ and $(3,2)$ modes}
\begin{figure}[t]
	\center
	\includegraphics[width=0.48\textwidth]{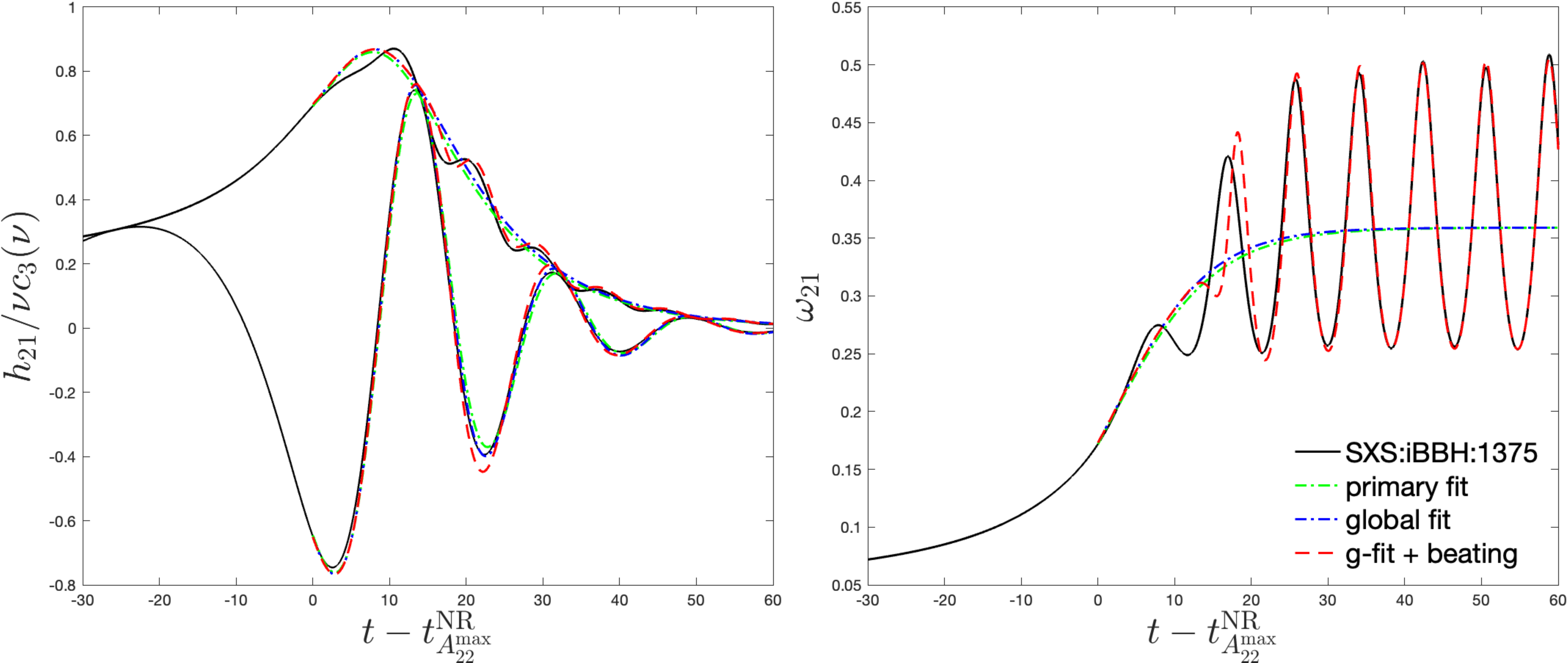}
	\includegraphics[width=0.48\textwidth]{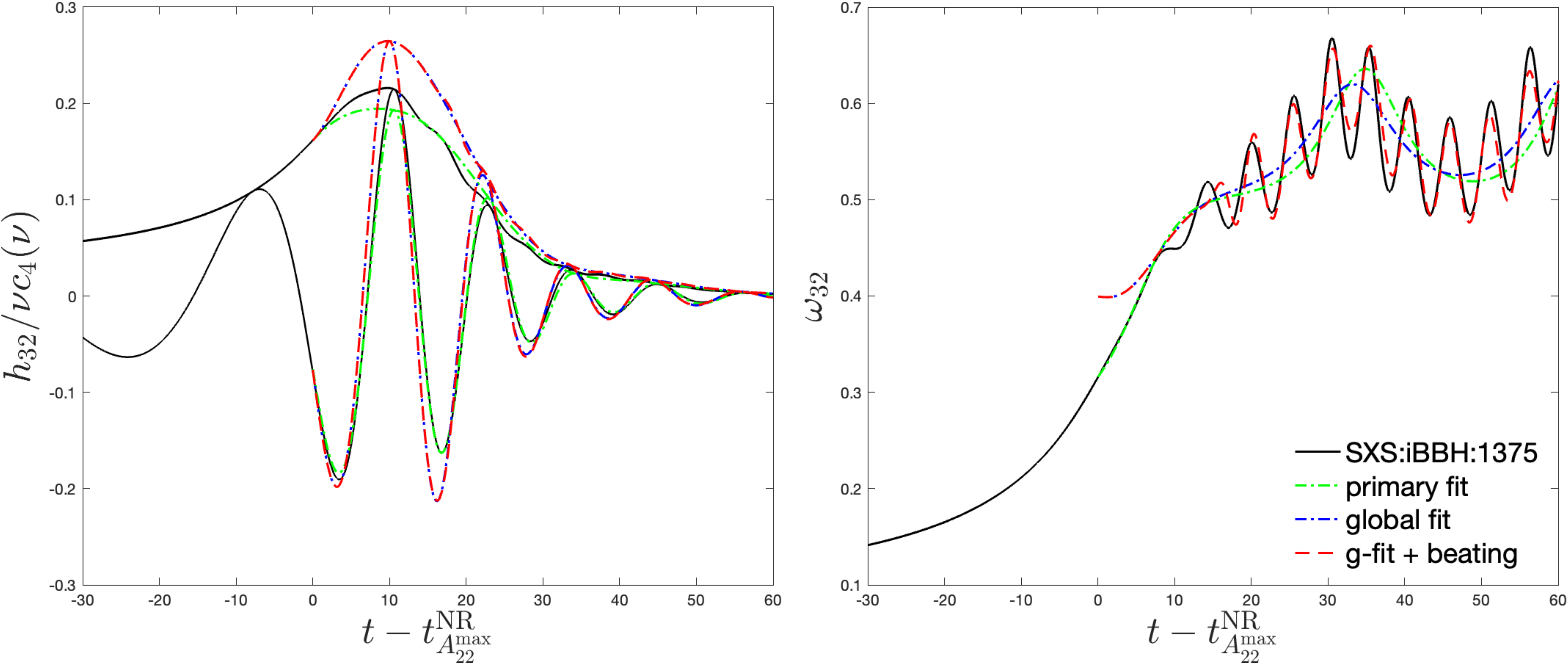}
	\caption{\label{fig:mix}Configuration $(8,-0.9,0)$, beating (mode-mixing) between different QNMs.
	Top panel: mode $(2,1)$ with and without the interference between co-rotating and counter-rotating
	QNMs. Bottom panel: same effect for the $(3,2)$ mode superposed to the modulation due to
	the mixing between $(2,2)$ and $(3,2)$ QNMs.}
\end{figure}
Now that we have seen the changes brought by the improvement of the $(2,1)$ mode, 
we turn our attention to other physical elements that may affect the accuracy of the 
higher modes in our model. In particular, we will focus on mode-mixing. 
There are two kinds of mode-mixing that occur when considering the ringdown 
of rotating black holes. 
The first one is due to the contribution of the counterrotating QNMs,
i.e. to the modes with $m<0$. The late behavior of this effect can be modeled
considering only the fundamental co-rotating and counter-rotating QNMs, as detailed 
in Ref.~\cite{Damour:2007xr,Nagar:2006xv,Bernuzzi:2010ty,Albanesi:2023bgi}. 
The second kind of mixing occurs between modes with same $m$ but different $\l$,
as noted in Ref.~\cite{Kelly:2012nd}. This occurs because the waveform modes are 
written using the spherical harmonic  basis ${}_{-2}Y_\lm$, while the natural 
base for the QNMs is the spheroidal harmonic one, ${}_{-2}S_{\lm n}$.
We model this mode-mixing for the (3,2) and (4,3) modes as done in \texttt{SEOBNR{\_}v5}, 
following the approach outlined in Sec.~IIIC of Ref.~\cite{Pompili:2023tna}, whose main ideas 
are recalled here (see also Ref.~\cite{Taracchini:2014zpa}). Reliable relations between spheroidal 
${}^S h_\lm$ and spherical $h_\lm$ modes can be obtained neglecting the contributions of the overtones
and considering only contribution of the $\l'<\l$ modes. Once that the relations are found, the numerical
spheroidal modes are extracted from the spherical ones and fitted 
using the usual templates of Eqs.~\eqref{eq:Abar},~\eqref{eq:Phbar}. The coefficients found
with the primary fits of the spheroidal modes are then fitted over the parameter space, so that 
the spheroidal ringdown can be reconstructed from the global fits. Then
to obtain the spherical modes it is sufficient to invert the spheroidal-spherical relations.

The inclusion of these effects in the complete EOB model requires several modifications and 
testing of the {\tt TEOBResumS} infrastructure that we postpone to future work. 
Here we just test the NR-informed ringdown waveform of Ref.~\cite{Pompili:2023tna} in a significative case. 
We consider the NR simulation \texttt{BBH:1375} of the SXS 
catalog~\cite{Boyle:2019kee}, that corresponds to a BBH with $(q,\chi_1,\chi_2) = (8,-0.9,0)$.
Figure~\ref{fig:mix} shows the waveform and the frequency (black) for the (2,1) and (3,2) modes of the aforementioned 
configuration, compared with different analytical prescriptions. Note that in this figure the multipole 
are normalized with $c_{\ell + \epsilon} = X_2^{\ell+\epsilon-1} + (-)^m X_1^{\ell+\epsilon-1}$, 
where $\epsilon$ is the parity of the mode (0 if $\l+m$ is even, 1 otherwise).
From the analytical point of view, we consider different prescriptions. We start by showing in green the waveform
obtained with the primary fits; for the (2,1) multipole it is simply obtained fitting the numerical 
spherical mode, while for the (3,2) spherical mode is obtained combining the fit of the spheroidal
(3,2) mode and of the spherical (2,2) mode, as discussed above. We also show in blue the waveform obtained 
with the global fits of Ref.~\cite{Pompili:2023tna} (see Appendix C and D therein). 
While for the (2,1)mode  this waveform is comparable to the one obtained directly
with the primary fit, the (3,2) waveform obtained from the global fits is less accurate for
what concerns the amplitude. This seems to suggest that in this specific region of the parameter space 
the global fits of Ref.~\cite{Pompili:2023tna} do not provide precise values and might need some improvements. 
However, despite this inaccuracy the waveform is still qualitatively reliable, and in particular
is able to catch the frequency modulation due to the $m=2$ mode-mixing. Finally, we also show in red the waveform
in which we have included the beating between co-rotating and counter-rotating fundamental QNMs; the coefficients 
needed to reproduce this effect are extracted from the numerical waveform (see Ref.~\cite{Albanesi:2023bgi} 
for more details). In this case, we are also able to reproduce the higher-frequency oscillations in the multipole 
frequency, both in the (2,1) and (3,2) multipoles. Such proof-of-principle study thus indicates that, provided 
a number of NR waveforms well placed in the parameter space, all mode-mixing effects can be modeled
at a reasonable level of accuracy.

\section{Conclusions}
\label{sec:conclusion}

The basic guiding principle behind the construction of robust and flexible EOB models is to explore, 
one by one, each physical element entering the construction of the model~\cite{Damour:2007xr}.
At the dawn of the development of EOB models this guiding principle was followed carefully
because of the need of understanding in detail the effect of each analytical choice and its impact
on describing accurately the physics of the plunge and merger. However, in the rush of constructing 
waveform models with higher and higher NR faithfulness, the original attitudes have 
progressively lost importance. In particular, the attitude of using automatized calibration procedures 
involving several parameters at the same time~\cite{Bohe:2016gbl,Cotesta:2018fcv,Pompili:2023tna}
may eventually hide the importance of each analytical element of the model. Although some recent 
studies in this direction were attempted recently (see e.g.~\cite{Rettegno:2019tzh,Nagar:2021xnh,Khalil:2022ylj}) 
the role of some building blocks of the procedure of NR-informing EOB models 
(and in particular \TEOBResumS{}) was not spelled out systematically so far. 
The understanding of the impact of each building block within the EOB construction can be
rephrased as {\it understanding waveform analytic systematics}.
To make well--precise statements around the concept of waveform systematics within the EOB
formalism, we focused on some pivotal building blocks of the models and illustrated how a careful 
mastery of their properties is important to yield improved waveform accuracy. 
We mainly considered {\it nonspinning} binaries and analyzed the importance of either the 
NR-tuning of the EOB flexibility parameters or the impact of high-order PN terms in the 
EOB (radial) potentials.
Our findings can be summarized as follows:
\begin{itemize}
\item[(i)] We have shown that a more precise way of NR-informing the effective 5PN-function $a_6^c(\nu)$ 
              allows us to improve the performance of \TEOBResumS{} (on nonspinning SXS data) of at least a factor
              2 with respect to current state-of-the-art: the maximum value of the EOB/NR unfaithfulness 
              (on the Advanced LIGO PSD) is lowered to $\sim 5\times 10^{-4}$. We remark that this is obtained by
              NR tuning the {\it single} function $a_6^c(\nu)$.
\item[(ii)] In order to better understand the potentialities of the model, we have also attempted to NR-tune
               the time-interval $\Delta t_{\rm NQC}$ that, loosely speaking, defines the merger location on
               the EOB time-axis. This procedure is similar to the one routinely adopted in {\tt SEOBNR} models~\cite{Bohe:2016gbl},
               although the precise understanding of the impact of each element was not spelled out 
               so far. For the illustrative $q=1$ case, we found that the tuning of $\Delta t_{\rm NQC}$
               allows one to reduce the EOB/NR phase difference of about a factor two during plunge
               and merger (see Fig.~\ref{fig:tuningDeltaTnqc}). Unfortunately, this also yields a slight, though
               noticeable, degradation of the phasing performance of the model during the inspiral.
               For this reason, we generally advocate to {\it avoid} using $\Delta t_{\rm NQC}$ as NR-informed
               parameter. In addition, the advantages in terms of EOB/NR unfaithfulness look so small 
               with respect to tuning only $a_6^c$ that it does not seem worth to introduce such 
               complication in the \TEOBResumS{} model. We stress, however, that in principle one 
               {\it should} tune  $\Delta t_{\rm NQC}$ in a way that it is compatible with the test-mass 
               case (see e.g.~\cite{Damour:2012ky}). Given the subtlety of this approach, we defer it
               to forthcoming studies.
\item[(iii)] We then explored whether the use of recently obtained (quasi-complete) 5PN information
                in the $D$ and $Q$ EOB potentials helps to obtain a model as flexible and as accurate 
                as the standard one (with or without iteration on next-to-quasi-circular waveform amplitude corrections).
                We highlighted that the performance of a (nonspinning) EOB waveform model crucially
                depends on the NR-faithful modelization of noncircular effects during the plunge
                up to merger. The new, and a priori unexpected, finding is that NR-informed 
                NQC corrections, that are purely phenomenological, are practically equivalent
                to analytically known noncircular effects entering the noncircular sector of the
                Hamiltonian. The fact that the NR-informed NQC iterations are crucial
                aspects of the model suggests that, possibly, genuine noncircular effects
                in the waveform (e.g. the leading-order ones as implemented in the eccentric model~\cite{Chiaramello:2020ehz}) 
                could be useful to improve the EOB/NR agreement further  during the late plunge up to merger.
\item[(iv)] Thanks to our enhanced understanding of noncircular effects in strong field (and in particular of
                 the radial force ${\cal F}_r$), we present a slightly modified version of the eccentric 
                 model of Ref.~\cite{Nagar:2021xnh}, although for the moment only limited to nonspinning binaries.
                 This model presents slightly improved matches with the SXS eccentric simulations available
                 as well as an excellent EOB/NR agreement for the scattering angle for strong-field configurations.
                 The extension to spins is postponed to future work as it requires an improvement of the description
                 of spin-orbit interaction for large, positive, spins, as already pointed out in Ref.~\cite{Nagar:2021xnh}.           
\item[(v)] We then obtained several new versions of \TEOBResumS{} model for spin-aligned binaries that 
                rely on the newly determined $a_6^c$ and (essentially) only differ by the NR-informed effective 
                N$^3$LO function $c_3$. Eventually, by changing only $a_6^c$ and $c_3$ with respect to the 
                \TEOBResumS{} default choices~\cite{Nagar:2020pcj,Riemenschneider:2021ppj}  we obtain a highly 
                NR-faithful new model, dubbed \TEOBResumSvfourthreetwo{}, that has maximal EOB/NR unfaithfulness 
                ${\cal F}_{\rm EOB/NR}^{\rm max}\sim 10^{-3}$ for the $(2,2)$ mode all over the publicly available SXS catalog. 
                This seems to suggest that  \TEOBResumSvfourthreetwo{} is currently the EOB waveform model with the
                lowest EOB/NR unfaithfulness, compatible with the {\tt SEOBNRv5} model~\cite{Pompili:2023tna}.
                It must be noted, however, that some of the NR datasets of Ref.~\cite{Pompili:2023tna} for large mass ratios are
                not public, so that we could not perform an actual apple-with-apple comparison.
                To comply with the spirit of this paper, we also kept all incremental steps that brought us to obtain \TEOBResumSvfourthreetwo{}, 
                i.e. we kept track of four different EOB models, summarized in Table~\ref{tab:models_vs_c3}, that always have 
                ${\cal F}_{\rm EOB/NR}^{\rm max}\lesssim 4\times 10^{-3}$ and the differ from details in the representation of
                the $c_3$ function. Each one of these models can be used in injection/recovery studies  to explore how such 
                small differences in EOB/NR mismatches impact the unbiased recovery of injected parameters.
                We finally want to remark that the values of the NR-informed parameters were obtained using simple procedures.
                For example, the first-guess $c_3$ values, were obtained only by visual inspection of the time-domain phasings and 
                then similarly fitted using simple fitting functions. Notably, the number of NR datasets used to do so is only composed
                by 8 nonspinning and 47 spin-aligned datasets, for a total of 55. This amounts to only the $10.3\%$ of the public 
                SXS waveform catalog. We also note that the calculation of mismatches only occurs as a validation check {\it a posteriori} of the model.
                We remark that this is different from other approaches~\cite{Bohe:2016gbl} were the NR-calibration is done on the 
                EOB/NR mismatches themselves.  This seems to indicate that the analytical structure of \TEOBResumS{} (in all its avatars) 
                is robust and a relatively moderate amount of NR-information is needed to determine the dynamical parameters\footnote{Note that 
                a separate issue is the modeling of post-peak (ringdown) waveform. Here we rely on the model of Ref.~\cite{Nagar:2020pcj},
                that was obtained using a large number of simulations for convenience, despite the evident redundancy of many datasets.}.
\item[(vi)] We also validated the \TEOBResumSvfourthreetwo{} model against two NR surrogate models, \nrsurqeight{} and \nrsurqfifteen{},
                i.e. up to mass ratios $q=15$. When focusing on the $(\ell, |m|) = (2,2)$ mode, our results reflect the previous 
                investigations performed against NR directly. The mismatches with \nrsurqfifteen{} are at most $\sim 10^{-3}$,
                despite no NR simulations were used to inform the model for $q\geq 10$. This finding is also consistent with previous
                NR validation of \TEOBResumS{}, in the nonspinning case only, for intermediate-mass-ratio binaries~\cite{ Nagar:2022icd}, 
                notably up to $q=100$.
                When higher modes are additionally included, the performance of the model degrades for systems with high mass ratio
    	       	and large spin magnitudes. Appropriate modeling of the $(2,1)$ and $(3,2)$ modes, especially, appears critical (notably, for anti-aligned spins) 
    	       	in order to obtain mismatches below $10^{-2}$. This is in line with previous findings using \TEOBResumS{}~\cite{Nagar:2020pcj}.
    	       	The accurate modelization of $(2,1)$, $(3,2)$, as well as $(4,3)$, modes during the plunge up to merger is known 
		to be the current Achilles' heel of the model. This is fundamentally related to the current implementation of NQC waveform corrections that do not
    	       	work for those configurations where the orbital frequency $\Omega$ crosses zero. This is a standard feature that occurs for 
    	       	certain configurations with anti-aligned spins and that is present also in the test-mass limit~\cite{Harms:2014dqa}.
    	       	Note however that in {\tt SEOBNR} models~\cite{Cotesta:2018fcv,Pompili:2023tna} this problem is efficiently 
    	       	avoided because the ringdown of each $(\ell,m)$ mode is matched at the peak of the $(2,2)$ mode and not at
    	       	the peak of the $(\ell,m)$ waveform mode itself. 
		As a pragmatical solution to this long-standing issue, we implemented within \TEOBResumS{} the $(2,1)$ mode 
		of the {\tt SEOBNRv5} model, precisely in the form described in Ref.~\cite{Pompili:2023tna}.
		This by itself is sufficient, as shown in Fig.~\ref{fig:teob_surrogate_summary_HM_21improved}, to
		have more than the $98\%$ of the total-mass-maximized unfaithfulness to lie below the $3\%$ threshold 
		when comparing to the surrogate models.
		In this respect, we also remind the reader that of the important features of the NR $(2,1)$ waveform amplitude, 
		i.e. that can develop a zero during the late inspiral for certain special configurations of the spins~\cite{Cotesta:2018fcv}, 
    	       	is naturally accounted for by the \TEOBResumS{}, resummed, waveform (see~\cite{Nagar:2020pcj}, Fig.~15), 
    	       	without the need of resorting to NR calibration (note that this seems partly necessary  for the {\tt SEOBNR} 
    	       	models~\cite{Cotesta:2018fcv,Pompili:2023tna}).
                 Moreover,  in the spirit of understanding the impact of the ringdown modelization on the global performance
                 of the model, in Figs.~\ref{fig:teob_surrogate_violin} and~\ref{fig:teob_surrogate_violin_M} we explore the 
                 relevance of using different representations for the ringdown  of the $(4,4)$ and $(3,3)$ modes.
	         We also preliminary investigate the impact of mode mixing in the ringdown of \TEOBResumS{}, adopting
	         the modelization of Ref.~\cite{Pompili:2023tna} and improving it further by additionally incorporating co-rotating
	         and counter-rotating QNMs (see Fig.~\ref{fig:mix}) in the spirit of Ref.~\cite{Albanesi:2023bgi}.
	         The complete modelization of QNMs mode mixing in the (3,2), and possibly also $(4,3)$, modes all over
	         the parameter space is however postponed to future work.	         					  	   
\item[(vii)] Finally, our detailed analysis of the nonspinning case in Fig.~\ref{fig:raw_phasing} indicates that, assuming
               that NR is {\it exact}, \TEOBResumSvfourthreetwo{} has to gain {\it only} between 0.1 and 0.2~rad at merger 
               to obtain EOB/NR unfaithfulnesses {\it below} $10^{-4}$ for present (and future) ground-based detectors.
               In particular, we remark that our simple study highlights that this phasing loss only occurs in the last orbit before 
               merger. Although some studies claim that the accuracy of NR simulations should be improved further 
               by at least one order of magnitude in terms of unfaithfulness~\cite{Purrer:2019jcp}, 
               improving the EOB phasing by 0.2~rad around merger looks like an easy task as it seems to be 
               mostly controlled by the NR-informed functions than by high-order contributions to the EOB potentials. 
               In this respect, we recall that the NR-informed NQC corrections, that are key to correctly shape the 
               EOB waveform at merger, are currently not very sophisticated, as one only imposes the 
               EOB/NR consistency between $(\omega_{22},\dot{\omega}_{22},A_{22},\dot{A}_{22})$, but there are {\it no} conditions 
               imposed on second-order time derivatives. Preliminary investigations in the test-mass limit~\cite{Albanesi:2023bgi} 
               suggest that additional noncircular corrections (either NR-informed or built-in due to the use of the native 
               generic Newtonian prefactor in the waveform) may play an important role in improving the analytic 
               waveform behavior up to merger. This is discussed extensively in Ref.~\cite{Albanesi:2023bgi}.      
\end{itemize}
The improvements introduced in \TEOBResumS{-\tt GIOTTO} in this work are automatically available to calculate spin 
precessing BBH waveforms~\cite{Akcay:2020qrj,Gamba:2021ydi}, 
binary neutron star waveforms~\cite{Bernuzzi:2014owa,Akcay:2018yyh,Gamba:2022mgx}, and black-hole--neutron-star waveforms~\cite{Gonzalez:2022prs}. 

\acknowledgements
We are grateful to J.~Yoo, V.~Varma, M.~Giesler, M.~Scheel, C.~J.~Haster, H.~P.~Pfeiffer, L.~Kidder and M.~Boyle for 
sharing with us the SXS simulations presented in Ref.~\cite{Yoo:2022erv}.
P.~R. is supported by the Italian Minister of University and Research (MUR) via the PRIN
2020KB33TP, {\tt Multimessenger astronomy in the Einstein Telescope Era (METE)}.
P.~R. and A.~A. acknowledge the hospitality of the Institut des Hautes Etudes Scientifiques
where part of this work was done.
R.~G. is supported by the Deutsche Forschungsgemeinschaft (DFG) under Grant No.
406116891 within the Research Training Group RTG 2522/1.
 A.~A. has been supported by the fellowship Lumina Quaeruntur No.
LQ100032102 of the Czech Academy of Sciences.
S.~B. acknowledges support by the EU H2020 under ERC Starting Grant, no. BinGraSp-714626 and
by the EU Horizon under ERC Consolidator Grant, no. InspiReM-101043372.
The present research was also 
partly supported by the {\it “2021 Balzan Prize for Gravitation: Physical and Astrophysical Aspects”}, 
awarded to Thibault Damour. We are also grateful to P.~Micca for inspiring suggestions. 
Calculations were performed on the Tullio server at INFN, Torino.

\noindent \TEOBResumS{} is developed open source and publicly available at
{\footnotesize \url{https://bitbucket.org/eob_ihes/teobresums/src/master/}} .
The code uses semantic versioning and the versions (\texttt{v?.?.?})
used in this work correspond to the code tags on the master branch.
\TEOBResumS{} can also be installed via 
\verb#pip install teobresums#. The code is interfaced to state-of-art
gravitational-wave data-analysis pipelines: 
\href{https://github.com/matteobreschi/bajes}{bajes}~\cite{Breschi:2021wzr}, \href{https://git.ligo.org/lscsoft/bilby}{bilby}~\cite{Ashton:2018jfp} and \href{https://pycbc.org/}{pycbc}~\cite{Biwer:2018osg}.

\bibliography{refs20231210.bib,local.bib}

\end{document}